\documentclass[singlespace, 10pt, vi,twoside]{mitthesis}
\usepackage{amsmath, amssymb, amsthm, amsfonts, bbm, fancyhdr}
\usepackage[noautoscale]{youngtab}
\usepackage{color, pstricks, amscd, revsymb}
\usepackage{graphicx, html, perpage}
\usepackage[frame,line,arrow,matrix,tips]{xy}

\CompilePrefix{xygui-}
\MakePerPage{footnote}

\newtheorem{theorem}{Theorem}[chapter]
\newtheorem{corollary}[theorem]{Corollary}
\newtheorem{lemma}[theorem]{Lemma}
\newtheorem{observation}[theorem]{Observation}
\newtheorem{proposition}[theorem]{Proposition}

\newtheorem{claim}[theorem]{Claim}
\newtheorem{fact}[theorem]{Fact}

\newtheorem{defi}[theorem]{Definition}
\newtheorem{eg}[theorem]{Example}

\newcommand{\bra}[1]{\langle #1 |}
\newcommand{\ket}[1]{| #1 \rangle}
\newcommand{\braket}[2]{\langle #1 | #2 \rangle}
\newcommand{\dblbraket}[1]{\mbox{$\langle #1 | #1 \rangle$}}
\newcommand{\proj}[1]{| #1 \>\< #1 |}
\newcommand{\oprod}[1]{\l| #1 \r\rangle\!\!\l\langle #1 \r|}
\newcommand{\smfrac}[2]{\mbox{$\frac{#1}{#2}$}}
\newcommand{\upto}[1]{\stackrel{#1}{\approx}}

\def\half{\smfrac{1}{2}}
\def\bmath#1{\mbox{\boldmath$#1$}}
\def\bi{\begin{itemize}}
\def\ei{\end{itemize}}
\def\be{\begin{equation}}
\def\ee{\end{equation}}
\def\bea{\begin{eqnarray}}
\def\eea{\end{eqnarray}}
\def\ben{\begin{eqnarray*}}
\def\een{\end{eqnarray*}}
\def\non{\nonumber}
\def\>{\rangle}
\def\<{\langle}
\def\l{\left}
\def\r{\right}
\def\la{\leftarrow}
\def\ra{\rightarrow}
\def\da{{\!\downarrow}}
\def\ot{\otimes}
\def\geqslant{\geq}
\def\leqslant{\leq}

\def\lbL{ \left[\rule{0pt}{2.4ex}\right. }
\def\rbL{ \left.\rule{0pt}{2.4ex}\right] }
\def\lbm{ \left[\rule{0pt}{2.1ex}\right. }
\def\rbm{ \left.\rule{0pt}{2.1ex}\right] }

\def\eps{\epsilon}
\def\g{\gamma}

\def\bl{\bar{\lambda}}
\def\va{\vec{a}}
\def\vb{\vec{b}}
\def\vx{\vec{x}}
\def\pfail{p_{\text{fail}}}
\def\st{\text{~s.t.~}}

\def\cA{{\cal A}}		 \def\cB{{\cal B}}		 
\def\cC{{\cal C}}
\def\cD{{\cal D}}		 \def\cE{{\cal E}}		 

\def\cG{{\cal G}}		 \def\cH{{\cal H}}		 
\def\cI{{\cal I}}
		 \def\cK{{\cal K}}		 
\def\cL{{\cal L}}

\def\cM{{\cal M}}		 \def\cN{{\cal N}}		 
\def\cO{{\cal O}}
\def\cP{{\cal P}}		 \def\cQ{{\cal Q}}		 
\def\cR{{\cal R}}
\def\cS{{\cal S}}		 \def\cT{{\cal T}}		 
\def\cU{{\cal U}}
		 		 
\def\cX{{\cal X}}

\def\bp{{\bf p}}
\def\bq{{\bf q}}
\def\br{{\bf r}}
\def\bL{{\bf L}}
\def\bP{{\bf P}}
\def\bQ{{\bf Q}}
\def\bR{{\bf R}}
\def\bS{{\bf S}}
\def\bT{{\bf T}}

\def\bbC{\mathbb{C}}
\def\bbE{\mathbb{E}}
\def\bbL{{\mathbb{L}}}
\def\bbM{{\mathbb{M}}}
\def\bbN{{\mathbb{N}}}
\def\bbP{{\mathbb{P}}}

\def\bbR{\mathbb{R}}
\def\bbZ{\mathbb{Z}}
\def\one{\openone}
\def\ccG{G}

\def\cnot{\textsc{cnot}}
\def\dcnot{\textsc{dcnot}}
\def\swap{\textsc{swap}}

\def\Ucg{U_{\text{CG}}}
\def\Ugpe{U_{\text{GPE}}}
\def\Usch{U_{\text{Sch}}}
\def\Usel{U_{\text{sel}}}
\def\Uqft{U_{\text{QFT}}}
\def\Uxoxo{U_{\text{XOXO}}}

\newcommand{\fig}[1]{Fig.~\ref{fig:#1}}
\newcommand{\eq}[1]{Eq.~(\ref{eq:#1})}
\newcommand{\peq}[1]{(Eq.~\ref{eq:#1})}
\newcommand{\eqs}[2]{Eqns.~(\ref{eq:#1}) and (\ref{eq:#2})}
\newcommand{\chap}[1]{Chapter~\ref{chap:#1}}
\newcommand{\chaps}[2]{Chapters~\ref{chap:#1} and \ref{chap:#2}}
\newcommand{\chaprange}[2]{Chapters~\ref{chap:#1}-\ref{chap:#2}}
\newcommand{\sect}[1]{Section~\ref{sec:#1}}
\newcommand{\sects}[2]{Sections~\ref{sec:#1} and \ref{sec:#2}}
\newcommand{\thm}[1]{Theorem~\ref{thm:#1}}
\newcommand{\prop}[1]{Proposition~\ref{prop:#1}}
\newcommand{\lem}[1]{Lemma~\ref{lemma:#1}}
\newcommand{\cor}[1]{Corollary~\ref{cor:#1}}
\newcommand{\defn}[1]{Definition~\ref{def:#1}}
\newcommand{\mscite}[1]{\cite{#1}}

\DeclareMathOperator{\diag}{diag}
\DeclareMathOperator{\End}{End}
\DeclareMathOperator{\GL}{GL}
\DeclareMathOperator{\Hom}{Hom}
\DeclareMathOperator{\id}{id}
\DeclareMathOperator{\poly}{poly}
\DeclareMathOperator{\rank}{rank}
\DeclareMathOperator{\Sch}{Sch}
\DeclareMathOperator{\sgn}{sgn}
\DeclareMathOperator{\Span}{Span}
\DeclareMathOperator{\spec}{spec}
\DeclareMathOperator{\tr}{Tr}
\DeclareMathOperator{\vol}{Vol}

\DeclareMathOperator{\A}{A}
\DeclareMathOperator{\B}{B}
\def\HA{\cH_{\A}}
\def\HB{\cH_{\B}}

\def\nqudits{(\bbC^d)^{\ot n}}
\newcommand{\iso}[1]{\stackrel{#1}{\cong}}
\def\dom{\bbZ^d_{++}}
\def\interlaces{\precsim}
\def\reverselaces{\succsim}
\def\*{\star}
\def\ext{\supseteq}
\def\reduction{\stackrel{{}_{\scriptstyle *}}{\geq}}
\def\tilde{\widetilde}
\def\bar{\overline}
\newcommand{\abs}{{\rm abs}}
\newcommand{\rel}{{\rm rel}}
\newcommand{\aux}{{\rm aux}}
\newcommand\app[1]{{\cal A}^{#1}}

\DeclareMathOperator{\cbf}{cbit(\ra)}
\DeclareMathOperator{\cbb}{cbit(\la)}

\DeclareMathOperator{\cc}{coherent\ bit}

\DeclareMathOperator{\rqbsf}{remote\ qubits(\ra)}

\def\qq{[q\, q]}
\def\qtq{[q\ra q]}
\def\qbq{[q\la q]}

\def\ctc{[c\ra c]}
\def\cbc{[c\la c]}
\def\cc{[c\,c]}
\def\cof{[\![c \ra c]\!]}
\def\cob{[\![c \la c]\!]}
\def\coftau{[\![c \ra c : \tau]\!]}
\def\ctctau{[c \ra c : \tau]}
\def\qtqtau{[q \ra q : \tau]}

\DeclareMathOperator{\CC}{CC}

\DeclareMathOperator{\CCE}{CCE}
\DeclareMathOperator{\QQE}{QQE}

\DeclareMathOperator{\CE}{CE}
\DeclareMathOperator{\QE}{QE}
\DeclareMathOperator{\CoCoE}{C\!_{\rm o}\!C\!_{\rm o}\!E}
\DeclareMathOperator{\CoQE}{C\!_{\rm o}\!CQE}
\DeclareMathOperator{\QCoE}{QC\!_{\rm o}\!E}

\def\gab{\gamma_{a' \!, b'}^{a,b}}
\def\gmany{
 |\g_{a \hspace*{-0.2ex} ' \! \oplus x \hs, b \hspace*{-0.1ex} ' \! \oplus y}
    ^{\hspace*{0ex} a \oplus x \hs, \hspace*{0.1ex} b\oplus y}
 \>} 
\def\hs{\hspace*{-0.2ex}}

\def\lbL{ \left[\rule{0pt}{2.4ex}\right. }
\def\rbL{ \left.\rule{0pt}{2.4ex}\right] }
\def\lbm{ \left[\rule{0pt}{2.1ex}\right. }
\def\rbm{ \left.\rule{0pt}{2.1ex}\right] }
\def\lpm{ \left(\rule{0pt}{2.1ex}\right. }
\def\rpm{ \left.\rule{0pt}{2.1ex}\right) }

\def\w{\ar@{-}[l]}
\def\n{*-{}\w}
\def\nw{*-{~}\w}
\def\nt{*-{~}}
\def\b{*={\bullet}}

\def\a#1{\save []="#1" \restore}
\def\op#1{*+[F]{\rule[-0.2ex]{0ex}{2.1ex}#1}}	

\newbox{\ssbox}

\def\gspace#1{*+{\rule[-0.2ex]{0ex}{2.1ex}%
	\setbox\ssbox=\hbox{$#1$}%
	\hspace*{\wd\ssbox}}}
	
\def\gnqubit#1#2{\gspace{#1}
		 \save [].[#2]!C="qq"*[F]\frm{}\restore
		 \save "qq"*[]{#1} \restore}

\font\gensymbols=drgen10
\def\male{{\gensymbols\char"1A}}
\def\female{{\gensymbols\char"19}}

\begin{document}
%
%
%
%
%
%
%
\title{Applications of coherent classical communication and the Schur
transform to quantum information theory}

\author{Aram Wettroth Harrow}
\prevdegrees{B.S., Massachusetts Institute of Technology (2001)}
\department{Department of Physics}
\degree{Doctor of Philosophy in Physics}
\degreemonth{September}
\degreeyear{2005}
\thesisdate{August 4, 2005}


\supervisor{Isaac L. Chuang}{Associate Professor of
Electrical Engineering and Computer Science, and Physics}

\chairman{Thomas J. Greytak}{Professor of Physics}

\maketitle



\cleardoublepage
\setcounter{savepage}{\thepage}
\begin{abstractpage}
Quantum mechanics has led not only to new physical theories, but also
a new understanding of information and computation.  Quantum
information not only yields new methods for achieving classical tasks
such as factoring and key distribution but also suggests a completely
new set of quantum problems, such as sending quantum information over
quantum channels or efficiently performing particular basis changes on
a quantum computer.  This thesis contributes two new, purely quantum,
tools to quantum information theory---coherent classical communication
in the first half and an efficient quantum circuit for the Schur
transform in the second half.

The first part of this thesis (Chapters
\ref{chap:shannon}-\ref{chap:family}) is in fact built around two
loosely overlapping themes.  One is quantum Shannon theory, a broad
class of coding theorems that includes Shannon and Schumacher data
compression, channel coding, entanglement distillation and many
others.  The second, more specific, theme is the concept of using
unitary quantum interactions to communicate between two parties.  We
begin by presenting new formalism: a general framework for Shannon
theory that describes communication tasks in terms of fundamental
information processing resources, such as entanglement and classical
communication.  Then we discuss communication with unitary gates and
introduce the concept of {\em coherent classical communication}, in
which classical messages are sent via some nearly unitary process.  We
find that coherent classical communication can be used to derive
several new quantum protocols and unify them both conceptually and
operationally with old ones.  Finally, we use these new protocols to
prove optimal trade-off curves for a wide variety of coding problems
in which a noisy channel or state is consumed and two noiseless
resources are either consumed or generated at some rate.

The second half of the thesis (Chapters
\ref{chap:schur}-\ref{chap:sch-Sn}) is based on the Schur transform,
which maps between the computational basis of $\nqudits$ and a basis
(known as the {\em Schur basis}) which simultaneously diagonalizes the
commuting actions of the symmetric group $\cS_n$ and the unitary group
$\cU_d$.  The Schur transform is used as a subroutine in many quantum
communication protocols (which we review and further develop), but
previously no polynomial-time quantum circuit for the Schur transform
was known.  We give such a polynomial-time quantum circuit based on
the Clebsch-Gordan transform and then give algorithmic connections
between the Schur transform and 
the quantum Fourier transform on $\cS_n$.

\end{abstractpage}


\cleardoublepage

\section*{Acknowledgments}
Very little of the work in this thesis, or indeed that I have done
throughout my time in grad school, would have been possible without
the support of countless colleagues, collaborators, advisors and
friends.

I first want to thank Ike Chuang for guiding me in quantum
information from even before I entered grad school and for doing such
a good job of alternatingly pushing me, supporting me and turning me
loose, always with an eye toward my growth as a scientist.  I am also
indebted to Eddie Farhi for his encouragement and for innumerable
discussions, as well as for teaching me quantum mechanics in the first
place.  Thanks to Peter Shor for, among other things, sharing so many
of his unpublished ideas with me, including a crucial improvement to
remote state preparation in the summer of 2001 that led to my first
research result in grad school.

I am deeply grateful to Charlie Bennett and Debbie Leung for being
wonderful collaborators, mentors and friends throughout my time in
grad school.  In fact, I'm indebted to the group at IBM Yorktown for
most of my research direction, and I would to thank Nabil Amer, Guido
Burkard, Igor Devetak, David DiVincenzo, Roberto Oliveira, Barbara
Terhal and especially John Smolin for everything I learned from
working with them.

I am have been fortunate to have so many productive travel opportunities.
Thanks to Michael Nielsen for inviting me to the U. of Queensland,
where I enjoyed great discussions with Mick Bremner, Chris Dawson, Jen
Dodd, Henry Haselgrove and many other researchers.  Thanks also to
John Preskill at Caltech, Keiji Matsumoto at ERATO and Noah Linden at
the Newton Institute for making these trips possible for me.

Most of my work in grad school has been collaborative and I am
grateful to my many collaborators for teaching me, sharing their ideas
with me and helping me improve my own ideas.  In particular, the work
in this thesis was done in collaboration with Dave Bacon, Charlie
Bennett, Ike Chuang, Igor Devetak, Debbie Leung, John Smolin and
Andreas Winter.  I have also had valuable collaborations with Ken
Brown, Patrick Hayden, Seth Lloyd, Hoi-Kwong Lo, Michael Nielsen (and
many others at UQ), Roberto Oliveira, Ben Recht and Barbara Terhal.

Besides the colleagues I have mentioned so far, I want to thank Herb
Bernstein, Carl Caves, Andrew Childs, Matthias Christandl, Andrew
Cross, Chris Fuchs, Masahito Hayashi, Denes Petz and Krysta Svore for
many valuable discussions.  Thanks to Nolan Wallach for crucial
discussions on the Schur transform and in particular for the idea
behind \sect{QFT-from-Sch}.

Most of my work in grad school was funded by a QuaCGR grant (ARO
contract DAAD19-01-1-06) for which I am grateful to Henry Everitt, 
Mark Heiligman, the NSA and ARDA.

\vspace{0.3in}

This thesis is dedicated to my parents and to my teachers:
in particular, to Mike Masterson, Mrs. Thomas, Mr. Steidle, Will Repko,
Susan Schuur and Gian-Carlo Rota.


\pagestyle{fancy}
\fancyhead[LE,RO]{\thepage}
\fancyhead[LO]{\leftmark}
\fancyhead[RE]{\rightmark}
\fancyfoot[CE,CO]{}

\tableofcontents

\addtocounter{chapter}{-1}
\chapter{Introduction}
\label{chap:intro}

\section{Motivation and context}
{\em Classical theories of information and computation:}
Though it may seem like a recent phenomenon, computation---the
manipulation, storage and transmission of information---has long been
one of the most central features of human civilization.  Markets of
buyers and sellers perform distributed computations to optimize the
allocation of scarce resources, natural languages carefully balance
the goals of reducing redundancy while correcting errors, and legal
systems have long sought reliable algorithms of justice that are
accurate and efficient even when implemented with unreliable
components.  Although these examples cannot be totally separated from
human intelligence, they all rely on an impersonal notion of
information that has two crucial attributes.  First, information can
be abstracted away from any particular physical
realization; it can be photocopied, memorized, dictated, transcribed
and broadcast, always in principle largely preserving the original
meaning. 
Likewise an abstract algorithm for processing information can be performed
equivalently using pencil and paper or with digital circuits, as long
as it is purely mechanical and makes no use of human insight or
creativity.  Though the particular features and efficiency of each
model of computation may differ, the class of problems they can solve
is the same.\footnote{For two very different perspectives on these
ideas, see {\em Cybernetics} (1948) by N.~Weiner and {\em The
Postmodern Condition} (1979) by J.-F.~Lyotard.}

These ideas of computation and information were expressed in their
modern forms by Turing and Church in 1936\cite{Turing36,Church36} and
Shannon in 1948\cite{Shannon48}, respectively.  Turing described a
hypothetical machine meant to be able to perform any purely mechanical
computation, and indeed every method of computation so far devised can
be simulated by a Turing machine.  Moreover, most practical algorithms
used today correspond to the class of problems that a Turing machine
can solve given a random number generator and running time bounded by
a polynomial of the input size.  While Turing showed the fungibility
of computation, Shannon proved that information is fungible, so that
determining whether any source can be reliably transmitted by any
channel reduces, in the limit of long strings, to calculating only two
numbers: the information content of the source and the information
capacity of the channel.

The abstract theories of Turing and Shannon have been extraordinarily
successful because they have happened to match the
information-processing technology within our reach in the
20$^{\text{th}}$ century; Shannon capacities are nearly achievable by
practical codes and most polynomial time algorithms are feasible on
modern computers.
 However, our knowledge of quantum mechanics
is now forcing us to rethink our ideas of information and computation,
just as relativity revised our notions of space and time.  The state
of a quantum mechanical system has a number of properties which cannot
be reduced to the former, classical, notion of information.

{\em The challenge from quantum mechanics:}
The basic principles of quantum mechanics are simple to state
mathematically, but hard to understand in terms we are familiar with
from classical theories of physics and information.  A quantum system
with $d$ levels (e.g. an electron in the $p$ orbital of an atom, which
can be in the $p_x$, $p_y$ or $p_z$ states) has a state
described by a unit vector $\ket{\psi}$ that belongs to a
$d$-dimensional complex vector space.  Thus, an electron could be in the
$p_x$ or $p_y$ state, or in a linear combination of the two, known in
chemistry as a hybrid orbital, or in quantum mechanics as a {\em
superposition}.  Systems combine via the tensor
product, so the combined state space of $n$ $d$-level systems is
$d^n$-dimensional.
A measurement with $K$ outcomes is given by a
collection of matrices $\{M_1,\ldots,M_K\}$ such that outcome $k$ has
probability $\bra{\psi}M_k^\dag M_k\ket{\psi}$ (here $\bra{\psi}$ is the
Hermitian conjugate of $\ket{\psi}$) and results in the normalized
output state $M_k\ket{\psi}/\sqrt{\bra{\psi}M_k^\dag M_k\ket{\psi}}$;
any measurement is possible (on a finite-dimensional system) as long as
it satisfies the normalization 
condition $\sum_{k=1}^K M_k^\dag M_k = \one$.  The possible forms of
time evolution are entirely described by the constraints of
normalization and linearity; they correspond to maps from $\ket{\psi}$
to $U\ket{\psi}$, where $U$ is a unitary operator ($U^\dag U=\one$).

These principles bear a number of resemblances to classical wave mechanics,
and at face value may not appear particularly striking.  However,
they have dramatic implications when quantum systems are used to store
and manipulate information.
\begin{itemize}
\item {\em Exponentially long descriptions:} While $n$ copies of a
classical system require $O(n)$ bits to describe, $n$ copies of a
comparable quantum system cannot be accurately described with fewer
than $\exp(O(n))$ bits.  This is a direct consequence of the tensor
product structure of composite quantum systems, in which $n$ two-level
systems are described by a unit vector in a $2^n$-dimensional complex
vector space.   On the other hand, the largest classical message that
can be reliably encoded in such a system is $n$ bits
long\cite{Holevo73}.  This 
enormous gap cannot be explained by any classical model of
information, even when probabilistic or analog models are considered.

\item {\em Nonlocal state descriptions:}  Another consequence of
applying the tensor product to state spaces is that a
composite system $AB$ can be in an {\em entangled state} that cannot be
separated into a state of system $A$ and a state of system $B$.  While
correlated probability distributions have a similar property, an
entangled quantum system differs in that the system as a whole can be
in a definite state, while its parts still exhibit (correlated)
randomness.  Moreover, measuring entangled states yields correlations
that cannot be obtained from any classical correlated random
variable\cite{Peres93}, though they nevertheless do not permit
instantaneous communication between $A$ and $B$.

\item {\em Reversible unitary evolution:}
Since time evolution is unitary, it is always reversible.
(Measurement is also reversible once we include the measuring
apparatus; see \cite{Peres93} or \sect{prelim} of this thesis for
details.)  As an immediate consequence, quantum information can never
be deleted, only rearranged, perhaps into a less accessible form.  An
only slightly more complicated argument can prove that it is
impossible to copy an arbitrary quantum state\cite{WZ82}, unless we
know that the state belongs to a finite set that is perfectly
distinguishable by some measurement.

This contrasts sharply with one of classical information's defining
properties, its infinite reproducibility.  The idea of {\em
possessing} information takes on an entirely new meaning when
referring to quantum information, one that we are only barely
beginning to appreciate (e.g. see \cite{Preskill99,GC01}).

\item {\em Complementary observables:}
Another way to prove that quantum information cannot be cloned is via
the {\em uncertainty principle}, which holds that complementary
observables, such as position and momentum, cannot be simultaneously
measured; observing one necessarily randomizes the other.  The reason
this implies no-cloning is that making a perfect copy of a particle
would allow the position of one and the momentum of the other to be
measured, thereby inferring both quantities about the original system.

Even though the uncertainty principle describes limitations of quantum
information, quantum cryptography turns this into a {\em strength} of
quantum communication, by using uncertainty to hide information from
an eavesdropper.  The idea is to encode a random bit in one of two
randomly chosen complementary observables, so that without knowing how
the bit is encoded, it is impossible to measure it without risking
disturbance.  This can detect any eavesdropper, no matter how
sophisticated, and even if the quantum information is sent through
completely insecure channels.  Combining this process with public
classical communication can be used to send unconditionally secure
messages\cite{BB84}.

\item {\em Interference of amplitudes:}
In the two-slit experiment, two beams of light from point sources
(such as slits cut into a screen) overlap on a screen, but instead of
simply adding, yield alternating bands of constructive and destructive
interference.  One insight of quantum mechanics is that particles are
waves with complex amplitudes, so that interference is still found in
the two-slit experiment with single photons, electrons, or even
molecules.  Measurement breaks the quantum coherence which makes this
possible, so observing which slit an electron passes through, no
matter how gently this is performed, completely eliminates the
interference effect.

The power of interference would be dramatically demonstrated by building
a large-scale quantum computer and using it to solve classical
problems.  Such a computer could interfere different branches of a
computation in much the same way that different paths of an electron
can interfere.
\end{itemize}

These examples are significant not only because they expand the range
of what is efficiently computable, but because they force us to revise
the logical terms with which we understand the world around us.  We
can no longer say that an electron either went through one path or the
other, or that a quantum computer took a particular computational
path or that Sch\"{o}dinger's cat must be either alive or dead.  At
one point, this suggested that quantum theory needed to revised, but
now a consensus is emerging that it is instead classical logic that
needs to be rethought.

{\em The operational approach to quantum information:}
Unfortunately, ever since quantum mechanics was first articulated
seventy years ago, it has been difficult to give a clear philosophical
interepretation of quantum information.  In the last 10-20 years,
though, a good deal of progress has been made by thinking about
quantum information {\em operationally}, and studying how
information-processing tasks can be accomplished using quantum
systems.  At the same time, we would like to study quantum information
in its own right, preferably by abstracting it away from any
particular physical realization.

This operational-yet-abstract approach to quantum information is best
realized by the idea of quantum computation.  While classical
computers are based on bits, which can be either 0 or 1, quantum
computers operate on quantum bits or {\em qubits}, which are 2-level
quantum systems.  Each state of a quantum memory register (a
collection of $n$ qubits, hence with $2^n$ states) has its own complex
amplitude.  Performing an elementary quantum gate corresponds to
multiplying this (length $2^n$) vector of amplitudes by a unitary
matrix of size $2^n\times 2^n$.  If we prepare an input with nonzero
amplitude in many different states, we can run a computation in
superposition on all of these input states and then interfere their
output amplitudes, just as the amplitudes of differents paths of an
electron can interfere.  Certain problems appear to lend themselves
well to this approach, and allow us to observe constructive
interference in ``correct'' branches of the computation and
destructive interference in ``incorrect'' branches; needless to say,
this technique is completely impossible on classical probabilistic
computers.  For example, Shor's algorithm\cite{Shor94} is able to use
interference to factor integers on a quantum computer much faster than
the best known classical algorithm can.  

Other applications use the fact that amplitudes can add linearly,
while probability (or intensity) is proportional to amplitude squared.
This is used in Grover's algorithm\cite{Grover96} to search a database
of $N$ items with time $O(\sqrt{N})$, or in the more colorful
application of ``interaction-free measurement,'' which can safely
detect a bomb that will explode if it absorbs a single photon.  Here
the idea is to constructively interfere $N$ photons, each of amplitude
$1/N$, while randomizing the phase that the bomb sees, so that the bomb
experiences a total intensity of $N\cdot(1/N)^2=1/N$, which can be
made arbitrarily small (see \cite{RG02} and references therein).

{\em Purely quantum problems in quantum information:}
So far all of the examples of the power of quantum information
describe goals that are defined entirely in terms of classical
information (sharing secret bits, unstructured search, factoring
integers) but are more efficiently achieved using quantum information
processing resources; we might call these hybrid classical-quantum
problems.

As our understanding of quantum information has improved, we have also
begun to study information processing tasks which are purely quantum;
for example, we might ask at what rate a noisy quantum channel can
reliably transmit quantum messages.  In fact, it is even possible to
think of classical information entirely as a special case of quantum
information, a philosophy known as the ``Church of the Larger Hilbert
Space''\footnote{This term is due to John Smolin.} which
\sect{prelim} will explain in detail.  The two main contributions of
this thesis involve such ``purely quantum'' tasks, in which both the
problem and the solution are given in terms of quantum information.
Before explaining them, we will discuss the fields of research that
give them context.


\begin{description}
\item[Quantum information theory] (or more specifically, {\em
quantum Shannon theory}) seeks a quantitative understanding of how
various quantum and classical communication resources, such as noisy
channels or shared correlation, can be used to simulate other
communication resources.  The challenge comes both from the much
richer structure of quantum channels and states, and from the larger
number of communication resources that we can consider; for example,
channels can be classical or quantum or can vary continuously between
these possibilities.  Moreover, (quantum) Shannon theory studies
asymptotic capacities; we might ask that $n$ uses of channel send
$n(C-\delta_n)$ bits with error $\epsilon_n$, where
$\delta_n,\epsilon_n\ra 0$ as $n\ra\infty$.  Since the state of $n$
quantum systems generally requires $\exp(O(n))$ bits to describe, the
set of possible communication strategies grows quite rapidly as the
number of channel uses increases.

While early work (such as \cite{Holevo73,BB84}) focused on using
quantum channels to transmit classical messages, the last ten years
have seen a good deal of work on the task of sending quantum
information for its own sake, or as part of a quantum computer.  The
main contribution of the first half of this thesis is to show that
many tasks previously thought of in hybrid classical-quantum terms
(such as using entanglement to help a noisy quantum channel send
classical bits) are better thought of as purely quantum communication
tasks.   We will introduce a new tool, called {\em coherent classical
communication}, to systematize this intuition.  Coherent classical
communication is actually a purely quantum communication resource; the
name indicates that it is obtained by modifying protocols that use
classical communication so that they preserve quantum coherence
between different messages.  We will find that coherent classical
communication, together with a rigorous theory of quantum information
resources, will give quick proofs of a wide array of optimal quantum
communication protocols, including several that have not been seen
before.

\item[Quantum complexity theory] asks how long it takes quantum
computers to solve various problems.  Since quantum algorithms include
classical algorithms as a special case, the interesting question is
when quantum algorithms can perform a task faster than the best
possible or best known classical algorithm.  The ultimate goal here is
generally to solve classical problems (factoring, etc.) and the
question is the amount of classical or quantum resources required to
do so.

When considering instead ``purely quantum'' algorithms, with quantum inputs
and quantum outputs, it is not immediately apparent what application
these algorithms have.  However, at the heart of Shor's factoring algorithm,
and indeed almost all of the other known or suspected exponential
speedups, is the quantum Fourier transform: a procedure that maps
quantum input $\sum_x f(x)\ket{x}$ to quantum output $\sum_x
\hat{f}(x)\ket{x}$, where $\hat{f}$ is the Fourier transform of the
function $f$.  Such a procedure, which Fourier transforms the amplitudes
of a wavefunction rather than an array of floating point numbers,
would not even make sense on a classical computer; complex
probabilities do not exist and global properties of a probability
distribution (such as periodicity) cannot be accessed by a single
sample.  Likewise, Grover's search algorithm can be thought of as an
application of quantum walks\cite{Szegedy04}, a versatile quantum
subroutine that is not only faster than classical random walks, but
again performs a task that would not be well-defined in terms of
classical probabilities.  These quantum subroutines represent the core
of quantum speedups, as well as the place where our classical
intuition about algorithms as logical procedures breaks down.  Thus,
finding new nontrivial purely quantum algorithms is likely to be the
key to understanding exactly how quantum computing is more powerful
than the classical model.

The second half of this thesis is based on the Schur transform, a
purely quantum algorithm which, like the quantum Fourier transform,
changes from a local tensor power basis to a basis that reflects the
global properties of the system.  While the Fourier transform involves
the cyclic group (which acts on an $n$-bit number by addition), the
Schur transform is instead based on the symmetric and unitary groups,
which act on $n$ $d$-dimensional quantum systems by permuting them and
by collectively rotating them.  The primary contribution of this
thesis will be an efficient quantum circuit implementing the Schur transform.
As a purely quantum algorithm, the Schur transform does not directly
solve any classical problem.  However, it is a crucial subroutine for
many tasks in quantum information theory, which can now be efficiently
implemented on a quantum computer using our methods.  More
intriguingly, an efficient implementation of the Schur transform
raises the hope of finding new types of quantum speedups.
\end{description}

%
%
%

This section has tried to give a flavor of why quantum information is
an interesting subject, and of the sort of problems that this thesis
contributes to.  In the next section, we will set out the
contributions of this thesis more precisely with a detailed technical
summary.

\section{Summary of results}\label{sec:intro-summary}
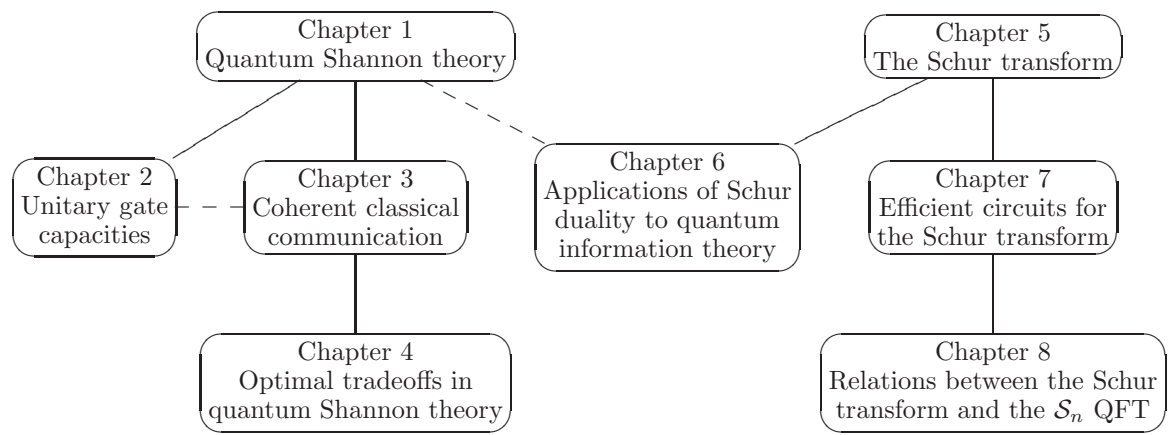
\begin{figure}[ht]
\begin{center}
\leavevmode\xymatrix@C=8pt{
& *+[F-:<9pt>]\txt{Chapter 1\\Quantum Shannon theory}
& ~~~~ & *+[F-:<9pt>]\txt{Chapter 5\\The Schur transform} \\
*+[u][u][F-:<9pt>]{\txt{Chapter 2\\Unitary gate\\capacities}}
\ar@{-}[ur] &
*+[F-:<9pt>]{\txt{Chapter 3\\Coherent classical\\communication}}
\ar@{--}[l]\ar@{-}[u]
&  *+[F-:<9pt>]\txt{Chapter 6\\Applications of Schur\\duality to
quantum\\information theory} 
\ar@{-}[ur] \ar@{--}[ul]
& *+[F-:<9pt>]\txt{Chapter 7\\Efficient circuits for\\the Schur transform}
\ar@{-}[u]\\
& *+[F-:<9pt>]\txt{Chapter 4\\Optimal tradeoffs in\\quantum Shannon theory}
\ar@{-}[u] & &
*+[F-:<9pt>]\txt{Chapter 8\\Relations between the Schur\\transform and the
 ${\cal S}_n$ QFT}\ar@{-}[u]
}
\caption{Dependencies between different chapters of this thesis.  The
solid lines indicate that one chapter depends on another, while the
dashed lines mean a partial dependence: \sect{schur-qit-apps} has
references to some of the protocols in \sect{known} and \chap{ccc} is
motivated by and extends the results of \chap{unitary}.}
\label{fig:dependencies}
\end{center}
\end{figure}

This thesis is divided into two halves: \chaprange{shannon}{family}
discuss information theory and \chaprange{schur}{sch-Sn} are on the
Schur transform.  The first chapter of each half is mostly background
and the other chapters are mostly new work, though some exceptions to
this rule will be indicated.  A diagram of how the chapters depend on
one another is given in \fig{dependencies}.

\begin{description}
\item[\chap{shannon}] introduces a rigorous framework for
concisely stating coding theorems in quantum Shannon theory.  The
key idea, which has long been tacitly understood but not spelled out
explicitly, is that communication protocols in quantum information
theory can be thought of as {\em inequalities} between asymptotic
information processing {\em resources}.  Channel coding, for example,
says that a noisy channel is at least as useful for communication as
the use of a noiseless channel at a particular rate.  This chapter
rigorously defines and proves the sort of claims we would like to take
for granted (e.g., that resources inequalities are transitive) in
\sect{resources}, goes on to prove some more advanced properties of
resource inequalities in \sect{general-inequalities} and then
summarizes many of the key results of quantum Shannon theory in terms
of this new formalism in \sect{known}. \chap{shannon} also lays out
various definitions and notation used in the rest of the thesis, and
in particular gives a detailed description of how the various
purifications we use make up the Church of the Larger Hilbert Space
(in \sect{prelim}).  This chapter, as well as \chap{family}, is based
on joint work with Igor Devetak and Andreas Winter, which is in the
process of being turned into a paper\cite{DHW05}.

\item[\chap{unitary}] applies this resource formalism to the problem
of communication using a unitary gate that couples two
parties. Unitary gates are in some ways more complicated than one-way
quantum channels because they are intrinsically bidirectional, but in
other ways they are simpler because they do not interact with the
environment.  The main results of this chapter are capacity formulae
for entanglement creation and one-way classical communication using
unlimited entanglement, as well as several relations among these and
other capacities.  We will see that most of these results are
superseded by those in the next chapter; the capacity formulae will be
simultaneously generalized while the relations between capacities will
be explained in terms of a deeper principle.  However this chapter
helps provide motivation, as well as basic tools, for the results that
follow.  It is based on \cite{BHLS02} (joint work with Charles
Bennett, Debbie Leung and John Smolin), though the original manuscript
has been rewritten in order to use the resource formalism of
\chap{shannon} (which has greatly simplified both definitions and
proofs) and to add new material.

\item[\chap{ccc}] introduces the concept of {\em coherent classical
communication}, a new communication primitive that can be thought of
either as classical communication sent through a unitary channel, or
as classical communication in which the sender gets the part of the
output that 
normally would go to the environment.  This provides an efficient (and
in fact, usually optimal) link from a wide variety of
classical-quantum protocols (teleportation, super-dense coding, remote
state preparation, HSW coding, classical capacities of unitary gates,
and more in the next chapter) to purely quantum protocols that often
would be much more difficult to prove by other means (super-dense
coding of quantum states, quantum capacities of unitary gates, etc.).

This chapter describes some of the general properties of coherent
communication, showing how it is equivalent to standard resources and
proving conditions under which classical-quantum protocols can be made
coherent.  After describing how the examples in the last paragraph can
all be fruitfully made coherent, we apply these results to find the
tradeoff between the rates of classical communication and entanglement
generation/consumption possible per use of a unitary gate.

Most of the material in this chapter is based on \cite{Har03}, with a
few important exceptions.  The careful proofs of the converse of
\thm{Ce-cap} (which showed that unlimited back communication does not
improve unitary gate capacities for forward communication or
entanglement generation) and of coherent remote state preparation are
new to the thesis.
The full bidirectional version of \thm{bidi-ccc}
(showing that sending classical communication through unitary channels
is as strong as coherent classical communication) and the discussion
of bidirectional rate regions in \sect{u-bidi-caps} are both from
\cite{HL04}, which was joint work with Debbie Leung.  Finally, the
formal rules for when classical communication can be made coherent
were sketched in \cite{DHW03} and will appear in the present form in
\cite{DHW05}, both of which are joint work with Igor Devetak and
Andreas Winter.

\item[\chap{family}] uses coherent classical communication from
\chap{ccc}, the resource formalism from \chap{ccc} and a few other
tools from quantum Shannon theory (mostly derandomization and measurement
compression) to (1) derive three new communication protocols, (2)
unify them with four old protocols into a family of related resource
inequalities and (3) prove converses that yield six different optimal
tradeoff curves for communication protocols that use a noisy channel
or state to produce/consume two noiseless resources, such as classical
communication, entanglement or quantum communication.

At the top of the family are two purely quantum protocols that can be
related by exchanging states with channels: the ``mother'' protocol
for obtaining pure entanglement from a noisy state assisted by a
perfect quantum channel, and the ``father'' protocol for sending
quantum information through a noisy channel assisted by entanglement.
Combining the parent protocols with teleportation, super-dense coding
and entanglement distribution immediately yields all of the other
``child'' protocols in the family.  The parents can in turn be
obtained from most of the children by simple application of coherent
classical communication.  It turns out that all of the protocols in
the family are optimal, but since they involve finite amounts of two
noiseless resources the converses take the form of two-dimensional
capacity regions whose border is a tradeoff curve.

This chapter is based on joint work with Igor Devetak and Andreas
Winter\cite{DHW03,DHW05}.  Most of the results first appeared in
\cite{DHW03}, though proofs of the converses and more careful
derivations of the parent protocols will be in \cite{DHW05}.

\item[\chap{schur}] begins the part of the thesis devoted to the Schur
transform.  Schur duality is a way of relating the representations
that appear when the unitary group $\cU_d$ and the symmetric group
$\cS_n$ act on $\nqudits$.  Schur duality implies the existence of a
{\em Schur basis} which simultaneously diagonalizes these
representations; and the unitary matrix relating the Schur basis to
the computational basis is known as the {\em Schur transform}.

The chapter begins by describing general properties of group
representations, such as how they combine in the Clebsch-Gordan
transform and how the Fourier transform decomposes the regular
representation, using the language of quantum information.  Then we go
on to describe the Schur transform, explain how it can be used to
understand the irreps of $\cU_d$ and $\cS_n$, and give an idea of how
Schur duality can generalized to other groups.

None of the material in this chapter is new (see \cite{GW98} for a
standard reference), but a presentation of this form has not appeared
before in the quantum information literature.  A small amount of the
material has appeared in \cite{BCH04} and most will later appear in
\cite{BCH05a,BCH05b}, all of which are joint work with Dave Bacon and
Isaac Chuang.

\item[\chap{sch-qit}] describes how Schur duality can be applied to
quantum information theory in a way analogous to the use of the method
of types in classical information theory.  It begins by reviewing the
classical method of types in \sect{classical-types} (following
standard texts\cite{CT91,CK81}) and then collects a number of facts
that justify the use of Schur duality as a quantum method of types in
\sect{quantum-types} (following \cite{GW98,Hayashi:02e,CM04}).
\sect{schur-qit-apps} then surveys a wide variety of information
theory results from the literature that are based on Schur duality.
This section will appear in \cite{BCH05a} and a preliminary version
was in \cite{BCH04} (both joint with Dave Bacon and Isaac Chuang).

The only new results of the chapter are in \sect{normal-form}, which
gives a way to decompose $n$ uses of a memoryless quantum channel in
the Schur basis, and shows how the components of the decomposition can
be thought of as quantum analogues of joint types.

\item[\chap{sch-algo}] turns to the question of computational
efficiency and gives a $\poly(n,d,\log 1/\eps)$ algorithm that
approximates the Schur transform on $\nqudits$ up to accuracy $\eps$.

The main idea is a reduction from the Schur transform to the
Clebsch-Gordan transform, which is described in \sect{schur-circuit}.
Then an efficient circuit for the Clebsch-Gordan transform is given in
\sect{cg-construct}.  Both of these algorithms are made possible by
using {\em subgroup-adapted bases} which are discussed in \sect{gz}.

\sect{schur-circuit} first appeared in \cite{BCH04} and the rest of
the chapter will soon appear in \cite{BCH05a}.  Again, all of this
work was done together with Dave Bacon and Isaac Chuang.

\item[\chap{sch-Sn}] explores algorithmic connections between the
Schur transform and the quantum Fourier transform (QFT) over $\cS_n$.
We begin by presenting {\em generalized phase estimation}, in which
the QFT is used to measure a state in the Schur basis, and then
discuss some generalizations and interpretations of the algorithm.
Then we give a reduction in the other direction, and show how a
variant of the standard $\cS_n$ QFT can be derived from one
application of the Schur transform.

Generalized phase estimation was introduced in the earlier versions of
\cite{BCH04}, and will appear along with the other
results in this chapter in \cite{BCH05b} (joint with Dave Bacon and
Isaac Chuang).
\end{description}

{\em Recommended background:} This thesis is meant to be
understandable to anyone familiar with the basics of quantum computing
and quantum information theory.  The textbook by Nielsen and
Chuang\cite{NC00} is a good place to start; Chapter 2 (or knowledge of
quantum mechanics) is essential for understanding this thesis,
Chapters 9 and 11 (or knowledge of the HSW theorem and related
concepts) are necessary for the first half of the thesis, and Sections
4.1-4.4, 5.1-5.2 and 12.1-12.5 are recommended.  The first six
chapters of Preskill's lecture notes\cite{Preskill98} are another
option.  Both \cite{NC00} and \cite{Preskill98} should be accessible
to anyone familiar with the basics of probability and linear algebra.
Further pointers to the literature are contained in Chapters
\ref{chap:shannon} and \ref{chap:schur}, which respectively introduce the
information theory background used in the first half of the thesis and
the representation theory background used in the second half.
\chapter{Quantum Shannon theory}\label{chap:shannon}

Two communicating parties, a sender (henceforth called Alice)
and a receiver (Bob), usually have, in a mathematical theory of communication,
a predefined goal like the perfect transmission of a classical
message, but at their disposal are only imperfect
resources\footnote{The term is used here in an everyday sense; later
in this chapter we make it mathematically precise.}
like a noisy channel. This is Shannon's channel coding
problem~\cite{Shannon48}: allowing the parties arbitrary
local operations (one could also say giving them local resources
for free) they can perform encoding and decoding of the
message to effectively reduce the noise of the given channel.
Their performance is measured by two parameters: the error
probability and the number of bits in the message, and quite naturally
they want to minimize the former while maximizing the latter.

In Shannon theory, we are particularly interested in the case that
the channel is actually a number of independent realizations of the
same noisy channel and that the message is long: the efficiency
of a code is then measured by the rate, i.e., the ratio of
number of bits in a message by number of channel uses.
And in particular again, we ask for the asymptotic regime of
arbitrarily long messages and vanishing error probability.

Note that not only their given channel, but also the goal of the
parties, noiseless communication, is a resource: the channel
which transmits one bit perfectly (it is ``noisy'' in the extreme
sense of zero noise), for which we reserve the special symbol
$[c\ra c]$ and call simply a \emph{cbit}.
Thus coding can be described more generally as
the conversion of one resource into another, i.e.,
simulation of the target resource by using the given resource
together with local processing. For a generic noisy channel,
denoted $\{c\ra c\}$, we express such an asymptotically faithful
conversion of rate $R$ as a \emph{resource inequality}
$$\{c\ra c\} \geq R [c\ra c],$$
which we would like to think of as a sort of chemical reaction,
and hence address the left hand side as \emph{reactant resource(s)}
and the right hand side as \emph{product resource(s)} with $R$ the
conversion ratio between these two resoures. 
In the asymptotic
setting, $R$ can be any real number, and the maximum $R$ is the
(operational) capacity of the channel --- to be precise: to transmit
information in the absence of other resources.

Obviously, there exist other useful or desirable resources, such as
perfect correlation in the form of a uniformly random bit
(abbreviated \emph{rbit}) known
to both parties, denoted $[c \, c]$, or more generally some noisy correlation.
In quantum information theory, we have further resources:
noisy quantum channels and quantum correlations between the
parties. Again of particular interest are the noiseless unit
resources; $[q\ra q]$ is an ideal quantum bit
channel (\emph{qubit} for short), and
$[q \, q]$ is a unit of maximal entanglement, a two-qubit
singlet state (\emph{ebit}).  The study of asymptotic conversion rates
between the larger class of quantum information-theoretic resources is
known as {\em quantum Shannon theory} and is the main focus of this
half of the thesis.

\medskip

To illustrate the goals of quantum Shannon theory, it is instructive
to look at the conversions permitted by the unit resources $[c\ra c]$,
$[q\ra q]$ and $[q \, q]$, where resource inequalities are finite and
exact: the following inequalities always refer to a specific integral
number of available resources of a given type, and the protocol
introduces no error. We mark such inequalities by a $*$ above the
$\geq$ sign.  For example, it is always possible to use a qubit to
send one classical bit, $[q\ra q] \reduction [c\ra c]$, and to
distribute one ebit, $[q\ra q] \stackrel{{}_{\scriptstyle *}}{\geq}
[q\, q]$; the latter is referred to as entanglement distribution (ED).

More inequalities are obtained by combining resources.
Super-dense coding~\cite{BW92} is a coding protocol
to send two classical bits using one qubit and one ebit:
\begin{equation*}
  [q\ra q]  + [q\, q] \stackrel{*}{\geq} 2[c\ra c].  \tag{SD}
\end{equation*}
Teleportation~\cite{BBCJPW98} is expressed as
\begin{equation*}
  2[c\ra c] + [q\, q] \stackrel{*}{\geq}  [q\ra q].  \tag{TP}
\end{equation*}
In~\cite{BBCJPW98} the following argument was used
that the ratio of $1:2$ between
$[q\ra q]$ and $[c\ra c]$ in these protocols is optimal, even with
unlimited entanglement, and even asymptotically: assume, with $R>1$,
$[q\ra q] + \infty [q\, q] \geq 2 R [c\ra c]$; then chaining this
with (TP) gives $[q\ra q] + \infty [qq] \geq R [q\ra q]$. Hence by
iteration $[q\ra q] + \infty [q\, q] \geq R^k [q\ra q] \geq R^k [c\ra c]$
for arbitrary $k$, which can make $R^k$ arbitrarily large,
and this is easily disproved. Analogously,
$2[c\ra c] + \infty [q\, q] \geq R[q\ra q]$, with $R>1$, gives, when
chained with (SD), $2[c\ra c] + \infty [q\, q] \geq 2R[c\ra c]$,
which also easily leads to a contradiction. In a similar way, the optimality
of the one ebit involved in both (SD) and (TP) can be seen.

While the above demonstration looks as if we did nothing but introduce
a fancy notation for things understood perfectly well otherwise,
in this chapter we want to make the case for a systematic
theory of resource inequalities. We will present a framework
general enough to include most two-player setups,
specifically designed for the asymptotic memoryless regime.
There are three main issues there: first, a suitably flexible
definition of a \emph{protocol}, i.e., a way of combining resources
(and with it a mathematically precise notion of a resource inequality);
second, a justification of the composition (chaining) of resource
inequalities; and third, general tools to produce new protocols
(and hence resource inequalities) from existing ones.

The benefit of such a theory should be clear then: while it
does not mean that we get coding theorems ``for free'', we \emph{do}
get many protocols by canonical modifications from others,
which saves effort and provides structural insights into
the logical dependencies among coding theorems. As the above
example shows, we also can relate (and sometimes actually
prove) the converses, i.e.~the statements of optimality,
using the resource calculus.

\bigskip\noindent
The remainder of this chapter will systematically develop the
resource formalism of quantum Shannon theory, and will show how it can
concisely express and relate many familiar results in quantum information.
\begin{description}
  \item[Section \ref{sec:prelim}]
    covers the preliminaries and
    describes several complementary formalisms
    for quantum mechanics, which serve diverse purposes in 
    the study of quantum information processing. Here also
    some basic facts are collected.

  \item[Section \ref{sec:resources}]
    sets up the basic communication scenario
    we will be interested in. It contains definitions and
    basic properties of so-called finite resources, and
    how they can be used in protocols.
    Building upon these we define asymptotic resources and
    inequalities between them, in such a way as to ensure 
    natural composability properties.

  \item[Section \ref{sec:general-inequalities}]
    contains a number of general and useful resource inequalities.

  \item[Section \ref{sec:known}]
    compiles most of the hitherto discovered 
    coding theorems, rewritten as resource inequalities.

   \item[Section \ref{sec:res-discuss}]
	concludes with a discussion of possible extensions to the
	resource formalism developed in the rest of the chapter.
\end{description}

The following three chapters will apply this formalism to develop new
coding results.
\begin{description}
\item[\chap{unitary}] will examine the communication and
entanglement-generating capacities of bipartite unitary gates.  It is
primarily based on \cite{BHLS02} (joint work with Charles Bennett,
Debbie Leung and John Smolin).
\item[\chap{ccc}] develops the idea of {\em coherent classical
communication} and applies it to a variety of topics in quantum
Shannon theory, following the treatment of \cite{Har03} and
\cite{HL04} (joint work with Debbie Leung).
\item[\chap{family}] shows how coherent classical communicaton can be
used to derive new quantum protocols and unify old ones into a family
of resource inequalities.  This chapter, as well as the present one,
are based on \cite{DHW03,DHW05} (joint work with Igor Devetak and
Andreas Winter).
\end{description}

\section{Preliminaries}
\label{sec:prelim}

This section is intended to introduce notation and ways of
speaking about quantum mechanical information scenarios.
We also state several key lemmas needed for the technical proofs.
Most of the facts and the spirit of this section can be
found in~\cite{Holevo01b}; a presentation slightly
more on the algebraic side is~\cite{WinterPhD}, appendix A.

\subsection{Variations on formalism of quantum mechanics}
\label{sec:formalism}
\index{Church of the Larger Hilbert Space|(}

We start by reviewing several equivalent formulations of quantum
mechanics and discussing their relevance for the study of quantum
information processing. As we shall be using several of them in
different contexts, it is useful to present them in a systematic way.
The main two observations are, first, that a classical random variable
can be identified with a quantum systems equipped with a preferred
basis, and second, that a quantum Hilbert space can always be extended
to render all states pure (via a reference system) and all operations
unitary (via an environment system) on the larger Hilbert space.

Both have been part of the quantum information processing 
folklore for at least a decade (the second of course
goes back much farther: the GNS construction, Naimark's
and Stinespring's theorems, see \cite{Holevo01b}),
and roughly correspond to
the ``Church of the Larger Hilbert Space'' viewpoint. 

Based on this hierarchy of embeddings  
${\rm C(lassical) \Rightarrow Q(uantum) \Rightarrow P(ure)}$,
in the above sense,
we shall see how the basic ``CQ'' formalism of
quantum mechanics gets modified to (embedded into)
CP, QQ, QP, PQ and PP formalisms. (The second letter refers
to the way quantum information is presented; the first,
how knowledge about this information is presented.)
We stress that from an
operational perspective they are all equivalent --- however, which
formalism is the most useful depends on the context.

\medskip
Throughout the thesis we shall use labels such as $A$ 
(similarly, $B$, $C$, etc.) 
to denote not only a particular quantum system but also the 
corresponding Hilbert space (which is also denoted $\cH_A$) and to
some degree even 
the set of bounded linear operators on that Hilbert space (also
denoted $\cL(\cH_A)$ or $\cL(A)$).  If $\ket{\psi}$ is a pure state,
then we will sometimes use $\ket{\psi}$ to denote the density matrix
$\oprod{\psi}$.
When talking about tensor products of spaces, we will
habitually omit the tensor sign, so $A\ot B = AB$, etc.
Labels such as $X$, $Y$, etc. will be used for
classical random variables. For simplicity, all spaces and
ranges of variables will be assumed to be finite.

\index{CQ formalism|(}
\paragraph{The CQ formalism.} This formalism is
the most commonly used one in the literature, as it captures most of
the operational features of a ``Copenhagen'' type quantum mechanics,
which concerns itself more with the behavior of quantum systems than
their meaning, reserving ontological statements about probabilities,
measurement outcomes, etc. for classical systems.  The postulates of
quantum mechanics can be classified into static and dynamic ones.  The
static postulates define the static entities of the theory, while the
dynamic postulates describe the physically allowed evolution of the
static entities.  In defining classes of static and dynamics
entities, we will try to highlight their (quantum)
information-theoretic significance.
 
The most general static entity is an \emph{ensemble} of
quantum states $(p_x, \rho_x)_{x \in \cX}$.
The probability distribution $(p_x)_{x \in \cX}$ is defined on some set $\cX$
and is associated with the random variable $X$.
The $\rho_x$ are density operators (positive Hermitian 
operators of unit trace) on
the Hilbert space of a quantum system $A$.  
The state of the quantum system $A$ is 
thus correlated with the classical index random variable $X$.
We refer to $XA$ as a hybrid classical-quantum system, 
and the ensemble $(p_x, \rho_x)_{x \in \cX}$ is the ``state'' 
of $XA$. 
We will occasionally refer to a classical-quantum 
system  as a ``$\{ c \, q \}$ {entity}''. 
Special cases of
$\{ c \, q \}$ entities are $\{ c \}$ entities 
(``classical systems'', i.e. random variables) 
and $\{ q \}$ entities (quantum systems).

The most general dynamic entity would be a map between two $\{ c \,q \}$
entities (hence, and throughout the thesis, we describe dynamics
in the Schr\"odinger picture). Let us highlight only a few special cases:

The most general map from a $\{ c \}$ entity to a $\{ q \}$ 
entity is a \emph{state preparation map} or a ``$\{ c \rightarrow q \}$
entity''. It is defined by a \emph{quantum alphabet} $(\rho_x)_{x \in \cX}$
and maps the classical index $x$ to the quantum state $\rho_x$.

Next we have a $\{ q \rightarrow c \}$ entity, a 
\emph{quantum measurement}, defined by a 
positive operator-valued measure (POVM) $(M_x)_{x \in \cX}$, where 
$M_x$ are positive operators satisfying $\sum_x M_x = \one$,
with the identity operator $\one$ on the underlying Hilbert space.
The action of the POVM $(M_x)_{x \in \cX}$ on  some
quantum system $\rho$ results in the random variable defined
by the probability distribution $(\tr \rho M_x)_{x \in \cX}$ on $\cX$.
POVMs will be denoted with roman capitals:
$L$, $M$, $N$, $P$,  etc.

A $\{q \rightarrow q \}$ entity is a \emph{quantum operation}, a
completely positive and trace
preserving (CPTP) map $\cN: A \rightarrow B$, described (non-uniquely)
by its \emph{Kraus representation}:  a set of operators
$\{ N_x \}_{x \in \cX}$, 
$\sum_x N_x^\dagger N_x = \one^B$, 
whose action is given by
$$
\cN(\rho) = \sum_x N_x \rho N_x^\dagger. 
$$ 
(When referring to operators, we use $\dagger$ for the adjoint,
while $*$ is reserved for the complex conjugate.  In
Chapters~\ref{chap:schur}-\ref{chap:sch-algo}, we will also apply $*$
to representation spaces to indicate the dual representation.)
A CP map is defined as above, but with the weaker restriction 
$\sum_x A_x^\dagger A_x \leq \one^B$, and by itself is unphysical
(or rather, it includes a postselection of the system).
Throughout, we will denote CP and CPTP maps by calligraphic letters:
$\cL$, $\cM$, $\cN$, $\cP$, etc. A special CPTP map
is the identity on a system $A$, $\id^A:A\rightarrow A$,
with $\id^A(\rho)=\rho$. More generally, for an isometry
$U:A\rightarrow B$, we denote --- for once deviating from the
notation scheme outlined here --- the corresponding CPTP map by the
same letter: $U(\rho) = U\rho U^\dagger$.

A $\{q \rightarrow c q \}$ entity is an \emph{instrument} $\bbP$,
described by an ordered set of CP maps $(\cP_x)_x$
that add up to a CPTP map.
$\bbP$ maps a quantum state $\rho$ to the
ensemble $(p_{x}, \cP_{x}(\rho)/p_x)_x$, with $p_x = \tr \cP_{x}(\rho)$.
A special case of an instrument is one in which $\cP_x = p_x \cN_x$,
and the $\cN_x$ are CPTP; it is equivalent to 
an ensemble of CPTP maps, $(p_x, \cN_x)_{x \in \cX}$. 
Instruments will be denoted by blackboard style capitals:
$\bbL$, $\bbM$, $\bbN$, $\bbP$, etc.

A $\{cq \rightarrow q \}$ entity is given by an ordered set 
of CPTP maps $(\cN_{x})_x$, and maps the ensemble
$(p_x, \rho_x)_{x \in \cX}$ to $\sum_x p_x \cN_x(\rho_x)$.  By
contrast, a $\{c,q\ra q\}$ map saves the classical label, mapping
$(p_x, \rho_x)_{x \in \cX}$ to $(p_x, \cN_x(\rho_x))_{x\in\cX}$.

In quantum information theory the CQ formalism
is used for proving direct coding theorems 
of a part classical -- part quantum nature, 
such as the HSW theorem \cite{Holevo98,SW97}. In addition, 
it is most suitable for computational purposes.

\medskip
For two states, we write $\rho^{ RA} \ext \sigma^{A}$ to mean
that the state $\sigma^{A}$ is a \emph{restriction} of 
$\rho^{RA}$, namely 
$\sigma^{A} = \tr_{\!R} \rho^{RA}$. The subsystem $R$ is
possibly null (which we write $R = \emptyset$), i.e., a $1$-dimensional
Hilbert space.
Conversely, $\rho^{RA}$ is called an
\emph{extension} of $\sigma^{A}$. 
Furthermore, if $\rho^{RA}$ is pure it is called a 
\emph{purification} of $\sigma^{R}$.
The purification is unique up to a local isometry on $R$:
this is an elementary consequence of the Schmidt decomposition
(discussed in \sect{u-schmidt}).
These notions carry over to dynamic entities as well.
For two quantum operations $\cA: A \rightarrow
B E$ and $\cB: A \rightarrow B$ we write $\cA \ext \cB$ 
if $\cB = \tr_{\!E} \circ \cA$. If $\cA$ is an isometry, 
is is called an \emph{isometric extension} of $\cB$, 
and is unique up to an isometry on $E$ --- this and the
existence of such a \emph{dilation} are known as
Stinespring's theorem~\cite{Sti55}.

Observe that we can safely represent noiseless quantum evolution by
isometries between systems (whereas quantum mechanics demands
\emph{unitarity}): this is because our systems are all finite, and we
can embed the isometries into unitaries on larger systems. Thus we
lose no generality but gain flexibility.
\index{CQ formalism|)}

\index{CP formalism|(}
\paragraph{The CP formalism.} In order to define the CP formalism, 
it is necessary to review an alternative representation
of the CQ formalism that involves fewer primitives. For instance,

\begin{itemize}
\item $\{q\}$. A quantum state  $\rho^{A}$ is referred to by 
   its purification $\ket{\phi}^{AR}$.
\item $\{c \, q\}$, $\{c \rightarrow q\}$. 
  The ensemble $(p_x, \rho_x^{A})_x$ 
  [resp.~quantum alphabet  $(\rho_x^{A})_x$] is
  similarly seen as restrictions of a pure state ensemble 
  $(p_x, \ket{\phi_x}^{AR})_x$ [resp.~quantum alphabet 
  $(\ket{\phi_x}^{AR})_x$].
\item $\{ q \rightarrow q \}$. A  CPTP map $\cN: A \rightarrow 
  B$ is referred to by its isometric extension $U_\cN:  A \rightarrow B E$.
\item $\{ q \rightarrow c \}$.
  A  POVM $(M_x)_x$ on the system $A$ is equivalent to 
  some isometry  $U_M: A \rightarrow A {E_X}$,
  followed by a von Neumann measurement of the system $E_X$ in 
  basis $\{ \ket{x}^{E_X} \}$, and discarding $A$.
\item $\{ q \rightarrow c \, q \}$.
  An instrument $\bbP$ is equivalent to some isometry 
  $U_\bbP: A \rightarrow B {E} {E_X}$,
  followed by a von Neumann measurement of the system $E_X$ 
  in basis $\{ \ket{x}^{E_X} \}$, and discarding $E$.
\item $\{ c, q \rightarrow  q \}$ The ensemble
  of CPTP maps $(p_x, \cN_x)_x$ is identified with the
  ensemble of isometric extensions $(p_x, U_{\cN_x})_x$.
\end{itemize}

In this alternative representation of the CQ formalism 
all the quantum static entities are  thus seen as restrictions 
of pure states and 
all quantum dynamic entities are combinations of
performing isometries, von Neumann measurements,
and discarding auxiliary subsystems. 
The \emph{CP formalism} is characterized by never discarding (tracing out)
the  auxiliary subsystems (reference systems,
environments, ancillas); 
they are kept in the description of our system.
As for the auxiliary subsystems that get (von-Neumann-) measured,
without loss of generality they may be discarded:
the leftover state of such a subsystem may be set
to a standard state $\ket{0}$
(and hence  decoupled from the rest of the system)
by a local unitary conditional upon the measurement outcome. 

The CP formalism is mainly used in quantum information theory
for proving direct coding theorems 
of a quantum nature, such as the quantum channel
coding theorem (see e.g.~\cite{Devetak03}).
\index{CP formalism|)}

\paragraph{The QP formalism.} 
\index{QP formalism|(}
The QP formalism  differs from CP in that 
the classical random variables, i.e. classical systems,
are embedded into quantum systems,
thus enabling a unified treatment of the two.
 
\begin{itemize}
\item $\{ c \}$. The classical random variable $X$ is identified with
  a dummy quantum system $X$ equipped with preferred basis $\{ \ket{x}^X \}$,
  in the state $\sigma^{X} = \sum_x p_x \proj{x}$.
  The main difference between random variables and
  quantum systems is that random variables exist
  without reference to a particular physical implementation,
  or a particular system ``containing'' it. 
  In the QP formalism this is reflected in the fact that
  the state $\sigma^{X}$ remains intact under the ``copying''
  operation $\bar{\Delta}: X \rightarrow X X'$, with Kraus representation
  $\{ \ket{x}^{X}  \ket{x}^{X'} \bra{x}^{X}  \}$. In this way, instances of
  the same random variable may be contained in 
  different physical systems. 
\item $\{ c \, q \}$. An ensemble $(p_x, \ket{\phi_x}^{AR})_x$
  is represented by a quantum state
  $$
  \sigma^{XAR} = \sum_x p_x \proj{x}^X \otimes \phi_x^{AR}.
  $$
\item $\{c \rightarrow q \}$. A state preparation map 
  $(\ket{\phi_x}^{AR})_x$ is given by the isometry
  $\sum_x \ket{\phi_x}^{AR}\ket{x}^{X} \bra{x}^{X}$, 
  followed by tracing out $X$.
\item $\{c q \rightarrow q \}$. The ensemble of isometries
  $(p_x, U_x)$ is represented by the controlled isometry
  $$
  \sum_x \proj{x}^X \otimes U_x.
  $$
\item $\{ q \rightarrow c \}, \{ q \rightarrow c \, q \}$. 
  POVMs and instruments are treated as in the CP picture, 
  except that the final von Neumann measurement is 
  replaced by a completely dephasing operation 
  $\bar{\id}: {E_X} \rightarrow {X}$, defined
  by the 
  Kraus representation $\{ \ket{x}^X \bra{x}^{E_X} \}_x$.
\end{itemize}

The QP formalism is mainly used in quantum information theory
for proving converse theorems. 
\index{QP formalism)|}

\paragraph{Other formalisms.}
The QQ formalism is obtained from the QP formalism by 
tracing out the auxiliary systems, and is also
convenient for proving converse theorems. In this 
formalism the primitives are 
general quantum states (static) and quantum operations 
(dynamic).

The PP formalism involves further ``purifying'' the 
classical systems in the QP formalism; it is distinguished
by its remarkably simple structure: all of quantum 
information processing is described in terms of 
isometries on pure states. There is 
also a PQ formalism, for which we don't see much use; 
one may also conceive of hybrid formalisms, such as 
QQ/QP, in which some but not all auxiliary systems are 
traced out. One should remain flexible.
We will usually indicate, however, which formalism we are using as we
go along.
\index{Church of the Larger Hilbert Space)|}

\subsection{Quantities, norms, inequalities, and miscellaneous notation}

For a state $\rho^{R A}$ and quantum operation 
$\cN: A \rightarrow B$ we identify, somewhat sloppily,
$$
\cN(\rho) := (\id^{R} \otimes \cN) \rho^{RA}.
$$
With each state $\rho^{B}$, one may associate a quantum
operation that appends the state to the input,
namely $\app{\rho} : {A} \rightarrow {AB}$,
defined by
$$
\app{\rho} (\sigma^{A}) =  \sigma^{A} \otimes \rho^{B}.  
$$
The state $\rho$ and the operation $\app{\rho}$ are
clearly equivalent in an operational sense. 

\medskip
\index{entropy}
\index{mutual information|see{entropy}}
\index{coherent information|see{entropy}}
Given some state, say $\rho^{XAB}$, one may define the usual entropic
quantities with respect to it.  Recall the definition of the von
Neumann entropy $H(A) = H(A)_\rho = H(\rho^A) = -\tr (\rho^A \log
\rho^A)$, where $\rho^A = \tr_{\!XB} \,{\rho}^{XAB}$.  When we
specialize to binary entropy this becomes $H_2(p) := -p\log p -
(1-p)\log p$. Throughout this thesis $\exp$ and $\log$ are base 2.
Further define the conditional entropy \cite{CA95}
$$H(A|B) = H(A|B)_\rho = H(AB) - H(B),$$
the quantum mutual information \cite{CA95}
$$I(A;B) = I(A;B)_\rho = H(A) + H(B) - H(AB),$$ 
the coherent information \cite{Sch96,SN96}
$$I(A\,\rangle B) = -H(A|B)=H(B)-H(AB),$$
and the conditional mutual information
$$I(A;B|X) =  H(A|X)+H(B|X)-H(AB|X).$$
Note that the conditional mutual information is always
non-negative, thanks to strong subadditivity~\cite{LR73}.

It should be noted that conditioning on 
classical variables (systems) amounts to averaging.
For instance, for a state of the form
$$
\sigma^{XA} = \sum_x p_x \proj{x}^X \otimes \rho_x^{A},
$$
$$
H(A|X)_\sigma = \sum_x p_x H(A)_{\rho_x}.
$$
We shall freely make use of standard identities for these
entropic quantities, which are formally identical to the
classical case (see Ch.~2 of \cite{CT91} or Ch.~1 of \cite{CK81}).
One such identity is the so-called chain rule for mutual information,
$$ I(A;BC) = I(A;B|C) + I(A;C),$$
and using it we can derive an identity will later be useful:
\be I(X;AB) = H(A) + I(A\> BX) - I(A;B) + I(X;B).
\label{eq:trip-entropy}\ee

We shall usually work in situations where the underlying state
is unambiguous, but as shown above, we can emphasize the state
by putting it in subscript.

\medskip
\index{distance!betweeen states}
We measure the distance between  
two quantum states $\rho^A$ and $\sigma^A$ by the trace norm,
$$
\| \rho^A - \sigma^A \|_1,
$$
where $\| \omega \|_1 = \tr\sqrt{\omega^\dagger\omega}$;
for Hermitian
operators this is the sum of absolute values of the eigenvalues.
If $\|\rho^A - \sigma^A\|_1\leq \epsilon$, then we sometimes write
that $\rho\upto{\eps}\sigma$.
An important property of the trace distance is its
monotonicity under quantum operations $\cN$:
$$
\| \cN(\rho^A) - \cN(\sigma^A) \|_1 \leq \| \rho^A - \sigma^A \|_1.
$$
In fact, the trace distance is operationally connected to the
distinguishability of the states: if $\rho$ and $\sigma$ have
uniform prior, Helstrom's theorem~\cite{Hel76} says that the
maximum probability of correct identification of the state by
a POVM is $\frac{1}{2}+\frac{1}{4}\| \rho - \sigma \|_1$.


 The trace distance induces a metric on density matrices under which
the von Neumann entropy is a continuous function.  This fact is known
as Fannes' inequality\cite{Fannes73,Nielsen00a}.
\begin{lemma}[Fannes] \label{lemma:fannes}
For any states $\rho^A,\sigma^A$ defined on a system $A$ of dimension
$d$, if $\|\rho^{A} - \sigma^{A} \|_1 \leq \epsilon$ then
\be
 | H(A)_\rho - H(A)_\sigma | \leq 
\epsilon\log d + \eta(\epsilon)\ee
where $\eta(\epsilon)$ is defined (somewhat unconventionally) to be
$-\epsilon\log\epsilon$ if $\epsilon\leq 1/e$ or $(\log e)/e$ otherwise.
\end{lemma}

Fannes' inequality leads to the following useful corollary:
\begin{lemma}
\label{lemma:fano}
  For the quantity $I(A \, \> B)$ defined on
  a system $AB$ of total dimension $d$, if
  $\|\rho^{AB} - \sigma^{AB} \|_1 \leq \epsilon$ then
  $$
  | I(A \, \> B)_\rho - 
  I(A \, \> B)_\sigma | \leq \eta'(\epsilon) + K \epsilon \log d,
  $$ 
  where $\lim_{\epsilon \rightarrow 0} \eta'(\epsilon) = 0$
  and $K$ is some constant. The same holds for
  $I(A;B)$ and other entropic quantities.
  \qed
\end{lemma}

Define a distance measure between  
two quantum operations $\cM, \cN: A_1 A_2 \rightarrow B$ 
with respect to some state
$\omega^{A_1}$
by 
\be
\| \cM - \cN \|_{\omega^{A_1}} := 
\max_{{\zeta}^{R A_1A_2} \ext \omega^{A_1}}
\bigl\| (\id^{R} \otimes \cM)\zeta^{R A_1A_2}
        - (\id^{R} \otimes \cN)\zeta^{R A_1A_2} \bigr\|_1.
\label{eq:lung}
\ee
The maximization may, w.l.o.g., be performed over pure
states $\zeta^{R A_1A_2}$. This is due
to the monotonicity of trace distance under the
partial trace map.
Important extremes are when $A_1$ or $A_2$ are null.
The first case measures absolute closeness between the two 
operations (and in fact, $\|\cdot\|_\emptyset$ is the
dual of the cb-norm\cite{KW03}),
while the second measures how similar they are relative
to a particular input state.
\eq{lung} is written more succinctly as
$$
\| \cM - \cN \|_\omega := \max_{\zeta \ext \omega}
                            \| (\cM - \cN) \zeta \|_1.
$$
We say that $\cM$ and $\cN$ are 
$\epsilon$-close with respect to
$\omega$ if
$$
\| \cM - \cN \|_{\omega} \leq \epsilon.
$$
Note that $\|\cdot\|_\omega$ is a norm only if $\omega$ has full
rank; otherwise, different operations can be at distance $0$.
If $\rho$ and $\sigma$ are $\epsilon$-close
then so are $\app{\rho}$ and $\app{\sigma}$ (with respect to
$\emptyset$, hence every state).

Recall the definition of
the fidelity of two density operators with respect to 
each other:
$$
F(\rho, \sigma) = \| \sqrt{\rho} \sqrt{\sigma} \|^2_1
 = \l(\tr\sqrt{\sqrt{\sigma}\rho\sqrt{\sigma}}\r)^2.
$$
For two pure states $\ket{\phi}$, $\ket{\psi}$ this amounts to
$$
F(\proj{\phi}, \proj{\psi}) = |\langle \phi \ket{\psi} |^2.
$$
We shall need the following relation between 
fidelity and the trace distance~\cite{FG97}
\be
1 - \sqrt{F(\rho, \sigma)} \leq  \frac{1}{2} \| \rho - \sigma \|_1
                                    \leq \sqrt{1 - F(\rho, \sigma)},
\label{eq:fid-trace}
\ee
the second inequality becoming an equality for pure states.
Uhlmann's theorem~\cite{Uhlmann76,Jozsa94} states that, for
any fixed purification $\proj{\phi}$ of $\sigma$,
$$
F(\rho, \sigma)  = \max_{\proj{\psi} \ext \rho} 
F(\proj{\psi}, \proj{\phi}).
$$
As the fidelity is only defined between two states living on
the same space, we are, of course, implicitly maximizing over 
extensions $\proj{\psi}$ that live on the same space as
$\proj{\phi}$.

\begin{lemma}
  \label{lemma:purify-dist}
  If $\| \rho - \sigma \|_1 \leq \epsilon$
  and $ \sigma' \ext \sigma$, then there exists
  some $ \rho' \ext \rho$ for which
  $\| \rho' - \sigma' \|_1 \leq 2 \sqrt{\epsilon}$.
\end{lemma}
\begin{proof}
  Fix a purification $\proj{\phi}^{ABC} \ext {\sigma'}^{AB} \ext \sigma^A$.
  By Uhlmann's theorem, there exists some 
  $\proj{\psi}^{ABC} \ext \rho^A$ such that
  $$ 
  F(\proj{\psi}, \proj{\phi}) = F(\rho, \sigma) \geq 1 - 2 \epsilon,
  $$
  using also \eq{fid-trace}
  Define ${\rho'}^{AB} = \tr_{\!C}\proj{\psi}^{ABC}$.
  By the monotonicity of trace distance under the partial trace map
  and \eq{fid-trace}, we have
  $$
  \| \rho' - \sigma' \|_1 \leq \| \proj{\psi}-  \proj{\phi}\|_1
                          \leq                2 \sqrt{\epsilon},
  $$
  as advertised.
\end{proof}

\begin{lemma}
\label{lemma:props}
  The following statements hold for density operators
  $\omega^{A}$, ${\omega'}^{A A'}$, $\sigma^{A}$, $\rho^{A'}$,
  $\Omega^{A_1 }$,
  and quantum operations $\cM', \cN': A A' B \rightarrow C$,
  $\cM, \cN: AB \rightarrow C$, $\cK, \cL:  A'B' \rightarrow C'$,
  and $\cM_i, \cN_i: A_i  A_i^{*} \rightarrow A_{i+1} \hat{A}_{i+1}$.

\begin{enumerate}
\item If $\omega' \ext \omega$ then
      $\| \cM' - \cN' \|_{\omega'} \leq  \| \cM' - \cN' \|_{\omega}$.
\item $\| \cM - \cN \|_{\omega}
        \leq \| \cM - \cN \|_{\sigma} + 2 \sqrt{\|\omega - \sigma \|_1}$.
\item $\| \cM \otimes \cK - \cN \otimes \cL \|_{\omega \otimes \rho}
        \leq \| \cM - \cN \|_{\omega} + \| \cK - \cL \|_{\rho}$.
\item $\|\cM_k \circ \dots \circ \cM_1 - \cN_k \circ \dots \circ \cN_1 \|_{\Omega}
         \leq \sum_i \|\cM_i - \cN_i \|_{ (\cM_{i-1} \circ \dots \circ \cM_1)(\Omega)}$. 
\end{enumerate}
\end{lemma}
\begin{proof}
Straightforward.
\end{proof}

\medskip\noindent
Finally, if we have systems $A_1$, $A_2$, \ldots, $A_n$,
we use the shorthand $A^n = A_1 \dots A_n$.
Also, the set $\{1,\ldots,d\}$ is denoted $[d]$.

\section{Information processing resources}
\label{sec:resources}
In this section, the notion of a
information processing resource will be rigorously introduced.
Unless stated otherwise, we shall be using the QQ formalism 
(and occasionally the QP formalism) in order
to treat classical and quantum entities in a unified way.

\subsection{The distant labs paradigm}
\label{sec:distantlabs}

The communication scenarios we will be interested involve two or more
separated parties.  Each party is allowed to perform arbitrary local
operations in his or her lab for free. On the other hand, non-local
operations (a.k.a. \emph{channels}) are valuable resources.   In this
thesis, we consider the following parties:
\begin{itemize}
\item {\em Alice} ($A$)
\item {\em Bob} ($B$):  Typically quantum Shannon theory considers
only problems involving communication from Alice to Bob.  This means
working with channels from Alice to Bob 
(i.e. of the form $\cN:A'\ra B$) and arbitrary states $\rho^{AB}$
shared by Alice and Bob.  However, the next two chapters will also
consider some bidirectional communication problems.
\item {\em Eve} ($E$): In the CP and QP formalisms, we purify noisy
channels and states by giving a share to the environment.  Thus, we
replace $\cN:A'\ra B$ with the isometry $U_\cN:A'\ra BE$ and replace
$\rho^{AB}$ with $\psi^{ABE}$.\footnote{For our purposes, we can think
of Eve as a passive environment, but other work, for example on private
communication\cite{Devetak03,BM04}, treats Eve as an active participant who
is trying to maximize her information.  In these settings, we introduce
 private environments for Alice and Bob $E_A$ and $E_B$, so that
they can perform noisy operations locally without leaking information
to Eve.} We consider a series of operations equivalent when they
differ only by a unitary rotation of the environment.
\item {\em Reference} ($R$):  Suppose Alice wants to send an ensemble of 
states $\{p_i,\ket{\alpha_i}^{A}\}$ to Bob with average density
matrix $\rho^A = \sum_i p_i \alpha_i^A$.  We would like to give a lower
bound on the average fidelity of this transmission in terms only of
$\rho$.  Such a bound can be accomplished (in the CP/QP formalisms) by
extending $\rho^A$ to a pure state $\ket{\phi}^{AR}\supseteq \rho^A$ and
finding the fidelity of the resulting state with the original state
when $A$ is sent through
the channel and $R$ is left untouched\cite{BKN98}.  Here the reference
system $R$ is introduced to guarantee that transmitting system $A$
preserves its entanglement with an arbitrary external system.  Like
the environment, $R$ is always inaccessible and its properties are not
changed by local unitary rotations.  Indeed the only freedom in
choosing  $\ket{\phi}^{AR}$ is given by a local unitary rotation on $R$.
\item {\em Source} ($S$)  In most coding problems Alice can choose how
she encodes the message, but cannot choose
the message that she wants to communicate to Bob; it can be thought of
as externally given.  Taking this a step further, we can identify the
source of the message as another protagonist ($S$), who begins a
communication protocol by telling Alice which message to
send to Bob.  Introducing $S$ is useful in cases when the Source does
more than simply send a state to Alice; for example in distributed
compression, the Source distributes a bipartite state to
Alice and Bob.
\end{itemize}

To each party corresponds a class of 
quantum or classical systems which they control or have access
to at different times. The systems corresponding to Alice 
are labeled by $A$ (for example, $A'$, $A_1$, $X_A$, etc.),
while Bob's systems are labeled by $B$.
When two classical systems, such as $X_A$ and $X_B$, have
the same principal label it means that they are instances of
the same random variable. In our example, $X_A$ is Alice's copy
and $X_B$ is Bob's copy of the random variable $X$.




\medskip
We turn to some important examples of quantum states and operations.
Let $A$, $B$, $A'$, $X_A$ and $X_B$ be $d$-dimensional systems
with respective distinguished bases
$\{ \ket{x}^{A} \}, \{ \ket{x}^{B} \}$, etc.
The standard maximally entangled state on $AB$ is given by 
$$
\ket{\Phi_d}^{AB} = \frac{1}{\sqrt{d}} \sum_{x = 1}^d 
                                        \ket{x}^A \ket{x}^B.
$$
The decohered, ``classical'', version of this state is 
$$
\bar{\Phi}_d^{X_A X_B} = \frac{1}{{d}} \sum_{x = 1}^d 
\proj{x}^{X_A} \otimes \proj{x}^{X_B},
$$
which may be viewed as two maximally correlated random variables
taking values on the set $[d] = \{ 1, \dots , d \}$.  The local
restrictions of either of these states is the maximally mixed state
$\tau_d := \frac{1}{d} \one_d$. (We write $\tau$ to remind us that it
is also known as the \emph{tracial state}.)  Define the identity
quantum operation $\id_d: A' \rightarrow B$ by the isometry $\sum_x
\ket{x}^{B} \bra{x}^{A'}$ (Note that this requires fixed bases of $A'$
and $B$!).  It represents a perfect quantum channel between the
systems $A'$ and $B$.  Its classical counterpart is the completely
dephasing channel $\bar{\id}_d: X_{A'} \rightarrow X_B$, given in the
Kraus representation by $\{\ket{x}^{X_B}\bra{x}^{X_{A'}}\}_{x\in[d]}$. It
corresponds to a perfect classical channel in the sense that it
perfectly transmits random variables, as represented by density
operators diagonal in the preferred basis. The channel $\bar{\Delta}_d
: X_{A'} \rightarrow X_A X_B $ with Kraus representation
$\{\ket{x}^{X_B} \ket{x}^{X_A} \bra{x}^{X_{A'}}\}_{x\in[d]}$ is a
variation on  
$\bar{\id}_d$ in which Alice first makes a (classical) copy of the
data before sending it through the classical channel.  The two
channels are essentially interchangeable.  In \chap{ccc} we will
discuss the so-called \emph{coherent} channel $\Delta_d: A'
\rightarrow AB$, given by the isometry $\sum_x \ket{x}^{A}\ket{x}^{B}
\bra{x}^{A'}$ which is a coherent version of the noiseless classical
channel with feedback, $\bar{\Delta}_d$. Here and in the following,
``coherent'' is meant to say that the operation preserves coherent
quantum superpositions.

The maximally entangled state 
$\ket{\Phi_d}^{AB}$ and perfect quantum channel $\id_d: A' \rightarrow B$ 
are locally basis covariant:
$(U \otimes U^*) \ket{\Phi_d}^{AB} = \ket{\Phi_d}^{AB}$
and $U^\dagger \circ \id_d \circ U = \id_d$ for
any unitary $U$. On the other hand, $\bar{\Phi}_d$, $\bar{\id}_d$,
$\bar{\Delta}_d$ and $\Delta_d$ are all locally basis-dependent.

\subsection{Finite resources}
\label{sec:finite-res}

In this subsection we introduce ``finite'' or 
``non-asymptotic'' resources. They can be either static or dynamic,
but strictly speaking, thanks to the appending maps $\app{\rho}$,
we only need to consider dynamic ones.

\begin{defi}[Finite resources]
  \label{defi:finite-resource}
  A finite \emph{static resource} 
  is a quantum state $\rho^{AB}$.
  A  finite \emph{dynamic resource} is an ordered
  pair $(\cN : \omega)$, where the $\cN: A'B' \rightarrow AB$
  is an operation, with Alice's and Bob's input systems decomposed
  as $A' = A^\abs A^\rel$, $B' = B^\abs B^\rel$,
  and $\omega^{A^\rel B^\rel}$ is a so-called  \emph{test state}.
\end{defi}

The idea of the resource character of states and channels (static
and dynamic, resp.) ought be clear.
The only thing we need to explain is why we assume that $\cN$ 
comes with a test state (contained in the ``relative'' systems $A^\rel
B^\rel$): for finite resources it serves only
a syntactic purpose --- the operation ``expects'' an extension
of $\omega$ as input, which will play a role for the
definition of (valid) protocols below.  The test state may not
comprise the entire input to $\cN$, in which case the remainder of the
input comes from the systems $A^\abs B^\abs$.

If $A^\rel B^\rel = \emptyset$, we identify $(\cN : \omega)$ with the
\emph{proper} dynamic resource $\cN$.  Note that $\app{\rho}$ is
always a proper dynamic resource, as it has no inputs.

A resource $(\cN: \omega)$ is called \emph{pure} if 
$\cN$ is an isometry. It is called \emph{classical} 
if $\cN$ is a  $\{ c \rightarrow c \}$ entity
and $\omega$ is a $\{c \}$ entity (though they may
be expressed in the QQ formalism).

We define a distance measure between 
two dynamic resources $(\cN: \omega)$ and $(\cN' : \omega)$ with
the same test state as
$$
 \| (\cN' : \omega) - (\cN : \omega) \| := \|\cN' - \cN \|_{\omega}.
$$

A central notion is that of comparison between resources:
we take the operational view that one finite resource,
$(\cN_1 : \omega_1)$, is stronger than another, $(\cN_2 : \omega_2)$,
in symbols $(\cN_1 : \omega_1) \reduction (\cN_2 : \omega_2)$,
if it the former can be used to perfectly simulate the latter.  We
demand first that there exist local operations $\cE_A :
A_2'\rightarrow A_1'$ 
and $\cD_A : A_1 \rightarrow A_2$ for Alice, and
$\cE_B : B_2'\rightarrow B_1'$
and $\cD_B : B_1 \rightarrow B_2$ for Bob, such that
\be
 \cN_2 = (\cD_A \otimes \cD_B) \cN_1 (\cE_A \otimes \cE_B);
\ee
and second that the simulation be {\em valid}, meaning that
for every $\zeta_1 \supset \omega_1$,
\be\zeta_2 := (\cE_A \otimes \cE_B)\zeta_1 \supset \omega_2.\ee
When this occurs, we also say that $(\cN_2 : \omega_2)$ {\em reduces
to} $(\cN_1 : \omega_1)$.

Two important properties of this relation are that
\begin{enumerate}
\item It is {\em transitive}; i.e. if 
$(\cN_1:\omega_1)\reduction (\cN_2:\omega_2)$ and 
$(\cN_2:\omega_2)\reduction (\cN_3:\omega_3)$, then
$(\cN_1:\omega_1)\reduction (\cN_3:\omega_3)$.
\item It is continuous; i.e. if $(\cN_1:\omega_1)\reduction
(\cN_2:\omega_2)$, then for any channel $\cN_1'$ there
exists $\cN_2'$ such that $(\cN_1':\omega_1)\reduction
(\cN_2':\omega_2)$ and $\|\cN_2' - \cN_2\|_{\omega_2}
\leq \|\cN_1' - \cN_1\|_{\omega_1}$.
\end{enumerate}

The tensor product of states naturally extends
to dynamic resources:
$$
(\cN_1 : \omega_1) \otimes (\cN_2 : \omega_2) :=
(\cN_1 \otimes \cN_2 : \omega_1 \otimes \omega_2). 
$$
However, contrary to what one might expect
$(\cN_1 \otimes \cN_2 : \omega_1 \otimes \omega_2)
\reduction (\cN_1 : \omega_1)$ holds if and only if $\omega_1^{A_1B_1}$ can be
perfectly mapped with local operations to a state
$\omega^{A_1B_1A_2B_2}$ such that $\omega^{A_1B_1} = \omega_1$ and
$\omega^{A_2B_2} = \omega_2$.  Thus, we will almost always consider
resoures where the test state $\omega$ is a product state.
Nevertheless, nontrivial examples exist when the tensor product is
stronger than its component resources; for example, $(\cN:\omega)^{\ot
2}$ when $\omega$ is a classically correlated state.

A more severe limitation on these resource comparisons is that they do
not allow for small errors or inefficiencies.  Thus, most resources
are incomparable and most interesting coding theorems do not yield
useful exact resource inequalities.  We will address these issues in
the next section when we define asymptotic resources and asymptotic
resource inequalities.

\medskip
Resources as above are atomic primitives: ``having'' such a resource
means (given an input state) the ability to invoke the operation
(once). When formalizing the notion of ``having'' several resources,
e.g., the choice from different channels, it would be too restrictive
to model this by the tensor product, because it gives us just another
resource, which the parties have to use in a sort of ``block code''.
To allow for --- finite --- recursive depth (think, e.g., of feedback,
where future channel uses depend on the past ones) in using the resources,
we introduce the following:

\begin{defi}[Depth-$\ell$ resources]
  \label{defi:depth-l-res}
  A finite depth-$\ell$ {resource} is an unordered collection
  of, w.l.o.g., dynamic resources 
  $$(\cN :  \omega)^{\ell} := 
      \bigl( (\cN_1 : \omega_1), \ldots,
             (\cN_{\ell} : \omega_{\ell}) \bigr).
  $$
  Both static and dynamic resources are identified with depth-$1$ resources.
  To avoid notational confusion, for $\ell$ copies of the same dynamic
  resource, 
  $\bigl( (\cN :  \omega),  \dots, (\cN : \omega) \bigr)$, we reserve
  the notation $(\cN : \omega)^{\times \ell}$.
\end{defi}

  The definition of the distance measure
  naturally extends to the case of
  two depth-$\ell$ resources:
  $$
   \| (\cN' : \omega)^\ell - (\cN : \omega)^\ell \| 
       := \min_{\pi\in\cS_{\ell}, \omega_j = \omega_{\pi(j)}\forall j}
\sum_{j\in[\ell]} \|(\cN_j':\omega_j) - (\cN_{\pi(j)}:\omega_{\pi(j)})\|.
  $$
Here $\cS_\ell$ is the set of permutations on $\ell$ objects; we need
  to minimize over it to reflect the fact that we're free to use
  depth-$\ell$ resources in an arbitrary order.

  To \emph{combine} resources
  there is no good definition of a tensor product (which operations
  should we take the products of?), but we can take tensor {\em
  powers} of a resource:
  $$ \l((\cN : \omega)^{\ell}\r)^{\ot k} :=
\bigl((\cN_1 :\omega_1)^{\ot k}, \ldots, 
(\cN_\ell:\omega_\ell)^{\ot k}\bigr).$$

  The way we combine a depth-$\ell$ and a depth-$\ell'$ resource is by
concatenation: let
  $$
    (\cN : \omega)^{\ell} + (\cN' : \omega')^{\ell'}
      := \bigl( (\cN_1 :\omega_1 ), \ldots, (\cN_\ell    :\omega_\ell    ),
                (\cN_1':\omega_1'), \ldots, (\cN_{\ell'}':\omega_{\ell'}') \bigr).
  $$

\medskip
We now have to extend the concept of one resource simulating another
to depth-$\ell$; at the same time we will introduce the notions
of approximation that will become essential for the asymptotic
resources below.

\begin{defi}[Elementary protocols]
\label{def:protocol}
An \emph{elementary protocol} ${\bf P}$ 
takes a depth-$\ell$ finite resource
$(\cN : \omega)^\ell$ to a depth-$1$ finite resource.
Given $\cN_i: A_i' B_i' \rightarrow A_i B_i$ and
test states ${\omega_i}^{A_i^\rel B_i^\rel}$, $i = 1 \ldots \ell$,  
${\bf P}[(\cN : \omega )^\ell]$ is a
finite depth-$1$ resource $(\cP: \Omega^{A^\rel B^\rel})$,
with a quantum operation $\cP: A' B' \rightarrow AB$,
which is constructed as follows:\footnote{
  We use diverse notation to emphasize the role of the systems in question. 
  The primed systems, such as $A'_i$, are channel inputs.
  The systems with no superscript, such as $B_i$, are channel outputs. 
  Some systems are associated with Alice's sources 
  (e.g. $A_i^\rel$) and Bob's possible side information about those
  sources (e.g. $B_i^\rel$). Furthermore, there
  are auxiliary systems, such as $A_i^\aux$.}
\begin{enumerate}
\item  select a permutation $\pi$ of the integers $\{1,\ldots,\ell\}$;
  \item perform local operations
    $\cE_0: A'\rightarrow A_0 A_0^\aux$
    and $\cE'_0: B' \rightarrow B_0 B_0^\aux$;
  \item repeat, for $i = 1, \dots, \ell$,
    \begin{enumerate}
      \item $\!\!\!_i \,$  perform local isometries
        $\cE_i: A_{i-1} A_{i-1}^\aux \rightarrow A'_{i} A_{i}^\aux$ and 
        $\cE'_i: B_{i-1} B_{i-1}^\aux \rightarrow B'_{i} B_{i}^\aux$;
      \item $\!\!\!_i \,$  apply the operation $\cN_{\pi(i)}$, mapping
    $A_i'B_i'$ to $A_iB_i$;
    \end{enumerate}
  \item perform local operations
    $\cE_{\ell+1}:  A'_\ell A_\ell^\aux \rightarrow A$ and 
    $\cE'_{\ell+1}: B'_\ell B_\ell^\aux \rightarrow B$.
\end{enumerate}
We allow the arbitrary permutation of the resources $\pi$ so that
depth-$\ell$ resources do not have to be used in a fixed order.
Denote by
$\cP_i$
the operation of performing the protocol up to, but not including,
step 3.$\! (b)_i$.
Define $\hat{\cP}_i$ to be $\cP_i$ followed
by a restriction onto $A_i^\rel B_i^\rel$.
The protocol ${\bf P}$ is called \emph{$\eta$-valid}
on the input finite resource $(\cN : \omega )^l$ if
the conditions
$$
\| \hat{\cP}_i(\Omega) - \omega_{\pi(i)}^{A_i^\rel B_i^\rel} \|_1 \leq
\eta 
$$
are met for all $i$.
\end{defi}

\begin{defi}[Standard protocol]
Define the \emph{standard protocol} ${\bf S}$, which is a $0$-valid
elementary protocol 
on a depth-$\ell$ finite resource $(\cN : \omega)^\ell$, by 
$$
 {\bf S} [(\cN : \omega)^l]
    = ( \bigotimes_{i=1}^l \cN_i : \bigotimes_{i=1}^\ell \omega_i).
$$
\end{defi}
That is, this protocol takes a list of resources, and
flattens them into a depth-$1$ tensor product.
\par\medskip

Whenever $(\cN : \omega) \reduction  (\cN' : \omega')$,  there is a natural
protocol ${\bf R}$, 
which is $0$-valid on $(\cN : \omega)$,
 \emph{implementing} the reduction:
$$
{\bf R} [(\cN : \omega) ] =  (\cN' : \omega'),
$$
which we write as
$$
{\bf R}: (\cN : \omega) \stackrel{{}_{\scriptstyle *}}{\geq} (\cN' : \omega').
$$

For resources with depth $>1$, 
$(\cN:\omega)^\ell=((\cN_1:\omega_1),\ldots,(\cN_\ell:\omega_\ell))$ and
$(\cN':\omega')^{\ell'} = ((\cN'_1:\omega'_1),\ldots,
(\cN'_{\ell'}:\omega'_{\ell'}))$,
we say that $(\cN:\omega)\reduction (\cN':\omega')$ if there exists an
injective function $f:[\ell']\ra[\ell]$ such that for all
$i\in[\ell']$, $(\cN_{f(i)}:\omega_{f(i)}) \reduction
(\cN'_i:\omega'_i)$.  In other words, for each $(\cN_i':\omega_i')$
there is a unique $(\cN_j:\omega_j)$ that reduces to $(\cN_i':\omega_i')$.
Note that this implies $\ell \geq \ell'$.  Again
there is a natural $0$-valid protocol ${\bf R}$ implementing the reduction.

The next two lemmas help justify aspects of our definition of a
protocol---$\eta$-validity and the fact that outputs are
depth-1---that will later be crucial in showing how protocols may be
composed.

First we show why $\eta$-validity is important.
In general we want our distance measures for states to satisfy the triangle
inequality, and to be nonincreasing under quantum operations.  These
properties guarantee that the error of a sequence of quantum
operations is no more than the sum of errors of each individual
operation (cf. part 4 of \lem{props} as well as \cite{BV93}).
However, this assumes that we are using the same distance measure
throughout the protocol; when working with relative resources, a small
error with respect to one input state may be much larger for a
different input state.  Thus, for a protocol to map approximately
correct inputs to approximately correct outputs, we need the
additional assumption that the protocol is $\eta$-valid.

\begin{lemma}[Continuity]
\label{lemma:continuity}
If some elementary protocol 
${\bf P}$ 
is $\eta$-valid  on $ [(\cN : \omega)^\ell]$ and 
$$
\|(\cN :\omega)^\ell  - (\cA: \omega)^\ell \| \leq \epsilon,
$$
then
$$
\| {\bf P} [(\cN : \omega)^\ell] - 
{\bf P} [(\cA : \omega)^\ell] \| 
\leq l (\epsilon + 2 \sqrt{\eta})
$$
and ${\bf P} [(\cA : \omega)^\ell]$ is 
$(\eta + \ell (\epsilon + 2 \sqrt{\eta}))$-valid.

\end{lemma}
\begin{proof}
Let $(\cP: \Omega) = {\bf P} [(\cN : \omega)^\ell]$
and $(\cP': \Omega) = {\bf P} [(\cA : \omega)^\ell]$.
By definition \ref{def:protocol}, $\cP$ is of the form
$$
\cP =
  \cE_{\ell+1} \circ \cN_\ell \circ \cE_\ell \circ \dots \circ \cN_1 \circ \cE_1  
$$
and similarly for $\cP'$.
The $\eta$-validity condition reads, for all $i$,
$$
\| \hat{\cP}_i(\Omega) - \omega_i \|_1 \leq \eta.
$$
By part 3 of \lem{props},
$$
\| \cP - \cP' \|_\Omega \leq \sum_i \|\cA_i - \cN_i\|_{\cP_i(\Omega)}.
$$
By part 1 of \lem{props},
$$
\|\cA_i - \cN_i\|_{\cP_i(\Omega)} \leq \|\cA_i - \cN_i
\|_{\hat{\cP}_i(\Omega)}.
$$
By part 2 of \lem{props} and 
$\eta$-validity
$$
\| \cA_i - \cN_i \|_{\hat{\cP}_i(\Omega)}
\leq \| \cA_i - \cN_i \|_{\omega_i} + 2 \sqrt{\eta}
$$
Hence
$$
\| \cP - \cP' \|_\Omega \leq \ell (\epsilon + 2 \sqrt{\eta}),
$$
which is one of the statements of the lemma.
To estimate the validity of  ${\bf P}$  on 
$[(\cA : \omega)^\ell]$, note that one obtains in the same way as above, 
for all $i$,
$$
\| \hat{\cP}_i - \hat{\cP}'_i \|_\Omega \leq \ell (\epsilon + 2 \sqrt{\eta}).
$$
Combining this with the $\eta$-validity condition via the
triangle inequality  finally gives
$$
\| \hat{\cP}'_i (\Omega) - \omega_i \|_1
      \leq \eta + \ell (\epsilon + 2 \sqrt{\eta}),
$$
concluding the proof.
\end{proof}

We note that we do not have a concept of what it means to turn a
depth-$\ell$ resource into a depth-$\ell'$ resource; instead, our
basic concept of simulation produces a depth-$1$ resource.
I.e., we can formulate what it means that a depth-$\ell$ resource
simulates the standard protocol of a depth-$\ell'$ resource.

The following lemma states that the standard
protocol is basically sufficient to generate any other,
under some i.i.d.-like assumptions.

\begin{figure}
\centerline{ {\scalebox{0.50}{\includegraphics{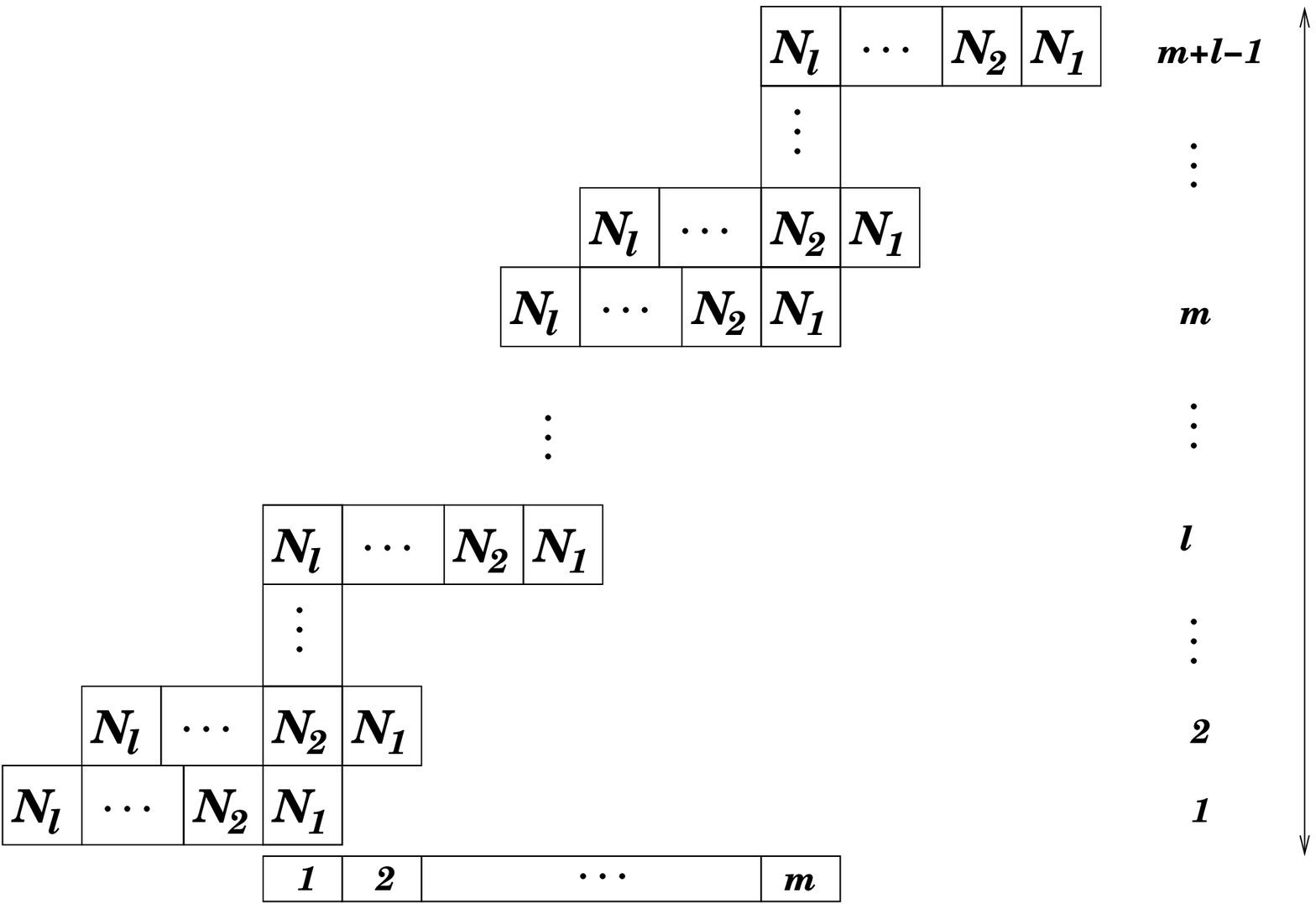}}}}
\caption{The sliding protocol.  We would like to simulate $\bP$, which
uses $\cN_1, \ldots, \cN_\ell$ consecutively, but we are only
given $\cN_1 \ot \ldots \cN_\ell$.  The horizontal blocks represent
uses of $\cN_1 \ot \ldots \ot \cN_\ell$ and stacking them vertically
indicates how we perform them consecutively with the output of one
block becoming the input of the block above it (i.e. time flows from
the bottom to the top).  Thus $m+l-1$ consecutive uses of $\cN_1 \ot
\ldots \ot \cN_\ell$ can simulate $m$ copies of $\bP$.}
\label{fig:slide}
\end{figure}

\begin{lemma}[Sliding]
\label{lemma:sliding}
If for some depth-$\ell$ finite resource
$(\cN : \omega)^\ell = ((\cN_1 : \omega_1), \dots, (\cN_\ell : \omega_\ell))$
and quantum operation $\cC$,
\be
\| (\cC : \bigotimes_i \omega_i  )  -  
     {\bf S}[(\cN : \omega)^\ell] \| \leq \epsilon,
\label{cab''}
\ee
then for any integer $m \geq 1$ and for any  
$\eta$- valid protocol ${\bf P}$ on $(\cN :  \omega)^\ell$, 
there exists a 
$( (m+\ell-1) (\epsilon + 2 \sqrt{\eta}) + \eta)$-valid protocol 
${\bf P}'$ on 
$(\cC:  \bigotimes_i \omega_i)^{\times (m + \ell - 1)}$,
such that
$$ 
\|{\bf P}'[ (\cC:  \bigotimes_i \omega_i)^{\times (m + \ell - 1)}] - 
({\bf P}[ 
(\cN :  \omega)^\ell])^{\otimes {m}} \|
\leq (m+\ell-1) (\epsilon + 2 \sqrt{\eta}).
$$
\end{lemma}
\begin{proof}
Denoting by ${\bf P}'$ the sliding protocol (see \fig{slide})
it is clear that 
$$
{\bf P}'[({\bf S}[ (\cN: \omega)^\ell])^{\times (m + \ell - 1)}] = 
({\bf P}[(\cN :  \omega)^\ell])^{\otimes {m}}.
$$
The result follows from \lem{continuity}.
\end{proof}

The sliding protocol shows how working with depth-1 resources is not
overly restrictive.  Another difficulty with resources is that
relative resources are only guaranteed to work properly when given the
right sort of input state.  Here we show that using shared randomness,
some of the standard relative resources can be ``absolutized,''
removing the restriction to a particular input state.
\begin{lemma}
\label{lemma:absolutizy}
For a operation $\cN: A' \rightarrow AB$ which is either the
perfect quantum channel
$\id_d$, the coherent channel 
$\Delta_d$ or the perfect classical channel $\bar{\id}_d$, 
there exists a $0$-valid protocol ${\bf P}$ such that
$$
{\bf P}[\bar{\Phi}^{X_A  X_B}, (\cN : \tau^{A'})] = 
\cN \otimes \app{\bar{\Phi}^{X_A  X_B}},
$$
where $\dim X_A =  (\dim A')^2$, and $\tau^{A'}$ is
the maximally mixed state on $A'$.
\end{lemma}
\begin{proof}
Consider first the case of $\cN$ being either $\id_d$ or 
the coherent channel $\Delta_d$.
The main observation is that there exist a set of 
unitary operations $\{U_x\}_{x \in [d^2]}$ (the
generalized Pauli, or discrete Weyl, operators) such that,
for any state $\rho$ living on a $d$-dimensional Hilbert space, 
\be
d^{-2} \sum_x U_x \rho U^\dagger_x = \tau_d, 
\label{randomo}
\ee
with $\tau_d$ being the maximally mixed state on that space.

Let Alice and Bob share the common randomness state
$$
\bar{\Phi}^{X_A  X_B} =  d^{-2} \sum_{x = 1}^{d^2} \proj{x}^{X_A} 
\otimes \proj{x}^{X_B},
$$
where $d:= \dim A'$.
Consider an arbitrary input state $\ket{\phi}^{R A'}$,
possibly entangled between Alice and a reference system $R$.
Alice performs the conditional unitary 
$\sum_x \proj{x}^{X_A} \otimes U_x^{A'}$, yielding a state
whose restriction to $A'$ is maximally mixed.
She then applies the operation $\cN$ (this is $0$-valid!),
which gives the state
$$
 d^{-2} \sum_{x = 1}^{d^2} \proj{x}^{X_A} 
\otimes \proj{x}^{X_B} \otimes (  \cN \circ U_x^{A'}) {\phi}^{R A'}.  
$$
In the case of the $\id_d$ channel, Bob simply applies the
conditional unitary $\sum_x \proj{x}^{X_B} \otimes (U^{-1}_x)^{B}$.
In the case of the $\Delta_d$ channel Alice must also perform
$$
\sum_x \proj{x}^{X_A} \otimes (U^{-1}_x)^{A}.
$$ 
Either way, the final state is 
$$
\bar{\Phi}^{X_A  X_B} \otimes \cN ({\phi}^{R A'}),
$$
as advertised.

The case of the perfect classical channel $\bar{\id}_d$ is a
classical analogue of the above. The observation here is
that there exists a set of $d$ unitaries 
$\{U_x\}_{x \in [d]}$ (all the cyclic permutations of the basis vectors),
each member of which commutes with $\Delta$,
such that (\ref{randomo}) holds 
for any state $\rho$ diagonal in the preferred basis.
Now Alice first applies a local $\Delta$ (diagonalizing the input),
before proceeding as above.
This concludes the proof.
\end{proof}

Observe that in the above lemma, the final output of $\cN$ is
uncorrelated with the shared randomness that is used.  In the QQ
formalism, this is immediately apparent from the tensor product
between $\cN$ and $\app{\bar{\Phi}^{X_A  X_B}}$.  Thus we say that the
shared randomness is (incoherently) {\em decoupled} from the rest of
the protocol. 

Now consider the case when $\cN=\bar{\id}_d$ and use the QP formalism,
so $\cN$ is a map from $A$ to $BE$.   If we condition on a particular
message sent by Alice, then the randomness is no longer
decoupled from the composite $BE$ system.  This is the problem of
reusing the key in a one-time pad: if the message is not uniformly
random, then information about the key leaks to Eve.

On the other hand, if $\cN$ is $\Delta_d$ or $\id_d$ then the shared
randomness is decoupled even from the environment.  This stronger form
of decoupling is called {\em coherent decoupling}.  Below we give
formal definitions of these notions of decoupling.\footnote{The notion
of an ``oblivious'' protocol for remotely preparing quantum states is
similar to coherent decoupling, but applies instead to quantum
messages\cite{LS02}.}

\begin{defi}[Incoherent decoupling]
\label{def-incoh-decouple}
Consider a protocol ${\bf P}$ on 
$((\bar{\cN} : \bar{\omega})^\ell,({\cN} : {\omega})^{\ell'})$,
where $(\bar{\cN} : \bar{\omega})^\ell$ is classical.  
Recall that in the QQ formalism 
classical systems are unchanged under
the copying operation $\bar{\Delta}$.  This means we can consider an
equivalent protocol in which the systems associated with the
classical resource $(\bar{\cN} : \bar{\omega})^\ell$ are copied into a
composite classical system $Z$, which
includes all the copies of all the random variables involved.
Let ${\bf P'}$ be the modified version of ${\bf P}$ which retains
$Z$ in the final state. Now 
$\cP' := {\bf P'}[((\bar{\cN} : \bar{\omega})^\ell,({\cN} : {\omega})^{\ell'}) ] 
\ext \cP$
takes a particular extension $\Upsilon^{R A' A^*B^*} \ext 
\Omega^{A^* B^*}$
to some state $\sigma^{Z R A B  A^* B^*}$.

We say that the classical resource 
$(\bar{\cN} : \bar{\omega})^\ell$
is $\epsilon-$\emph{incoherently decoupled} 
(or just $\epsilon-$\emph{decoupled}) with respect to 
the protocol ${\bf P}$ on 
$((\bar{\cN} : \bar{\omega})^\ell,({\cN} : {\omega})^{\ell'})$ if 
for any $\Upsilon^{R A' A^* B^*}$ the state 
$\sigma^{Z R A B  A^* B^*}$ satisfies
\be
\| \sigma^{Z R A B  A^* B^*} - \sigma^{Z} \otimes 
\sigma^{R A B  A^* B^*} \|_1 \leq \epsilon.
\label{otto}
\ee
\end{defi}

We describe separately how classical resources used in the input and
the output of a protocol may be coherently decoupled.
\begin{defi}[Coherent decoupling of input resources]
\label{def:coh-decoupling-input}
Again, consider a protocol ${\bf P}$ on 
$((\bar{\cN} : \bar{\omega})^\ell,({\cN} : {\omega})^{\ell'})$,
where $(\bar{\cN} : \bar{\omega})^\ell$ is classical.  
Now we adopt a QP view in which all non-classical states are
purified and all channels are isometrically extended.
Again, we define a classical system $Z$ which contains copies of all
the classical variables associated with the resource $({\cN} :
{\omega})^{\ell'}$. 
The final state of the protocol is then some 
$\sigma^{Z R A B  A^* B^* E}$. 
We say that the classical resource $(\bar{\cN} : \bar{\omega})^\ell$
is $\epsilon-$\emph{coherently decoupled} 
with respect to the protocol ${\bf P}$ on 
$((\bar{\cN} : \bar{\omega})^\ell,({\cN} : {\omega})^{\ell'})$
if for any $\Upsilon^{R A' A^* B^*}$ the final state 
$\sigma^{Z R  A B  A^* B^* E}$ satisfies
$$
\| \sigma^{Z R  A B  A^* B^* E} - \sigma^{Z} \otimes 
\sigma^{R  A B  A^* B^* E} \|_1
 \leq \epsilon.
$$
\end{defi}

\begin{defi}[Coherent decoupling of output resources]
\label{def:coh-decoupling-output}
Remaining within the QP formalism, let ${\bf P}$ be a protocol mapping
$({\cN} : {\omega})^{\ell}$ to $(\bar{\cP}_1\ot\cP_2 : 
\bar{\Omega}_1^{A_1} \ot \Omega_2^{A_2B_2})$; i.e. the tensor
product of a classical resource
$(\bar{\cP}_1 : \bar{\Omega}_1^{A_1})$ and a quantum resource
$(\cP_2 : \Omega_2^{A_2B_2})$.  Define $Z$ to consist of copies of
$A_1B_1$ together with all the other classical resources associated
with $\bar{\cP}_1$, such as outputs (if different from $A_1$) and
inputs other than $A_1$ (if any).

We now say that the classical resource  
$(\cP_1 : \Omega_1 )$ is $\epsilon-$\emph{coherently decoupled} 
with respect to the protocol ${\bf P}$ on 
$({\cN} : {\omega})^{\ell}$
if 
$$
\| \sigma^{Z Q} - \sigma^{Z} \otimes \sigma^{ Q} \|_1
 \leq \epsilon,
$$
where now $Q$ comprises all the quantum systems involved (including
environments and reference systems).
\end{defi}

We will give some applications of decoupling in
\sect{general-inequalities}, but its primary utility will be seen in
\chaps{ccc}{family}.

One simple example of decoupling is when a protocol involves several
pure resources (i.e. isometries) and one noiseless classical resource.
In this case, decoupling the classical resource is rather easy, since
pure resources don't involve the environment.  However, it is possible
that the classical communication is correlated with the ancilla system $Q$
that Alice and Bob are left with.  If $Q$ is merely discarded, then
the cbits will be incoherently decoupled.  To prove that coherent
decoupling is in fact possible, we will need to carefully account for
the ancillas produced by the classical communication.  This will be 
accomplished in \sect{ccc-proofs}, where we prove that classical
messages sent through isometric channels can always be coherently
decoupled.

\subsection{Asymptotic resources}

\begin{defi}[Asymptotic resources]\label{def:asy-resource}
An  \emph{asymptotic resource} $\alpha$ is defined by 
a sequence of finite depth-$\ell$ resources $(\alpha_n)_{n = 1}^\infty$, 
where $\alpha_n$ is w.l.o.g. of the form
$\alpha_n = (\cN_n : \omega_n )^\ell := 
((\cN_{n,1} : \omega_{n,1}), (\cN_{n,2} : \omega_{n,2}), \ldots,
 (\cN_{n,\ell} : \omega_{n,\ell})) $, such that
\begin{itemize}
\item
\be \alpha_n \reduction \alpha_{n-1}
\quad\text{for all} n;
\label{eq:asy-monotone}\ee
\item for any $\delta>0$, any integer $k$ 
and all sufficiently large $n$,
\be
\alpha_{\lfloor  n(1 + \delta) \rfloor} \reduction
(\alpha_{\lfloor n/k \rfloor})^{\otimes k} \reduction
\alpha_{ \lfloor  n(1 - \delta) \rfloor}.
\label{eq:asy-quasi-iid}\ee
We sometimes refer to this as the requirement that a resource be
``quasi-i.i.d.''
\end{itemize}
\end{defi}
Denote the set of asymptotic resources by ${\cR}$.

Given two resources $\alpha = (\alpha_n)_{n = 1}^\infty$ 
and $\beta = (\beta_n)_{n = 1}^\infty$, if 
$\alpha_n \reduction \beta_n$ for all sufficiently large $n$,
then we write $\alpha \reduction \beta$. 
We shall use the following convention:
if $\beta = (\cN_n)_n$, where all $\cN_n$ are proper dynamic resources
and $\gamma = (\omega_n)_n$, where all $\omega_n$ are proper static
resources, 
then $(\beta: \gamma)   := (\cN_n : {\omega_n})_n$.   Note that
typically $\omega_n$ is product state, so the resource $\gamma$
reduces to the null resource $\emptyset$; however this is no problem
as long as we are interested in $\gamma$ only as a test state for $\beta$.

Our next goal is to define what it means to
simulate one (asymptotic) resource by another.

\begin{defi}[Asymptotic resource inequalities]
\label{def:asy-ineq}
A resource \emph{inequality} $\alpha \geq \beta$ holds
between two resources $\alpha = (\alpha_n)_n$
and $\beta = (\beta_n)_n$
if for any $\delta > 0$ there exists an integer $k$
such that for any $\epsilon>0$ there exists $N$ such that for all
$n\geq N$
there exists 
an $\epsilon$-valid protocol ${\bf P}^{(n)}$ on 
$(\alpha_{\lfloor n/k \rfloor})^{\times k}$ (i.e. $k$ sequential uses
of $\alpha_{\lfloor n/k \rfloor}$)
for which
$$
\| {\bf P}^{(n)} [ (\alpha_{\lfloor n/k \rfloor})^{\times k}]
   - {\bf S}[\beta_{\lfloor (1- \delta)n \rfloor}] \|  \leq \epsilon.
$$

$\alpha$ is called the input resource, $\beta$ is
called the output resource, $\delta$ is the inefficiency (or sometimes
the fractional inefficiency) and $\epsilon$ (which bounds both the
validity and the error) is called the accuracy (or sometimes just the
error).
\end{defi}

At first glance it may seem that we are demanding rather little from
asymptotic resource inequalities: we allow the depth of the input
resource to grow arbitrarily, while requiring only a depth-1 output.
However, later in this section we will use tools like the sliding
lemma to show that this definition is nevertheless strong enough to
allow the sort of protocol manipulations we would like.  

Also, for resources that consist entirely of states
one-way channels, it is never necessary to use protocols with depth
$>1$.  Thus, we state here a ``flattening'' lemma that will later be
useful in proving converses; i.e. statements about when certain
resource inequalities are impossible.

\begin{lemma}[Flattening]\label{lemma:flattening}
Suppose $\alpha\geq\beta$ and $\alpha$ is a ``one-way'' resource,
meaning that it consists
entirely of static resources ($\app{\rho}$) and dynamic resources
which leave nothing on Alice's side (e.g. $\cN^{A'\ra BE}$).  Then for
any $\epsilon,\delta>0$ for sufficiently large $n$ there is an
$\epsilon$-valid protocol ${\bf P}^{(n)}$ on $\alpha_n$ such that
$$ \| {\bf P}^{(n)}[\alpha_n] - {\bf
S}[\beta_{\lfloor(1-\delta)n\rfloor}] \| \leq \epsilon.$$
\end{lemma}
\begin{proof}
To prove the lemma, it will suffice to convert a protocol on
$(\alpha_{\lfloor n/k\rfloor})^{\times k}$ to a protocol on
$(\alpha_{\lfloor n/k\rfloor})^{\otimes k}$. Then we can use the fact
that $\alpha_{\lfloor n(1+\delta)\rfloor} \reduction (\alpha_{\lfloor
n/k\rfloor})^{\otimes k}$ and the lemma follows from a suitable
redefinition of $n$ and $\delta$.

Since $\alpha$ is a one-way resource, any protocol that uses it can be
assumed to be of the following form: first Alice applies all of the
appending maps, then she does all of her local operations, then she
applies all of the dynamic resources, and finally Bob does his
decoding operations.  The one-way nature of the protocol means that
Bob can wait until all of Alice's operations are finished before he
starts decoding.  It also means that Alice can apply the dynamic
resources last, since they have no outputs on her side, so none of her
other operations can depend on them.  Finally, the appending maps can
be pushed to the beginning because they have no inputs.
Thus $(\alpha_{\lfloor n/k\rfloor})^{\times k}$ can be simulated using
$(\alpha_{\lfloor n/k\rfloor})^{\otimes k}$, completing the proof.
\end{proof}

\begin{defi}[i.i.d. resources]
A  resource $\alpha$ 
is called \emph{independent and identically distributed  (i.i.d.)} if 
$\alpha_n = (\cN^{\otimes n}: \omega^{\otimes n})$ for some state
$\omega$ and operation $\cN$.
We use shorthand notation $\alpha =  \<\cN: \omega\>$.
\end{defi}

We shall use the following notation for \emph{unit} asymptotic resources:
\begin{itemize}
\item ebit $[q \, q]:= \<\Phi_2 \>$
\item rbit $[c \, c] := \<\bar{\Phi}_2 \>$
\item qubit $[q \rightarrow q] := \< \id_2 \>$
\item cbit  $[c \rightarrow c] := \<\bar{\id}_2 \>$
\item cobit $\cof := \< \Delta_2 \>$ (cobits will be explained in \chap{ccc})
\end{itemize}
In this thesis, we tend to use symbols for asymptotic resource
inequalities (e.g. ``$\<\cN\>\geq C\ctc$'') and words for finite protocols
(e.g. ``$\cN^{\ot n}$ can be used to send $\geq n(C-\delta_n)$ cbits
with error $\leq \epsilon_n$'').  However, there is no formal reason
that they cannot be used interchangeably.

We also can define versions of the dynamic resources with respect to
the standard ``reference'' state $\tau_2^{A'} = \one_2^{A'}/2$: a
qubit in the maximally mixed state.  These are denoted as follows:
\begin{itemize}
\item $[q \rightarrow q : \tau] := \< \id_2 : \tau_2 \>$
\item $[c \rightarrow c : \tau] := \<\bar{\id}_2  : \tau_2\>$
\item $\coftau := \<\Delta_2  : \tau_2\>$
\end{itemize}

\begin{defi}[Addition]
The addition operation $+: \cR \times \cR \rightarrow \cR$ is defined
for $\alpha = (\alpha_n)_n$, 
$\alpha_n =  
((\cN_{n,1}  : \omega_{n,1}),  \dots, (\cN_{n,l}  : \omega_{n,l}) ) $,
and $\beta = (\beta_n)_n$, 
$\beta_n = 
((\cN'_{n,1}  : \omega'_{n,1}),  \dots, (\cN'_{n,l'}  : \omega'_{n,l'}) )$,
 as
$\alpha + \beta = (\gamma_n)_n$ with 
$$
\gamma_n = (\alpha_n, \beta_n):= 
((\cN_{n,1}  : \omega_{n,1}),  \dots, (\cN_{n,l}  : \omega_{n,l}),
(\cN'_{n,1}  : \omega'_{n,1}),  \dots, (\cN'_{n,l'}  : \omega'_{n,l'}) ).
$$
\end{defi}

Closure is trivially verified. It is also easy to see
that the operation $+$ is associative and commutative. 
Namely,
\begin{enumerate}
\item ${\alpha}  + {\beta} = {\beta} + {\alpha}$ 
\item $({\alpha}  + {\beta})  + \gamma =
{\alpha}  + ({\beta}  + \gamma)$ 
\end{enumerate}

\begin{defi}[Multiplication]
The multiplication operation 
$\cdot: \cR \times \mathbb{R}_+ \rightarrow \cR$ 
is defined for any positive real number $z$ and resource 
$\alpha = (\alpha_n)_n$
by $z \alpha  = (\alpha_{\lfloor z n \rfloor})_n$.
\end{defi}

Of course, we need to verify that $\cR$ is indeed closed under
multiplication. Define
$\beta := z \alpha$, so that $\beta_n = \alpha_{\lfloor zn \rfloor}$.
We know, for all sufficiently large $n$, that
$$
\alpha_{\lfloor  \lfloor zn \rfloor(1 + \delta) \rfloor}   \reduction
  (\alpha_{\lfloor \lfloor zn \rfloor/k \rfloor})^{\otimes k}
 \reduction  \alpha_{\lfloor  \lfloor zn \rfloor(1 - \delta) \rfloor}.
$$ 
We need to prove
$$
\alpha_{ \lfloor z\lfloor  n(1 + \delta') \rfloor\rfloor}   \reduction 
 (\alpha_{ \lfloor z \lfloor n/k \rfloor\rfloor})^{\otimes k}
 \reduction  \alpha_{ \lfloor z \lfloor  n(1 - \delta') \rfloor\rfloor},
$$ 
which is true for the right $\delta'$.


\medskip
\begin{defi}[Asymptotic decoupling]
\label{def:asy-decouple}
Consider a resource inequality of the form $\alpha + \gamma \geq
\beta$, or $\alpha \geq \gamma $, where $\gamma$ is a classical
resource, and $\alpha$ and $\beta$ are quantum resources.  In either
case, if in the definition above, for each sufficiently large $n$
we also have that $\gamma_n$ is $\epsilon-$(coherently) decoupled with
respect to ${\bf P}^{(n)}$, then we say that $\gamma$ is (coherently)
decoupled in the resource inequality.
\end{defi}

The central purpose of our resource formalism is contained in the
following ``composability'' theorem, which states that resource
inequalities can be combined via concatenation and addition.  In other
words, the source of a resource (like cbits) doesn't matter; whether
they were obtained via a quantum channel or a carrier pigeon, they can
be used equally well in any protocol that takes cbits as an input.  A
well-known example of composability in classical information theory is
Shannon's joint source-channel coding theorem which states that a
channel with capacity $\geq C$ can transmit any source with entropy
rate $\leq C$; the coding theorem is proved trivially by composing
noiseless source coding and noisy channel coding.

\begin{theorem}[Composability]
\label{thm:composability}
For resources in $\cR$:
\begin{enumerate}
\item   if $\alpha \geq  \beta$ and $\beta \geq \gamma$ then
$\alpha \geq \gamma$
\item  if $\alpha \geq \beta$ and $  \gamma \geq \varepsilon$ then
$\alpha +  \gamma \geq  \beta + \varepsilon $
\item  if $\alpha \geq  \beta$ then $z \alpha \geq z \beta$
\end{enumerate}
\end{theorem}

\begin{proof}
\begin{enumerate}\item  Fix $\delta > 0$.
Then there exist $k, k'$, 
such that for any $\epsilon$ and sufficiently large $n$ 
\be
\| {\bf P}_1 [(\alpha_{\lfloor n(1 -  \delta)/(mkk') \rfloor})^{\times k}]
- {\bf S}[\beta_{\lfloor n(1 - 2\delta)/(mk') \rfloor}] \| \leq \epsilon,
\label{eq:compose-a1}
\ee
\be
\| {\bf P}_2 [(\beta_{\lfloor n(1 -  2\delta)/(mk') \rfloor})^{\times k'}]
- {\bf S}[\gamma_{\lfloor n(1 - 3 \delta)/m \rfloor}]
 \|  \leq \epsilon, 
\label{eq:compose-a2}
\ee
\be
\gamma_{\lfloor n(1- 3 \delta)/m \rfloor }^{\otimes m}  
\reduction \gamma_{\lfloor n(1- 4 \delta)\rfloor}, 
\label{eq:compose-a5}
\ee
with $m \geq k' l/\delta$, where $l$ is the depth of $\beta$, and 
where ${\bf P}_1$ and ${\bf P}_2$ are both $\epsilon$-protocols. 
Equation (\ref{eq:compose-a5})
implies the existence of a reduction protocol 
$$
\bR_1:  {\bf S}[\gamma_{\lfloor n(1 - 3 \delta)/m \rfloor}]^{\otimes m}
\reduction 
\bS[\gamma_{\lfloor n(1 - 4 \delta) \rfloor}].
$$
By \eq{compose-a2}
\be
\| {\bf P}_2 [(\beta_{\lfloor n(1 - 2 \delta)/(mk') \rfloor})^{\times k'}]
^{\otimes m}
- {\bf S}[\gamma_{\lfloor n(1 - 3 \delta)/m \rfloor}]^{\otimes m}
 \| \leq m \epsilon.
\label{eq:compose-pedva}
\ee
Define
$
\iota = {\bf P}_1 [(\alpha_{\lfloor n(1 - \delta)/(mkk') \rfloor})^{\times k}]
^{\otimes k'},
$
which, by \eq{compose-a1}, satisfies 
\be
\| \iota - 
{\bf S}[(\beta_{\lfloor n(1 - 2 \delta)/(mk') \rfloor})^{\times k'}] \|
 \leq k' \epsilon.
\label{eq:compose-a1a}
\ee
Let 
$\epsilon' = ( m + k' l- 1) ( k' \epsilon + 2 \sqrt{\epsilon}) + \epsilon$.
We shall exhibit an $\epsilon'$-valid protocol ${\bf P}_3$
such that 
\be
\| {\bf P}_3 [\alpha_n]
- {\bf S}[\gamma_{\lfloor n(1 - 4 \delta) \rfloor}] 
\|  
\leq  \epsilon'  + m\epsilon.
\label{eq:compose-p3-claim}\ee
\sloppypar{By \eq{asy-quasi-iid}, there is a reduction $\bR'$ from
the initial finite resource  
$\alpha_n$ to 
$(\alpha_{\lfloor n(1 - \delta)/(mkk') \rfloor})^{\times
\lfloor mkk'(1 + \delta) \rfloor}$, which in turn 
suffices to implement $\iota^{\times m + k' l - 1}$.
By the Sliding Lemma (\ref{lemma:sliding}) and \eq{compose-a1a}, there 
exists some $\epsilon'$-valid protocol ${\bf P}'$
such that }
$$
\|   {\bf P}'[\iota^{\times  m + k' l - 1}] - 
{\bf P}_2 [(\beta_{\lfloor n(1 -  \delta)/(mk') \rfloor})^{\times k'}]^{\otimes m} 
\| \leq \epsilon'.
$$
Now we claim that the protocol $\bP_3 := \bR \circ \bP' \circ
\bP_1^{\ot k} \ot \bR'$ satisfies \eq{compose-p3-claim}.  Indeed
$\bP_3[\alpha_n] = {\bf R} \circ {\bf P}'[\iota^{\times m + k' l - 1}]$
maps $\alpha$ to  $\gamma$ with
inefficiency $\delta' \leq 4\delta + 1/m \leq 5\delta$, depth $\leq
mkk'(1+\delta) \leq k(k')^2l(1 + 1/\delta)$ (where $k,k'$ depend only
on $\delta$) and error $\epsilon'' \leq \epsilon' + m\epsilon$.  Since
$\delta'\ra 0$ as depth increases and $\epsilon''\ra 0$ as
$n\ra\infty$, this satisfies our definition of an asymptotic protocol.

\item We begin with the standard quantifiers from our definition of a
resource inequality: $\forall \delta>0, \exists
k,k', \forall \epsilon>0, \exists N, \forall n\geq N$
\be
\| {\bf P}_1 [(\alpha_{\lfloor n/(kk') \rfloor})^{\times k}]
- {\bf S}[\beta_{ \lfloor n(1 -  \delta) / k' \rfloor }] \|  \leq \epsilon,
\ee
\be
\| {\bf P}_2 [(\gamma_{\lfloor n/(kk') \rfloor})^{\times k'}]
- {\bf S}[\varepsilon_{ \lfloor n(1 -  \delta) / k \rfloor }] \|
  \leq \epsilon,
\ee
\be
 {\bf R}_1:  (\beta_{\lfloor n(1 - \delta)/k' \rfloor})^{\otimes k'} 
\reduction  
 \beta_{\lfloor  n (1- 2 \delta) \rfloor },
\ee
\be
 {\bf R}_2 : (\varepsilon_{\lfloor n(1 - \delta)/k \rfloor})^{\otimes k} 
\reduction  
 \varepsilon_{\lfloor  n (1- 2 \delta) \rfloor },
\ee
where ${\bf P}_1$ and ${\bf P}_2$ are both $\epsilon$-protocols. 
Hence the depth-$(k + k')$ ${(k + k')\epsilon}$-protocol ${\bf P}_3$  
given by
$$
{\bf R}_1 \circ
{\bf P}_1 [(\alpha_{\lfloor n/(kk') \rfloor})^{\times k}]^{\otimes k'} 
\otimes
{\bf R}_2 \circ {\bf P}_2 [(\gamma_{\lfloor n/(kk') \rfloor})^{\times k'}]^{\otimes k},
 $$
satisfies
\be
\| {\bf P}_3 [((\alpha+\gamma)_{\lfloor n/(kk') \rfloor})^{\times kk'}]
- {\bf S}[(\beta+\varepsilon)_{ \lfloor n(1 - 2 \delta) \rfloor }] \|
 \leq (k + k') \epsilon.
\ee

\item The proof is trivial.
\end{enumerate}
\end{proof}

It is worth noting that our definitions of resources and resource
inequalities were carefully chosen with the above theorem in mind; as
a result the proof exposes most of the important features of our
definitions.  (It is a useful exercise to try changing aspects of our
definitions to see where the above proof breaks down.)  By contrast,
the remainder of this section will establish a number of details about
the resource formalism that mostly depend only on
\eqs{asy-monotone}{asy-quasi-iid} and not so much on the details of
how we construct protocols and resource inequalities.

\begin{defi}[Equivalent resources]
Define an equivalence between resources $\alpha \equiv \beta$
iff $\alpha \geq  \beta$ and $\beta \geq \alpha$.
\end{defi}

\begin{eg}
It is easy to see that $R [q \, q] \equiv ( \Phi_{D'_n} )_n$ with 
$D'_n = {\lfloor 2^{n R} \rfloor}$. 
\end{eg}

\begin{lemma}\label{lemma:mult-RI}
For resources in  ${\cR}$:
\begin{enumerate}
\item $(zw) {\alpha} \equiv z(w {\alpha})$ 
\item $z({\alpha}  + {\beta}) = 
z {\alpha}  + z {\beta}$
\item $(z + w) {\alpha} \equiv  z {\alpha} + w {\alpha}$
\end{enumerate}
\end{lemma}
\begin{proof}
\begin{enumerate}
\item The $\geq$ is
trivial, since $\lfloor zw n \rfloor  \geq 
\lfloor z \lfloor w n \rfloor \rfloor$.
The $\leq$ follows from
$\lfloor zw n \rfloor \leq zwn 
\leq \lfloor z \lfloor w n \rfloor \rfloor + z + 1$.

\item Immediate from the definitions.

\item
Let  $k = \lfloor zm \rfloor$ and $k' = \lfloor wm \rfloor$, where $m$
is a parameter we will choose later.

For any $\delta$ and sufficiently large $n$ (depending on $\delta$ and
$m$),
$$
\alpha_{\lfloor z n(1 + 2\delta) \rfloor} \reduction 
(\alpha_{\lfloor\lfloor z n(1 + \delta)  \rfloor /k  \rfloor })^{\otimes k},
$$
$$
\alpha_{\lfloor w n(1 + 2 \delta) \rfloor} \reduction 
(\alpha_{\lfloor\lfloor w n(1 + \delta) \rfloor /k'  \rfloor })^{\otimes k'},
$$
$$
\alpha_{\lfloor (z+w) n(1 +2 \delta) \rfloor} \reduction 
(\alpha_{\lfloor\lfloor (z+w) n(1 + \delta) \rfloor /(k + k')  \rfloor })^{\otimes (k + k')},
$$
$$
(\alpha_{\lfloor\lfloor z n(1 - \delta) \rfloor /k  \rfloor })^{\otimes k}
\reduction \alpha_{\lfloor z n(1 -2 \delta) \rfloor},
$$
$$
(\alpha_{\lfloor\lfloor w n(1 - \delta) \rfloor /k'  \rfloor })^{\otimes k'}
\reduction \alpha_{\lfloor w n(1 -2 \delta) \rfloor},
$$
$$ 
(\alpha_{\lfloor\lfloor (z+w) n(1 - \delta) \rfloor /(k + k')  \rfloor })^{\otimes (k + k')}
\reduction \alpha_{\lfloor (z+w) n(1 - 2 \delta) \rfloor}.
$$
Observe:
\ben
|zn -  kn/m| & \leq  & n/m \\
|\lfloor zn \rfloor -  \lfloor kn/m \rfloor| & \leq  & n/m + 1\\
|\lfloor zn \rfloor -  k\lfloor n/m \rfloor| & \leq  & n/m + k + 2\\
|\lfloor \lfloor zn \rfloor/k \rfloor -  \lfloor n/m \rfloor| & \leq  & n/(km) + 2 + 2/k.
\een
Thus, for sufficiently large $n$ and an appropriate 
choice of $m$,
$$
\lfloor \lfloor  zn(1 + \delta) \rfloor/k \rfloor \geq \lfloor n/m \rfloor \geq 
\lfloor \lfloor  zn (1 - \delta)\rfloor/k \rfloor.
$$ 
Analogously,
$$
\lfloor \lfloor  wn(1 + \delta)  \rfloor/k \rfloor \geq \lfloor n/m \rfloor \geq 
\lfloor \lfloor  wn(1-\delta) \rfloor/k \rfloor
$$ 
and
$$
\lfloor \lfloor  (w + z)n (1 + \delta)\rfloor/(k + k') \rfloor \geq \lfloor n/m \rfloor \geq 
\lfloor \lfloor  (w + z)n (1 - \delta)\rfloor/(k + k') \rfloor.
$$
 Let us start with the $\leq$ direction.
\ben
\alpha_{\lfloor z n(1 + 2\delta) \rfloor} \otimes
\alpha_{\lfloor w n(1 + 2\delta) \rfloor}\
&  \reduction & 
(\alpha_{\lfloor\lfloor z n(1 + \delta)  \rfloor /k  \rfloor })^{\otimes k}
\otimes 
(\alpha_{\lfloor\lfloor w n(1 + \delta)  \rfloor /k'  \rfloor })^{\otimes k'}\\
&  \reduction & (\alpha_{ \lfloor n/m \rfloor})^{\otimes k} \otimes
(\alpha_{ \lfloor n/m \rfloor})^{\otimes k'} \\
& = & (\alpha_{ \lfloor n/m \rfloor})^{\otimes (k + k')} \\
&  \reduction & ( \alpha_{\lfloor \lfloor (w + z)n (1 - \delta)\rfloor/(k + k')\rfloor})
^{\otimes (k + k')}\\
&  \reduction & \alpha_{\lfloor (z+w) n(1 - 2 \delta) \rfloor}.
\een
The $\geq$ direction is proven similarly.
\end{enumerate}
\end{proof}

\begin{defi}[Equivalence classes of resources]
Denote by $\tilde{\alpha}$ the equivalence class of $\alpha$,
i.e. the set of all $\alpha'$ such that $\alpha' \equiv \alpha$.
Define $\tilde{\cR}$ to be the set of
equivalence classes of resources in $\cR$.
Define the relation $\geq$ on $\tilde{\cR}$ by
$\tilde{\alpha} \geq \tilde{\beta}$ iff $\alpha' \geq \beta'$
for all $\alpha' \in \tilde{\alpha}$ and 
$\beta' \in \tilde{\beta}$.
Define the operation $+$ on $\tilde{\cR}$ such that
$\tilde{\alpha} + \tilde{\beta}$ is the union
of $\tilde{\alpha' + \beta'}$ over  all 
$\alpha' \in \tilde{\alpha}$ and $\beta' \in \tilde{\beta}$.
Define the operation $\cdot$ on $\tilde{\cR}$ such that
$z \tilde{\alpha}$ is the union
of $\tilde{z \alpha'}$ over  all 
$\alpha' \in \tilde{\alpha}$.
\end{defi}

\begin{lemma}
\label{ate}
For resources in $\cR$:
\begin{enumerate}
\item  $\tilde{\alpha} \geq \tilde{\beta}$ iff 
$\alpha \geq \beta$
\item $\tilde{\alpha} + \tilde{\beta} = \tilde{\alpha + \beta}$
\item $z \tilde{\alpha} = \tilde{z \alpha}$
\end{enumerate}
\end{lemma}
\begin{proof}
Regarding the first item: it suffices to show the
``if'' direction. Indeed, for any  $\alpha' \in \tilde{\alpha}$ and 
$\beta' \in \tilde{\beta}$
$$
\alpha' \geq \alpha \geq \beta \geq \beta',
$$
by \thm{composability}.
Regarding the second item: it suffices to show
that if  $\alpha' \equiv \alpha$, $\beta' \equiv \beta$
then $\alpha' + \beta' \equiv \alpha + \beta$.
This follows from \thm{composability}.
Similarly, for the  third item it suffices to show
that if  $\alpha' \equiv \alpha$ then
$z \alpha' \equiv z \alpha$, which is true by 
\thm{composability}.
\end{proof}

We now state a number of additional properties of $\tilde{\cR}$, each
of which can be easily verified.
\begin{theorem}
The relation $\geq$  forms a
partial order on the set $\tilde{\cR}$:
\begin{enumerate}
 \item $\tilde{\alpha} \geq \tilde{\alpha}$ (reflexivity) 
 \item if $\tilde{\alpha} \geq  \tilde{\beta}$ and 
$\tilde{\beta} \geq \tilde{\gamma}$ then
$\tilde{\alpha} \geq \tilde{\gamma}$  (transitivity) 
 \item  if $\tilde{\alpha} \geq  \tilde{\beta}$ and 
$\tilde{\beta} \geq \tilde{\alpha}$ 
 then $\tilde{\alpha} = \tilde{\beta}$ (antisymmetry) 
\end{enumerate}
\qed
\end{theorem}

\begin{theorem}
The following properties hold for the set $\tilde{\cR}$
with respect to $+$ and multiplication by positive real numbers.
\begin{enumerate}
\item $(zw) \tilde{\alpha} = z(w \tilde{\alpha})$ 
\item $(z + w) \tilde{\alpha} =  z \tilde{\alpha} + w \tilde{\alpha}$
\item $z( \tilde{\alpha}  + \tilde{\beta}) = 
z \tilde{\alpha}  + z \tilde{\beta}$
\item $1 \,  \tilde{\alpha} =  \tilde{\alpha}$
\end{enumerate}
\qed
\end{theorem}

\begin{theorem}
For equivalence classes in $\tilde{\cR}$:
\begin{enumerate}
\item if $\tilde{\alpha}_1 \geq  \tilde{\alpha}_2$ and 
$\tilde{\beta}_1 \geq \tilde{\beta}_2$ then
$\tilde{\alpha}_1 + \tilde{\beta}_1 \geq \tilde{\alpha}_2 + \tilde{\beta}_2 $
\item if $\tilde{\alpha} \geq \tilde{\beta}$ then 
$z \tilde{\alpha} \geq z \tilde{\beta}$
\end{enumerate}
\qed
\end{theorem}

{\bf Warning:}
Lemma \ref{ate} has essentially allowed us
to replace resources with their equivalence classes
and $\equiv$ with $=$. Henceforth we shall 
equate the two, and drop the $\sim$ superscript.
The one exception to this rule is when writing relative resources as
$(\beta : \gamma)$ where $\beta$ is a proper dynamic resource and
$\gamma$ is a proper static resource; in this case replacing $(\beta :
\gamma)$ with its equivalence class is well-defined, but replacing
$\beta$ and $\gamma$ with their equivalence classes wouldn't make sense.

\section{General resource inequalities}
\label{sec:general-inequalities}

In this section, we describe several resource inequalities that will
serve as useful basic tools for manipulating and combining other
resource inequalities.

\begin{lemma}
\label{lemma:relatif}
Let $\beta$ and $\beta'$ be proper dynamic resources,
and $\gamma$ and $\gamma'$ static test resources.
The following  resource inequalities hold:
\begin{enumerate}
\item $\beta   \geq   (\beta : \gamma)  $
\item $ (\beta : \gamma)  + \gamma  \geq  \beta(\gamma) $
\item if $\gamma \ext \gamma'$ then 
$(\beta : \gamma') \geq (\beta : \gamma) $ 
\item
$ \beta : \gamma  + \beta' : (\beta \gamma) 
 \geq   (\beta' \circ  \beta) : \gamma.$
\end{enumerate}
\end{lemma}
\begin{proof}
Immediate from definitions.
\end{proof}

\begin{lemma}[Closure]\label{lemma:closure}
For resources in $\cR$,
if $w_0 >0$ and $w \alpha \geq \beta $ for every $w > w_0 $ 
then $w_0 \alpha  \geq \beta$.
\end{lemma}
\begin{proof}
The statement is equivalent to
$$
w_0 \alpha \geq (1 - \delta) \beta, \,\,\, \forall \delta > 0, 
$$
which by definition implies the statement for $\delta = 0$.
\end{proof}

The case of $w_0=0$ is special and corresponds to the use of a
sublinear amount of a resource.
\begin{defi} [Sublinear $o$ terms]
We write
$$
\alpha + o \gamma \geq \beta 
$$
if for every $w>0$
$$
\alpha + w \gamma \geq \beta.
$$
\end{defi}

At the other extreme we might consider the case when we are allowed an
unlimited amount of some resource, typically when proving converse
theorems.
\begin{defi}[$\infty$ terms]
We write
$$\alpha + \infty\gamma \geq \beta$$
if for any $\delta>0$, there exists $k$ such that for any $\epsilon>0$
there exists $n_1,n_2$ and a $\epsilon$-valid protocol $\bP$
satisfying
$$\l\| \bP[(\alpha_{\lfloor n_1/k\rfloor} + \gamma_{\lfloor
n_2/k\rfloor})^{\times k}] - \beta_{\lfloor(1-\delta)n\rfloor}
\r\| \leq \epsilon.$$
\end{defi}
This means that we can use an amount of $\gamma$
that increases arbitrarily quickly with $n$.
Note that $\infty\gamma$ cannot be defined as a resource, since it
violates \eq{asy-quasi-iid}.

\begin{defi}[Negative terms]
For any $z<0$, define the statement
$$\alpha + z \gamma \geq \beta$$
to mean that 
$$\alpha \geq \beta + (-z) \gamma.$$
Similarly, $\alpha \geq \beta + z \gamma$ means that
$\alpha + (-z)\gamma \geq \beta$.
\end{defi}
Again $-\gamma$ is obviously not a resource, but the above definition
lets us treat it as such.

We now return to sublinear terms.
In general we cannot neglect sublinear resources; e.g. in entanglement
dilution, they are both necessary\cite{HL02,HW02} and
sufficient\cite{LP99}.  However, this situation only occurs when
they cannot be generated from the other resources being used in the
protocol.

\begin{lemma}[Removal of $o$  terms]
\label{lemma:noo}
For $\alpha, \beta , \gamma \in \cR$,
if 
\ben
\alpha + o \gamma &  \geq &  \beta \\ 
z \alpha &\geq & \gamma 
\een
for some real $z > 0$,
then 
$$
\alpha \geq \beta.
$$
\end{lemma}
\begin{proof}
For any $w > 0$
$$
(1 + zw) \alpha  \geq \alpha + w \gamma \geq \beta, 
$$
and the lemma follows by the Closure Lemma (\ref{lemma:closure}).
\end{proof}

One place that sublinear resources often appear is as catalysts,
meaning they are used to enable a protocol without themselves being
consumed.  Repeating the protocol many times reduces the cost of the
catalyst to sublinear:

\begin{lemma}[Cancellation]
\label{lemma:cancel}
For $\alpha, \beta , \gamma \in {\cR}$, if
$$
 \alpha +  \gamma \geq \beta +  \gamma ,
$$
then $\alpha + o \gamma \geq  \beta$.
\end{lemma}
\begin{proof}
Combine $N$ copies of the inequality (using part 1 of \thm{composability}) to
obtain  
$$
\gamma + N \alpha \geq \gamma + N \beta.
$$
Divide by $N$:
$$N^{-1}\gamma  + \alpha \geq N^{-1}\gamma + \beta \geq \beta.
$$
As $N^{-1}$ is arbitrarily small, the result follows.
\end{proof}

Often we will find it useful to use shared randomness as a catalyst.
The condition for this to be possible is that the randomness be
incoherently decoupled:
\begin{lemma} [Recycling common randomness]
\label{lemma:rcr}
If $\alpha$ and $\beta$ are resources for which
$$
\alpha + R \, [c \, c] \geq \beta, 
$$
and the $[c \, c]$ is incoherently decoupled in
the above resource inequality (RI), then 
$$
\alpha +  o \, [c \, c] \geq \beta.
$$
\end{lemma}
\begin{proof}
Since $[c \, c]$ is asymptotically independent
of the $\beta$ resource, by definitions \ref{def-incoh-decouple} and 
\ref{def:asy-decouple} 
it follows that
$$
\alpha + R \, [c \, c] \geq \beta + R \, [c \, c].
$$
An application of the cancellation lemma (\ref{lemma:cancel})
yields the desired result.
\end{proof}

\begin{corollary}
\label{cor:purim}
If $\alpha \geq [c \, c]$ and $\beta$ is pure then
$$
\alpha + R \, [c \, c] \geq \beta 
$$
can always be derandomized to
$$
\alpha \geq \beta. 
$$
\end{corollary}
\begin{proof}
It suffices to notice that for a  pure  output resource $\beta$, 
equation (\ref{otto}) is automatically satisfied. 
\end{proof}

The following theorem tells us that in proving channel coding theorems
one only needs to consider the case where the input state is maximally
mixed.  A similar result was shown in \cite{BKN98} (see also
\cite{KW03,Yard05a}), though with quite different techniques and
formalism.

\begin{theorem} [Absolutization]
\label{thm:absolutize}
The following resource inequalities hold:
\begin{enumerate}
\item $\qtqtau  \geq  [q \rightarrow q]$
\item $\coftau  \geq  [q \rightarrow qq] $
\item $\ctctau  \geq  [c \rightarrow c] $
\end{enumerate}
\end{theorem}
\begin{proof}
The lemma is a direct consequence of \lem{absolutizy}.
We shall prove case 1., as the proofs of 2. and 3. are identical.
By \lem{absolutizy}, we know that
$$
{[q \rightarrow q] :[\tau] }  + 2 [c \, c] \geq  [q \rightarrow q]
+ 2 [c \, c].
$$
By the cancellation lemma,
$$
{[q \rightarrow q] :[\tau] }  + o [c \, c] \geq  [q \rightarrow q].
$$
Since 
$$ 
{[q \rightarrow q] :[\tau] } \geq  [c \, c],
$$
by \lem{noo} the $o$ term can be dropped, and we are done.
\end{proof}

Finally, we note how convex combinations of static resources can be thought
of as states conditioned on classical variables.
\begin{theorem}
\label{thm:alfav}
Consider some static i.i.d. resource 
$\alpha = \< \sigma \>$, where
$$
\sigma^{A X_A B X_B} = 
\sum_x p_x \proj{x}^{ X_A} \otimes \proj{x}^{ X_B} 
 \otimes  \rho^{AB}_x.
$$
Namely, Alice and Bob share an ensemble of bipartite states,
and they both have the classical information about which state they hold.
Denote $\alpha_x = \<\rho_x \>$.
Then
$$
\alpha \geq \sum_x p_x \alpha_x.
$$
\end{theorem}
\begin{proof}
Recall the notion of the typical set\cite{CT91,CK81} $\cT$ such that
for any $\epsilon, \delta > 0 $ and sufficiently
large $n$, $p^{\otimes n}(\cT) \geq 1 - \epsilon$ and
for any $x^n \in \cT$,
$$
| n_x - p_x n| \leq \delta n,
$$
where $n_x$ is the number of occurrences of the symbol $x$ in 
$x^n$. Then
$$
\left\| \sigma^{\otimes n} - \sum_{x^n \in \cT} 
\proj{x^n}^{ X_A} \otimes p^{\ot n}(x^n)
\proj{x^n}^{ X_B} \otimes \rho_{x^n} \right\|_1
\leq \epsilon.
$$
The state that we want to simulate is $\bS[\sum_x p_x \alpha_x] =
(\omega_n)_n$ with  
$$
\omega_n = \bigotimes_x \rho_x^{\otimes \lfloor p_x n \rfloor}.
$$
For any $x^n \in \cT$ there is, clearly, a unitary 
$U_{x^n}^{A} \otimes U_{x^n}^{B}$ that maps $\rho_{x^n}$ to 
$\omega_{([1 - \delta] n - 1)} \otimes \hat{\rho}_{x^n}$ 
exactly for some state $\hat{\rho}_{x^n}$. Performing 
$$
\left(\sum_{x^n} \proj{x^n}^{ X_A} \otimes U_{x^n}^{A}\right) \otimes 
\left(\sum_{x^n} \proj{x^n}^{ X_B} \otimes U_{x^n}^{B}\right) 
$$
and tracing out subsystems thus brings $\sigma^{\otimes n}$
$\epsilon$-close to $\omega_{([1 - \delta] n - 1)}$.
Hence the claim.
\end{proof}

In fact, the above result could be strengthened to the equality
\be \alpha = \sum_x p_x \alpha_x + H(X_A)_{\sigma} [cc], \ee
but we will not need this fact, so leave the proof as an exercise for
the reader.  However, we will
show how a similar statement to \thm{alfav} can be made about
relative resources. 

\begin{theorem} 
\label{thm:betav}
Consider some channel $\cN$ with input Hilbert space $A$ and a state
$\sigma$ of the form
$$
\sigma^{R A X_A X_B} = 
\sum_x p_x \proj{x}^{ X_A} \otimes \proj{x}^{ X_B} 
 \otimes  \phi^{R A}_x.
$$
Namely, Alice has an ensemble of states $\ket{\phi_x}$,
and both parties have the classical information identifying the state.
Then
$$
\sum_x p_x \<\cN: \phi_x^A\> \geq \<\cN: \sigma\>.
$$
\qed
\end{theorem}
\begin{proof}
We will only give an outline of the simulation procedure; the proof of
correctness is essentially the same as for the last theorem.  Given
$\sigma^{\ot m}$ with $m = (1-\delta)n - 1$, Alice will locally
prepare $\hat{\rho}^{x^m}$ conditioned on $x^m$ from $\sigma^{\ot m}$
(which is possible since $\phi_x^{RA}$ can be locally prepared by
Alice), perform the inverse of the map $U^A_{x^m}$ from the
last theorem and then apply $\cN^{\ot n}$.
\end{proof}

\section{Known coding theorems 
 expressed as resource inequalities} 
\label{sec:known}

There have been a number of quantum and classical coding theorems 
discovered to date, typically along with so-called converse theorems
which prove that the coding theorems cannot be improved upon. The
theory of resource inequalities has been 
developed to provide an underlying unifying principle. 
This direction was initially suggested in \cite{DW03a}.

We shall state theorems such as Schumacher compression, the classical reverse 
Shannon theorem,  the instrument compression theorem, 
the classical-quantum Slepian-Wolf theorem,
the HSW theorem, and CR concentration  as resource inequalities.
Then we will show how some of these can be used as building
blocks, yielding transparent and concise proofs of some derivative results.

We shall work within the QQ formalism.

\paragraph{Schumacher compression.}
The quantum source compression theorem was proven by Schumacher in
\cite{JS94,Sch95}. 
Given a quantum state $\rho^{A'}$, define $\sigma^{B} := \id^{ A'
\rightarrow B}(\rho^{A'})$.  Then the 
following resource inequality (RI) holds:  
\be
(H(B)_\sigma + \delta) [q \rightarrow q] \geq  
\<\id^{ A' \rightarrow B}  : \rho^{A'} \>
\label{eq:schu}
\ee
if and only if $\delta\geq 0$.

Note that this formulation simultaneously
expresses both the coding theorem and the converse theorem.

\paragraph{Entanglement concentration.} The problem 
of entanglement concentration was solved in \cite{BBPS96}, and is,
in a certain sense, a static counterpart to Schumacher's 
compression theorem.
Entanglement concentration can be thought of as a coding theorem which
says that given a pure bipartite quantum state $\ket{\phi}^{AB}$ 
the following RI holds: 
\be
\< \phi^{A B} \> \geq H(B)_\phi \,[q \, q].
\label{eq:E-concentration}\ee

The reverse direction is known as \emph{entanglement dilution} \cite{BBPS96},
and thanks to Lo and Popescu \cite{LP99} it is known that
\be
H(B)_\phi \,[q \, q] + o\, [c \ra c] \geq \< \phi^{AB} \>.
\label{eq:E-dilution}\ee

Were it not for the $o\,[c\ra c]$ term, we would have the equality
$\<\phi^{AB}\>=H(B)_\phi \, \qq$.  However, it turns out
that the $o [c \ra c]$ term cannot be avoided\cite{HL02,HW02}.  This
means that the strongest equality we can state has a sublinear amount
of classical communication on both sides:
\be
H(B)_\phi \,[q \, q] + o\,\ctc = \< \phi^{AB}\> + o\,\ctc
.\label{eq:BBPS-equality}\ee

Note how \eq{BBPS-equality} states the converse in a form that is in
some ways stronger than \eq{schu}, since it implies the transformation
is not only optimal, but also asymptotically reversible.  We can also
state a converse when more classical communication is allowed, though
no longer as a resource equality:
$$ \< \phi^{A B} \> + \infty\,\ctc 
\geq (H(B)_\phi -\delta) \,[q \, q]$$
iff $\delta\geq 0$; and similarly for entanglement dilution.

\paragraph{Shannon compression.} Shannon's classical compression
theorem was proven in \cite{Shannon48}.  Given a classical state
${\rho}^{X_A}$ and defining
$${\sigma}^{X_B} = \bar{\id}^{ X_A \rightarrow X_B}({\rho}^{X_A}),$$
Shannon's theorem says that
\be
(H(X_B)_{{\sigma}} + \delta) [c \rightarrow c] \geq  
\<\bar{\id}^{ X_A \rightarrow X_B} : {\rho}^{X_A} \>, 
\label{eq:shannon-compression}
\ee
if and only if $\delta\geq 0$.

\paragraph{Common randomness concentration.}
This is the classical analogue of entanglement concentration,
and a static counterpart to Shannon's compression theorem.
It states that, if Alice and Bob have a copy of the same random
variable $X$, embodied in the classical bipartite state
$$
\rho^{X_A X_B} = \sum_x p_x \proj{x}^{X_A} \otimes \proj{x}^{X_B},
$$
then
\be
\< \rho^{X_A X_B} \> \geq H(X_B)_\rho \, [c \, c].
\label{eq:crc}
\ee
Incidentally, common randomness dilution can do without the $o$ term:
$$
H(X_B)_\rho \, [c \, c] \geq \< \rho^{X_A X_B} \>.
$$

Thus we obtain a simple resource equality:
$$H(X_B)_\rho \,\ctc = \<\rho^{X_AX_B}\>.$$

\paragraph{Classical reverse Shannon theorem (CRST).} This theorem
was proven in \cite{BSST01,Winter:02a}, and it generalizes
Shannon's compression theorem to compress probability distributions
of classical states instead of pure classical states.  
Given a classical channel $\bar{\cN}: X_{A'} \rightarrow Y_B$ and a
classical state ${\rho}^{X_{A'}}$, the CRST states that
\be
I(X_A; Y_B)_\sigma [c \rightarrow c] + H(X_A |Y_B)_\sigma [c \, c] \geq  
\<\bar{\cN} : \rho^{X_{A'}} \>, 
\label{eq:crst}
\ee
where 
$$
{\sigma}^{X_A Y_B} = \bar{\cN} \circ 
\bar{\Delta}^{X_{A'} \rightarrow X_{A'} X_{A}} 
(\rho^{X_{A'}}).
$$
Moreover, given a modified classical channel $\bar{\cN}': 
X_{A'} \rightarrow Y_A Y_B$ which also provides Alice with a copy
of the channel output,
$$
\bar{\cN}' =  \bar{\Delta}^{Y_B  \rightarrow Y_A Y_B}   \circ \bar{\cN}, 
$$
 the following stronger RI also holds:
\be
I(X_A; Y_B)_\sigma [c \rightarrow c] + H(X_A| Y_B)_\sigma [c \, c] \geq  
\<\bar{\cN}'  : \rho^{X_{A'}} \>, 
\label{eq:crst2}
\ee
In fact, this latter RI can be reversed to obtain the equality
\be
I(X_A; Y_B)_\sigma [c \rightarrow c] + H(X_A| Y_B)_\sigma [c \, c] =
\<\bar{\cN}'  : \rho^{X_{A'}} \>.
\ee

However, in the case without feedback, the best we can do is a
tradeoff curve between cbits and rbits, with \eq{crst} representing
the case of unlimited randomness consumption.  The full tradeoff
will be given by an RI of the following form
$$a \,\ctc + b \,\cc \geq \<\bar{\cN}  : \rho^{X_{A'}} \>$$
where $(a,b)$ range over some convex set $CR(\bar{\cN})$.  It can be
shown\cite{Wyner75, BW05} that $(a,b)\in CR(\bar{\cN})$ iff there
exist channels 
$\bar{\cN}_1:X_{A'}\ra W_{C'}, \bar{\cN}_2:W_{C'}\ra Y_B$ such that 
$\bar{\cN}=\bar{\cN}_2 \circ \bar{\cN}_1$
and $a \geq I(X_A;W_C)_\omega, b \geq I(X_AY_B;
W_C)_\omega$, where 
$$\omega^{X_AW_CY_B} := 
\bar{\cN}_2 \circ \bar{\Delta}^{W_{C'} \rightarrow W_{C'} W_{C}} 
\circ \bar{\cN}_1 \circ
\bar{\Delta}^{X_{A'} \rightarrow X_{A'} X_{A}} 
(\rho^{X_{A'}}).
$$

\paragraph{Classical compression with quantum side information.}
This problem  was solved in \cite{DW02, WinterPhD}, 
and is a generalization of Shannon's classical compression
theorem in which Bob has quantum side information 
about the source. 
Suppose Alice and Bob are given an ensemble  
$$
{\rho}^{X_A B} = \sum_x p_x \proj{x}^{X_A} \otimes \rho_x^{B}, 
$$ 
and Alice wants to communicate $X_A$ to Bob, which would give them the
state 
$$
{\sigma}^{X_B B} := \bar{\id}^{X_A \rightarrow X_B}({\rho}^{X_A B}).
$$
To formalize this situation, we use the Source as one of the
protagonists in the 
protocol, so that the coding theorem inputs a map from the Source
to Alice and Bob $\<\bar{\id}^{S_X\ra X_A}\ot \id^{S_B\ra B}:\rho^{S_X
S_B}\>$ and outputs a map from the Source entirely to Bob.  The coding
theorem is then
\be
\< \bar{\id}^{S_X\ra X_A}\ot \id^{S_B\ra B}:\rho^{S_XS_B}\> +
(H(X_B | B)_{{\sigma}}+\delta) [c \rightarrow c] \geq  
\<\bar{\id}^{S_X \rightarrow X_B} \ot \id^{S_B\ra B}:\rho^{S_XS_B}\>,
\label{eq:cqsw}
\ee
which holds iff $\delta\geq 0$.  This formulation ensures that we work
with well-defined resources instead of using the natural-seeming, but
incorrect $\<\id^{X_A\ra X_B}:\rho^{X_AB}\>$ (which violates
\eqs{asy-monotone}{asy-quasi-iid}). 

Of course, with no extra resource cost Alice could keep a copy of
$X_A$.

\paragraph{Instrument compression theorem.}
This theorem was proven in \cite{Winter01}, 
and is a generalization of the CRST.
Given a remote instrument ${\bf T}: 
A' \rightarrow {A} X_B$,
and a quantum state $\rho^{A'}$, the 
following RI holds:
\be
I(R; X_B)_\sigma [c \rightarrow c] + 
H(X_B| R)_\sigma [c \, c] \geq \< {\bf T} : \rho^{A'} \>,
\label{eq:ict}
\ee
where 
$$
{\sigma}^{R A X_B} = {\bf T}(\psi^{R A'})
$$
and $\proj{\psi}^{R X_A} \ext \rho^{X_A}$.
Moreover, given a modified remote instrument
 which also provides Alice with a copy of the instrument output,
$$
{\bf T}'  =  \bar{\Delta}^{X_B  \rightarrow X_A X_B} \circ {\bf T}, 
$$
the  RI still holds:
\be
I(R; X_B)_\sigma [c \rightarrow c] + 
H(X_B| R)_\sigma [c \, c] \geq \< {\bf T}' : \rho^{A'} \>.
\label{eq:ict2}
\ee
Only this latter RI is known to be optimal (up to a trivial
substitution of $\ctc$ for $\cc$); indeed 
\be
a [c \rightarrow c] + b [c \, c] \geq \< {\bf T}' : \rho^{A'} \>.
\ee
iff $a\geq I(R;X_B)_\sigma$ and $a+b \geq H(X_B)_\sigma$.

By contrast, only the communication rate of \eq{ict} is known to be
optimal; examples are known in which less randomness is necessary.

\paragraph{Remote state preparation (RSP)}
Instrument compression can be thought of as a generalization of the CRST
from $\{c\ra c\}$ channels to $\{q\ra c\}$ channels.  In contrast,
remote state preparation (proved in \cite{BHLSW03}) generalizes the
CRST to $\{c\ra q\}$ channels. 

Let $\cE=\sum_i p_i \oprod{i}^{X_A} \ot \oprod{\psi_i}^{AB}$ be an
ensemble of bipartite states.  Define the corresponding $\{c\ra q\}$
channel $\cN_\cE$ by
\be \cN_\cE(\ket{i}\!\bra{j}^{X_A}) = \delta_{ij}
\oprod{i}^{X_A} \ot \oprod{\psi_i}^{AB}.
\label{eq:RSP-map}\ee
  This means that $\cN_\cE$
measures the input in the standard basis and maps outcome $i$ to the
joint state $\psi_i^{AB}$.  Thus, $\cE = \cN_\cE(\cE^{X_A})$, where
$\cE^{X_A}$ is the classical input state
$\sum_i p_i\oprod{i}^{X_A}$.

The coding theorem of RSP states that
\be
I({X_A};B)_\cE [c\ra c] + H(B) \qq \geq \<\cN_\cE : \cE^{X_A}\>,
\label{eq:RSP}\ee
meaning that Alice can use the resources on the LHS to prepare a
sequence of states $\ket{\psi_{i_1}}\cdots\ket{\psi_{i_n}}$ of her
choosing, with high fidelity on average if she chooses $i^n$ according
to $p^{\ot n}$.  Note that since Alice
holds the purification of Bob's state, this is stronger than the
ability to simulate a $\{c\ra q\}$ channel that gives Bob mixed
states.  The cbit cost is optimal in either case, since HSW coding
(\eq{hsw}, below)
yields $\<\cN_\cE : \cE^{X_A}\> \geq I({X_A};B)_\cE [c\ra c]$ even if Alice's half
of $\psi_i^{AB}$ is discarded.  However, the entanglement cost of \eq{RSP}
is only known to be optimal for the setting when Alice holds the
purification of Bob's output.  Determining the minimal resources
necessary to perform {\em visible mixed-state data compression} has
been a long-standing open problem in quantum information
theory.\cite{Barnum01,KI01,Winter:02a}

Ref.~\cite{BHLSW03} also proved a stronger ``single-shot'' version of
RSP, the simplest form of which is that $n(1+o(1))$ cbits and $n$
ebits can be used to prepare an arbitrary $n$ qubit state.  It it
interesting to note that this does not form an asymptotic resource (as
given in \defn{asy-resource}) because it fails to satisfy
\eq{asy-quasi-iid}.\footnote{This has a number of interesting
implications.  For example, ``single-shot'' RSP is not amenable to the
sort of cbit-ebit tradeoffs that are possible in the ensemble
case\cite{Devetak01,HJW02,BHLSW03}.  In fact, the $\exp(n)$ cbit cost
for simulating single-shot RSP of $n$ qubits is one of the few known
examples where infinite, or super-linear, resources are useful.  Also,
the RSP capacities of channels appear to be different for single-shot
and ensemble RSP\cite{Leung-private-04}.}

\paragraph{Teleportation and super-dense coding.} 
Teleportation \cite{BBCJPW98} and super-dense coding \cite{BW92} 
are finite protocols, and we have discussed them already in the
introduction.
In a somewhat weaker form they may be written as resource 
inequalities. Teleportation (TP):
\be
 2\,[c \rightarrow c] + [q \, q] \geq [q \rightarrow q].
\label{eq:tp}
\ee
Super-dense coding (SD):
\be
[q \rightarrow q] + [q \, q] \geq 2  \,[c \rightarrow c].
\label{eq:sd}
\ee
Finally, entanglement distribution:
\be
[q \rightarrow q] \geq  [q \, q].
\ee
All of these protocols are optimal (we neglect the precise
statements), but composing them with each other (e.g. trying to
reverse teleportation by using super-dense coding) is wasteful.  We
will give a resolution to this problem in \chap{ccc} by using
coherent classical communication.


\paragraph{Holevo-Schumacher-Westmoreland (HSW) theorem.}
The direct part of this theorem was proven in \cite{Holevo98, SW97} and
the converse in \cite{Holevo73}. Together they say that 
given a quantum channel ${\cN}: A' \rightarrow B$, 
for any ensemble  
$$
{\rho}^{X_A A'} = \sum_x p_x \proj{x}^{X_A} \otimes \rho_x^{A'} 
$$ 
the following RI holds:  
\be
\<\cN : \rho^{A'} \>  \geq
(I(X_A; B)_\sigma -\delta) [c \rightarrow c], 
\label{eq:hsw}
\ee
iff $\delta\geq 0$, where 
$$
{\sigma}^{X_A B} = {\cN}^{A'\ra B}(\rho^{X_A A'}).
$$


\paragraph{Shannon's noisy channel coding theorem}
This theorem was proven in \cite{Shannon48} and today can be understood
as a special 
case of the HSW theorem.
One version of the theorem says that
given a classical channel $\bar{\cN}: X_{A'} \rightarrow Y_B$
and any classical state ${\rho}^{X_{A'}}$ the 
following RI holds:  
\be
\<\bar{\cN} : \rho^{X_{A'}} \>  \geq
(I(X_A; Y_B)_\sigma-\delta) [c \rightarrow c], 
\label{shan2a}
\ee
iff $\delta\geq 0$ and where 
\be
{\sigma}^{X_A Y_B} := \bar{\cN} \circ 
\bar{\Delta}^{X_{A'} \rightarrow X_{A'}X_{A}} 
(\rho^{X_A}).
\label{eq:shannon-sigma-def}\ee

If we optimize over all input states, then we find that
\be \<\bar{\cN}\> \geq C \ctc \ee
iff there exists an input ${\rho}^{X_{A'}}$ such that $C\geq
I(X_A;Y_B)_\sigma$, with $\sigma$ given by \eq{shannon-sigma-def}.


\paragraph{Entanglement-assisted capacity theorem.}
This theorem was  proven in \cite{BSST01,Holevo01a}.
The direct coding part of the theorem says that, 
given a quantum channel ${\cN}: A' \rightarrow B$, 
for any quantum state $\rho^{A'}$
the following RI holds:  
\be
 \< \cN : \rho^{A'} \> + H(R)_\sigma [q \, q]
\geq I(R; B)_\sigma \,[c  \rightarrow c],
\label{eq:eac},
\ee
where 
$$
{\sigma}^{R B} = {\cN}(\psi^{R A'})
$$
for an arbitrary $\psi$ satisfying  $\proj{\psi}^{R A'} \ext
\rho^{A'}$.

The only converse proven in \cite{BSST01,Holevo01a} was for the case of
infinite entanglement: they found that $\<CN\> + \infty\qq \geq
C\ctc$ iff $C\leq I(R; B)_\sigma$ for some appropriate $\sigma$.
\mscite{Shor04} gave a full solution to the tradeoff problem for
entanglement-assisted classical communication which we will present an
alternate derivation of in \sect{NCE-toff}.


\paragraph{Quantum capacity (LSD) theorem.}
This theorem was conjectured in \cite{Sch96,SN96},
a heuristic (but not universally accepted) proof given
by Lloyd~\cite{Lloyd96}
and finally proven by Shor~\cite{Shor02} and
with an independent method by Devetak~\cite{Devetak03}.
The direct coding part of the theorem says that, 
given a quantum channel ${\cN}: A' \rightarrow B$, 
for any quantum state $\rho^{A'}$
the following RI holds:  
\be
 \< \cN : \rho^{A'} \> 
\geq (I(R\,\rangle B)_\sigma-\delta) \,[q  \rightarrow q],
\label{eq:lsd}
\ee
iff $\delta\geq 0$ and 
where 
$$
{\sigma}^{R B} = {\cN}(\psi^{R A'})
$$
for any $\psi^{RA'}$ satisfying $\proj{\psi}^{R A'} \ext \rho^{A'}$.

\paragraph{Noisy super-dense coding theorem.} 
This theorem was  proven in \cite{HHHLT01}.
The direct coding part of the theorem says that, 
given a bipartite quantum state $\rho^{A B}$, 
the following RI holds:  
\be
   \< \rho^{AB} \> + 
H(A)_\rho \, [q \rightarrow q]  
\geq  I(A; B)_\rho \,[c \rightarrow c].
\label{eq:nsd}
\ee
A converse was proven in \cite{HHHLT01} only for the case when an
infinite amount of $\<\rho^{AB}\>$ is supplied, but we will return to
this 
problem and provide a full trade-off curve in \sect{nSD-toff}.

\paragraph{Entanglement distillation.}
The direct coding theorem for one-way entanglement distillation
is embodied in the \emph{hashing inequality}, proved in~\cite{DW03b,DW03c}:
given a bipartite quantum state $\rho^{A B}$, 
\be
\<\rho^{AB}\> + I(A;E)_\psi \, [c \rightarrow c] 
                      \geq I(A\,\rangle B)_\psi \,[q \, q],
\label{eq:hashing}
\ee
where 
$\proj{\psi}^{A B E} \ext \rho^{A B}$.

Again, the converse was previously only known for the case when an
unlimited amount of classical communication was
available\cite{Sch96,SN96, DW03b,DW03c}.  In \sect{distill-toff} we
will give an expression for the full trade-off curve.

\paragraph{Noisy teleportation.} This RI was 
discovered in \cite{DHW03}. 
Given a bipartite quantum state $\rho^{A B}$,
\ben
\label{eq:ntp}
\<\rho^{AB}\> + I(A;B)_\rho \, [c \rightarrow c] 
 \geq  I(A\,\rangle B)_\rho \,[q \rightarrow q].
\een 
Indeed, letting
$\proj{\psi}^{A B E} \ext \rho^{A B}$,
\ben
\<\rho^{AB}\> + I(A;B)_\psi \, [c \rightarrow c] 
& = & \<\rho^{AB} \> + I(A;E)_\psi \, [c \rightarrow c] + 
2 I(A\,\rangle B)_\psi [c \rightarrow c] \\
& \geq &  I(A\,\rangle B)_\psi \,[q \, q]  + 
2 I(A\,\rangle B)_\psi [c \rightarrow c] \\
& \geq & I(A\,\rangle B)_\psi \,[q \rightarrow q].
\een 
The first inequality follows from \eq{hashing} and
the second from teleportation.

\paragraph{Classical-quantum communication trade-off
           for remote state preparation.} 
The main coding theorem of \cite{HJW02} has
two interpretations. Viewed as a statement
about quantum compression with classical side information,
it says that, given an ensemble
$$
{\rho}^{X_{A'} A'} = \sum_x p_x \proj{x}^{X_{A'}} \otimes \rho_x^{A'}, 
$$
for any classical channel $\bar{\cN}: X_{A'} \rightarrow Y_B$,
the following RI holds:
\be
\label{eq:cqrsp}
 H(B|Y_B)_\sigma [q \rightarrow q] + I(X_A;Y_B)_\sigma [c \rightarrow c]
\geq \<\id^{A' \rightarrow B} : \rho^{X_{A'}  A'} \>.
\ee
where
$$
\sigma^{X_A Y_B B} = 
((\bar{\cN}^{X_{A'}\ra Y_B} \circ \bar{\Delta}^{X_{A'}\ra X_{A'} X_{A}} )
 \otimes \id^{A' \rightarrow B}) \rho^{X_{A'}  A'}.
$$
Conversely, if $a\qtq + b \ctc \geq \<\id^{A'\ra B}:\rho^{X_{A'}A'}\>$
then there exists a classical channel $\bar{\cN}: X_A \rightarrow Y_B$
with corresponding state $\sigma$ such that $a\geq H(B|Y_B)_\sigma$
and $b\geq I(X_A;Y_B)_\sigma$.

We shall now show how the proof from \cite{HJW02}
may be written very succinctly
in terms of previous results.
Define $\bar{\cN}' = \bar{\Delta}^{Y_{B} \rightarrow Y_{A} Y_{B}}
\circ \bar{\cN}$. By the Classical Reverse Shannon Theorem
(\eq{crst2}) and part 3 of \lem{relatif}, 
$$
I(X_A; Y_B)_\sigma [c \rightarrow c] + H(X_A| Y_B)_\sigma [c \, c] \geq  
\<\bar{\cN}'  : \rho^{X_{A'} A'} \>. 
$$
On the other hand, Schumacher compression (\eq{schu}) and \thm{betav} 
imply
$$
H(B| Y_B)_\sigma [q \rightarrow q] \geq
\<\id^{ A' \rightarrow B}  : \bar{\cN}' (\rho^{X_A  A'}) \>.
$$
Adding the two equations and invoking part 2 of \lem{relatif} gives
$$
 H(B|Y_B)_\sigma [q \rightarrow q] + I(X_A;Y_B)_\sigma [c \rightarrow c]
+ H(X_A| Y_B)_\sigma [c \, c]
\geq \<\id^{A' \rightarrow B} : \rho^{X_{A'}  A'} \>.
$$
Finally, derandomizing via \cor{purim}
gives the desired result \peq{cqrsp}.

The result of \cite{HJW02} may be also viewed as a statement about 
remote state preparation.  Suppose we are given a classical state
$\rho^{X_{A''}}$ and a $\{c \rightarrow q \}$ map
$\cN_\cE' : X_{A''} \rightarrow B$, $\cN_\cE' = \id^{A' \rightarrow B} \circ 
\cN_\cE$, where $\cN_\cE$ has Kraus representation 
$\{ \ket{\phi_x}^{A^*A'} \bra{x}^{X_{A''}} \}_x$.  Then for any
classical channel $\bar{\cN}: X_{A} \rightarrow Y_B$, 
the following RI holds:
\be
\label{eq:cqrsp2}
 H(B|Y_B)_\sigma [q \rightarrow q] + I(X_A;Y_B)_\sigma [c \rightarrow c]
\geq \<\cN_\cE' : \rho^{X_{A''}} \>,
\ee
where $\sigma^{X_A Y_B B}$ is defined as above and
$$
\rho^{X_{A'}  A^* A'} = (\cN_\cE \circ 
\bar{\Delta}^{X_{A''} \rightarrow X_{A''} X_{A'}})\rho^{X_{A''}}  .
$$
This follows from adding (\eq{cqrsp}) to
\ben
\<\id^{A' \rightarrow B} : \rho^{X_{A'}  A'} \> & \geq & 
\<\id^{A' \rightarrow B} : \rho^{X_{A'} A^*  A'} \>  + 
\< (\cN_\cE \circ 
\bar{\Delta}^{X_{A''} \rightarrow X_{A''} X_{A'}}) : \rho^{X_{A''}}\> \\
& \geq &   \<\cN_\cE' : \rho^{X_{A''}} \>.
\een
The first inequality follows from part 3 of \lem{relatif} and
the locality of the map $\cN_\cE$.
The second is an application of part 4 of \lem{relatif}.

\paragraph{Common randomness distillation.}
This theorem was originally proven in \cite{DW03a}.
Given an ensemble 
$$
\rho^{X_A B} = \sum_x p_x \proj{x}^{X_A} \otimes \rho_x^{B}, 
$$ 
the
following RI holds:
\be
\< \rho^{X_A B} \> + H(X_A|B)_\rho [c \rightarrow c] 
 \geq  H(X_A)_\rho [c \, c].
\label{crd}
\ee
Armed with our theory of resource inequalities, the
proof becomes extremely simple.
\ben
\< \rho^{X_A B} \> + H(X_A|B)_\rho [c \rightarrow c] 
& \geq & \< \rho^{X_A B} \> + 
\< \bar{\Delta}^{X_A \rightarrow X_A X_B} : \rho^{X_A B} \> \\
& \geq & \< \rho^{X_A X_B B} \>\\
& \geq & \< \rho^{X_A X_B}\>\\
& \geq & H(X_A)_\rho [c \, c].
\een
The first inequality is by classical compression with quantum side
information (\eq{cqsw}), the second by \lem{relatif},
part 2, and the fourth by common randomness concentration (\eq{crc}).

\section{Discussion}\label{sec:res-discuss}
This chapter has laid the foundations of a formal approach to quantum
Shannon theory in which the basic elements are asymptotic resources
and protocols mapping between them.  Before presenting applications of
this approach in the next three chapters, we pause for a moment to
discuss the limitations of our formalism and possible ways it may be
extended.

The primary limitation is that our approach is most successful when
considering one-way communication and when dealing with only one noisy
resource at a time.  These, and other limitations, suggest a number of
ways in which we might imagine revising the notion of an asymptotic
resource we have given in \defn{asy-resource}.  For example, if we
were to explore unitary and/or bidirectional resources more carefully,
then we would need to reexamine our treatments of depth and of
relative resources.  Recall that in \defn{asy-ineq} we (1) always
simulate the depth-1 version of the output resource, (2) are allowed
to use a depth-$k$ version of the input resource where $k$ depends
only on the target inefficiency and not the target error.  These
features were chosen rather delicately in order to guarantee the
convergence of the error and inefficiency in the Composability Theorem
(\ref{thm:composability}), which in turn gets most of its depth
blow-up from the double-blocking of the Sliding Lemma
(\ref{lemma:sliding}).  However, it is possible that a different model
of resources would allow protocols which deal with depth differently.
This won't make a difference for one-way resources due to the
Flattening Lemma (\ref{lemma:flattening}), but there is evidence that
depth is an important resource in bidirectional
communication\cite{Klauck01}; on the other hand, it is unknown how
quickly depth needs to scale with $n$.

Relative resources are another challenge for studying bidirectional
communication.  As we discussed in \sect{finite-res}, if $\rho^{AB}$
cannot be locally duplicated then $\<\cN : \rho^{AB}\>$ fails to
satisfy \eq{asy-monotone} therefore is not a valid resource.  The
problem is that being able to simulate $n$ uses of a channel on $n$
copies of a correlated or entangled state is not necessarily stronger
than the ability to simulate $n-1$ uses of the channel on $n-1$ copies
of the state.  The fact that many bidirectional problems in classical
information theory\cite{Shannon61} remain unsolved is an indication
that the quantum versions of these problems will be difficult.  On the
other hand, it is possible that special cases, such as unitary gates
or Hamiltonians, will offer simplifications not possible in the
classical case.

Another challenge to our definition of a resource comes from unconventional
``pseudo-resources'' that resemble resources in many ways but fail to
satisfy the quasi-i.i.d. requirement (\eq{asy-quasi-iid}).  For
example, the ability to remotely prepare an arbitrary $n$ qubit state
(in contrast with the ensemble version in \eq{RSP}) cannot be
simulated by the ability to remotely prepare $k$ states of
$n(1+\delta)/k$ qubits each.  There are many fascinating open
questions surrounding this ``single-shot'' version of RSP; for
example, is the RSP capacity of a channel ever greater than its quantum
capacity?\footnote{Thanks to Debbie Leung for suggesting this
question.}  Another example comes from the 
``embezzling states'' of \cite{vanDam03}.  The $n$-qubit embezzling state
can be prepared from $n$ cbits and $n$ ebits (which are also
necessary\cite{HW02}) and can be used as a resource for entanglement
dilution and for simulating noisy quantum channels on
non-i.i.d. inputs\cite{BDHSW05}; however, it also cannot be prepared
from $k$ copies of the $n(1+\delta)/k$-qubit embezzling state.
These pseudo-resources are definitely useful and interesting, but it
is unclear how they should fit into our resource formalism.

Other extensions of the theory will probably require less
modification.  For example, it will not a priori be hard to extend the
theory to multi-user scenarios.  Resources and capacities can even be
defined in non-cooperative situations pervasive in cryptography (see
e.g.~\cite{WNI03}), which will mostly require a more careful
enumeration of different cases.  We can also consider privacy to be a
resource.  Our definitions of decoupled classical communication are a
step in this direction; also there are expressions for the private 
capacity of quantum channels\cite{Devetak03} and states\cite{DW03b},
and there are cryptographic versions of our Composability
Theorem\cite{BM04,Unruh04}.

\chapter{Communication using unitary interactions}\label{chap:unitary}

In this chapter, we approach bipartite unitary interactions through
the lens of quantum Shannon theory, by viewing them as a two-way
quantum channels.  For example, we might try to find the classical
communication capacity of a $\cnot=\ket{0}\bra{0}\otimes I +
\ket{1}\bra{1}\otimes \sigma_x$ with control qubit in Alice's
laboratory and target qubit in Bob's laboratory.  
More generally, we will fix a bipartite gate
$U\in \cU_{d\times d} = \cU_{d^2}$ and investigate the rate at which $U$ can
generate entanglement, send classical or quantum messages and so on.

This work can be applied both to computation (in a model where local
operations are easy and interactions are expensive) and to the rest of
Shannon theory, which will be our primary focus in the next two chapters.
Most other work on bipartite unitary gates has been more concerned
with computational issues, but in \sect{u-background} we survey the
literature with an eye toward information theory applications.  The
main results of this chapter are the capacities of a bipartite unitary
gate to create entanglement (in \sect{u-e-cap}) and to send classical
messages in one direction when assisted by an unlimited amount of
entanglement (in \sect{u-c-cap}).  Along the way, we also establish
some easily computable bounds on and relations between these
capacities (in \sects{u-e-cap}{u-c-cap}) and discuss these capacities for
some interesting specific gates in \sect{u-examples}.
We conclude with a summary and discussion in \sect{u-discuss}.

{\em Bibliographical note:}
Except where other works are cited, most of the results in this
chapter are from \cite{BHLS02} (joint work with Charles Bennett,
Debbie Leung and John Smolin).  However, this thesis reformulates them
in the formalism of \chap{shannon}, which allows many of the
definitions, claims and proofs to be greatly simplified.


\section{Background}\label{sec:u-background}
\subsection{Survey of related work}\label{sec:u-related}

The nonlocal strength of unitary interactions was first discussed
within a model of communication complexity, when Nielsen introduced
the Schmidt decomposition of a unitary gate (described below) as a
measure of its nonlocality\cite{Nielsen98}.  The idea of studying a
gate in terms of nonlocal invariants---parameters which are unchanged
by local unitary rotations---was first applied to two-qubit gates by
\cite{Makhlin00}, which found that the nonlocal properties of these
gates are completely described by three real parameters.  Later these
invariants would be interpreted by \cite{Khaneja01, KC01} as
components of a useful general decomposition of two-qubit gates: for
any $U\in\cU_{2\times 2}$, there exist $A_1,A_2,B_1,B_2\in\cU_2$ and
$\theta_x,\theta_y,\theta_z \in (-\frac{\pi}{4},\frac{\pi}{4}]$ such
that
\begin{equation}
U=(A_1\ot B_1)e^{i(\theta_x \sigma_x\ot \sigma_x+\theta_y \sigma_y\ot
\sigma_y +\theta_z \sigma_z\ot\sigma_z)}(A_2\ot B_2)
\label{eq:magic-basis}.\end{equation}
This fact has a number of useful consequences, but the only one we
will use in this work is that the nonlocal part $\exp(i\sum_j \theta_j
\sigma_j \ot \sigma_j)$ is symmetric under exchange of Alice and Bob,
implying that $\<U\>=\<\swap U \swap\>$ for any $U\in\cU_{2\times 2}$.
This symmetry no longer holds\cite{Bennett02a} (and similar
decompositions generally do not exist) for $\cU_{d\times d}$ with
$d>2$.

Other early work considered the ability of unitary gates to
communicate and create entanglement.  \cite{Collins00} showed that
$\<\cnot\> + [qq] \geq [c\ra c] + [c\la c]$ and \cite{Chefles00}
proved that $2\log d([c\ra c]+[c\la c]+[qq]) \geq \<U\>$ for any
$U\in\cU_{d\times d}$.  The first discussion of asymptotic capacity was in
\cite{DVCLP01}, which found the rate at which Hamiltonians can
generate entanglement.  Their technique would be adopted mostly
unchanged by \cite{Leifer03,BHLS02} to find the entanglement capacity
of unitary gates.  In general it is difficult to exactly calculate the
entanglement capability of Hamiltonians and gates, but \cite{CLVV03}
finds the rate at which two-qubit Hamiltonians of the form $H=\alpha
\sigma_x\ot \sigma_x + \beta \sigma_y\ot\sigma_y$ can generate
entanglement.

Instead of reducing gates and Hamiltonians to standard resources, such
as cbits and ebits, one can consider the rates at which Hamiltonians
and gates can simulate one another.  The question of when this is
possible is related to the issue of computational universality, which
we will not review here; rather, we consider optimal simulations in
which fast local operations are free.  \cite{Bennett02a} found the
optimal rate at which a two-qubit Hamiltonian can simulate another, if
the time evolution is interspersed by fast local unitaries that do not
involve ancilla systems.  \cite{VC02b} showed that adding local
ancilla systems improves this rate, but that classical communication
does not.  Hamiltonian simulation is further improved when we allow
the ancilla systems to contain entanglement that is used
catalytically\cite{VC02a}.

The question of optimally generating two-qubit unitary interactions
using a given nonlocal Hamiltonian was solved (without ancillas) in
\cite{VHC01} and the proof was greatly simplified in \cite{HVC02}.  A
more systematic approach to the problem was developed in
\cite{Khaneja01}, which considers systems of many qubits and applies
its techniques to nuclear magnetic resonance.  Recently, generic
gates on $n$ qubits were shown by \cite{Nielsen05} to require one- and
two-qubit Hamiltonians to be applied for $\cO(\exp(n))$ time.
Hopefully this work will lead to useful upper bounds on the strengths
of Hamiltonians, which so far have been difficult to obtain.

Finally, one can also consider the reverse problem of simulating a
nonlocal Hamiltonian or gate using standard resources such as cbits
and ebits.  This problem has so far resisted optimal solutions, except
in a few special cases, such as Gottesman's\cite{Got98} simulation of
the \cnot\ using $[c\ra c]+[c\la c]+[qq]$.  For general $d\times d$
unitary gates, a simple application of teleportation yields $2\log
d([c\ra c]+[c\la c] + [qq]) \geq \<U\>$ \,\cite{Chefles00} (and see
also \prop{U-TP-bound}).
Unfortunately this technique cannot be used to efficiently simulate
evolution under a nonlocal Hamiltonian for time $t$, since allowing
Alice and Bob to intersperse fast local Hamiltonians requires breaking
the simulated action of $H$ into $t/\epsilon$ serial uses of
$e^{-iH\eps}$ for $\eps\ra 0$.  This ends up requiring classical
communication on the order of $t^2$ in order to achieve constant
error.  However, we would like the cost of a simulation to be linear
in the time the Hamiltonian is applied, so that we can discuss {\em
simulation rates} that are asymptotically independent of the time the
Hamiltonian is applied.  If classical communication is given for free,
then \cite{CDKL01} shows how to simulate a general Hamiltonian for time
$t$ using $\cO(t)$ entanglement.  This result was improved by
Kitaev\cite{Kit04}, who showed how to use $\cO(t) ([q\ra q]+[q\la q])$
to simulate a Hamiltonian for time $t$.  However, though these
constructions are efficient, their rates are far from optimal.

\subsection{Schmidt decompositions of states and operators}
\label{sec:u-schmidt}
Here we review the familiar Schmidt decomposition of bipartite quantum
states\cite{Peres93,NC00}, and explain the analogous, but less
well-known, operator Schmidt decomposition for bipartite
operators\cite{Nielsen98}. \index{Schmidt decomposition}

\begin{proposition}[Schmidt decomposition]
Any bipartite pure state $\ket{\psi}\in\cH_A \ot
\cH_B$ can be written as $\ket{\psi}=\sum_{i=1}^m \sqrt{\lambda_i}
\ket{\alpha_i}_{\A} \ket{\beta_i}_{\B}$, where $\lambda_i>0$, $\sum_i
\lambda_i=1$ (i.e. $\lambda$ is a probability distribution with full
support) and $\ket{\alpha_i}\in\cH_{\A}$ and
$\ket{\beta_i}\in\cH_{\B}$ are orthogonal sets of vectors
(i.e. $\braket{\alpha_i}{\alpha_{i'}} =
\braket{\beta_i}{\beta_{i'}}=\delta_{ii'}$).  Since these vectors are
orthogonal, $m \leq \min(\dim\HA,\dim\HB)$.

Furthermore, the Schmidt rank $\Sch(U):=m$ is unique, as are the
Schmidt coefficients $\lambda_i$, up to a choice of ordering.
Therefore unless otherwise specified we will take the $\lambda_i$ to
be nonincreasing.  Also, for any other decomposition
$\ket{\psi}=\sum_{i=1}^l \ket{\alpha_i'}_A
\ket{\beta_i'}_B$ (with $\ket{\alpha_i'},\ket{\beta_i'}$ not
necessarily orthogonal or 
normalized), we must have $l \geq \Sch(U)$.
\end{proposition}

Our proof follows the approach of \cite{NC00}.

\begin{proof}
The key element of the proof is the singular value decomposition
(SVD).  Choose orthonormal bases $\{\ket{j}\}_{1\leq j \leq d_A}$ and
$\{\ket{k}\}_{1 \leq k \leq d_B}$ for
$\HA$ and $\HB$ respectively, where $d_A=\dim\HA$ and $d_B=\dim\HB$.
Then $\ket{\psi}$ can be written as $\ket{\psi} = \sum_{j,k}
a_{jk}\ket{j}_A\ket{k}_B$, where $a$ is a $d_A\times d_B$ matrix.
The SVD states that there exists a set of positive numbers
$\sqrt{\lambda_1},\ldots,\sqrt{\lambda_m}$ and
isometries $u:\bbC^m\ra\bbC^{d_B}$ and $v:\bbC^{d_A}\ra\bbC^m$ such
that $a = u \cdot \sqrt{\diag(\vec{\lambda})} \cdot v$.  Let
$\ket{\alpha_i}_{\A}:= 
\sum_j u_{ji} \ket{j}_{\A}$ and $\ket{\beta_i}_{\B} := \sum_k
v_{ik}\ket{k}_{\B}$.  Since $u$ and $v$ are isometries, it follows that
$\{\ket{\alpha_i}\}$ and $\{\ket{\beta_i}\}$ are orthonormal sets.

To prove the second set of claims, note that the Schmidt coefficients
are just the singular values of the matrix $a_{jk}=\bra{\psi}\cdot
\ket{j}\ket{k}$; since singular values are unique, so are Schmidt
coefficients.  Finally if $\ket{\psi}=\sum_{i=1}^l \ket{\alpha_i'}_A
\ket{\beta_i'}_B$, then $a_{jk} = \sum_{i=1}^l \braket{\alpha_i'}{j}
\braket{\beta_i'}{k}$ and $\Sch(\psi) = \rank a \leq l$.
\end{proof}

{\em Schmidt decomposition and entanglement manipulation:} The Schmidt
coefficients are central to the study of bipartite pure state
entanglement.  For example, two states can be transformed into one
another via local unitary transformations if and only if they have the
same Schmidt coefficients.  Thus, we usually choose entanglement
measures on pure states to be functions only of their Schmidt
coefficients.  

Moreover, the intuitive requirement that entanglement be
nonincreasing under local operations and classical communication (LOCC) is equivalent to the mathematical requirement
that entanglement measures be Schur-concave functions of a state's
Schmidt coefficients.  (A function $f:\bbR^n\ra\bbR$ is Schur-concave
iff $v\prec w \Rightarrow f(v)\geq f(w)$\cite{Bhatia97}).   The proof
is as follows:
Suppose a bipartite pure state
$\ket{\psi}$ can be transformed by LOCC into $\ket{\varphi_i}$ with
probability $p_i$ (i.e. the state $\sum_i p_i
\ket{ii}\bra{ii}_{A_1B_1} \otimes
\ket{\varphi_i}\bra{\varphi_i}_{A_2B_2}$).  Then \cite{Nielsen99a}
showed that this transformation is possible if and only if there exist
$\vec{\lambda}$ and $\vec{\mu}_i$ such that $\vec{\lambda}=\sum_i p_i
\vec{\mu}_i$ where $\vec{\lambda}$ is the set of
Schmidt coefficients of $\ket{\psi}$ (ordered arbitrarily) and the
$\vec{\mu_i}$ are the Schmidt coefficients for $\ket{\varphi_i}$
(again in an arbitrary ordering).  As a consequence, if
$E(\ket{\psi})$ is an entanglement measure that is a Schur-concave
function of the Schmidt coefficients of
$\ket{\psi}$, then the expectation of $E$ is nonincreasing under LOCC;
i.e. $E(\ket{\psi})
\geq \sum_i p_i E(\ket{\varphi_i})$\cite{Nielsen99a,Nielsen99b}.  This
general principle unifies many results about entanglement not
increasing under LOCC.  If we take $E$ to be the standard entropy of
entanglement $E(\psi)=H(\tr_B \psi)$, then we find that its
expectation doesn't increase under LOCC; similarly for the min-entropy
$E_\infty(\psi)=-\log \| \tr_B
\psi\|_\infty$.  Since $(E_0(\psi))^\alpha=\Sch(\psi)^\alpha$ is Schur-concave
for all $\alpha\geq 0$, we also find that Schmidt number has zero
probability of increasing under any LOCC transformation (of course,
this result follows more directly from the relation $\vec{\lambda}=\sum_i p_i
\vec{\mu}_i$).

{\em Operator-Schmidt decomposition:}  A similar Schmidt decomposition
exists for bipartite linear operators $M\in
\cL(\HA\ot\HB)$\footnote{We use Nielsen's definition of operator
Schmidt number from \cite{Nielsen98}.  In \cite{Terhal00}, Terhal and
Horodecki defined an alternative notion of Schmidt number for
bipartite density matrices which we will not use.}.  Define the
Hilbert-Schmidt inner product on $\cL(\cH)$ by $(X,Y):=\tr X^\dag Y /
\dim\cH$ 
for any $X,Y\in\cL(\cH)$.  For example, a complete orthonormal basis
for the space of one-qubit operators is the set of Pauli
matrices, $\{I,X,Y,Z\}$.

Let $d_A=\dim\HA$ and $d_B=\dim\HB$.  Then any $M\in\cL(\HA\ot\HB)$
can be Schmidt decomposed into
\be M = \sum_{i=1}^{\Sch(M)} \sqrt{\lambda_i} A_i \otimes B_i\ee
where $\Sch(M)\leq \min(d_A^2,d_B^2)$, $\tr A_i^\dag A_j =
d_A\delta_{ij}$ and $\tr B_i^\dag B_j = \delta_{ij}$.  Normalization
means that $\tr M^\dag M = d_Ad_B
\sum_i\lambda_i$; typically $M$ is unitary, so $\sum_i \lambda_i =
\tr M^\dag M / d_Ad_B = 1$.

A simple example is the \cnot\ gate which has operator-Schmidt
decomposition
\begin{equation}
\cnot=\frac{1}{\sqrt2}\,\oprod{0}\ot I + \frac{1}{\sqrt{2}}\,
\oprod{1}\ot X
\end{equation}
and hence has Schmidt coefficients $\{1/\sqrt2,1/\sqrt2\}$, and
$\Sch(\cnot)=2$.  The \swap\ gate for qubits has operator-Schmidt
decomposition
\begin{equation}
\swap=\frac{1}{4}\l(I\ot I+ X\ot X + Y\ot Y + Z\ot Z\r)
\end{equation}
and hence $\Sch(\swap)=4$.  

Most facts about the Schmidt decomposition for bipartite states carry
over to bipartite operators: in particular, if $M=A_1\otimes B_1+
\ldots + A_m\otimes B_m$, then $m\geq \Sch(M)$.  This implies a useful
lemma (originally due to \cite{Nielsen98}, but further discussed in
\cite{Nielsen03a}):
\begin{lemma}[Submultiplicity of Schmidt number]
\label{lemma:u-sch-mult}
Let $U$ and $V$ be bipartite operators and $\ket{\psi}$ a bipartite state.
Then
\begin{enumerate}
\item $\Sch(UV) \leq \Sch(U)\Sch(V)$
\item $\Sch(U\ket{\psi}) \leq \Sch(U)\Sch(\ket{\psi})$
\end{enumerate}
\end{lemma}
\begin{proof}
If $U=\sum_i \sqrt{u_i} A_i\otimes B_i$ and $V=\sum_j \sqrt{v_j}
C_j\otimes D_j$ are Schmidt decompositions, then $UV = \sum_{i,j}
\sqrt{u_iv_j} A_iC_j \otimes B_iD_j$ is a decomposition of $UV$ into
$\Sch(U)\Sch(V)$ terms.  Therefore $\Sch(UV)\leq\Sch(U)\Sch(V)$.

Claim (b) is similar.  If
$\ket{\psi}=\sum_j\sqrt{\lambda_j}\ket{a_j}\otimes \ket{b_j}$, then
$U\ket{\psi} = \sum_{i,j} \sqrt{u_i\lambda_j} A_i\ket{a_j} \otimes
B_i\ket{b_j}$ is a decomposition with $\Sch(U)\Sch(\ket{\psi})$ terms.
Thus $\Sch(U\ket{\psi})\leq\Sch(U)\Sch(\ket{\psi})$.
\end{proof}

Part (b) of the above Lemma provides an upper bound for how quickly
the Schmidt number of a state can grow when acted on by a
bipartite unitary gate.  It turns out that this bound is saturated when
the gate acts on registers that are maximally entangled with local
ancilla systems.  This is proven
by the next Lemma, a simple application of the Jamiolkowski
state/operator isomorphism\cite{Jamiolkowski72} that was first pointed
out by Barbara Terhal in an unpublished comment.
\begin{lemma}\label{lemma:jam-schmidt}
Given Hilbert spaces $\A,\A',\B,\B'$ with $d_A:=\dim\A=\dim\A'$ and
$d_B:=\dim\B=\dim\B'$, let $M\in\cL(\HA\ot\HB)$ have Schmidt
decomposition $M=\sum_i \sqrt{\lambda_i} A_i\ot B_i$.  Then the state
$\ket{\Phi(M)} := (M_{AB} \ot I_{A'B'}) \ket{\Phi_{d_A}}_{AA'}
\ket{\Phi_{d_B}}_{BB'}$ also has Schmidt coefficients $\{\lambda_i\}$.
\end{lemma}

\begin{proof}
If we define $\ket{a_i} = (A_i \ot I)\ket{\Phi_{d_A}}$ and
$\ket{b_i} = (B_i \ot I)\ket{\Phi_{d_B}}$, then 
$\ket{\Phi(M)}$ can be written as 
\be \ket{\Phi(M)} = \sum_i \sqrt{\lambda_i}
\ket{a_i}\ket{b_i}.\label{eq:jam-schmidt}\ee  
Note that $\braket{a_i}{a_j} =  \tr (A_i^\dag A_j \otimes I)
\Phi_{d_A} = \tr A_i^\dag A_j / d_A = \delta_{ij}$ and similarly
$\braket{b_i}{b_j} = \delta_{ij}$.  Thus \eq{jam-schmidt} is a Schmidt
decomposition of $\ket{\Phi(M)}$, and since the Schmidt coefficients
are unique, $\ket{\Phi(M)}$ has Schmidt coefficients $\{\lambda_i\}$.
\end{proof}

\section{Entanglement capacity of unitary gates}\label{sec:u-e-cap}
In this section, we investigate the entanglement generating capacity
of a unitary interaction.  Fix a gate $U\in\cU_{d\times d}$ (the
generalization to $d_A\times d_B$ is straightforward) and let $\<U\>$
denote the corresponding asymptotic resource.\footnote{Note that our
definition of $\<U\>$ differs slightly from the definition in
\cite{BHLS02}; whereas \cite{BHLS02} allowed $n$ sequential uses of $U$
interspersed by local operations (i.e. the depth $n$ resource
$U^{\times n}$), we follow \defn{asy-ineq} and allow only $(U^{\ot
n/k})^{\times k}$ where $k$ depends only on the target inefficiency
and not the desired accuracy of the protocol.} Then we define the {\em
entanglement capacity} $E(U)$ 
to be the largest $E$ such that 
\be \<U\> \geq E [qq] \ee
For a bipartite pure state $\ket{\psi}^{AB}$, we will also use
$E(\ket{\psi})$ to indicate the entropy of entanglement of
$\ket{\psi}$; i.e. $H(A)_\psi = H(B)_\psi$ in the language of the last
chapter. 

We will start by stating some easily computable bounds on $E(U)$, then
prove a general expression for the capacity and conclude by discussing
some consequences. 

{\em Simple bounds on entanglement capacity:}  We can establish some
useful bounds on $E(U)$ merely by knowing the
Schmidt coefficients of $U$.   The following
proposition expresses these bounds.

\begin{proposition}\label{prop:U-nonzero-E}
If $U$ has Schmidt decomposition $U = \sum_i \sqrt{d_Ad_B\lambda_i}
A_i \otimes B_i$, then
\be H(\lambda) = \sum_i -\lambda_i \log \lambda_i
\leq E(U) \leq H_0(\lambda) = \log\Sch(U). \ee
\end{proposition}

\begin{proof}
\sloppypar{The lower bound follows from \lem{jam-schmidt} and entanglement
concentration (recall from \eq{E-concentration}  that if $\psi$ is a
bipartite pure 
state then $\<\psi\> \geq E(\psi)[qq]$).  Thus, $\<U\> \geq
\<\Phi(U)\> \geq H(\lambda)[qq]$, where $\Phi(U)$ is defined as in
\eq{jam-schmidt} and we have used the fact that
$E(\Phi(U))=H(\lambda)$.}

To prove the upper bound, we use
\lem{u-sch-mult} to show that $n$ uses of $U$ together with LOCC can
generate only mixtures of pure states with Schmidt number $\leq
\Sch(U)^n = \exp(nH_0(\lambda))$.  Since approximating
$\ket{\Phi}^{\ot nE(1-\delta)}$ to accuracy $\epsilon$ requires a
mixture of pure states with expected Schmidt number $\geq
(1-\epsilon)\exp(nE(1-\delta))$, asymptotically we must have
$H_0(\lambda)\geq E(U)$.
\end{proof}

As a corollary, any nonlocal $U$ has a
nonzero $E(U)$.  A similar, though less quantitative, result holds for
communication as well.

\begin{proposition}\label{prop:U-nonzero-CC}
If $U$ is nonlocal then $\<U\> \geq C[c\ra c]$ for some $C>0$.
\end{proposition}
We state this here since we will need it for the proof of
the next theorem, but defer the proof of \prop{U-nonzero-CC} until
\sect{u-c-cap} so as to focus on entanglement generation in this
chapter.


{\em General formula for entanglement capacity:} The main result on
the entanglement capacity is the following method of expressing it in
terms of a single use of $U$:

\begin{theorem}\label{thm:U-E-cap}
\be E(U) = \Delta E_U := \sup_{\ket{\psi}\in\cH_{\A\A'\B\B'}} E\l(
(U_{AB} \otimes I_{A'B'})\ket{\psi}\r) - E(\ket{\psi})
\label{eq:U-E-cap}\ee
where the supremum ranges over Hilbert spaces $A',B'$ of any finite
dimension.
\end{theorem}
In other words, the asymptotic entanglement capacity $E(U)$ is equal
to the largest {\em single-shot} increase of entanglement $\Delta
E_U$, if we are allowed to start with an arbitrary pure (possibly
entangled) state.  This result was independently obtained in
\cite{Leifer03} and is based on a similar result for Hamiltonians in
\cite{DVCLP01}.  Here we restate the proof of \cite{BHLS02} in the
language of asymptotic resources.  

\begin{proof}

{\em $E(U)\leq \Delta E_U$ [converse]:} Consider an arbitrary protocol
that uses $U$ $n$ times in order to generate $\approx_\epsilon
{\Phi}^{\otimes nE(U)(1-\delta)}$.  We will prove a stronger result,
in which even with unlimited classical communication $U$ cannot
generate more than $\Delta E_U$ ebits per use.  Since communication is
free, we assume that instead of discarding subsystems, Alice and Bob
perform complete measurements and classically communicate their
outcomes.  Thus, we always work with pure states.

Since LOCC cannot increase expected entanglement and Alice and Bob
start with a product state, their final state must have expected
entanglement $\leq n\Delta E_U$.  However, by Fannes' inequality
(\lem{fannes}) the output state must have entanglement $\geq
nE(U)(1-\delta)(1-\epsilon)-\eta(\epsilon)$.  Thus $\forall
\epsilon,\delta>0$ we can choose $n$ sufficiently large that
$E(U)(1-\delta)(1-\epsilon) - \eta(\epsilon)/n \leq \Delta E_U$,
implying that $E(U)\leq \Delta E_U$.\footnote{A more formal (and
general) version of this argument will also appear in the proof of
\thm{Ce-cap} in \sect{U-Ce-cap}.}

{\em $E(U) \geq \Delta E_U$ [coding theorem]:} Assume $\Delta E_U >0$;
otherwise the claim is trivial.  Recall from
\eqs{E-concentration}{E-dilution} our formulation of entanglement
concentration $\<\psi\> \geq E(\psi) [qq]$ and dilution $E(\psi) [qq]
+ o[c\ra c]
\geq \<\psi\>$.  
Then
\be \<U\> + E(\psi)[qq] \geq 
\<U\> + o [c\ra c] + E(\psi)[qq] \geq 
\<U\> + \<\psi\> \geq \< U(\psi) \>
\geq E(U\ket{\psi}) [qq], \ee
where we have used \prop{U-nonzero-CC} in the first inequality,
entanglement dilution in the second inequality, and entanglement
concentration in the last inequality.  
Using the Cancellation Lemma (\ref{lemma:cancel}), we find that $\<U\>
+ o [qq] \geq E(U\ket{\psi}) - E(\ket{\psi}) [qq]$, and the sublinear
$[qq]$ term can be removed due to
\prop{U-nonzero-E} and the fact that $\Delta E_U>0$ implies
$\Sch(U)>1$.  Thus $\<U\> \geq E(U\ket{\psi}) -
E(\ket{\psi}) [qq]$ for all $\psi$.  Taking the supremum over $\psi$
and using the Closure Lemma (\ref{lemma:closure}) yields the desired
result.
\end{proof}

The problem of finding $E(U)$ is now reduced to calculating the
supremum in \eq{U-E-cap}.  To help understand the properties of
\eq{U-E-cap}, we now consider a number of possible variations on it,
as well as some attempts at simplification.

\begin{itemize}
\item {\em Restricting the size of the ancilla appears hard:}  Solving
\eq{U-E-cap} requires optimizing over ancilla systems $\A'$ and $\B'$
of unbounded size.  Unfortunately, we don't know if the supremum is
achieved for any finite dimensional ancilla size, so we can't give an
algorithm with bounded running time that reliably approximates $E(U)$.
On the one hand, we know that ancilla systems are sometimes necessary.
The two-qubit \swap\ gate can generate no entanglement without
entangled ancillas, and achieves its maximum of 2 ebits when acting on
$\ket{\Phi}_{\A\A'}\ket{\Phi}_{\B\B'}$; a separation that is in a
sense maximal.  On the other hand, some gates, such as $\cnot$, can
achieve their entanglement capacity with no ancillas.  Less trivially,
\cite{CLVV03} proved that two-qubit Hamiltonians of the form $H=\alpha
X\otimes X + \beta Y\otimes Y$ can achieve their entanglement capacity
without ancilla systems, though this no longer holds when a
$Z \otimes Z$ term is added.

It is reasonable to assume that even when ancilla are necessary, it
should suffice to take them to be the same size as the input systems.
Indeed, no examples are known where
achieving the entanglement capacity requires $\dim\A'>\dim\A$ or
$\dim\B'>\dim\B$.  On the other hand, there is no proof that the
capacity is achieved for any finite-dimensional ancilla; we cannot
rule out the possibility that there is only an infinite sequence of
states that converges to the capacity.

\item{\em Infinite dimensional ancilla don't help:} Though we cannot put an
upper bound on the necessary dimensions of $\A'$ and $\B'$, we can
assume that they are finite dimensional.  In other words, we will show
that $\Delta E_U$
is unchanged if we modify
the $\sup$ in \eq{U-E-cap} to optimize over
$\ket{\psi}\in\cH_{\A\A'\B\B'}$ s.t. $E(\psi)<\infty$ and
$\dim\cH_{\A'} = \dim\cH_{\B'} = \infty$.  Denote this modified
supremum by $\Delta E_U'$.  We will prove that $\Delta E_U=\Delta
E_U'$.  

First, we state a useful lemma.
\begin{lemma}\label{lemma:inf-dim}
 Any bipartite state $\ket{\psi}$ with
$E(\psi)<\infty$ can be approximated by a series of states
$\ket{\varphi_1},\ket{\varphi_2},\ldots$, each with finite Schmidt
number and obeying
$\|\psi-\varphi_n\|_1 \log\Sch(\varphi_n) \ra 0$ as $n\ra\infty$.  (In
other words, the error converges to zero faster than
$1/\log\Sch(\varphi_n)$.)
\end{lemma}
\begin{proof}
Schmidt decompose $\ket{\psi}$ as $\ket{\psi}=\sum_{i=1}^\infty
\sqrt{\lambda_i} \ket{i}\ket{i}$ and define the normalized state
$\ket{\varphi_n} = 
\sum_{i=1}^n \sqrt{\lambda_i}\ket{i}\ket{i}/
\sqrt{\sum_{i=1}^n\lambda_i}$.  Let $\delta_n :=
\smfrac{1}{2}\|\psi-\varphi_n\|_1 =
\sum_{i>n} \lambda_i$.  Now, use the fact that $E(\psi)<\infty$ and
$\lambda_n \leq 1/n$ to obtain
\be
E(\psi) - \sum_{i=1}^{n} \lambda_i \log (1/\lambda_i)
= \sum_{i=n+1}^{\infty} \lambda_i \log (1/\lambda_i)
\geq \sum_{i=n+1}^{\infty} \lambda_i \log (1/\lambda_n)
= \delta_n \log(1/\lambda_n)
\geq \delta_n \log n
\ee
Since the term on the left converges to 0 as $n\ra\infty$, we also
have that $\delta_n\log n\ra 0$ as $n\ra \infty$.  Using
$n=\Sch(\varphi_n)$ and $\delta_n = \smfrac{1}{2}\|\psi-\varphi_n\|_1$,
our desired result follows.
\end{proof}

Now  $\forall \epsilon>0, \exists 
\ket{\psi}\in\cH_{\A\A'\B\B'}$ with $\dim \cH_{A'}=\dim \cH_{B'}=\infty$
such that $E(U\ket{\psi}) - E(\ket{\psi}) > \Delta E_U' -
\epsilon$.  By \lem{inf-dim}, we can choose $\ket{\varphi}$ with
$\Sch(\varphi)< \infty$ (and thus can belong to $\cH_{\A\A'\B\B'}$
with $\dim\A', \dim\B' < \infty$) such that
$\|\psi-\varphi\|_1 \log\Sch(\varphi) \leq \epsilon$.
By Fannes' inequality (\lem{fannes}), $|E(\psi)-E(\varphi)|\leq \epsilon +
\eta(\epsilon)/\log \Sch\varphi$ and
$|E(U\ket{\psi})-E(U\ket{\varphi})|\leq (\epsilon + 
\eta(\epsilon))(1 + (\log \Sch(\varphi))/(\log\Sch(U)))$ (since
$\Sch(U\ket{\varphi}) \leq \Sch(U)\Sch(\varphi)$).
Combining these, we
find that $E(U\ket{\varphi})-E(\ket{\varphi}) \ra \Delta E_U'$ as
$\epsilon\ra 0$, implying that $\Delta E_U = \Delta E_U'$. 

\item{\em Sometimes it helps to start with entanglement:} Subtracting
one entropy from another in \eq{U-E-cap} is rather ugly; it would be
nice if we could eliminate the second term (and at the same time
restrict $\dim \A'\leq \dim\A$ and $\dim \B'\leq \dim\B$) by
maximizing only over product state inputs.  However, this would result
in a strictly lower capacity for some gates.  This is seen most
dramatically for Hamiltonian capacities, for which
$\smfrac{d}{dt}E(e^{-iHt}\ket{\alpha}\ket{\beta}) =0$ for any
$\ket{\alpha}\in \cH_{AA'}, \beta\in\cH_{\B\B'}$, due to the quantum
Zeno effect: after a small amount of time $t$, the largest Schmidt
coefficient is $1-\cO(t^2)$.  The same principle applies to the gate
$U=e^{-iHt}$ for $t$ sufficiently small: the entanglement capacity is
$\cO(t)$ (because $\cO(1/t)$ uses of $U$ give a gate far from the
identity with $\cO(1)$ entanglement capacity), though the most
entanglement that can be created from unentangled inputs by one use of
$U$ is $\cO(t^2\log (1/t))$.

As a corollary, the lower bound of \prop{U-nonzero-E} is not tight for
all gates.

\item{\em Mixed states need not be considered:}  We might also
try optimizing over density matrices rather than pure states.  For
this to be meaningful, we need to replace the entropy of entanglement
with a measure of mixed-state entanglement\cite{BDSW96}, such as
entanglement of 
formation $E_f(\rho) := \min\{\sum_i p_i E(\psi_i) : \rho=\sum_i p_i
\psi_i\}$, entanglement cost $E_c(\rho) := \inf_m
\frac{1}{m}E_f(\rho^{\otimes m}) = \inf\{e : e[qq] + \infty [c\ra c]
\geq \<\rho\>\}$, or distillable entanglement
$D(\rho) := \sup\{e :\<\rho\> + \infty [c\ra c] + \infty
[c\la c] \geq e [qq]\}$ \cite{BDSW96}.

We claim that $\Delta E_U = \sup_\rho E_f(U(\rho))-E_f(\rho)
= \sup_\rho E_c(U(\rho))-E_c(\rho)
= \sup_\rho D(U(\rho))-E_c(\rho)$.  To prove this for $E_f$, decompose
an arbitrary $\rho^{AB}$ into pure states as $\rho=\sum_i p_i \psi_i$ s.t. $E_f(\rho)=\sum_i
p_i E(\psi_i)$.  Now we use the convexity of $E_f$ to show that
$E_f(U(\rho)) = E_f(\sum_i p_iU(\psi_i)) \leq \sum_i p_i
E(U(\psi_i))$, implying that  
$$E_f(U(\rho))-E_f(\rho) \leq \sum_i p_i
\l[E(U(\psi_i)) - E(\psi_i)\r] \leq \max_i E(U(\psi_i)) - E(\psi_i).$$
Thus, any increase in $E_f$ can be achieved by a pure state.

A similar, though slightly more complicated, argument applies for
$E_c$.  For any $\epsilon>0$ and any 
$\rho$, there exists $m$ sufficiently large that $E_c(\rho) +
\epsilon \geq
\frac{1}{m}\sum_i p_i E(\psi_i)$ for some $\{p_i,\psi_i\}$
such that $\rho^{\otimes m} = \sum_i p_i\psi_i$.  Using first the
definition of $E_c$ and then convexity, we have $E_c(U(\rho))
\leq \smfrac{1}{m}E_f(U(\rho)^{\otimes m}) \leq \smfrac{1}{m} 
\sum_i p_i E(U^{\otimes m}(\psi_i))$.  Thus,
\begin{gather*}
E_c(U(\rho))-E_c(\rho)-\epsilon \leq \frac{1}{m}\sum_i p_i \l[
E(U^{\otimes m}(\psi_i)) - E(\psi_i)\r] \leq \max_i (E(U^{\otimes
m}(\psi_i)) - E(\psi_i))/m
\\ \leq \max_i \max_{j\in
\{1,\ldots,m\}} E((U^{\otimes j}\otimes I^{\otimes m-j})(\psi_i)) -
E((U^{\otimes j-1}\otimes I^{\otimes m-j+1})(\psi_i)) \leq \Delta
E_U.
\end{gather*}
  This proof implicitly uses the fact that $E(U)$ is (sub)additive;
i.e. $E(U^{\otimes 2}) = 2E(U)$.

Finally, $\Delta E_U = \sup_\rho D(U(\rho)) - E_c(\rho)$ because of
the $E_c$ result from the last paragraph and the fact that $D(\rho)
\leq E_c(\rho)$.  This case corresponds to the operationally
reasonable scenario of paying $E_c(\rho)[qq]$ for the input state and
getting $D(U(\rho))[qq]$ from the output state.  Of course, this case
also follows from the fact that classical communication
doesn't help entanglement capacity.
\end{itemize}

{\em Contrasting the entanglement capacity of unitary gates and noisy
quantum channels:}  The problem of generating entanglement with a
unitary gate turns out to have a number of interesting differences
from the analogous problem of using a noisy quantum channel to share
entanglement.  Here we survey some of those differences.
\begin{itemize}
\item{\em Free classical communication doesn't help:}  In the proof of
the converse of \thm{U-E-cap}, we observed that unlimited classical
communication in both directions doesn't increase the entanglement
capacity.  For noisy quantum channels, it is known that forward
communication doesn't change the entanglement
capacity\cite{BDSW96}, though in some cases back
communication can improve the capacity (e.g. back communication
increases the capacity of the 50\% erasure channel from zero to $1/2$)
and two-way communication appears to further improve the
capacity\cite{BDSS04}.
\item{\em Quantum and entanglement capacities appear to be different:}
A noisy quantum channel $\cN$ has the same capacity to send quantum
data that it has to generate entanglement (i.e. $\<\cN\>\geq Q[q\ra
q]$ iff $\<\cN\>\geq Q[qq]$)\cite{BDSW96}, though with free classical
back communication this is no longer thought to hold\cite{BDSS04}.
Since unitary gates are intrinsically bidirectional, we might instead
ask about their total quantum capacity $Q_+(U) := \max\{Q_1 + Q_2 :
\<U\> \geq Q_1 [q\ra q] + Q_2[q\la q]\}$ and ask whether it is equal
to $E(U)$.  All that is currently known is the bound $Q_+(U)\leq
E(U)$, which is saturated for gates like \cnot\ and \swap.  However,
in
\sect{CnE-example}, I will give an example of a gate that
appears to have 
$Q_+(U) < E(U)$, though this conjecture is supported only by heuristic
arguments.
\item{\em Entanglement capacities are strongly additive:} For any two
bipartite gates $U_1$ and $U_2$, we have $E(U_1\ot U_2) \geq E(U_1) +
E(U_2)$, since we can always run the optimal entanglement generating
protocols of $U_1$ and $U_2$ in parallel.  On the other hand,
$E(U_1\ot U_2) = \sup_\psi E((U_1\ot U_2)\ket{\psi})-E(\ket{\psi}) =
\sup_\psi \l[E((U_1\ot U_2)\ket{\psi}) - E((U_1\ot I)\ket{\psi})\r] +
\l[ E((U_1\ot I)\ket{\psi}) - E(\ket{\psi})\r] \leq \Delta E_{U_2} +
\Delta E_{U_1} = E(U_2) + E(U_1)$.  Thus $E(U_1\ot U_2) =E(U_1) +
E(U_2)$.

In contrast, quantum channel capacities (equivalently either for
quantum communication or entanglement generation) appear to be
superadditive\cite{SST01}.
\item{\em Entanglement capacities are always nonzero:}  If $U$ is a
nonlocal gate (i.e. cannot be written as $U=U_A\ot U_B$), then
according to \prop{U-nonzero-E}, $E(U)>0$.  On the other hand,
there exist nontrivial quantum channels with zero entanglement
capacity: classical channels cannot create entanglement and bound
entangled channels cannot be simulated classically, but also cannot
create any pure entanglement.
\end{itemize}

\section{Classical communication capacity}\label{sec:u-c-cap}
Nonlocal gates can not only create entanglement, but can also send
classical messages both forward (from Alice to Bob) and backwards
(from Bob to Alice).  Therefore, instead of a single capacity, we need
to consider an achievable classical rate region.  Define $\CC(U) :=
\{(C_1,C_2) : \<U\> \geq C_1 [c\ra c] + C_2 [c\la c]\}$.  Some useful special
cases are the forward capacity $C_\ra(U) = \max\{C_1:
(C_1,0)\in\CC(U)\}$, backward capacity $C_\la(U) = \max\{C_2:
(0,C_2)\in\CC(U)\}$ and bidirectional capacity $C_+(U) = \max\{C_1+C_2:
(C_1,C_2)\in\CC(U)\}$.  (By \lem{continuity} $\CC(U)$ is a closed
set, so these maxima always exist.)

We can also consider the goal of simultaneously transmitting classical
messages and generating entanglement.  Alternatively, one might want
to use some entanglement to help transmit classical messages.  We
unify these scenarios and others by considering the three-dimensional
rate region $\CCE(U):= \{(C_1,C_2,E) : \<U\> \geq C_1 [c\ra c] + C_2
[c\la c] + E [qq]\}$.  When some of $C_1,C_2$ and $E$ are negative, it
means that the resource is being consumed; for example, if $E<0$ and
$C_1,C_2\geq 0$, then the resource inequality $\<U\> + (-E) [qq] \geq
C_1 [c\ra c] + C_2 [c\la c]$ represents entanglement-assisted
communication.  Some useful limiting capacities are $C_\ra^E(U) :=
\max\{C_1 : (C_1,0,-\infty)\in\CCE(U)\}$, $C_\la^E(U) :=
\max\{C_2 : (0,C_2,-\infty)\in\CCE(U)\}$ and $C_+^E(U) :=
\max\{C_1+C_2 : (C_1,C_2,-\infty)\in\CCE(U)\}$.

\begin{figure}[ht]
$$
\setlength{\unitlength}{0.8mm}
\begin{picture}(172,62)(-50,0)
\put(100,0){\textcolor{red}{.}}
\put(-100,0){\textcolor{red}{.}}
\put(-50,0){\textcolor{red}{.}}
\put(10,10){\vector(1,0){45}}
\put(10,10){\vector(0,1){45}}
\put(10,51){\line(1,-1){41}}
\qbezier(10,40)(25,40)(29,29)
\qbezier(29,29)(32,20)(33,10)
\put(26,35){\circle*{2}}
\put(33,10){\circle*{2}}
\put(10,40){\circle*{2}}
\put(56,9){\makebox{$C_1$}}
\put(8,56){\makebox{$C_2$}}
\put(40,21){\makebox{$C_1 \! + \! C_2 \! = C_+$}}
\put(32,5){\makebox{$C_\ra$}}
\put(3,42){\makebox{$C_\la$}}
\end{picture}$$
\caption{
Example of a possible achievable rate region $\CC(U)$, with the
limiting capacities of $C_\ra, C_\la$ and $C_+$ indicated.}
\label{fig:unit-cap-example}
\end{figure}
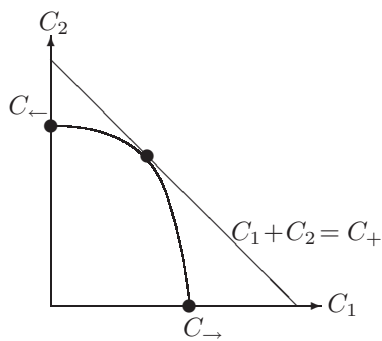

To get a sense of what these capacity regions can look like,
 \fig{unit-cap-example} contains
 a schematic diagram for the achievable region $\CC(U)$ and the
 definitions of the various capacities when we set $E=0$.  We present
 all the {\em known} properties and intentionally show the features
 that are not ruled out, such as the asymmetry of the region, and the
 nonzero curvature of the boundary.

There are much simpler examples -- the unassisted achievable region
for {\sc cnot} and {\sc swap} are similar triangles with vertices
$\{(0,0), (0,1), (1,0)\}$ and $\{(0,0), (0,2), (2,0)\}$ respectively
(see \sect{u-2qubit-examples}).

In general, little is known about the unassisted achievable region of
$(C_1,C_2)$ besides the convexity and the monotonicity of its
boundary.
The most perplexing question is perhaps whether the region has
reflective symmetry about line $C_1 = C_2$, which would imply that
$C_\ra(U) = C_\la(U)$.
\eq{magic-basis} shows that any two-qubit gate or Hamiltonian
is locally equivalent to one with Alice and Bob interchanged, so that
the achievable region is indeed symmetric.
In higher dimensions, on the other hand, \cite{Bennett02a} shows that there are
Hamiltonians (and so unitary gates) that are intrinsically asymmetric.
However, it remains open whether the achievable rate pairs are
symmetric, or more weakly, whether $C_\ra=C_\la$.

The rest of this section is as follows:
\begin{description}
\item[\sect{U-CCE-general}] proves some basic facts about the achievable
classical communication region.  Then we establish some bounds on
communication rates similar to, but weaker than, the bounds on
entanglement rate in \prop{U-nonzero-E}.
\item[\sect{U-lglass}] proves a capacity
formula for $C_\ra^E(U)$ (or equivalently $C_\la^E(U)$) that parallels
the formula in \thm{U-E-cap}.  This formula will be improved in the
next chapter when we introduce coherent classical communication.
\item[\sect{U-relations}] discusses relations between the classical
communication and the entanglement generation capacities of unitary
gates. 
\item[\sect{U-bidi-challenges}] explores the difficulties involved in
proving capacity theorems for bidirectional communication.
\end{description}

\subsection{General facts about the achievable classical communication
rate region}\label{sec:U-CCE-general}
We begin with some basic facts about $\CCE$.
\begin{itemize}
\item {\em Monotonicity: If $(C_1,C_2,E)\in\CCE(U)$ then $(C_1-\delta_1,
C_2-\delta_2, E-\delta_3)\in\CCE(U)$ for any
$\delta_1,\delta_2,\delta_3 \geq 0$.}  This is because we can always
choose to discard resources.

\item {\em Convexity: $\CCE(U)$ is a convex set.}  This follows from
time-sharing (part 2 of \thm{composability} and part 3 of \lem{mult-RI}.

\item {\em Classical feedback does not help: If
$(C_1,C_2,E)\in\CCE(U)$, then $(C_1,0,E)\in\CCE(U)$ and
$(0,C_2,E)\in\CCE(U)$.}  We mention this fact now, but defer the proof
until \chap{ccc}.

Combining this with monotonicity and the fact that classical feedback
doesn't improve entanglement capacity, we obtain as a corollary that
$\CCE(U) \subseteq [-\infty, C_\ra^E(U)] \times [-\infty, C_\la^E(U)]
\times [-\infty, E(U)] \subseteq [\infty,2\log d] \times [\infty,2\log
d] \times [\infty,2\log d]$.  This second inclusion depends on
\prop{U-TP-bound}, proven below.

\item {\em No more than $E(U^\dag)$ ebits are ever needed: If
$(C_1,C_2,E)\in\CCE(U)$, then $(C_1,C_2,-E(U^\dag))\in\CCE(U)$.}
A proof of this will be sketched in \sect{U-relations}, and it also
follows from \thm{bidi-ccc} in the next chapter.

\item {\em Shared randomness does not help: If $\<U\> + \infty [cc]
\geq C_1[c\ra c] + C_2[c\la c] + E[qq]$, then
$(C_1,C_2,E)\in\CCE(U)$.}

This is due to a standard derandomization argument (further
developed in \cite{CK81,DW05}).  Let $r$ denote the shared randomness and
let $x:=(a,b)$ run over all possible messages sent by Alice and Bob with
$n$ uses of $U$ (a set of size $\leq\exp(Cn)$ for $C:=C_1+C_2$).  If
$e_{x,r}$ is the corresponding probability of error, then our error-correcting
condition is that $\max_x
\bbE_r e_{x,r} \leq \epsilon$.  Now sample $m$ copies of the shared
randomness, $(r_1,\ldots,r_m)=:\vec{r}$, where $m$ is a parameter we
will choose later.  According to Ho\"effding's
inequality\cite{Hoeffding63}, we have 
\be \Pr_{\vec{r}}\l[ \frac{1}{m}\sum_{i=1}^m e_{x,r} \geq 2\epsilon
\r] \leq \exp(-m\epsilon^2/2), \ee
for any particular value of $x$.
We apply the union bound over all $\leq\exp(Cn)$ values of $x$ to
obtain
\be \Pr_{\vec{r}}\l[\max_x  \frac{1}{m}\sum_{i=1}^m e_{x,r} \geq 2\epsilon
\r] \leq \exp(Cn-m\epsilon^2/2). \ee
Thus, if we choose $m>2Cn/\epsilon^2$, then there exists a choice of
$\vec{r}$ with maximum error $\leq 2\epsilon$.  If Alice and Bob
preagree on $\vec{r}$, then they need only $\log m$ bits of shared
randomness to agree on which $r_i$ to use.  Since $\log m = \cO(\log n +
\log(1/\epsilon))$, this randomness can be generated by a negligible
amount of extra communication.
\end{itemize}

We now state an upper bound, originally due to \cite{Chefles00}.
\begin{proposition}\label{prop:U-TP-bound}
If $U\in \cU_{d\times d}$, then $C_\ra^E(U)\leq 2\log d$ and
$C_\la^E(U)\leq 2\log d$. 
\end{proposition}
\begin{proof}
The proof is based on simulating $U$ with teleportation: Alice
teleports her input to Bob using $2\log d[c\ra c] + \log d[qq]$, Bob
applies $U$ locally (and hence for free), and then Bob teleports
Alice's half of the state back using $2\log d[c\la c] + \log d[qq]$.
Thus we obtain the resource inequality
\be 2\log d\l([c\ra c] + [c\la c] + [qq]\r) \geq 
\log d \l([q\ra q] + [q\la q]\r) \geq \<U\>\ee
Allowing free entanglement and back communication yields $2\log d
[c\ra c] + \infty [q\la q] \geq C_\ra^E(U) [c\ra c]$.  
Causality\cite{Holevo73} implies that $C_\ra^E(U)\leq 2\log d$.  A
similar argument proves that $C_\la^E(U)\leq 2\log d$. 
\end{proof}

It is an interesting open question whether any good bounds on
classical capacity can be obtained as functions of a gate's Schmidt
coefficients, as we found with \prop{U-nonzero-E} for the case of
entanglement generation.

We now prove \prop{U-nonzero-CC}, which stated that any nonlocal $U$ has a
nonzero classical capacity.  An alternate proof can be found in
\cite{Beckman01}.

\begin{proof}[Proof of \prop{U-nonzero-CC}]
Let $E_0$ the amount of entanglement created by
applying $U$ to the $AB$ registers of
$|\Phi_d\>_{AA'}|\Phi_d\>_{BB'}$.  If $U$ is nonlocal, then $E_0>0$ according to \prop{U-nonzero-E}.

Alice can send a noisy bit to Bob with the following $t$-use protocol.
Bob inputs $|\Phi_d\>_{BB'}^{\otimes t}$ to all $t$ uses of $U$. 
To send ``$0$'' Alice inputs $|\Phi_d\>_{AA'}^{\otimes t}$ to share $t
E_0$ ebits with Bob, i.e. inputting a fresh copy of $|\Phi_d\>$ each
time.  To send ``$1$'', Alice inputs $|0\>_A$ to 
the first use of $U$, takes the output and uses it as the input to the
second use, and so on.  Alice only interacts a $d$-dimensional
register throughout the protocol, so their final entanglement is no more
than $\log d$.
Thus different messages from Alice result in very different amounts of 
entanglement at the end of the protocol.

Let $\rho_0$ and $\rho_1$ denote Bob's density matrices when Alice
sends 0 or 1 respectively.  Using Fannes'
inequality (\lem{fannes}), $t E_0 - \log d \leq \log d
\; \|\rho_{0}-\rho_{1}\|_1 + \frac{\log e}{e}$.  If we choose $t>(\log
d + \frac{\log e}{e})/E_0$, then $\rho_0 \neq \rho_1$ and Bob has a
nonzero probability of distinguishing $\rho_0$ from $\rho_1$ and thereby
identifying Alice's message.  Thus the $t$-use
protocol simulates a noisy classical channel with nonzero
capacity and $C_{\ra}(U) > 0$.  
\end{proof}

\subsection{Capacity theorem for entanglement-assisted one-way
classical communication}\label{sec:U-lglass}

We conclude the section with a general expression for $C_\ra^E(U)$.
Though we will improve it in \chap{ccc} to characterize the entire
one-way tradeoff region $\CE(U):=\{(C,E) : (C,0,E)\in\CCE(U)\}$, the
proof outlines useful principles which we will later use.

First, we recall some notation from our definition of remote state
preparation (RSP) in \sect{known}.  Let
\be \cE = \sum_i p_i \oprod{i}^{X_A} \ot \oprod{\psi_i}^{A_1A_2B_1B_2}
\label{eq:cE-bipartite}\ee
be an ensemble of bipartite states $\ket{\psi_i}$, where Alice holds
the index $i$, $U$ acts on $A_1B_1$ and $A_2,B_2$ are ancilla spaces.
Thus we can define $U(\cE)$ by
\be U(\cE) := \sum_i p_i \oprod{i}^{X_A} \ot (U^{A_1B_1} \ot
\one^{B_1B_2}) (\oprod{\psi_i}^{A_1A_2B_1B_2})
\ee
We will use $A$ to denote the composite system $A_1A_2$ and $B$ to
denote $B_1B_2$.  As in \sect{known}, define the $\{c\ra q\}$ channel
$\cN_\cE$ by $\cN_\cE(\oprod{i}) = \oprod{i} \ot \psi_i$, so that 
that $\cE = \cN_cE(\cE^{X_A})$.  Defining $\cN_{U(\cE)}$ similarly, we
can use \lem{relatif} to show that
\be \<\cN_\cE : \cE^{X_A}\> + \<U\> \geq
\<U \circ \cN_\cE : \cE^{X_A}\> = \<\cN_{U(\cE)} : \cE^{X_A}\>
.\ee

Recall from HSW coding (\eq{hsw}) that
\be \<\cN_\cE : \cE^{X_A} \> \geq I(X_A;B)_\cE \ctc,\ee
while RSP (\eq{RSP}) states that
\be I(X_A;B)_\cE \ctc + H(B) \qq \geq \<\cN_\cE : \cE^{X_A} \>.\ee

In the presence of free entanglement, these resource inequalities
combine to become an equality:
\be I(X_A;B)_\cE \ctc + \infty \qq = \<\cN_\cE : \cE^{X_A} \>
+ \infty \qq.\label{eq:RSP-HSW-eq}\ee
This remarkable fact can be thought of as a sort of reverse Shannon
theorem for $\{c\ra q\}$ channels, stating that when
entanglement is free (in contrast to the CRST, which requires free
rbits), any $\{c\ra q\}$ channel on a fixed source is equivalent to an 
amount of classical communication given by its capacity.

Recall the similar equality for partially entangled states in the
presence of a sublinear amount of classical communication:
$\<\psi^{AB}\> + o[c\ra c] = H(B)_\psi\qq + o[c\ra c]$.  By analogy
with entanglement generation in \thm{U-E-cap}, we will use the
resource equality in \eq{RSP-HSW-eq} to derive a capacity theorem for
classical communication in the presence of unlimited entanglement.

\begin{theorem}\label{thm:U-CE-cap}
\be C_\ra^E(U) = \Delta\chi_U := \sup_\cE 
\l[I(X_A;B)_{U(\cE)} - I(X_A;B)_{\cE}\r]
\label{eq:U-CE-cap}\ee
where the supremum is over all ensembles $\cE$ of the form in
\eq{cE-bipartite}. 
\end{theorem}
The proof closely follows the proof of \thm{U-E-cap}.

\begin{proof}We begin with the converse, proving that $C_\ra^E(U)\leq
\Delta\chi_U$.  Alice and Bob begin with a fixed input state, which
can be thought of as an ensemble $\cE_0$ with $I(X_A;B)_\cE=0$.
Local operations (which for simplicity, we can assume are all
isometries) cannot increase $I(X_A;B)$, so after $n$ uses of $U$ the
mutual information must be $\leq n\Delta\chi_U$.  (For a generalized
and more formal verson of this argument, see the proof of
\thm{Ce-cap} in \sect{U-Ce-cap}.)  The bound $C_\ra^E(U)\leq
\Delta\chi_U$ then follows from Fannes' inequality.

{\em Coding theorem:} For any ensemble $\cE$, we have $\<U\> +
I(X_A;B)_\cE[c\ra c] + \infty [qq] \geq \<U\> + \<\cN_\cE:\cE^{X_A}\>
+ \infty [qq] \geq \<\cN_{U(\cE)}:\cE^{X_A}\> + \infty [qq] \geq
I(X_A;B)_{U(\cE)}[c\ra c] + \infty [qq]$.  Using the Cancellation
Lemma (\ref{lemma:cancel}) and taking the supremum over $\cE$, we find
that $\<U\> + o[c\ra c] + \infty [qq] \geq
\Delta\chi_U [c\ra c] + \infty[qq]$.  Finally, we can use
\prop{U-nonzero-CC} and \lem{noo} to eliminate the sublinear
classical communication cost.
\end{proof}

Although the coding theorem is formally very similar to the coding
theorem for entanglement generation, its implementation looks rather
different.  Achieving the bound in \thm{U-E-cap} is rather
straightforward: 1) $n_1$ copies are created of some state
$\psi^{A_1A_2B_1B_2}$ 
s.t. $\Delta E_U\approx H(B)_{U(\psi)}-H(B)_{\psi}$,  2) $U^{\ot n_1}$
is applied to 
$\psi^{\ot n_1}$,
3) entanglement is concentrated from $(U\ket{\psi})^{\ot n_1}$, 4)
$\approx\! n_1H(B)_{\psi}$ ebits are used to recreate $\psi^{\ot n_1}$ and
$\approx\! n_1(H(B)_{U(\psi)}-H(B)_{\psi}) \approx n_1\Delta E_U$ ebits are set
aside as output, 5) steps 2-4 are repeated $n_2$ times to make the
cost of the catalyst vanish.  The coding scheme for
entanglement-assisted classical communication is similar, but has some
additional complications because different parts of the message are
not interchangeable.  The resulting protocol involves a peculiar
preprocessing step in which Alice runs through the entire protocol
backwards before $U$ is used for the first time; for this reason, we
call it the ``looking-glass protocol.'' \index{looking-glass protocol}
The procedure is as follows:
\begin{enumerate}
\item Choose an ensemble $\cE=\sum_i p_i \oprod{i} \ot \psi_i$ with
$I(X_A;B)_{U(\cE)}-I(X_A;B)_\cE \approx \Delta\chi_U$. 
\item The message is broken into $n_1$ blocks $M_1,\ldots,M_{n_1}$,
each of length $\approx\! n_2\Delta\chi_U$.  Initialize $R_{n_1}$ to be
an arbitrary string of length $\approx\! n_2I(X_A;B)_\cE$.
\item For $k=n_1,n_1-1,\ldots, 1$:
\begin{enumerate}
\item Encode the string $(R_k,M_k)$ ($\approx\! n_2I(X_A;B)_{U(\cE)}$
bits) into an element of $(U(\cE))^{\ot n_2}$, say $U\ket{\psi_{x_{k,1}}}
\ot \cdots \ot U\ket{\psi_{x_{k,n_2}}}$ for some $p$-typical string
$x_k^{n_2}$.  This is accomplished via HSW coding.
\item Alice now wishes to use RSP to send $\ket{\psi_{x_k^{n_2}}} :=
\ket{\psi_{x_{k,1}}} \ot \cdots \ot \ket{\psi_{x_{k,n_2}}}$ to
Bob.  She performs the RSP measurement on some shared entanglement and
obtains an outcome with $\approx\! n_2I(X_A;B)_\cE$ bits, which
she doesn't send to Bob directly, but instead stores in the register
$R_{k-1}$.
\end{enumerate}
\item Finally, Alice sends $R_0$ to Bob using $\approx\!
n_2I(X_A;B)_\cE [c\ra c]$.
\item For $k=1,\ldots,n_1$:
\begin{enumerate}
\item Bob uses $R_{k-1}$ to perform his half of RSP and reconstruct
his half of $\ket{\psi_{x_k^{n_2}}}$.
\item Alice and Bob apply $U$ $n_2$ times to obtain $\approx\! U^{\ot
n_2}\ket{\psi_{x_k^{n_2}}}$.
\item Bob performs HSW decoding to obtain $(M_k,R_k)$ with a high
probability of success.
\end{enumerate}
\end{enumerate}
It might seem that errors and inefficiencies from the many HSW and RSP
steps accumulate dangerously over the many rounds of the looking-glass
protocol.  In \cite{BHLS02}, the protocol was carefully analyzed and
the errors and inefficiency were shown to converge to zero.
However, the validity of the composite protocol follows even more
directly from the Composability Theorem (\ref{thm:composability});
remarkably, this permits a proof that is much more compact and
intuitive than even the description of the above protocol, let alone
a verification of its correctness.

As a corollary of \thm{U-CE-cap}, entanglement-assisted capacities are
additive (i.e.  $C_\ra^E(U_1 \ot U_2) = C_\ra^E(U_1) +
C_\ra^E(U_2)$).  The proof is basically the same as the proof that
$E(U)$ is additive.

Another corollary we can obtain is an optimal coding theorem for
entanglement-assisted one-way quantum communication: $Q_\ra^E(U) :=\max\{Q
: \<U\> +\infty[qq] \geq Q[q\ra q]\} = C_\ra^E(U)/2$.  This is because 
when entanglement is free, teleportation and super-dense coding imply
that 2 cbits are equivalent to 1 qubit.

\subsection{Relations between entanglement and classical communication
capacities}\label{sec:U-relations}
One of the most interesting properties of unitary gates as
communication channels is that their different capacities appear to be
closely related.  In this section we prove that $C_+(U) \leq E(U)$ and
then discuss some similar bounds.


\begin{proposition}\label{prop:U-CgE}
If $(C_1,C_2,E)\in\CCE(U)$ then  $E(U) \geq C_1 + C_2 + E$.
\end{proposition}
Using the fact that back communication does not improve capacities
(proved in the next chapter), we can improve this bound to $E(U) \geq
\max(C_1,0) + \max(C_2,0) + E$.

This claim is significant for two reasons.  First
is that it implies that it may be easier to connect different unitary
gate capacities than it has been to relate different capacities of
noisy channels.  It is directly useful in finding gate capacities and
raises the intriguing question of whether the converse
inequality of \prop{U-nonzero-CC} (that $E(U)>0 \Rightarrow
C_\ra(U)>0$) can be strengthened, and ultimately whether
$C_+(U)=E(U)$.

The fact that $C_+(U)\leq E(U)$ has a deeper implication as well,
which is that not all classical communication is created equal.  While
normally $[c\ra c]\not\geq [qq]$, a cbit sent through unitary means
{\em can} be converted into entanglement.  This suggests that using unitary
gates to communicate gives us something stronger than classical bits;
a resource that we will formally define in the next chapter as {\em
coherent bits} or cobits.  The consequences will be productive not
only for the study of unitary gate capacities, but also for many other
problems in quantum Shannon theory.

\begin{proof}[Proof of \prop{U-CgE}]
Assume for now that $E\geq 0$.
For any $n$, there is a protocol $\cP_n$ that uses $U$ $n$ times to
send $C_1^{(n)} \cbf + C_2^{(n)} \cbb$ and create $E^{(n)}$ ebits
with $C_1^{(n)} \geq n(C_1-\delta_n)$, $C_2^{(n)} \geq
n(C_2-\delta_n)$, $E^{(n)} \geq n(E-\delta_n)$ and error $\leq
\epsilon_n$, where $\delta_n,\eps_n\ra 0$ as $n\ra \infty$.
We analyze the protocol using the QP formalism, in which $\cP_n$ is an
isometry such that for any
$a\in\{0,1\}^{C_1^{(n)}}, b\in\{0,1\}^{C_2^{(n)}}$,
\bea
      \ket{\varphi_{ab}}
 & := & \cP_n |a\>_{A_1} |b\>_{B_1}
\non
\\
& {\rm and} & 
F \, ( \,|b\>_{A_1} |a\>_{B_1} |\Phi\>^{\ot E^{(n)}}_{A_2B_2}\,, 
\tr_{\!A_3\!B_3} 
|\varphi_{ab}\>\<\varphi_{ab}|_{A_{1,2,3}\!B_{1,2,3}} \,) 
= 1 - \epsilon_{ab} \geq 1-\epsilon_n 
\,.
\eea
for some $\epsilon_{ab}\leq \epsilon_n$.
By Uhlmann's Theorem\cite{Uhlmann76}, there exist normalized (though
not necessarily orthogonal) states
$\ket{\gamma_{ab}}$ and 
$\ket{\eta_{ab}}$ satisfying
\be \ket{\varphi_{ab}} = 
\sqrt{1-\epsilon_{n}}\ket{b}_{A_1}\ket{a}_{B_1}
\ket{\Phi}^{E^{(n)}}_{A_2B_2} \ket{\gamma_{ab}}_{A_3B_3} + \sqrt{\epsilon_n}
\ket{\eta_{ab}}_{A_{1,2,3}B_{1,2,3}}.
\ee
Note that we have changed $\epsilon_{ab}$ to $\epsilon_n$ by an
appropriate choice of $\ket{\eta_{ab}}$.  This will simplify the
analysis later.

To generate entanglement, Alice and Bob will apply $\cP_n$ to
registers $A_1B_1$ that are maximally entangled with local ancillas
$A_4B_4$; i.e. the states
$\ket{\Phi}^{\ot C_1^{(n)}}_{A_1A_4} = 2^{-C_1^{(n)}/2}\sum_a
\ket{a}_{A_1}\ket{a}_{A_4}$ and
$\ket{\Phi}^{\ot C_2^{(n)}}_{B_1B_4} = 2^{-C_2^{(n)}/2}\sum_b
\ket{b}_{B_1}\ket{b}_{B_4}$.
The resulting output state is
\be \ket{\bar{\varphi}_n}_{AB} = \sqrt{1-\epsilon_n}\ket{\psi_n}_{AB} +
\sqrt{\epsilon_n}\ket{\delta_n}_{AB},
\label{eq:UCE-output}\ee
where
\be
\ket{\psi_n}_{AB} = 
2^{-(C_1^{(n)}+C_2^{(n)})/2} \sum_{a,b} \ket{b}_{A_1}\ket{a}_{A_4}
\ket{a}_{B_1}\ket{b}_{B_4} \ket{\Phi}^{\ot E^{(n)}}_{A_2B_2}
\ket{\gamma_{ab}}_{A_3B_3}.\ee
 A similar expression exists for $\ket{\delta_n}_{AB}$, but it is not
needed, so we omit it.
Note that every Schmidt coefficient of $\ket{\psi_n}$ is $\leq
\exp(-(C_1^{(n)} + C_2^{(n)} + E^{(n)}))$, so 
 $E(\ket{\psi_n}) \geq C_1^{(n)} + C_2^{(n)} + E^{(n)}$.

We will use Fannes' inequality (\lem{fannes}) to relate
$E(\ket{\bar{\varphi}_n})$ to $E(\ket{\psi_n})$.  From \eq{UCE-output}, we have
$|\braket{\bar{\varphi}_n}{\psi_n}| 
\geq \sqrt{1-\epsilon_n}$.  Applying the relation between fidelity
and trace distance in \eq{fid-trace}, we find $\|\varphi_n - \psi_n
\|_1 \leq  
2\sqrt{\epsilon_n}$.  Also, $\ket{\bar{\varphi}_n}$ was created with
$n$ uses of $U$, so $\Sch(\ket{\bar{\varphi}_n}) \leq (\Sch(U))^n \leq
d^{2n}$.  Thus
\bea
 |E(\ket{\psi_n}) - E(\ket{\varphi_n})| &\leq&
\l(2n \log d\r)2\sqrt{\epsilon_n} + \eta(2\sqrt{\epsilon_n})
\non \\ E(\ket{\bar{\varphi}_n}) & \geq &
n\l(C_1 + C_2 + E - 3\delta_n - 4\sqrt{\epsilon_n}\log d - 
\frac{\eta(2\sqrt{\epsilon_n})}{n}\r)
\eea
Therefore as $n\ra \infty$, $\smfrac{1}{n}E(\ket{\bar{\varphi}_n}) \ra C_1 +
C_2 + E$.  Since 
$n\<U\> \reduction \<\bar{\varphi}_n\>$, it follows that $\<U\> \geq
(C_1 + C_2 + E)[qq]$.

We omit the quite similar proof of the $E<0$ case; however, note that
this case also follows from the more general \thm{bidi-ccc}, which
will be proved in \sect{ccc-proofs}.
\end{proof}

A similar bound exists for the entanglement-assisted capacity:
$C_+^E(U) \leq  E(U) + E(U^\dag)$.  This result is proved in
\cite{BS03b}, though some preliminary steps are found in
\cite{BHLS02,BS03a}.  Here we give a sketch of the argument and
explain its evolution through \cite{BS03a,BHLS02,BS03b}.

As in \prop{U-CgE}, Alice and Bob will input halves of maximally
entangled states into a communication protocol $\cP_n$ that uses $U$
$n$ times.  This creates  $\approx nC_+^E(U)$ ebits.  However, the
entanglement assistance leads to two additional complications.  First,
we need to bound the amount of entanglement that $\cP_n$
 uses to communicate.  Say that $\cP_n$ starts with
$E^{(n)}$ ebits.  Then its entanglement consumption is no greater than
$\max_{a,b} \l[E^{(n)} - E(\cP_n\ket{a}_A\ket{b}_B 
\ket{\Phi}_{AB}^{E^{(n)}})\r] \leq n E(U^\dag)$ (using $\Delta
E_{U^\dag}=E(U^\dag)$ from \thm{U-E-cap}).  Here $E(U^\dag)$ can 
be thought of as an {\em entanglement destroying capacity} of $U$ if
we recognize that unitarily disentangling a state is a nonlocal task.
For $U\in\cU_{2\times 2}$, we always have $E(U)=E(U^\dag)$, but for
$d>2$, numerical evidence suggests that equality no longer
holds\cite{CLS02}.  Since $\cP_n$ uses no more than 
$nE(U^\dag)$ ebits, we have $\<U\> + E(U^\dag) [qq] \geq C_+^E(U)[qq]$
and thus $E(U) \geq C_+^E(U) - E(U^\dag)$, implying the desired
result.  More generally, for any $(C_1,C_2,E)\in\CCE(U)$ this result implies
that $(C_1,C_2,-E(U^\dag))\in\CCE(U)$; i.e. more than $E(U^\dag)$ ebits
are never needed for any communication protocol.

The argument outlined above follows the presentation of \cite{BS03a}.
However, we also need to address the second problem introduced by free
entanglement. For the inefficiency caused by communication errors to
vanish as in
\prop{U-CgE}, we need to ensure that the logs of the Schmidt numbers
of the states we work with grow at most linearly with $n$.
Equivalently, we need to show that the parameter $E^{(n)}$ from the
previous paragraph can be chosen to be $\leq
Kn$ for some constant $K$.  In \cite{BHLS02}, the explicit
construction of \thm{U-CE-cap} was used to achieve this bound for
one-way communication, and thereby to prove the weaker result that
$C_\ra^E(U) \leq E(U) + E(U^\dag)$.

 Finally \cite{BS03b} proves an exponential bound on Schmidt rank for
general bidirectional protocols, by applying HSW coding in both
directions to $\cP_n$.  Specifically, for any input of Bob's, Alice
can consider $\cP_n$ to be a channel that communicates
$n(C_1-\delta_n)$ bits with error $\leq \epsilon_n$; such a channel
has HSW capacity $\approx n(C_1 -
\delta_n)(1-\epsilon_n) = nC_1 - o(n)$.  Similarly, Bob can code for a
channel to Alice that has capacity $nC_2 - o(n)$.  These block codes
require $k$ blocks of $\cP_n$ with $k\gg \exp(n)$, but now the total
error goes to zero as $k\ra \infty$, while the entanglement cost
$kE^{(n)}$ grows linearly with $k$.  So the desired capacity is
achieved by taking $k\ra\infty$ before $n$.  A refined version of this
argument will be presented in the proof of \thm{bidi-ccc} in
\sect{ccc-proofs}. 

Technically, HSW coding is not quite appropriate here, since Alice's
channel weakly depends on Bob's input and vice versa.  Thus, a small
modification of \cite{BS03b}'s proof is necessary.  The correct coding
theorem to use for bidirectional channels was given in 1961 by
Shannon\cite{Shannon61} and can be used to obtain the result claimed
in \cite{BS03b} (see also \cite{CLL05} for a generalization of
Shannon's 1961 result to noisy bidirectional quantum channels).
Unlike the HSW theorem and Shannon's original noisy 
channel coding theorem\cite{Shannon48}, the two-way coding theorem
only achieves low average error instead of low maximum error.  For
entanglement generation, average error is sufficient, but in the next
chapter we will show (in \thm{bidi-ccc}) that maximum error can also
be made small for bidirectional protocols.  In fact, the average error
and maximum error conditions appear to be asymptotically equivalent in
general, given some mild assumptions\cite{DW05,CK81}.

\subsection{Challenges for bidirectional communication}
\label{sec:U-bidi-challenges}
We conclude our discussion of classical communication using unitary
gates in this section, by reviewing attempts to extend \thm{U-E-cap}
to the case of bidirectional communication and pointing out the
difficulties that arise.

{\em There is no bidirectional analogue of HSW coding, even
classically.}  In \cite{Shannon61}, Shannon considers communication
with noisy bidirectional channels---a model in some ways simpler, but
in other ways more complex, than unitary gates---and establishes upper
and lower bounds that do not always coincide.  We briefly restate
those bounds here.  Define a bidirectional channel
$N(A_{\text{out}}B_{\text{out}} | A_{\text{in}}B_{\text{in}})$ where
$A_{\text{in}}$ is Alice's input, $B_{\text{in}}$ is Bob's input,
$A_{\text{out}}$ is Alice's output and $B_{\text{out}}$ is Bob's
output.  For any probability distribution on the inputs
$A_{\text{in}}B_{\text{in}}$, consider the rate pair
$I(A_{\text{in}};B_{\text{out}}|B_{\text{in}})[c\ra c] +
I(B_{\text{in}};A_{\text{out}}|A_{\text{in}})[c\la c]$.
\mscite{Shannon61} proves that this rate pair is
\begin{itemize}
\item achievable if we
maximize over {\em product} distributions on
$A_{\text{in}}B_{\text{in}}$ (i.e. $I(A_{\text{in}};B_{\text{in}})=0$)
; and
\item an upper bound if we maximize over arbitrary distributions on
$A_{\text{in}}B_{\text{in}}$ (i.e. if $\<N\> \geq C_1[c\ra c] +
C_2[c\la c]$, then there exists a joint distribution on
$A_{\text{in}}B_{\text{in}}$ such that $C_1 =
I(A_{\text{in}};B_{\text{out}}|B_{\text{in}})$ and $C_2 =
I(B_{\text{in}};A_{\text{out}}|A_{\text{in}})$).
\end{itemize}
Using the chain
rule\cite{CK81} we can rewrite these quantities suggestively as
$C_1=I(A_{\text{in}} ; B_{\text{in}}B_{\text{out}}) -
I(A_{\text{in}};B_{\text{in}})$ and $C_2=I(B_{\text{in}} ;
A_{\text{in}}A_{\text{out}}) - I(A_{\text{in}};B_{\text{in}})$.  In
this form, they resemble \eq{U-CE-cap}: for communication from Alice
to Bob we measure the difference
between the output correlation
$I(A_{\text{in}};B_{\text{in}}B_{\text{out}})$ and the input
correlation $I(A_{\text{in}};B_{\text{in}})$ and a similar expression
holds for communication from Bob to Alice.  This has led
\cite{BS03b} to conjecture that a bidirectional version of
$\Delta\chi_U$ (defined in \eq{U-CE-cap})
should describe the two-way classical capacity of a
unitary gate.  However, even in the classical case, Shannon's inner
and outer bounds on the capacity region (corresponding to uncorrelated
or correlated inputs respectively) are in general different.

This highlights the difficulties in coding for bidirectional
channels.  The messages both parties send may interfere with each
other, either positively or negatively.  The best known protocols
reduce the bidirectional channel to a pair of one-way channels for
which Alice and Bob code independently.  However, we cannot rule out
the case in which Alice and Bob use correlated channel
inputs to improve the rate.  

The same general concerns apply to quantum bidirectional channels,
including unitary gates, although not all of the corresponding bounds
have been proven.  Some promising steps towards this goal are in
\cite{Yard05a,Yard05b}, which derive capacity expressions for quantum
channels with two inputs and one output.


{\em Reversible RSP is not possible for all bidirectional ensembles.}
The crucial ingredient in the proof of \thm{U-CE-cap} was the
equivalence for any ensemble $\cE$ (given unlimited entanglement) between the induced $\{c\ra q\}$
map $\cN_\cE$ and the standard resource $I(X_A;B)_\cE [c\ra c]$.  Now
suppose $\cE$ is a bidirectional 
ensemble $\sum_{i,j} p_iq_j \oprod{i}^{X_A} \ot \oprod{j}^{Y_B} \ot
\ket{\psi_{ij}}^{AB}\}$ which has a corresponding $\{cc\ra qq\}$
channel $\cN_\cE$ mapping 
$\ket{i}^A\ket{j}^B$ to  $\ket{\psi_{ij}}^{AB}$.   To extend
\thm{U-CE-cap} to the bidirectional case, we would begin by trying to
find pairs $(C_1,C_2)$ such that $\<\cN_\cE\> + \infty [qq] = C_1[c\ra
c] + C_2[c\la c] + \infty [qq]$.  It turns out that there are
ensembles for which no such equivalence exists.  In fact, classical
communication cannot reversibly simulate any ensemble whose classical
capacity region is not just a rectangle.  The proof of this is
trivial: if $\<\cN_\cE\> + \infty [qq] = C_1[c\ra c] + C_2[c\la c] +
\infty [qq]$ then $\<\cN_\cE\> +\infty [qq] \geq R_1[c\ra c] +
R_2[c\la c]$ if and only if $R_1\leq C_1$ and $R_2\leq C_2$.

One simple example of an ensemble that cannot be reversibly simulated
is the ensemble corresponding to the AND channel: 
$\ket{\psi_{ij}}^{AB}=\ket{i\land j}^A\ket{i\land j}^B$, where
$i,j\in\{0,1\}$ and $i\land j$ is the logical AND operation.  Clearly
$(1,0)\in\CC(\text{AND})$ and $(0,1)\in\CC(\text{AND})$; i.e.  the AND
ensemble can send one bit from Alice to Bob or one bit from Bob to
Alice.  (The channel is effectively classical, so we need not consider
entanglement) If AND were reversibly simulatable, then we would expect
$(1,1)\in\CC(\text{AND})$.  However,
$(1,\epsilon)\not\in\CC(\text{AND})$ for any $\epsilon>0$.  Suppose
Bob sends zero with probability $p$ and one with probability $1-p$.
When Bob sends zero, the channel output is $\ket{00}$ regardless of
Alice's input.  Alice can only communicate to Bob
during the $1-p$ fraction of time that he sends one, so she can only
send $1-p$ bits to him.  Thus we must have $p=0$.  Since Bob always
sends one, he cannot communicate any information to Alice.

One might object to the AND example by pointing out that
simulating a relative resource is a more reasonable goal, since the
capacities of ensembles like AND vary with the probability
distribution of Alice and Bob's inputs.  In fact, even in the one-way
case the HSW/RSP equivalence in \eq{RSP-HSW-eq} is only proven for
relative resources.\footnote{Actually, the quantum reverse Shannon
theorem\cite{BDHSW05} gives a reversible simulation of {\em
unrelativized} $\{c\ra q\}$ channels, though this appears not to be
possible for general $\{q\ra q\}$ channels, or for the coherent
version of $\{c\ra q\}$ channels that we will consider in the next
chapter.}
However, one can construct ensembles where reversible simulation is
impossible even if the probability distribution of the input is fixed.
We construct one such ensemble (or channel) as follows:

Alice and Bob both input $m+1$ bit
messages, $(a_1,a_2)$ and $(b_1,b_2)$, where $a_1$ and $b_1$ are
single bits and $a_2$ and $b_2$ are $m$-bit strings.  The channel
$\cN$ computes the following string: $(a_1\oplus b_1, (a_1\oplus b_1
? a_2 : b_2))$ and gives Alice and Bob both a copy of it.  The
notation $(a_1\oplus b_1 ? a_2 : b_2)$ means that the channel outputs
$a_2$ if $a_1\oplus b_1=1$ and $b_2$ if $a_1\oplus b_1=0$.  We choose
the input probability distributions to be uniform for both parties.
Alice and Bob are allowed to agree upon any sort of block coding
protocol they wish as long as they still send each input approximately
the same number of times.

First, we argue that $(m,0),(0,m)\in\CC(\cN)$.
The protocol to achieve $(m,0)$ is as follows: Alice sets $a_1=0$ for
the first $n/2$ rounds and $a_1=1$ for the last $n/2$ rounds.
Likewise, Bob sets $b_1=0$ for the first $n/2$ rounds and $b_2=1$ for
the last $n/2$ rounds.  The other two registers are set uniformly at
random.  This satisfies the criteria of $p_i$ and $q_j$ being uniform,
although it is a very particular coding scheme.  Since $a_1 \oplus
b_1$ is always zero, it is always Alice's message $a_2$ which is
broadcast to both parties.  Thus, this transmits $m$ bits to Bob per
use of $\cN$.  If Bob instead sets $b_1=1$ for the first $n/2$ rounds
and $b_2=0$ for the last $n/2$ rounds, then the communication direction
is reversed.

If $\cN$ with uniformly distributed inputs had a rectangular rate
region, then $(m,m)$ would also be achievable.  However any achievable
$(C_1,C_2)$ must satisfy $C_1+C_2\leq m+2$, since there is a natural
multi-round simulation for $\cN$ that uses $m+2$ total cbits.
Choosing $m>2$ yields a non-rectangular rate region and hence a
channel that cannot be efficiently simulated, even with a fixed input
probability distribution.

Arguably, even this example does not go far enough, since we could
talk about simulating $\cN$ with respect to a bipartite test state
$\rho^{AB}$.  However, it is hard to define a corresponding asymptotic
resource; the natural choice of $\<\cN:\rho^{AB}\> = (\cN^{\ot n} :
\rho^{\ot n})_{n=1}^\infty$ violates \eq{asy-quasi-iid} since extra
input test states $\rho^{AB}$ can no longer be created for
free locally.  On the other hand, $\<\cN:\rho_1^A \ot \rho_2^B\>$ is a
well-defined resource for which there may be a reversible simulation,
but since it cannot contain any correlations between Alice and Bob it
is hard to imagine using it in a protocol analogous to the one in
\thm{U-CE-cap}. 

Combined, these facts mean that we are likely to need new methods and
possibly new ways of thinking about resources to find the two-way
capacity regions of unitary gates.
 
\section{Examples}\label{sec:u-examples}
There are only a handful of examples of unitary gates where any
capacities can be computed exactly.  On the other hand, some more
complicated gates appear to give separations between quantities like
$C_\ra$ and $C_\la$ or $C_+$ and $E$, though we will only be able to
offer incomplete proofs for these claims.   This section will describe
what is known about the capacities of all of these examples.  Many of
the results on the two-qubit gates $\swap$, $\cnot$ and $\dcnot$ are
taken from \cite{Collins00}.

\subsection{SWAP, CNOT and double CNOT}\label{sec:u-2qubit-examples}
We begin by reviewing three well-known gates in $\cU_{2\times 2}$.

{\bf SWAP:}
The \swap\ gate on two qubits is in a sense the strongest two qubit
gate; i.e. for any two-qubit $U$, $\<\swap\> = [q\ra q] + [q\la q]
\geq \<U\>$.  The proof follows the lines of \prop{U-TP-bound}: any
$U$ can be simulated by sending Alice's input to Bob using $[q\ra q]$,
Bob performing $U$ locally, and then Bob sending Alice's qubit back with
$[q \la q]$.  Thus, we would expect it to saturate all of the upper
bounds we have found on capacities.

In fact the capacity region is
\be \CCE(\swap) = \{(C_1,C_2,E):
C_1 \leq 2, C_2\leq 2, E\leq 2, \max(C_1,0)+\max(C_2,0)+E\leq 2\}.\ee
 The first two upper bounds follow from \prop{U-TP-bound} and the
last two upper bounds from 
$$ \max(C_1,0) + \max(C_2,0) + E \leq
E(\swap) \leq \log\Sch(\swap) = 2.$$
To show that this entire region is achievable, we can apply
\prop{U-CgE} to the single point $(2,2,-2)\in\CCE(\swap)$.  This in
turn follows from applying super-dense coding in both directions to
obtain $\swap + 2[qq] = [q\ra q] + [qq] + [q\la q] + [qq] \geq 2[c\ra
c] + 2 [c\la c]$.

There are more direct proofs for some of the other extreme points of
the capacity region; the interested reader should try the exercise of
finding a simple alternate proof of $\swap \geq 2[c\ra c]$ (see
Eq. (63) of \cite{Collins00} for the answer).

{\bf CNOT:}
It turns out that the capacity region of \cnot\ is exactly one half
the size of the capacity region for \swap: $\CCE(\cnot) =
\{(C_1,C_2,E): C_1 \leq 1, C_2\leq 1, E\leq 1,
\max(C_1,0)+\max(C_2,0)+E\leq 1\}$. 
(We will later see that this is no accident but rather a consequence
of the asymptotic equivalence $2\<\cnot\>=\<\swap\>$.)

The first three upper bounds follow from a simulation due to
Gottesman\cite{Got98}, 
\be [c\ra c]+[c\la c]+[qq] \geq \<\cnot\> \ee
and causality.  Then applying \prop{U-CgE} yields the last bound.

In terms of achievability, $\<\cnot\>\geq [c\ra c]$ is obvious, and
$\<\cnot\>\geq [c\la c]$ follows from $(H\ot H)\cnot(H\ot
H)\ket{0}\ket{b} = \swap\,\cnot\,\swap\ket{0}\ket{b} = \ket{b}\ket{b}$.
However, just as the entire \swap\ capacity region follows from
$(2,2,-2)$, the entire \cnot\ region follows from the inequality $\cnot +
[qq] \geq [c\ra c] + [c\la c]$, which is achieved by a protocol due to
\cite{Collins00}:
\be (Z^aH\otimes I)\textsc{cnot}(X^a\otimes Z^b)\ket{\Phi_2}_{AB} =
\ket{b}_A\ket{a}_B
\label{eq:e-asstd-cnot}
\ee

{\bf Double CNOT:}
The double CNOT is formed by applying two CNOTs consecutively: first
one with Alice's qubit as control and Bob's as target, and then one
with Bob's qubit as control and Alice's as target.  Equivalently we
can write $\dcnot=\swap\,\cnot\,\swap\,\cnot$.  For $a,b\in\{0,1\}$,
we have $\dcnot\ket{a}\ket{b} = \ket{b}\ket{a\oplus b}$.

The double \cnot\ seems weaker than two uses of a \cnot, but it turns out
to have the same capacity region as the \swap\ gate, or as
$(\cnot)^{\times 2}$:
\be \CCE(\dcnot)\!=\!\CCE(\swap)\!=\!\{(C_1,C_2,E):
C_1 \leq 2, C_2\leq 2, E\leq 2, \max(C_1,0)+\max(C_2,0)+E\leq 2\}.\ee
The upper bounds are the same as for \swap, and achievability is shown
in \cite{Collins00}.  Specifically, they give a protocol for the point
$(2,2,-2)\in\CCE(\dcnot)$, from which all other points follow.

{\bf Relations among SWAP, CNOT and Double CNOT:}
If we were to judge the strengths of the \swap, \cnot\ and \dcnot\
gates solely based on their capacity regions, then it would be
reasonable to conclude that 
\be \<\swap\> = 2\<\cnot\> = \<\dcnot\>.
\label{eq:swap-cnot-dcnot}\ee
However, it has been historically difficult to construct efficient
maps between these gates.  \cite{Collins00} has conjectured that
$2\cnot \not\geq \swap$, and since $2\cnot\geq\dcnot$, this would
imply that $\dcnot\not\geq\swap$.  Moreover, \cite{HVC02} shows that
\dcnot\ is takes less time than \swap\ to simulate using nonlocal
Hamiltonians, implying that it somehow has less nonlocal power.  A
cute side effect of coherent classical communication, which we will
introduce in the next chapter, will be a concise proof of
\eq{swap-cnot-dcnot}, confirming the intuition obtained from capacity
regions.

Of course, this simple state of affairs appears to be the
exception rather than the rule.  We now consider two examples of gates
whose capacity regions appear to be less well behaved.

\subsection{A gate for which $C_\la(U)$ may be less than $C_\ra(U)$}
\label{sec:u-xoxo}
In this section we introduce a gate $U_{\text{XOXO}}\in\cU_{d\times d}$
that appears to have $C_\la(U)< C_\ra(U)$ when $d$ is sufficiently large.
Define $\Uxoxo$ as follows:
\ben
\Uxoxo \ket{x0} &=& \ket{xx} \qquad\forall 0\leq x < d \\
\Uxoxo \ket{xx} &=& \ket{x0} \qquad\forall 0\leq x < d \\
\Uxoxo \ket{xy} &=& \ket{xy} \qquad\forall x\neq y \neq 0
\een
The first two lines are responsible for the gate's affectionate
nickname, ``XOXO.''   The $d=2$ case corresponds to a \cnot,
which is locally equivalent to a symmetric gate, though as $d$
increases $\Uxoxo$ appears to be quite asymmetric.

\subsubsection{Bounds on capacities for $\Uxoxo$}
If Alice inputs $\ket{a}$ and Bob inputs $\ket{0}$, then Bob will
obtain a copy of Alice's input $a$.  Thus $C_\ra(\Uxoxo) \geq \log d$.

Define $S_x\in \cL(\cH_B)$ by
$$
S_x\ket{y} = \left\{
\begin{array}{ll}
\ket{0} & \mbox{ if } x=y \\
\ket{x} & \mbox{ if } 0=y \\
\ket{y} & \mbox{ otherwise}
\end{array}\right.
$$

Then $\Uxoxo=\sum_x \ket{x}\bra{x} \otimes S_x$, so $\Sch(\Uxoxo) \leq
d$.  Thus 
$E(\Uxoxo) \leq \log d$.  Combining this with $C_\ra(\Uxoxo) \geq \log
d$ yields 
$ \log d \leq C_\ra(\Uxoxo) \leq C_+(\Uxoxo) \leq E(\Uxoxo) \leq \log
\Sch(\Uxoxo) \leq \log d$.
Thus these must all be equalities, and we have
$$ C_\ra(\Uxoxo) = C_+(\Uxoxo) = E(\Uxoxo) = \log \Sch(\Uxoxo) = \log d$$

These are the only capacities that know how to determine exactly.
However, we can bound a few other capacities.

Suppose Alice and Bob share a $d$-dimensional maximally entangled
state $\ket{\Phi_d}=\frac{1}{\sqrt{d}}\sum_x \ket{x}\ket{x}$.  Using
such a state Bob can communicate $\log d$ bits to Alice.  The protocol
is as follows.  Let $b\in\{0,\ldots,d-1\}$ be the message Bob wants to
send and let $\omega=\exp(2\pi i/d)$.  First Bob applies the unitary
transformation $\sum_x \omega^{bx} \ket{x}\bra{x}$ to his half of
$\ket{\Phi_d}$, leaving them with the state $\frac{1}{\sqrt{d}}\sum_x
\omega^{bx}\ket{x}\ket{x}$.  Then they apply the gate $\Uxoxo$ to
obtain the product state $\frac{1}{\sqrt{d}}\sum_x
\omega^{bx}\ket{x}\ket{0}$.  Alice can now apply the inverse Fourier
transform $\frac{1}{\sqrt{d}}\sum_{xy} \ket{x}\bra{y} \omega^{-xy}$ to
recover Bob's message.

Thus $C_\la^E(\Uxoxo) \geq \log d$.  This yields a lower bound for
$C_\la(\Uxoxo)$ as well, since one possible communication strategy for
Bob is to use $\Uxoxo$ once to create a copy of $\ket{\Phi_d}$ and a
second time to send $\log d$ bits to Alice, using up the copy of
$\ket{\Phi_d}$.

So $\frac{1}{2}\log d \leq C_\la(\Uxoxo) \leq \log d$.  We would like
to know whether $C_\la(\Uxoxo) < C_\ra(\Uxoxo) = \log d$.  We cannot
prove this expression asymptotically, but can show that if Alice and
Bob share no entanglement and are initially uncorrelated, Alice's
mutual information with Bob's message is strictly less than $\log d$
after a single use of $\Uxoxo$.

\subsubsection{Bounding the one-shot rate of $\Uxoxo$}
\begin{proposition} If Alice and Bob share no entanglement and input
uncorrelated states into $\Uxoxo$, Alice's mutual information with Bob's
message is less than $(1-\epsilon)\log d + \cO(1)$ for some
constant $\epsilon>0$.
\end{proposition}

\begin{proof}
Let $\alpha,\beta,\gamma$ be small positive parameters that we will
choose later.

\sloppypar{Assume Alice begins with fixed input $\ket{\psi^A}^A=\sum_i
a_i\ket{i}^{A_1} 
\sum_j A_{ij}\ket{j}^{A_2}$ where $\sum_i |a_i|^2 = \sum_j |A_{ij}|^2 =
1$ and $A$ denotes the composite Hilbert space $A_1A_2$.  Let
$R\subseteq \{0,\ldots,d-1\}$ be the set given by 
$$ R = \left\{ i : |a_i|^2>\alpha\right\}.$$
The normalization condition means that $|R|\leq 1/\alpha$.
}

Bob will signal to Alice with some ensemble
$\cE=\sum_x p_x \oprod{x}^{X_B} \ot \oprod{\psi^B_x}^B$.  We will
divide the indices $x$ into three sets $S_1, S_2$ and $S_3$, according
to various properties of the states 
$\ket{\psi^B_x}$.  Write one such state as $\sum_i b_i^{(x)}
\ket{i}^{B_1} \sum_j B_{ij}^{(x)} \ket{j}_{B^2}$, where again $B$
denotes the composite Hilbert space $B_1B_2$, $\Uxoxo$ acts on
$A_1B_1$ and $A_2B_2$ are ancilla systems.  Now define $S_1, S_2$ and
$S_3$ by 
\begin{align}
S_1 & =  \l\{x : |b_0^{(x)}|^2 \geq \beta\r\} &  \\
S_2 & =  \l\{x :  |b_0^{(x)}|^2 < \beta 
\mbox{ and } \sum_{i\in R} |b_i^{(x)}|^2 \geq \gamma\r\} \\
S_3 & =  \l\{x : |b_0^{(x)}|^2 < \beta 
\mbox{ and } \sum_{i\in R} |b_i^{(x)}|^2 < \gamma \r\}
\end{align}
Without loss of generality, we can introduce a second classical
register for Bob, $Y_B$, that records which of the $S_y$ the index $x$
belongs to.  If we also include Alice's fixed input state, then $\cE$
becomes 
\be \cE = \sum_{y\in\{1,2,3\}} \oprod{y}^{Y_B} \ot
\sum_{x\in S_y} \oprod{x}^{X_B} \ot \oprod{\psi_x}^{AB},\ee
where $\ket{\psi_x}:=\ket{\psi^A}\ket{\psi^B_x}$.  

After $U\!:=\Uxoxo$ is applied, the parties are left with the ensemble
$U(\cE) := (U^{A_1B_1} \ot I^{X_BY_BA_2B_2})(\cE)$.
The mutual information of Alice's state with Bob's message is given
by
\be\begin{split}
I(X_B ; A)_{U(\cE)} &= 
I(X_BY_B ; A)_{U(\cE)} =
I(X_B ; A | Y_B)_{U(\cE)} + I(A ; Y_B)_{U(\cE)} 
\\ & \leq
I(X_B ; A | Y_B)_{U(\cE)} + \log 3
\leq 
\max_{y\in\{1,2,3\}} I(X_B ; A)_{U(\cE_y)} + \log 3.
\end{split}\ee
Here we have defined the ensemble $\cE_y$, 
for $y\in \{1,2,3\}$ to be the ensemble $\cE$
conditioned on $Y_B=y$; i.e.
\be \cE_y := \l(\sum_{x\in S_y}p_x\r)^{-1}
\sum_{x\in S_y}p_x \oprod{x}^{X_B} \ot \oprod{y}^{Y_B}
\ot \oprod{\psi_x}^{AB}.\ee
Thus to prove our proposition it suffices to verify that $I(X_B ;
A)_{U(\cE_y)} < (1-\epsilon)\log d + \cO(1)$ for each choice of $y$.

For cases $y=1,2$ we will use the following two facts.
\begin{fact}\label{fact:proj-meas}
Let $\rho$ be a $d$-dimensional state and suppose that
$\tr \Pi\rho = p$ for some $k$-dimensional projector $\Pi$.  Then
measuring $\{\Pi, I-\Pi\}$ yields a state with entropy no greater than
\be
-k\frac{p}{k}\log\frac{p}{k}-(d-k) \frac{1-p}{d-k}\log\frac{1-p}{d-k}
= H_2(p) + p\log k + (1-p)\log (d-k)
< 1 + \log k + (1-p)\log d.\ee
Since $H(\rho) \leq H(\Pi\rho\Pi + (1-\Pi)\rho(1-\Pi))$ it follows
that $H(\rho)\leq (1-p)\log d + 1 + \log k$.
If we treat $k$ as a constant, then this is $(1-p)\log d + \cO(1)$.
\end{fact}

\begin{fact}\label{fact:input-entropy}
The mutual information of the output is bounded by entropy of Bob's
input as follows:
\be I(X_B ; A)_{U(\cE_y)} 
\leq I(X_B ; AB_1)_{U(\cE_y)}
= I(X_B ; AB_1)_{\cE_y} 
\leq H(AB_1)_{\cE_y} = H(B_1)_{\cE_y}.\ee
\end{fact}

We can now prove that $I(X_B;A)_{U(\cE_1)} < (1-\epsilon)\log d
+ \cO(1)$.  By the definition of $S_1$, we have
$\bra{0}\cE_1^{B_1}\ket{0} \geq \beta$.  Now we use first
Fact~\ref{fact:input-entropy} and then Fact~\ref{fact:proj-meas} (with
the projector $\Pi=\oprod{0}^{B_1}$) to obtain
\be I(X_B;A)_{U(\cE_1)}
\leq H(B_1)_{\cE_1}
< (1-\beta)\log d + 1.\ee
This last expression is $\leq (1-\epsilon)\log d + 1$ as long as
$\epsilon\leq\beta$.

The case of $y=2$ will yield to similar analysis.  Define
$\ket{i'}\in\cH_{B}$ 
by $\ket{i'}^B = \ket{i}^{B_1} \otimes \sum_j B_{ij}\ket{j}^{B_2}$.
Now define a projector $\Pi = \sum_{i\in R} \oprod{i'}^B$ so that
$\tr\Pi = |R|$ and $p:= \tr 
\bra{\psi_x^B}\Pi\ket{\psi_x^B} = \sum_{i\in R} |b_i|^2$.  Note that
$\tr\Pi \leq 1/\alpha$ 
and $p\geq\gamma$.  Now we can again use
Facts~\ref{fact:input-entropy} and \ref{fact:proj-meas} to bound
$I(X_B;A)_{U(\cE_2)} < 1 + \log 1/\alpha + (1-\gamma)\log d$.  This is $\leq
(1-\epsilon)\log d + \cO(1)$ if we choose $\epsilon\leq\gamma$.

Note that these two bounds are independent of $U$.  They simply say
that having a lot of weight in a small number of dimensions limits the
potential for communication.  Case $S_3$ is the interesting case.  Here
we will argue that if Bob inputs a state that is not zero and does not
match Alice's state well, he will not change Alice's state very much.

Suppose a particular input state can be expressed as
$$\ket{\psi} = \sum_{i,j,k,l} a_i b_j A_{ik} B_{il}
\ket{ijkl}^{A_1A_2B_1B_2}.$$
According to the definition of $S_3$, $|b_0|^2 < \beta$ and
$\sum_{i\in R} |b_i|^2 < \gamma$, where $R=\{i:|a_i|^2\geq\alpha\}$.

After one use of the nonlocal gate $U$, the new state is
$$ \ket{\psi'} := U\ket{\psi}
= \ket{\psi}
+ \sum_{i\neq 0}\sum_{k,l}
a_ib_iA_{ik}B_{il}\ket{i0kl}
+ a_ib_0A_{ik}B_{0l}\ket{iikl}
- a_ib_iA_{ik}B_{il}\ket{iikl}
- a_ib_0A_{ik}B_{0k}\ket{i0kl}
$$

Writing $\ket{\psi'}$ in this form is useful for bounding the state
change
\be\begin{split}
\l\| \ket{\psi'} - \ket{\psi} \r\|^2 &=
\sum_{i,k,l}|a_i|^2 |A_{ik}|^2 \l(|b_iB_{il} - b_0B_{0l}|^2
 + |b_0B_{0l} - b_iB_{il}|^2\r)
\\&= 2 \sum_{i,l} |a_i|^2 |b_iB_{il}-b_0B_{0l}|^2
\\&\leq 4\sum_{i,l} |a_i|^2
\l(|b_i|^2|B_{il}|^2+|b_0|^2|B_{0l}|^2\r) 
\\&=4\sum_i |a_i|^2 \l(|b_i|^2 + |b_0|^2\r)  
\\&= 4\sum_{i\in R} |a_ib_i|^2 + 4\sum_{i\not\in R} |a_ib_i|^2 +
4|b_0|^2
\end{split}\label{eq:Uxoxo-diff}\ee
where the inequality on the third line follows from the general bound
$|x-y|^2 \leq (|x|+|y|)^2 = 2(x^2+y^2) - (|x|-|y|)^2 \leq 2(x^2+y^2)$.

We can bound each of the three terms in \eq{Uxoxo-diff} separately.
First,
$$\sum_{i\in R}|a_ib_i|^2 \leq \sum_{i\in R}|b_i|^2 < \gamma$$
The second term is
$$\sum_{i\not\in R}|a_ib_i|^2 
\leq \sum_{i\not\in R}\l(\sum_{j\not\in R}|a_j|^2\r)|b_i|^2 
< \alpha \sum_{j\not\in R} |b_j|^2
\leq \alpha \sum_j |b_j|^2 = \alpha
$$
The third term is simply $|b_0|^2 < \beta$.

Thus $\l\| \ket{\psi'} - \ket{\psi} \r\|^2 < 4(\alpha + \beta +
\gamma)$.  In terms of fidelity, $F(\ket{\psi},\ket{\psi'}) =
|\braket{\psi}{\psi'}|^2 > 1 - 4(\alpha + \beta + \gamma)$. Converting
this to trace distance means that $\half\|\oprod{\psi}
-\oprod{\psi'}\|_1 < 2\sqrt{(\alpha + \beta + \gamma)}$.  Since this
holds for each element 
of $\cE_3$ and trace distance is convex it follows that
$\half\|\cE_3^A - U(\cE_3)^A\|_1 < 2\sqrt{(\alpha +
\beta + \gamma)}$.  Alice's system is initially in a pure state, so we
can do a Schmidt decomposition between $A$ and ${A'}$ and thus
assume that $\dim {A'}=d$.  This also means that
$H(\cE_3^A)=0$.  Using Fannes' inequality then yields
$I(X_B;A)_{U(\cE_3)} \leq H(A)_{U(\cE_3)} < 4\sqrt{(\alpha +
\beta + \gamma)}\cdot 2\log d + (\log e)/e \leq (1-\epsilon)\log d + \cO(1)$
as long as $\epsilon \leq 1-8\sqrt{(\alpha + \beta + \gamma)}$.

This proves our claim for any $\alpha,\beta,\gamma>0$ as long as
$\epsilon\leq \beta$, $\epsilon\leq \gamma$ and $\epsilon\leq 1 -
8\sqrt{(\alpha + \beta + \gamma)}$.  This clearly holds as
long as $\alpha,\beta,\gamma$ and $\epsilon$ are small enough.  The
largest value of $\epsilon$ possible is
$\sqrt{\sqrt{33}-\sqrt{32}}\approx 0.2962$,
when $\alpha \approx 0$ and $\beta=\gamma=\epsilon$.
\end{proof}

I suspect that the actual asymptotic capacity is closer to
 $\frac{1}{2}\log d +  \cO(1)$ for large values of $d$, but
more careful techniques will be required to prove this.

\subsection{A gate for which $C_+(U)$ may be less than $E(U)$}
\label{sec:CnE-example}

Another separation that appears plausible is between the total
classical capacity $C_+(U)$ and the entanglement capacity $E(U)$.
In this section we present an example of a gate $U$ for which it
appears that $C_+(U)<E(U)$, though, as with the last section, we
cannot actually prove this claim.

The gate is defined (for any $d$) as follows:
$U = I + \ket{\Phi_d}\bra{01} + \ket{01}\bra{\Phi_d} - \ket{01}\bra{01} -
\ket{\Phi_d}\bra{\Phi_d}$.   Obviously, $E(U)\geq\log d$, since
$U\ket{01}=\ket{\Phi_d}$.  This inequality is not quite tight
(i.e. $E(U)>\log d$ and probably $E(U)\approx \log d + O(1)$), but
this doesn't matter for the argument.

I conjecture that $C_+^E(U) = O(1) < \log d$ for large $d$.  However,
the only statement that can readily be proven is that, like the last
section, 
 a single use of $U$ for one-way communication on uncorrelated
product inputs can create strictly less than $\log d$ bits of mutual
information, for $d$ sufficiently large.

The proof is actually almost identical to the proof of the last
section, though slightly simpler.  If Alice and Bob input product
states, then the overlap of their states with $\ket{\Phi_d}$ is $\leq
1/\sqrt{d}$, so this portion of $U$ has little effect.  We divide
Alice's signal ensemble into a part with a large $\ket{0}$ component
(which has low entropy) and a part with a small $\ket{0}$ component
(which is nearly unchanged by the action of $U$).  As a result, the
total amount of information that Alice can send to Bob (or that Bob
can send to Alice) with one use of $U$, starting from uncorrelated
product states, is strictly less than the entanglement capacity.
However, this argument is far from strong enough to prove a separation
between asymptotic capacities.

\section{Discussion}\label{sec:u-discuss}
We conclude this chapter by restating its key results and discussing
some of the major open questions.  Most of the gate capacities can be
expressed in terms of the three-dimensional region
$\CCE(U):=\{(C_1,C_2,E) : \<U\> \geq C_1[c\ra c] + C_2[c \la c] +
E[qq]\}$.  The two coding theorems (\ref{thm:U-E-cap} and
\ref{thm:U-CE-cap}) establish that
\begin{itemize}
\item 
$\max\{E : \<U\>\geq E[qq]\} =: E(U) = \Delta E_U
 := \sup_{\psi} H(B)_{U(\psi)} - H(B)_{\psi}$
\item
$\max\{C : \<U\> + \infty [qq] \geq C [c\ra c]\} =: C_\ra^E(U) =
\Delta\chi_U 
:= \sup_\cE \chi(\tr_A U(\cE)) - \chi(\tr_A \cE)$
\end{itemize}
The key bounds (from Propositions \ref{prop:U-nonzero-E},
\ref{prop:U-nonzero-CC}, \ref{prop:U-TP-bound} and \ref{prop:U-CgE})
are
\begin{itemize}
\item $C_+(U) \leq E(U) \leq \log \Sch(U) \leq 2\log d$
\item $C_+^E(U) \leq \min(4\log d, E(U)+E(U^\dag))$
\item $C_\ra^E(U),C_\la^E(U) \leq 2\log d$
\item $E(U) \geq H(\lambda)[qq]$ where $\{\lambda_i\}$ are Schmidt
coefficients of $U$.
\item $C_\ra(U)\neq 0 \iff C_\la(U)\neq 0 \iff E(U)\neq 0 \iff
\Sch(U)\neq 1 \iff U$ is nonlocal.
\end{itemize}

These results suggest a number of open questions.
\bi\item Can
we find an upper bound on the dimension of the ancillas $A'B'$ that
are needed for an optimal input state for entanglement generation?
For entanglement-assisted classical communication, how large do the
dimensions of $A'B'$ need to be, and how many states are needed in the
optimal ensemble?
These are important for numerical studies of the capacities.  

\item Do there exist $U$ such that $C_\ra(U)\neq C_\la(U)$?  Note that
$U$ cannot be a two-qubit gate since the decomposition in
\eq{magic-basis} implies that two-qubit gates have symmetric
capacities.  I conjecture that $C_\ra(\Uxoxo)\neq C_\la(\Uxoxo)$ for
$\Uxoxo$ defined as in \sect{u-xoxo}.

\item Do there exist $U$ such that $C_\ra^E(U)\neq C_\la^E(U)$?  All
of the examples of gates in \sect{u-examples} satisfy $U=U^\dag$, but
unpublished work with Peter Shor proves that in this case the
entanglement-assisted capacity regions are fully symmetric.  It seems
plausible that this situation would hold in general, but no proof or
counterexample is known.

\item Do there exist $U$ for which $C_+(U) < E(U)$?  I conjecture that
this inequality holds for the gate defined in \sect{CnE-example}.

\item Is $E(U)=E(U^\dag)$?  Both quantities relate to how entangling a
nonlocal gate is.  However, 
we can only prove the equality when $U = U^T$, by using the fact $E(U)
= E(U^*)$\footnote{This
is because $\max_\psi E(U\ket{\psi}) - E(\ket{\psi}) = 
\max_\psi E(U\ket{\psi^*}) - E(\ket{\psi^*}) =
\max_\psi E(U^*\ket{\psi}) - E(\ket{\psi})$.}.
This generalizes the proof in Ref.~\cite{BS03a} for $2$-qubit gates
since $U = U^T$ for all $2$-qubit gates that are decomposed in the
form of \eq{magic-basis}.
Numerical work suggests that the equality does not hold for some $U$ 
in higher dimensions~\cite{CLS02}.  More generally, we can ask whether
$\CCE(U)=\CCE(U^\dag)$.  

\item Is $E(U)$ completely determined by the Schmidt coefficients of $U$?

\item It seems unlikely that classical capacity can be determined by Schmidt
coefficients alone, but can we derive better lower and upper bounds on
classical capacity based on the Schmidt coefficients of a gate?
Specifically, can we show that $C_+^E(U) \leq 2\log \Sch(U)$, or even
better, that $\log \Sch(U) \l([q\ra q] + [q\la q]\r) \geq \<U\>$?
Right now these inequalities are only known to be true when $\Sch(U)$
is maximal (i.e. equal to $d_Ad_B$ when $U\in\cU_{d_A\times d_B}$).
\ei

\chapter{Coherent classical communication}\label{chap:ccc}
\section{Introduction and definition}

One of the main differences between classical and quantum Shannon
theory is the number of irreversible, but optimal, resource
transformations that exist in quantum Shannon theory.  The highest
rate that ebits or cbits can be created from qubits is one-for-one:
$\qtq\geq \qq$ and $\qtq\geq \ctc$.  But the best way to
create qubits from cbits and ebits is teleportation: $2\ctc+\qq\geq
\qtq$.  These protocols are all asymptotically 
optimal---for example, the classical communication requirement of
teleportation cannot be decreased even if entanglement is free---but
composing them is extremely wasteful: $3\qtq \geq 2\ctc + \qq
\geq \qtq$.  This sort of irreversibility represents one of the
main challenges of quantum information theory: resources may be
qualitatively equivalent but quantitatively incomparable.

In this chapter we will introduce a new primitive resource: the {\em
coherent bit} or {\em cobit}.  To emphasize its connection with
classical communication, we denote the asymptotic resource (defined
below) by $\cof$.\footnote{Other work\cite{DHW05,Devetak05a} 
uses $[q\ra qq]$ to denote cobits, in order to emphasize their central
place among isometries from $A$ to $AB$.}
Coherent classical communication will
simplify and improve a number of topics in quantum Shannon theory:
\bi
\item We will find that coherently decoupled cbits can be described
more simply and naturally as cobits.
\item Replacing coherently decoupled cbits with cobits will make many
resource 
transformations reversible.  In particular, teleportation and
super-dense coding become each other's inverses, a result previously
only known when unlimited entanglement is allowed.
\item More generally, we find that many forms of irreversibility in
quantum Shannon theory are equivalent to the simple map $\cof\geq
\ctc$. 
\item We will expand upon \prop{U-CgE} to 
precisely explain how cbits are more powerful when they are sent
through unitary means.  This has a number of consequences for unitary
gate capacities.
\item In the next chapter, coherent classical communication will be
used to relate many of the different protocols in quantum
Shannon theory, give simple proofs of some existing protocols and
create some entirely new protocols.  These will allow us to determine
two-dimensional tradeoff curves for the capacities of channels and
states to create or consume cbits, ebits and qubits.
\ei

Coherent classical communication  can be defined in
two ways, which we later show to be equivalent.  

\bi
\item {\em Explicit definition in terms of finite resources:} 

\sloppypar{Fix a basis for $\bbC^d$: $\{\ket{x}\}_{x=0}^{d-1}$.
First, we recall from 
\sect{distantlabs} the definitions of quantum and classical
communication: $\id_d = \sum_x
\ket{x}^{B} \bra{x}^{A'}$ (a perfect quantum
channel), $\bar{\id}_d = \sum_x \ket{x}^{B}\ket{x}^{E} \bra{x}^{A'}$
(a perfect classical channel in the QP formalism) and
$\bar{\Delta}_d = \sum_x \ket{x}^A\ket{x}^{B}\ket{x}^{E} \bra{x}^{A'}$
(the classical copying operation in the QP formalism).  Then we define
a perfect coherent channel as
\be \Delta_d = \sum_{x=0}^{d-1}
\ket{x}^A\ket{x}^{B} \bra{x}^{A'}.\ee
It can be thought of as a
purification of a cbit in which Alice controls the environment, as a
sort of quantum analogue to a feedback channel.
The asymptotic resource is then given by $\cof := \<\Delta_2\>$.}

\item {\em Operational definition as an asymptotic resource:}
We can also define a cobit as a cbit sent through unitary, or more
generally isometric, means.  The approximate version of this statement
is that whenever a protocol creates coherently decoupled cbits
(cf.~\defn{coh-decoupling-output}), then a modified
version of the protocol will create cobits.  Later we will prove a precise
form of this statement,  known as ``Rule
O,'' because it describes how output cbits should be made coherent.

When $C$ {\em input cbits} are coherently decoupled
(cf.~\defn{coh-decoupling-input}) we instead find that replacing them
with $C$ cobits results in $C$ extra ebits being generated in the
output.  This input rule is known as ``Rule I.''  Both rules are
proved in \sect{ccc-proofs}.

The canonical example of coherent decoupling is when cbits are sent
using a unitary gate.  In \thm{bidi-ccc}, we show that cbits sent
through unitary means can indeed be coherently decoupled, and thereby
turned into cobits.
\ei

The rest of the chapter is organized as follows.
\begin{description}
\item[\sect{ccc-source}] will give some simple examples of how cobits
can be obtained.
\item[\sect{ccc-use}] will then describe how to use coherent
classical communication to make quantum protocols reversible and more
efficient.  It will conclude with a precise statement of Rules I and O.
\item[\sect{ccc-apps}] will apply these general principles to remote
state preparation\cite{BHLSW03}, which leads to new 
protocols for super-dense coding of quantum state\cite{HHL03} as well
as many new results for unitary gate capacities.
\item[\sect{ccc-proofs}] collects some of the longer proofs from the
chapter, in order to avoid interrupting the exposition of the rest of
the chapter. 
\item[\sect{ccc-discuss}] concludes with a brief discussion.
\end{description}

{\em Bibliographical note:} Most of this chapter is based on
\cite{Har03}, though in \sect{ccc-proofs} the proofs of Rules I and O
are from 
\cite{DHW05} (joint work with Igor Devetak and Andreas Winter).
and the proof of \thm{bidi-ccc} is from
\cite{HL04} (joint work with Debbie Leung).

\section{Sources of coherent classical communication}
\label{sec:ccc-source}

Qubits and cbits arise naturally from noiseless and dephasing channels
respectively, and can be obtained from any noisy channel by
appropriate coding \cite{Holevo98,SW97,Lloyd96,Shor02,Devetak03}.
Similarly, we will show both a 
natural primitive yielding coherent bits and a coding theorem that
can generate coherent bits from a broad class of unitary operations.

The simplest way to send a coherent message is by modifying
super-dense coding (SD).  In SD, Alice and Bob begin
with $\ket{\Phi_2}$ and want to use $\id_2$ to send a two bit
message $a_1a_2$ from Alice to Bob.  Alice encodes her message by
applying $Z^{a_1}X^{a_2}$ to her half of 
$\ket{\Phi_2}$ and then sending it to Bob, who decodes by applying
$(H\otimes I)\textsc{cnot}$ to the state, obtaining
$$(H\otimes I)\textsc{cnot}(Z^{a_1}X^{a_2}\otimes I)\ket{\Phi_2}
=\ket{a_1}\ket{a_2}$$
Now modify this protocol so that Alice starts with a quantum state
$\ket{a_1a_2}$ and applies $Z^{a_1}X^{a_2}$ to her half of
$\ket{\Phi_2}$ conditioned on her quantum input.  After she sends her
qubit and Bob decodes, they will be left with the state
$\ket{a_1a_2}^A\ket{a_1a_2}^B$.  Thus, 
\be \qtq + \qq \reduction 2 \cof \label{eq:SDC-cc}\ee

In fact, any unitary operation capable of classical communication is
also capable of an equal amount of coherent classical communication,
though in general this only holds asymptotically.  The following
theorem gives a general prescription for obtaining coherent
communication and proves part of the equivalence of the two
definitions of cobits given in the introduction.

\begin{theorem}\label{thm:bidi-ccc}
For any bipartite unitary or isometry $U$, if
\be
	\<U\>  \geqslant  C_1 \ctc + C_2 \cbc + E \qq
\label{eq:cbit-toff} 
\ee
for $C_1,C_2\geq 0$ and $E\in\bbR$ then
\be
	\<U\>  \geqslant  C_1 \cof + C_2 \cob + E \qq
\label{eq:cobit-toff} 
\ee
\end{theorem}
If we define $\CoCoE(U)=\{(C_1,C_2,E) : \<U\>  \geqslant  C_1 \cof + C_2
\cob + E \qq\}$, then this theorem states that $\CCE(U)$ and
$\CoCoE(U)$ coincide on the quadrant $C_1,C_2\geq 0$.

Here we will prove only the case where $C_2=0$, deferring the full
bidirectional proof to \sect{ccc-proofs}.  By appropriate coding (as
in \cite{BS03b}), we can reduce the one-way case of \thm{bidi-ccc} to
the following coherent analogue of HSW coding.

\begin{lemma}[Coherent HSW]\label{lemma:HSW}
Given a PP ensemble of bipartite pure states
\be \ket{\cE} = \sum_{x\in\cX}
\sqrt{p_x}\ket{x}^R\ket{x}^{X_A}\ket{\psi_x}^{AB}\ee
 and an isometry 
\be U_\cE = \sum_x \oprod{x}^{X_A} \ot \ket{\psi_x}^{AB}\ee
then 
\be \<U_\cE : \cE^{X_A}\> \geq I(X_A;B)_\cE\cof + H(B|X_A)\qq.\ee

\end{lemma}

\begin{proof}
A slightly modified form of HSW coding (e.g. \cite{Devetak03}) holds
that for any $\delta>0,\epsilon>0$ and 
every $n$ sufficiently large there exists a code $\cC\subset\cS^n$
with $|\cC|=\exp(n(I(X_A;B)_\cE-\delta))$, a decoding POVM
$\{D_{c^n}\}_{c^n\in\cC}$ with error $<\epsilon$ and a type $q$ with
$\|p-q\|_1\leqslant |\cX|/n$ such that every codeword $c^n:=c_1\ldots
c_n\in\cC$ (corresponding to the state $\ket{\psi_{c^n}}^{AB}:=
\ket{\psi_{c_1}}^{A_1B_1}\cdots \ket{\psi_{c_n}}^{A_nB_n}$)
has type $q$ (i.e. $\forall x, |\{c_j=x\}|=nq_x$).  By
error $<\epsilon$, we mean that for any $c^n\in\cC$,
$\bra{\psi_{c^n}}(I\otimes D_{c})\ket{\psi_{c^n}} > 1-\epsilon$.

Using Neumark's Theorem\cite{Peres93}, Bob can make his decoding POVM into
a unitary operation $U_D$ defined by $U_D\ket{0}\ket{\phi} = \sum_{c^n}
\ket{c^n} \sqrt{D_{c^n}}\ket{\phi}$.  Applying this to his half of a
codeword $\ket{\psi_{c^n}}$ will
yield a state within $\epsilon$ of $\ket{c^n}\ket{\psi_{c^n}}$, since
measurements with nearly certain outcomes cause almost no
disturbance\cite{Winter99}.

The communication strategy begins by applying $U_\cE$ to
$\ket{c^n}_{X_A}$ to obtain $\ket{c^n}^{X_A}\ket{\psi_{c^n}}^{AB}$.
Bob then decodes unitarily with $U_D$ to yield a state within
$\epsilon$ of $\ket{c^n}^{X_A}\ket{c^n}^{X_B}\ket{\psi_{c^n}}^{AB}$.
Since $c^n$ is of type $q$, Alice and Bob can coherently permute the
states of $\ket{\psi_{c^n}}$ to obtain a state within $\epsilon$ of
$\ket{c^n}_{X_A}\ket{c^n}_{X_B}\ket{\psi_1}^{\otimes nq_1}\cdots
\ket{\psi_{|\cX|}}^{\otimes nq_{|\cX|}}$.  Then they can apply
entanglement concentration\cite{BBPS96} to $\ket{\psi_1}^{\otimes
nq_1}\cdots \ket{\psi_{|\cX|}}^{\otimes nq_{|\cX|}}$ to obtain
$\approx nH(B|X_A)_\cE$ ebits without disturbing the coherent message
$\ket{c^n}_{X_A}\ket{c}_{X_B}$.
\end{proof}
This will be partially superseded by the full proof of \thm{bidi-ccc}.
However, it is worth appreciating the 
key ideas of the proof---making measurements coherent via Neumark's
Theorem and finding a way to decouple ancillas shared by Alice and
Bob---as they will appear again in the later proofs, but surrounded by
more mathematical details.

There are many cases in which no ancillas are produced, so we do not
need the assumptions of Lemma~\ref{lemma:HSW} that communication 
 occurs in large blocks.  For example, a {\sc cnot} can
transmit one coherent bit from Alice to Bob or one coherent bit from
Bob to Alice.  Recall the protocol given in \eq{e-asstd-cnot} for
$\cnot + \qq\reduction \ctc+\cbc$:
$ (Z^aH\otimes I)\textsc{cnot}(X^a\otimes Z^b)\ket{\Phi_2}^{AB} =
\ket{b}^A\ket{a}^B$. 
This can be made coherent by conditioning
the encoding on a quantum register $\ket{a}^{A'}\ket{b}^{B'}$, so that
\be\textsc{cnot} + \qq \reduction \cof + \cob
\label{eq:CNOT-power1}\ee

\section{Rules for using coherent classical communication}
\label{sec:ccc-use}

By discarding her state after sending it, Alice can convert coherent
communication into classical communication, so
$\cof\geqslant \ctc$.  Alice can also generate entanglement by
inputting a superposition of messages (as in \prop{U-CgE}), so
$\cof\geqslant \qq$.  The true power of coherent
communication comes from performing both tasks---classical communication
and entanglement generation---simultaneously.  This is possible
whenever the classical message sent is coherently decoupled,
i.e. random and nearly independent 
of the other states at the end of the protocol.

Teleportation \cite{BBCJPW98} satisfies these conditions, and indeed a
coherent version has already been proposed in \cite{BBC98}.  Given an
unknown quantum state $\ket{\psi}^{A}$ and an EPR pair
$\ket{\Phi_2}^{AB}$, Alice begins coherent teleportation not by a Bell
measurement on her two qubits but by unitarily rotating the Bell basis
into the computational basis via a {\sc CNOT} and Hadamard gate.  This
yields the state $\frac{1}{2}\sum_{ij}\ket{ij}^AX^iZ^j\ket{\psi}^B$.
Using two coherent bits, Alice can send Bob a copy of her register to
obtain $\frac{1}{2}\sum_{ij}\ket{ij}^A\ket{ij}^BX^iZ^j\ket{\psi}^B$.
Bob's decoding step can now be made unitary, leaving the state
$(\ket{\Phi_2}^{AB})^{\otimes 2}\ket{\psi}^B$.  In terms of resources,
this can be summarized as: $2\cof + \qq \reduction \qtq + 2\qq$.
Canceling the ebits on both sides (possible since $\cof\geq \qq$) gives
$2\cof \geq \qtq + \qq$.
Combining this relation with \eq{SDC-cc} yields the
equality\footnote{Our use of the Cancellation Lemma means that this
equality is only asymptotically valid.  \mscite{vanEnk05} proves a
single-shot version of this equality, but it requires that the two
cobits be applied in series, with local unitary operations in between.}
\be 2\cof = \qtq + \qq\label{eq:cc-equality}.\ee
This has two important implications.  First, teleportation and
super-dense coding are reversible so long as all of the classical
communication is left coherent.  Second, cobits are equivalent, as
resources, to the existing resources of qubits and ebits.  This means
that we don't need to calculate quantities such as the cobit capacity
of a quantum channel; coherent communication introduces a new tool for
solving old problems in quantum Shannon theory, and is not directly a
source of new problems.

Another protocol that can be made coherent is Gottesman's
method\cite{Got98} for simulating a distributed CNOT using one
ebit and one cbit in either direction.  At first
glance, this appears completely irreversible, since a CNOT can be used
to send one cbit forward or backwards, or to create one ebit, but no
more than one of these at a time.

Using coherent bits as inputs, though, allows the recovery of 2 ebits
at the end of the protocol, so $\cof + \cob + \qq \reduction
\<\cnot\> + 2\qq$, or using entanglement catalytically,
$\cof + \cob \geq \<\cnot\> + \qq$.  Combined with
\eq{e-asstd-cnot}, this yields another equality:
$$\<\cnot\>+\qq = \cof + \cob.$$
Another useful bipartite unitary gate is \textsc{swap}, which we
recall is equivalent to $\qtq + \qbq$.  Applying
\eq{cc-equality} then yields
$$2\<\cnot\>=1\<\swap\>$$
which explains the similar communication and entanglement capacities
for these gates found in the last chapter.  Previously, the most efficient
methods known to transform between these gates gave $3\<\cnot\>\geq
1\<\swap\> \geq 1\<\cnot\>$.

A similar argument can be applied to \dcnot.  Since $\<\dcnot\> + 2\qq
\geq 2\ctc + 2\cbc$, it follows (from \thm{bidi-ccc} or direct
examination) that $\<\dcnot\> + 2\qq \geq 2\cof + 2\cob$ and
that $\<\dcnot\> \geq \qtq + \qbq = \<\swap\>$.  Combining
this with \prop{U-TP-bound}, we find that $\<\dcnot\>=\<\swap\>$,
a surprising fact in light of the observation of \cite{HVC02} that
\dcnot\ is easier for some nonlocal Hamiltonians to simulate than
\swap.  In fact, by the same argument, any gate in $\cU_{d\times d}$
with $C_+^E(U)=4\log d$ must be equivalent to the $d\times d$ \swap\
gate.

The above examples give the flavor of when classical communication can
be replaced by coherent communication (i.e. ``made coherent.'')  In
general, we require that the classical message be (almost) uniformly
random and (almost) coherently decoupled from all other systems,
including the environment.  This leads us to two general rules regarding making
classical communication coherent.  When coherently-decoupled cbits are
in the input to a protocol, Rule I (``input'') says that replacing them with
cobits not only performs the protocol, but also has the side 
effect of generating entanglement.  Rule O (``output'') is simpler; it
says that if a protocol outputs coherently-decoupled cbits, then it
can be modified to instead output cobits.  
Once coherently decoupled cbits are replaced
with cobits we can then use \eq{cc-equality} to in turn replace
cobits with qubits and ebits.  Thus, while cobits are conceptually
useful, we generally start and finish with protocols involving the
standard resources of cbits, ebits and qubits.

Below we give formal statements of rules I and O, deferring their
proofs till the end of the chapter.

\begin{theorem}[Rule I]
If, for some quantum resources $\alpha, \beta \in {\cR}$,
$$
\alpha + R \, [c \rightarrow c : \tau] \geq \beta 
$$
and the classical resource $ R \, \ctctau$ is 
coherently decoupled  then
$$
\alpha + \frac{R}{2} \,[q \rightarrow q] \geq \beta + \frac{R}{2} \,[q \, q].
$$
\end{theorem}

{\bf Remark:} This can be thought of as a coherent version of
\lem{rcr}.

The idea behind the proof is that replacing $R [c\ra
c:\tau]$ with $R \coftau$ then gives an extra output of $R\qq$,
implying that $\alpha + R\coftau \geq \beta + R\qq$.  Then $\coftau$
can be replaced by $\half(\qtq + \qq)$ using \eq{cc-equality} and
\lem{noo}.  To prove this rigorously will require carefully accounting
for the errors, which we will do in \sect{ccc-proofs}.

\begin{theorem}[Rule O]
If, for some quantum resources $\alpha, \beta \in {\cR}$, 
$$
\alpha  \geq \beta + R \,[c \rightarrow c]
$$
and the classical resource is 
decoupled with respect to the RI then 
$$
\alpha  \geq \beta + \frac{R}{2} \, [q \, q] +  
\frac{R}{2} \, [q \rightarrow q].
$$
\end{theorem}

Here the proof is even simpler: $R\ctc$ in the output is replaced with
$R\cof$, which is equivalent to $\frac{R}{2}(\qtq+\qq)$.  Again, the
details are given in \sect{ccc-proofs}.

In the next chapter, we will show how Rules I and O can be used to
obtain a family of optimal protocols (and trade-off curves) for
generating cbits, ebits and qubits from noisy channels and states.
First, we show a simpler example of how a protocol can be made
coherent in the next section.

\section{Applications to remote state preparation and unitary gate
capacities}
\label{sec:ccc-apps}
\subsection{Remote state preparation}
Remote state preparation (RSP) is the task of simulating a $\{c\ra
q\}$ channel, usually using cbits and ebits.  In this section, we show
how RSP can be made coherent, not only by applying Rule I to the input
cbits, but also by replacing the $\{c\ra q\}$ channel by a coherent
version that will preserve superpositions of inputs.  Finally, we will
use this coherent version of RSP to derive the capacity of a unitary
gate to send a classical message from Alice to Bob while
using/creating an arbitrary amount of entanglement.

Begin by recalling from \sect{known} our definition of RSP.  Let
$\cE=\sum_i p_i \oprod{i}^{X_A} \ot \oprod{\psi_i}^{AB}$ be an 
ensemble of bipartite states and  $\cN_\cE:\oprod{i}^{X_A}
\ra \oprod{i}^{X_A} \ot \oprod{\psi_i}^{AB}$ the $\{c\ra q\}$ channel
such that $\cN(\cE^{X_A}) = \cE$.  The main coding theorem of
RSP\cite{BHLSW03} states that
\be
I(X_A ; B)_\cE [c\ra c] + H(B)_\cE \qq \geq \<\cN_\cE : \cE^{X_A}\>.
\label{eq:RSP2}\ee

We will show that the input cbits in \eq{RSP2} are coherently
decoupled, so that according to Rule I, replacing them with cobits
will perform the protocol and return some entanglement at the same
time.  This reduces the entanglement cost to $H(B) - I(X_A;B) =
H(B|X_A)$, so that
\be
I({X_A}:B)_\cE \cof + H(B|X_A)_\cE \qq \geq \<\cN_\cE : \cE^{X_A}\>.
\label{eq:cobit-rsp}\ee
In fact, we can prove an even stronger statement, in which not only is
the input coherently 
decoupled, but there is a sense in which the output is as well.
Define a coherent analogue of $\cN_\cE$, which we call $U_\cE$, by
\be U_\cE = \sum_i \oprod{i}^{X_A} \ot \ket{\psi_i}^{AB}.\ee
We also replace the QP ensemble $\cE$ with the (PP formalism) pure
state $\ket{\cE}$ given by
\be \ket{\cE} = \sum_i \sqrt{p_i} \ket{i}^R \ket{i}^{X_A}
\ket{\psi_i}^{AB}.\ee
We will prove that
\be
I({X_A}:B)_\cE \cof + H(B|X_A)_\cE \qq \geq \<U_\cE : \cE^{X_A}\>.
\label{eq:coherent-rsp}\ee
Since $\<U_\cE : \cE^{X_A}\> \geq \<\cN_\cE : \cE^{X_A}\>$, this
RI implies \eq{cobit-rsp}; in particular, the presence of
the reference system $R$ ensures that $\cE^{X_A}$ is the same in both
cases, even if the $\ket{\psi_i}$ are not all orthogonal.
Proving \eq{coherent-rsp} will require careful examination of the
protocol from \cite{BHLSW03}, so we defer the details until
\sect{ccc-proofs}.

{\bf Remark:} An interesting special case is when $H(A)_\cE=0$, so
that Alice is preparing pure states in Bob's lab rather than entangled
states.  In this case, $H(B|X_A)_\cE=0$ and \eq{cobit-rsp} becomes
simply
\be H(B)_\cE \cof \geq \<U_\cE : \cE^{X_A}\>.\ee
Thus, if we say (following \cite{BHLSW03}) that \eq{RSP2} means that
``1 cbit + 1 ebit $\geq$ 1 
remote qubit,'' then \eq{cobit-rsp} means that ``1 cobit $\geq$ 1
remote qubit.''  Here ``$n$ remote qubits'' mean the ability of Alice
to prepare an $n$-qubit state of her choice in Bob's lab, though we
cannot readily define an asymptotic resource corresponding to this
ability, since it would violate the quasi-i.i.d. condition
(\eq{asy-quasi-iid}).  Despite not being
formally defined as a resource, we can 
think of remote qubits as intermediate in strength between qubits and
cbits, just as cobits are; i.e. 1 qubit $\geq$ 1 remote qubit $\geq$ 1
cbit.  As resources intermediate between qubits and cbits, remote
qubits and cobits have complementary 
attributes: remote qubits share with qubits the ability to transmit
arbitrary pure states, though they cannot create entanglement, while
cobits can generate entanglement, but at first glance appear to only
be able to faithfully transmit the computational basis states to Bob.
Thus it is interesting that in fact 1 cobit $\geq$ 1 remote
qubit, and that (due to \cite{BHLSW03}) this map is optimal.

\eq{cobit-rsp} yields two other useful corollaries, which we state
in the informal language of remote qubits.

\begin{corollary}[RSP capacity of unitary gates]
If $U$ is a unitary gate or isometry with
 $\<U\>\geq C \ctc$  then $\<U\>\geq C \rqbsf$.
\end{corollary}

\begin{corollary}
{\bf (Super-dense coding of quantum states)}
\label{cor:sddc}
$\qtq + \qq \geq 2 \rqbsf$
\end{corollary}
More formally, we could say that if
 $H(B)_\cE \leq C$  for an ensemble $\cE$, then $\<U\> \geq \<U_\cE
 : \cE^{X_A}\>$, and similarly for \cor{sddc}.  We can also express
 \cor{sddc} entirely in terms of standard resources as
\be \half I(X_A ; B)_\cE \qtq  + \l(H(B)_\cE - \half I(X_A ; B)_\cE\r)
 \qq \geq \<U_\cE : \cE^{X_A}\>.\ee
Though this last expression is not particularly attractive, it turns
out to be optimal, and in fact to give rise to optimal trade-offs for
performing RSP with the three resources of cbits, ebits and
qubits\cite{AH03} (see also \cite{AHSW04} for a single-shot version of
the coding theorem).  We will find this pattern repeated many times in
the next chapter; by making existing protocols coherent and using
basic information-theoretic inequalities, we obtain a series of
optimal tradeoff curves.

Corollary~\ref{cor:sddc} was first proven directly in \cite{HHL03}
(see also \cite{AHSW04}) and
in fact, finding an alternate proof was the original motivation for
the idea of coherent classical communication.

{\em Coherent RSP:} Now, we explore the consequences of the stronger
version of coherent RSP in \eq{coherent-rsp}.  Just as RSP and HSW
coding reverse one another given free entanglement, coherent RSP
(\eq{coherent-rsp}) and coherent HSW coding (\lem{HSW}) reverse each
other, even taking entanglement into account.  Combining them gives
the powerful equality
\be
I({X_A}:B)_\cE \cof + H(B|X_A) \qq = \<U_\cE : \cE^{X_A}\>,
\label{eq:coh-RSP-HSW}\ee
which improves the original RSP-HSW duality in \eq{RSP-HSW-eq} by
eliminating the need for free 
entanglement.   This remarkable statement simultaneously implies
entanglement concentration, entanglement dilution, the HSW theorem and
remote state preparation and super-dense coding of entangled
states.\footnote{On the other hand, we had to use almost all of these
statements in order to prove the result!  Still it is nice to see them
all unified in a single powerful equation.  Also, recent work by
Devetak\cite{Devetak05a} further generalizes the equalities that can
be stated about isometries from $A$ to $AB$.}  

\subsection{One-way classical capacities of unitary gates}
\label{sec:U-Ce-cap}
Here we will use \eq{coh-RSP-HSW} to determine the 
capacity of a unitary gate $V$ to simultaneously send a classical message
and generate or consume entanglement at any finite rate.  The proof idea is
similar to one in \thm{U-CE-cap}; we will use the equivalence between
(coherent) ensembles and standard resources (cobits and ebits) to turn
a one-shot improvement in mutual information and expected entanglement
into an asymptotically efficient protocol.  Now that we have an
improved version of the duality between RSP and HSW coding, we obtain
a precise accounting of the amount of entanglement generated/consumed.

\begin{theorem}\label{thm:Ce-cap}
Define $\CE(V) := \{(C,E) : (C,0,E)\in\CCE(V)\}$ and 
\be \Delta_{I,E}(V) := \l\{ (C,E) : \exists \cE \st
I(X_A ; B)_{V(\cE)} - I(X_A ; B)_{\cE} \geq C
\text{ and }
H(B|X_A)_{V(\cE)} - H(B|X_A)_{\cE} \geq E\r\},\ee
where $\cE$ is an ensemble of bipartite pure states in $AB$ conditioned on a
classical register $X_A$.

Then $\CE(V)$ is equal to the closure of $\Delta_{I,E}(V)$.
\end{theorem}
Thus the asymptotic capacity using $-E$ ebits of assistance per use of
$V$ (or simultaneously outputting $E$ ebits) equals the largest
increase in mutual information possible with one use of $V$ if the
average entanglement decreases by no more than $-E$.  \thm{U-CE-cap}
proved this for $E=-\infty$ and our proof here is quite similar.  Note that
the statement of the theorem is the same whether we consider QP
ensembles $\cE$ or PP ensembles $\ket{\cE}$, though the proof
will use the coherent version of RSP in \eq{coherent-rsp}.

\begin{proof}
{\em  Coding theorem:}
Suppose there exists an ensemble $\cE$ with 
$C = I(X_A ; B)_{V(\cE)} - I(X_A ; B)_{\cE}$ and 
$E = H(B|X_A)_{V(\cE)} - H(B|X_A)_{\cE}$.
Then
\ben
\<V\> + \<U_\cE\> &\geq& \<U_{V(\cE)}\> \non\\
&\geq & I(X_A ; B)_{V(\cE)} \cof + H(B|X_A)_{V(\cE)} \qq
\\ &\geq & 
 \l(I(X_A ; B)_{V(\cE)} - I(X_A ; B)_{\cE}\r)\cof +
\l(H(B|X_A)_{V(\cE)} - H(B|X_A)_{\cE}\r)\qq + \,\<U_\cE\>
\een
Here the second RI used coherent HSW coding (Lemma~\ref{lemma:HSW}) and
the third RI used
coherent RSP (\eq{coherent-rsp}).  We now use the Cancellation Lemma
to show that 
$\<V\> \geq C\cof + E\qq$, implying that $(C,0,E)\in\CCE(V)$.

{\em Converse:}
We will actually prove a stronger result, in which Bob is allowed
unlimited classical communication to Alice.  Thus, we will show that
if $\<V\> + \infty \cbc \geq C \ctc + E \qq$, then there is a sequence
of ensembles $\{\tilde{\cE}_n\}$ with $(I(X_A ; B)_{V(\tilde{\cE}_n)}
- I(X_A ; B)_{\tilde{\cE}_n}, H(B|X_A)_{V(\tilde{\cE}_n)} -
H(B|X_A)_{\tilde{\cE}_n})$ converging to $(C,E)$ as $n\ra\infty$.
This will imply that $(C,E)$ is in the closure of $\Delta_{I,E}(V)$.

Let $Y:=Y_AY_B$ the cumulative record of all of Bob's classical
messages to Alice.  Using the QP formalism, we 
assume without loss of generality that Bob always transmits his full
measurement outcome (cf.~Section III of \cite{HL02}) so that Alice
and Bob always hold a pure state 
conditioned on $X_AY$; i.e. $H(AB|X_AY)=0$ and $H(A|X_AY)=H(B|X_AY) =
I(A\> BX_AY) = I(B \> AX_AY)$.  

First consider the case when $E>0$.  Fix a protocol which uses $V$ $n$
times to communicate $\geq n(C-\delta')$ bits and create $\geq
n(E-\delta')$ ebits with error $\leq \epsilon$.  They start with a
product state $\cE_0$ for which $I(X_A ; B)_{\cE_0}=0$ and $H(B |
X_A)_{\cE_0}=0$.  Denote their state immediately after $j$ uses of $V$
by ${\cE_j}$. (Without loss of generality, we assume that the $n$ uses
of $V$ are applied serially.)  Then by \lem{fano}, $I(X_A;B)_{\cE_n}
\geq n(C-\delta)$ and $H(B|X_AY)_{\cE_n} \geq n(E-\delta)$ where
$\delta=O(\delta+\epsilon)\ra 0$ as $n\ra \infty$.

Now define the ensemble $\tilde{\cE}_n=\smfrac{1}{n}
\sum_{j=1}^{n} \oprod{jj}^{Z_AZ_B} \ot V^\dag(\cE_j^{ABX_AY_AY_B})$.  We
think of $\hat{X}:=X_AY_AZ_A$ as the message variable and $\hat{B}:=
BY_BZ_B$ as representing Bob's system.  We will prove that
$I(\hat{X} ; \hat{B})_{V(\tilde{\cE}_n)}
- I(\hat{X} ; \hat{B})_{\tilde{\cE}_n} \geq C-\delta$ and that
$H(\hat{B}|\hat{X})_{V(\tilde{\cE}_n)} -
H(\hat{B}|\hat{X})_{\tilde{\cE}_n} \geq E-\delta$.

First consider the change in mutual information.  Since $Y_A=Y_B$ and
$Z_A=Z_B$ (as random variables), $I(\hat{X} ; \hat{B})_{\tilde{\cE}_n}
= I(X_AY_AZ_A ; BY_BZ_B)_{\tilde{\cE}_n} = I(X_A ; B |
YZ)_{\tilde{\cE}_n} + H(YZ)_{\tilde{\cE}_n}$ and similarly when we
replace $\tilde{\cE}_n$ with $V(\tilde{\cE}_n)$.  Since $V$ doesn't
act on $Y$ or $Z$, we have 
$H(YZ)_{\tilde{\cE}_n}=H(YZ)_{V(\tilde{\cE}_n)}$ and thus
\be\begin{split} I(\hat{X} ; \hat{B})_{V(\tilde{\cE}_n)}
&- I(\hat{X} ; \hat{B})_{\tilde{\cE}_n}
= I(X_A ; B | YZ)_{V(\tilde{\cE}_n)}
- I(X_A ; B | YZ)_{\tilde{\cE}_n}
\\&= \frac{1}{n}\sum_{j=1}^n I(X_A ; B | Y)_{\cE_{j}} - 
I(X_A ; B | Y)_{V^\dag(\cE_j)}
\\&= \frac{1}{n}\l(I(X_A ; B | Y)_{\cE_n} - I(X_A ; B | Y)_{\cE_0}\r)
+ \frac{1}{n}\sum_{j=1}^n\l(I(X_A ; B | Y)_{\cE_{j-1}} - 
I(X_A ; B | Y)_{V^\dag(\cE_j)}\r)
\\&\geq C - \delta
+ \frac{1}{n}\sum_{j=1}^n\l(I(X_A ; B | Y)_{\cE_{j-1}} - 
I(X_A ; B | Y)_{V^\dag(\cE_j)}\r)
\end{split}\ee
Recall that going from $\cE_{j-1}$ to $V^\dag(\cE_j)$ involves local
unitaries, a measurement by Bob and classical communication of the
outcome from Bob to Alice.  We claim that $I(X_A;B|Y)$ does not
increase under this process, meaning that the 
expression inside the sum on the last line is always nonnegative and
that $I(\hat{X} ; \hat{B})_{V(\tilde{\cE}_n)}
- I(\hat{X} ; \hat{B})_{\tilde{\cE}_n}\geq C-\delta$, implying our
desired conclusion.  To prove this,
write $I(X_A;B|Y)$ as $I(X_A;BY) - I(X_A;Y)$.  The $I(X_A;BY)$ term is
nonincreasing due to the data-processing inequality\cite{SN96}, while
$I(X_A;Y)$ can only increase since each round of communication only
causes $Y$ to grow.

Now we examine the change in entanglement.
\be\begin{split} H(\hat{B}|\hat{X})_{V(\tilde{\cE}_n)}
-& H(\hat{B}|\hat{X})_{\tilde{\cE}_n} = 
H(BY_BZ_B | X_AY_AZ_A)_{V(\tilde{\cE}_n)}
- H(BY_BZ_B | X_AY_AZ_A)_{\tilde{\cE}_n}
\\&=H(B | X_AYZ)_{V(\tilde{\cE}_n)}
- H(B | X_AYZ)_{\tilde{\cE}_n}
\\&= \frac{1}{n}\l(H(B|X_AY)_{\cE_n} - H(B|X_AY)_{\cE_0}\r)
 + \frac{1}{n}\sum_{j=1}^n
\l(H(B|X_AY)_{\cE_{j-1}} - H(B|X_AY)_{V^\dag(\cE_j)}\r)
\\ &\geq E - \delta
 + \frac{1}{n}\sum_{j=1}^n
\l(H(B|X_AY)_{\cE_{j-1}} - H(B|X_AY)_{V^\dag(\cE_j)}\r)
\end{split}\ee
 We would like to show that this last term is positive, or
equivalently that $H(B|X_AY)$ is at least as large for $\cE_{j-1}$ as
it is for $V^\dag(\cE_j)$.  This change from $\cE_{j-1}$ to
$V^\dag(\cE_j)$ involves local unitaries, a measurement by Bob and
another classical message from Bob to Alice, 
which we call $Y_j$.  Also, call the first $j-1$ messages $Y^{j-1}$.
Thus, we would like to show that $H(B|X_AY^{j-1})_{\cE_{j-1}} -
H(B|X_AY^{j-1}Y_j)_{V^\dag(\cE_j)} \geq 0$.  This can be expressed as an
average over $H(B)_{\cE_{j-1|x,y^{j-1}}} -
H(B|Y_j)_{V^\dag(\cE_{j|x,y^{j-1}})}$, where $\cE_{j-1|x,y^{j-1}}$
indicates that we have conditioned $\cE_{j-1}$ on $X_A=x$ and
$Y^{j-1}=y^{j-1}$.  This last quantity is positive 
because of principle that the average entropy of states output from a
projective measurement is no greater than the entropy of the
original state\cite{Nielsen99a}.  Thus $H(\hat{B}|\hat{X})_{V(\tilde{\cE}_n)}
- H(\hat{B}|\hat{X})_{\tilde{\cE}_n} \geq E-\delta$.

As $n\ra \infty$, $\delta\ra 0$, proving the theorem.

The case when $E\leq 0$ is similar.  We now begin with
$H(B|X_AY_A)_{\cE_0} \leq n(-E+\delta) = -n(E-\delta)$ and since $X_A$
and $Y$ are classical registers, finish with
$H(B|X_AY_A)_{\cE_n} \geq 0$.  Thus $H(B|X_AY_A)_{\cE_n} -
H(B|X_AY_A)_{\cE_0} \geq n(E-\delta)$.  The rest of the proof is the
same as the $E>0$ case.
\end{proof}

\subsection{Two-way cbit, cobit, qubit and ebit capacities of unitary
gates}
\label{sec:u-bidi-caps}
So far we have two powerful results about unitary gate capacity
regions: \thm{bidi-ccc} relates $\CCE$ and $\CoCoE$ in the
$C_1,C_2\geq 0$ quadrant and \thm{Ce-cap} gives an expression for
$\CE(U)$ in terms of a single use of $U$.  Moreover, the proof of
\thm{Ce-cap} also showed that backwards classical communication cannot
improve the forward capacity of a unitary gate.  This allows us to
extend \thm{bidi-ccc} to $C_1 \leq 0$ or $C_2 \leq 0$ as follows:

\begin{theorem}\label{thm:bidi-nnn}
For arbitrary real numbers $C_1,C_2,E$, 
\be
  (C_1,C_2,E) \in \CCE  \Longleftrightarrow
		(C_1,C_2,E{-}\min(C_1,0){-}\min(C_2,0)) \in 
	{\CoCoE} \,. 
\label{eq:thm12}\ee
\end{theorem}

This theorem is a direct consequence of the following Lemma, which
enumerates the relevant quadrants of the $(C_1,C_2)$ plane.

\begin{lemma}
For any bipartite unitary or isometry $U$ and $C_1,C_2 \geq 0$, 
\bea 
	C_2 \cbc + \<U\> & \geqslant & C_1 \ctc + E \qq
	\quad \quad {\rm iff}
\label{eq:1locc} 
\\
	\<U\> & \geqslant & C_1 \ctc + E \qq
	\quad \quad {\rm iff}
\label{eq:celo}
\\
	\<U\> & \geqslant & C_1 \cof + E \qq 
	\quad \quad {\rm iff}
\label{eq:ccelo}
\\
	C_2 \cob + \<U\> & \geqslant & C_1 \cof + (E{+}C_2) \qq
\label{eq:1loccc} 
\eea
and
\bea 
	C_1 \ctc + C_2 \cbc + \<U\> & \geqslant & E \qq 
	\quad {\rm iff}
\label{eq:2locc} 
\\
	\<U\> & \geqslant & E \qq 
	\quad {\rm iff}
\label{eq:elo}
\\
	C_1 \cof + C_2 \cob + \<U\> & \geqslant & (E{+}C_1{+}C_2) \qq 
\label{eq:2loccc} 
\eea
\end{lemma}

Basically, the rate at which Alice can send Bob cbits while
consuming/generating ebits is not increased by (coherent) classical
communication from Bob to Alice, except for a trivial gain of
entanglement when the assisting classical communication is coherent.

\begin{proof}
Combining (TP) and coherent SD (\eq{SDC-cc}) yields $2\ctc + \qq +
\qtq + \qq \reduction \qtq + 2\cof$.  Canceling the $\qtq$ from both sides
and dividing by two gives us
\be
	\ctc +  \qq \geqslant  \cof \,.
\label{eq:tp-sd}
\ee

For the first part of the lemma, recall from the proof of
\thm{Ce-cap} that free backcommunication does not improve the forward
capacity of a gate.  This means that \eq{1locc} $\Rightarrow$
\eq{celo}.  We obtain \eq{celo} $\Leftrightarrow$ \eq{ccelo} from
\thm{bidi-ccc} and \eq{ccelo} $\Rightarrow$ \eq{1loccc} follows from
$\cof\geq\qq$ and composability (\thm{composability}).  Finally,
\eq{1loccc} $\Rightarrow$ \eq{1locc} because of \eq{tp-sd}.

For the second part of the theorem, \eq{2locc} $\Rightarrow$ \eq{elo}
follows from \thm{U-E-cap}, \eq{elo} $\Rightarrow$ \eq{2loccc} is
trivial and \eq{2loccc} $\Rightarrow$ \eq{2locc} is a consequence of
\eq{tp-sd}.
\end{proof}

{\em Quantum capacities of unitary gates:} These techniques also allow
us to determine the quantum capacities of unitary gates.  Define $\QQE$
to be the region $\{(Q_1,Q_2,E): U \geqslant Q_1 \qtq
 + Q_2 \qbq +E \qq\}$, corresponding to two-way quantum
communication.  We can also consider coherent classical communication
in one direction and quantum communication in the other; let $\QCoE$ be
the region $\{(Q_1,C_2,E) : U \geqslant Q_1 \qtq + C_2\cob
+E \qq\}$ and define $\CoQE$ similarly.

As a warmup, we can use the equality $2\cof = \qtq + \qq$ to relate
C$\!_{\rm o}\!$E and QE, defined as 
${\rm C}\!_{\rm o}\!{\rm E}=\{(C,E) : (C,0,E)\in {\CoCoE} \}$ and 
$\QE=\{(Q,E) : (Q,0,E)\in \QQE\}$.  We claim that
\be
 (Q,E)\in\QE \Leftrightarrow (2Q, E-Q)\in {\rm C}\!_{\rm o}\!{\rm E} \,.
\label{eq:one-way-toff}
\ee
To prove \eq{one-way-toff}, choose any $(Q,E)\in\QE$.  
Then $\<U\> \geq Q \qtq + E\qq = 2Q \cof + (E-Q)\qq$, so 
$(2Q, E-Q)\in {\rm C}\!_{\rm o}\!{\rm E}$.  Conversely, if
$(2Q,E-Q)\in {\rm C}\!_{\rm o}\!{\rm E}$, then 
$U\geq 2Q \cof + (E-Q)\qq = Q \qtq + E \qq$, so $(Q,E)\in \QE$.

Note that the above argument still works if we add the same resource,
such as $Q_2\qbq$, to the right hand side of each resource inequality.
Therefore, the same argument that proved \eq{one-way-toff} also
establishes the following equivalences for bidirectional rate regions:
\be
\begin{array}{ccc}
(Q_1,Q_2,E)\in\QQE & \Longleftrightarrow & (2Q_1, Q_2, E-Q_1) \in{\CoQE} 
\\[2ex] \Updownarrow & & \Updownarrow
\\[2ex] (Q_1,2Q_2, E-Q_2) \in {\QCoE} & \Longleftrightarrow & 
	(2Q_1,2Q_2,E-Q_1-Q_2)\in {\CoCoE}
\end{array} .
\ee
Finally, \eq{thm12} further relates QQE, QCE, CQE, CCE, where QCE and
CQE are defined similarly to $\QCoE$ and $\CoQE$ but with incoherent
classical communication instead.

Thus once one of the capacity regions (say $\CoCoE$) is determined, all 
other capacity regions discussed above are determined.  The main open
problem that remains is to find an efficiently computable expression
for part of this capacity region.  \thm{Ce-cap} gives a formula for
the one-way cbit/ebit tradeoff that involves only a single use of the
unitary gate, but we still need upper bounds on the optimal ensemble
size and ancilla dimension for it to be practical.

\section{Collected proofs}
\label{sec:ccc-proofs}
In this section we give proofs that various protocols can be made
coherent.  We start with Rules I and O (from \sect{ccc-use}), which
gave conditions for when coherently decoupled cbits could be replaced
by cobits in asymptotic protocols.  Then we show specifically how
remote state preparation can be made coherent, proving
\eq{coherent-rsp}.  Finally, we show how two-way classical
communication from unitary operations can be made coherent, and prove
 \thm{bidi-ccc}. 

\subsection{Proof of Rule I}

In what follows we shall fix $\epsilon$ and consider a sufficiently
large $n$ so that the protocol is $\epsilon$-valid,
$\epsilon^2$-decoupled and accurate to within $\epsilon$.
  
Whenever the resource inequality features
$[c \rightarrow c]$ in the input this means that Alice performs
a von Neumann measurement on some subsystem $A_1$ of dimension $D
\approx \exp(n(R+\delta))$, 
the outcome of which she sends to Bob,
who then performs an unitary operation depending on the received information.

Before Alice's von Neumann measurement, the joint state of  $A_1$ and 
the remaining quantum system $Q$ is
$$
\sum_x \sqrt{p_x}\ket{x}^{A_1} \ket{\phi_x}^{Q},
$$
where by $\epsilon$-validity
$$
\sum_x |p_x - D^{-1}| \leq \epsilon. 
$$
Upon learning the measurement outcome $x$, Bob
performs some unitary $U_x$ on his part of $Q$, almost decoupling
it from $x$:
$$
\left\|\sum_x p_x \proj{x} \otimes \theta_x' -
\sum_x p_x \proj{x} \otimes \bar{\theta}' \right\|_1 
= \sum_x p_x  \| \theta_x' - \bar{\theta}'  \|_1 
\leq \epsilon^2,
$$
where $\ket{\theta_x'} = U_x \ket{\phi_x}$ and
$\bar{\theta}' = \sum_x p_x \theta_x$.  To simplify the analysis,
extend $Q$ to a larger Hilbert space on which there exist
purifications $\oprod{\bar{\theta}} \ext \bar{\theta}'$ and 
$\oprod{\theta_x} \ext \oprod{\theta_x'}$ such that (according to
\lem{purify-dist}) $\|\theta_x-\bar{\theta}\|_1 \leq 2
\sqrt{\|\theta_x'-\bar{\theta}'\|_1}$.  Then
\be
\sum_x p_x  \| \theta_x - \bar{\theta}  \|_1 \leq
\sum_x p_x  2\sqrt{\| \theta_x' - \bar{\theta}' \|_1} \leq
2 \sqrt{\sum_x p_x  \| \theta_x' - \bar{\theta}' \|_1} \leq
2\epsilon,\ee
where the second inequality uses the concavity of the square root.

If Alice refrains from the measurement and instead sends $A_1$ through
a \emph{coherent} channel, using $n(R+\delta)$ cobits, the resulting state is
$$
\sum_x \sqrt{p_x} \ket{x}^{A_1} \ket{x}^{B_1} \ket{\phi_x}^{Q}.
$$
Bob now performs the \emph{controlled} unitary 
$\sum_x \proj{x}^{B_1} \otimes  U_x$, giving rise to
$$
\ket{\Upsilon}^{A_1B_1Q} = 
\sum_x \sqrt{p_x} \ket{x}^{A_1} \ket{x}^{B_1} \ket{\theta_x}^Q.
$$
We may assume, w.l.o.g., that $\braket{\bar{\theta}}{\theta_x}$ is real
and positive for 
all $x$, as this can be accomplished by either Alice or Bob via
an $x$-dependent global phase rotation. 

We now claim that $\ket{\Upsilon}^{A_1B_1Q}$ is close to
$\ket{\Phi_D}^{A_1B_1} \ket{\bar{\theta}}^Q$.
Indeed
\be 
\bra{\Upsilon} \Gamma\rangle \ket{\bar{\theta}} =
\sum_x \sqrt{\frac{p_x}{D}} \braket{\theta_x}{\bar{\theta}}
\geq \sum_x \sqrt{\frac{p_x}{D}} 
\l(1 - \frac{1}{2}\| \theta_x - \theta \|_1\r), \ee
according to \eq{fid-trace}.
To bound this, we split the sum into two.
For the first term, we apply \eq{fid-trace} to the diagonal density
matrices $\sum_x p_x\proj{x}$ and $\sum_x D^{-1}\proj{x}$ to obtain
\be \sum_x \sqrt{\frac{p_x}{D}} \geq 1 - \frac{1}{2} \sum_x |p_x -
D^{-1}| \geq 1-\frac{\epsilon}{2} \ee
The second term is
\ben \sum_x \sqrt{\frac{p_x}{D}} \frac{1}{2}\| \theta_x - \theta \|_1
 &=& \sum_x \frac{1}{2}\l[p_x + \frac{1}{D} -
\bigl(\sqrt{p_x}-\sqrt{1/D}\bigr)^2\r]
 \frac{1}{2}\| \theta_x - \theta \|_1 \\
&\leq & \sum_x \frac{1}{2}\l(p_x + \frac{1}{D}\r)
 \frac{1}{2}\| \theta_x - \theta \|_1
\leq \sum_x \frac{1}{2}\l(2p_x + \l|p_x - \frac{1}{D}\r|\r)
 \frac{1}{2}\| \theta_x - \theta \|_1
\\ &\leq &\sum_x p_x  \frac{1}{2}\| \theta_x - \theta \|_1
+ \sum_x \l|p_x - \frac{1}{D}\r| \leq 2\epsilon. \een
Putting this together, we find that 
$$ \bra{\Upsilon} \Gamma\rangle \ket{\bar{\theta}} \geq 1-3\epsilon $$
and  by \eq{fid-trace},
$$
\|\Upsilon - \Phi_D \otimes \bar{\theta} \|_1 \leq \sqrt{6\epsilon}
$$ 
Finally, since tracing out subsystems cannot increase trace distance,
$$
\|\Upsilon^{A_1 B_1} - \Phi_D \|_1 \leq \sqrt{6\epsilon}
$$ 
Thus, the total effect of replacing cbits
cobits is the generation of a state close to
$\Phi_D$.  This analysis ignores the fact that the cobits are only
given up to an error $\epsilon$.  However, due to the triangle
inequality, this only enters in as an
additive factor, and the overall error of $\epsilon +
\sqrt{6\epsilon}$ is still asymptotically vanishing.  Furthermore,
this mapping preserves the $\epsilon$-validity of the original
protocol (with respect to the inputs of $\alpha$) since all we have
done to Alice's states is to add purifying systems and add phases,
which w.l.o.g.~we can assume are applied to these purifying systems.

We have thus shown
$$
\alpha + R \,\cof \geq \beta + R \,[q \, q].
$$
\eq{cc-equality} and Lemmas \ref{lemma:noo} and \ref{lemma:cancel}
give the desired result
$$
\alpha + \frac{R}{2} \,[q \rightarrow q] \geq \beta + \frac{R}{2} \,[q \, q].
$$
\qed\bigskip


\subsection{Proof of Rule O}
Again fix $\epsilon$ and consider a sufficiently
large $n$ so that the protocol is $\epsilon$-valid,
$\epsilon^2$-decoupled and accurate to within $\epsilon$.
Now the roles of Alice and Bob are somewhat interchanged.
Alice performs a unitary operation depending on the classical
message $x$ to be sent and Bob performs a von Neumann measurement on
some subsystem $B_1$ which almost always succeeds in reproducing the message.
Namely, if  we denote by $p_{x'|x}$ the probability of outcome $x'$ given 
Alice's message was $x$ then, for sufficiently large $n$,
$$
\frac{1}{D} \sum_x p_{x|x} \geq 1 - \epsilon.
$$
Again $D=\exp(n(R+\delta))$.
Before Bob's measurement, the state of $B_1$ and 
the remaining quantum system $Q$ is
$$
\sum_{x'} \sqrt{p_{x'|x}} \ket{x'}^{B_1} \ket{\phi_{xx'}}^{Q}.
$$
Based on the outcome $x'$ of his measurement, Bob
performs some unitary $U_{x'}$ on $Q$, 
leaving the state of $Q$ almost decoupled from $xx'$:
$$
\left\|\sum_{xx'} D^{-1} p_{x'|x} \proj{x} \otimes \proj{x'} \otimes
\theta_{xx'}' - \sum_{xx'} D^{-1} p_{x'|x} \proj{x} \otimes \proj{x'} 
\otimes \bar{\theta}'\right\|_1 \leq \epsilon^2,
$$
where $\ket{\theta_{xx'}'} = U_{x'} \ket{\phi_{xx'}}$ and
$\bar{\theta}' = D^{-1} \sum_{xx'} p_{x'|x} \theta_{xx'}'$. 
Observe, as before, that we can use \lem{purify-dist} to extend $Q$ so
that $\bar{\theta}\ext\bar{\theta}'$ and
$\theta_{xx'}\ext\theta_{xx'}'$ are pure states,
$\braket{\bar{\theta}}{\theta_{xx}}$ is real and positive and
$\|\theta_{xx'}-\bar{\theta}\|_1 \leq 
2\sqrt{\|\theta_{xx'}'-\bar{\theta}'\|_1}$.  Again we use the
concavity of $x\ra\sqrt{x}$ to bound
$$
D^{-1}\sum_{x} p_{x|x} \| \theta_{xx} - \bar{\theta}  \|_1 \leq
D^{-1}\sum_{xx'} p_{x|x'} \| \theta_{xx'} - \bar{\theta}  \|_1 \leq 2\epsilon.
$$

We now modify the protocol so that instead Alice performs
\emph{coherent} communication. Given a subsystem $A_1$ in the
state $\ket{x}^{A_1}$ she  encodes via 
\emph{controlled} unitary operations, yielding 
$$
\ket{x}^{A_1} \sum_{x'} \sqrt{p_{x'|x}} \ket{x'}^{B_1} \ket{\phi_{xx'}}^{Q}.
$$

Bob refrains from measuring $B_1$ and instead 
performs the \emph{controlled} unitary 
$\sum_{x'} \proj{x'}^{B_1} \otimes  U_{x'}$, giving rise to
$$
\ket{x}^{A_1}\ket{\Upsilon_x}^{B_1Q} = 
\ket{x}^{A_1} \left( \sum_{x'} \sqrt{p_{x'|x}} \ket{x'}^{B_1} 
\otimes \ket{\theta_{xx'}}^{Q} \right).
$$
We claim that this is a good approximation for $R\coftau +
\<\bar{\theta}\>$, and according to the correctness of 
the original protocol, $\bar{\theta}$ is close to the output of
$\beta_n$.
To check this, suppose Alice inputs $\ket{\Phi_D}^{RA_1}$ into the
communication protocol.  We will compare the actual state
$$\ket{\Upsilon}^{RA_1B_1Q} :=
D^{-\smfrac{1}{2}}\sum_x
\ket{x}^R\ket{x}^{A_1}\ket{\Upsilon_x}^{B_1Q}$$
  with the ideal state
$$\ket{\Phi_{\text{GHZ}}}^{RA_1B_1}\ot\ket{\bar{\theta}}^Q = 
D^{-\smfrac{1}{2}}\sum_x
\ket{x}^R\ket{x}^{A_1}\ket{x}^{B_1}\ket{\bar{\theta}}^Q.$$
  Their inner
product is 
\ben \bra{\Upsilon}\Phi_{\text{GHZ}}\rangle\ket{\bar{\theta}}&=&
\frac{1}{D}\sum_x
\sqrt{p_{x|x}}\braket{\theta_{xx}}{\bar{\theta}}
\geq
\frac{1}{D}\sum_x
p_{x|x}\braket{\theta_{xx}}{\bar{\theta}}
\geq
\frac{1}{D}\sum_x p_{x|x} 
\l(1 - \half\l\|\theta_{xx}-\bar{\theta}\r\|_1\r)
\\&\geq&
\frac{1}{D}\sum_x p_{x|x} 
 - \frac{1}{D}\sum_x \half\l\|\theta_{xx}-\bar{\theta}\r\|_1
\geq (1 - \epsilon)-\epsilon = 1-2\epsilon
\een
Thus, we can apply \eq{fid-trace} to show that
$$
\|\Upsilon - \Phi_{GHZ} \otimes \theta \|_1 \leq 2 \sqrt{\epsilon}.
$$ 

We have thus shown that
$$
\alpha \geq \beta + R \,\qtqtau.
$$
Using \thm{absolutize} and \eq{cc-equality}
gives the desired result
$$
\alpha  \geq \beta + \frac{R}{2} \, \qq +  \frac{R}{2} \, \qtq. 
$$
\qed\bigskip

\subsection{Proof of Coherent RSP \peq{coherent-rsp}}
To prove that RSP can be made coherent, we review the proof of
\eq{RSP2} from \cite{BHLSW03} and show how it needs to be modified.
We will assume knowledge of typical and conditionally typical
projectors; for background on them, as well as the operator Chernoff
bound used in the proof, see \cite{Winter99}.

The (slightly modified) proof from \cite{BHLSW03} is as follows.  Let
$\cE =\sum_i p_i 
\oprod{i}^{X_A} \ot \psi_i^{AB}$ be an ensemble of bipartite states,
for which we would like to simulate $\cN_\cE$ or $U_\cE$.  Alice is
given a string $i^n=(i_1,\ldots, i_n)$ and wants to prepare the joint
state $\ket{\psi_{i^n}}^{AB} :=
\ket{\psi_{i_1}}^{AB}\cdots\ket{\psi_{i_n}}^{AB}$.  
Let $Q_{i^n}$ be the empirical distribution of $i^n$, i.e. the
probability distribution on $i$ obtained by sampling from $i^n$.  We
assume that  
$\|p-Q_{i^n}\|_1\leq \delta$, and since our
simulation of $U_\cE$ will be used in some $\eta$-valid protocol, we
can do so with error $\leq \eta + \exp(-O(n\delta^2))$.  (Here
$\eta,\delta\ra 0$ as $n\ra\infty$.)  Thus, the protocol begins by
Alice projecting onto the set of $i^n$ with $\|p-Q_{i^n}\|_1\leq
\delta$, in contrast with the protocol in \cite{BHLSW03}, which
begins by having Alice measure $Q_{i^n}$ and send the result to Bob
classically. 

Define $\Pi_{\cE^B|i^n,\delta}^n$ to be the
conditionally typical projector for Bob's half of
$\ket{\psi_{i^n}}^{AB}$, and let $\Pi_{\cE^B\!,\delta}^n$ be the typical
projector for $n$ copies of $\cE^B$.  These projectors are defined in
\cite{Winter99}, which also proves that the subnormalized state
\be \ket{\psi'_{i^n}} = (\one \ot \Pi_{\cE^B\!,\delta}^n
\Pi_{\cE^B|i^n,\delta}^n) \ket{\psi_{i^n}},
\label{eq:rsp-typ-fidelity}\ee
satisfies $\dblbraket{\psi'_{i^n}} \geq 1-2\epsilon$, where
$\delta,\epsilon\ra 0$ as $n\ra\infty$.  This implies that
$\|\psi_{i^n} - \psi''_{i^n}\|_1 \leq 2\sqrt{\epsilon}$, where we
define the normalized state
$\ket{\psi''_{i^n}} := \ket{\psi'_{i^n}} / \sqrt{\dblbraket{\psi'_{i^n}}}$.
We will now write $\ket{\psi'_{i^n}}$ in a way which suggests
how to construct it.  Let $\ket{\Phi_D}^{AB}$ be a maximally entangled
state with $D := \rank \Pi_{\cE^B\!,\delta}^n$ and $\Phi_D^B = 
\Pi_{\cE^B\!,\delta}^n / D$.  (By contrast, \cite{BHLSW03} chooses
$\Phi$ to be a purification of $\Pi_{\sigma,\delta}^n$ with $\sigma :=
\sum_x Q_{i^n}(x) \psi_x^B$.)
Then $\ket{\psi'_{i^n}}$ can be
written as $(M_{i^n} \ot \one)\ket{\Phi_D}$ where $\tr M_{i^n}^\dag
M_{i^n} = D^{-1} 
\dblbraket{\psi'_{i^n}}$.  Thus, Alice will apply a POVM
composed of rescaled and rotated versions of $M_{i^n}$ to her half of
$\ket{\Phi_D}$, and after transmitting the measurement outcome $k$ to Bob,
he can undo the rotation and obtain his half of the correct state.
The cost of this procedure is $\log D$ ebits and $\log K$ ebits, where
we will later specify the number of POVM outcomes $K$.

We now sketch the proof that this is efficient.  From \cite{Winter99},
we find the bounds
\begin{align}
D = \rank \Pi_{\cE^B\!,\delta}^n & \leq \exp\l(n(H(B)_\cE+\delta)\r)\\
\tr_A \oprod{\psi'_{i^n}} &\leq
\exp\l(-n(H(B|X_A)_\cE + \delta)\r) \Pi_{\cE^B\!,\delta}^n
\end{align}
Combining these last two equations and \eq{rsp-typ-fidelity} with the
operator Chernoff 
bound\cite{Winter99} means that there exist a set of unitaries
$U_1,\ldots,U_K$ such that 
$\log K \leq n(I(X_A ; B)_\cE + 3\delta + o(1))$ and
whenever $\|Q_{i^n}-p\|_1\leq \delta$ we have
\be
(1-\epsilon)\frac{\Pi_{\cE^B\!,\delta}^n}{D} \leq
\frac{1}{K}\sum_{k=1}^K \frac{U_k^\dag M_{i^n}^\dag M_{i^n} U_k}
{\tr M_{i^n}^\dag M_{i^n}} \leq 
(1+\epsilon)\frac{\Pi_{\cE^B\!,\delta}^n}{D}.
\label{eq:rsp-randomization}\ee
These conditions mean that Alice can construct a POVM
$\{A_1^{(i^n)},\ldots,A_K^{(i^n)},A_{\text{fail}}^{(i^n)}\}$ with
\be\begin{split}
A_k^{(i^n)} &:= \frac{D}{\sqrt{K(1+\epsilon)\tr M_{i^n}^\dag M_{i^n}}}
M_{i^n} U_k^*\\
A_{\text{fail}}^{(i^n)} & := \sqrt{\Pi_{\cE^B\!,\delta}^n-\sum_k
A_k^\dag A_k} 
\end{split}\ee
According to \eq{rsp-randomization}, the ``fail'' outcome has
probability $\leq 2\epsilon$ of occurring when Alice applies this POVM
to half of $\ket{\Phi_D}$.  And since $(U_k^* \ot
\one)\ket{\Phi_D} = (\one \ot U_k^\dag)\ket{\Phi_D}$, if Alice sends
Bob the outcome $k$ and Bob applies $U_k$ then the residual state
will be $\ket{\psi''_{i^n}}$.

We now explain how to make the above procedure coherent.  First
observe that conditioned on not observing the ``fail'' outcome, the
residual state is completely independent of the classical message $k$.
Thus, we can apply Rule I.  However, a variant of Rule O is also
applicable, in that there is no need to assume the input $\ket{i^n}$
is a classical register.  Again conditioning on success,  the
only record of $i^n$ in the final state is the output state
$\ket{\psi''_{i^n}}$.  Thus, if Alice performs the POVM
\be A_k := \sum_{i^n} \oprod{i^n} \ot A_k^{(i^n)},\ee
(with $A_{\text{fail}}$ defined similarly) and Bob decodes using
\be \sum_k \oprod{k} \ot U_k\ee
then (conditioned on a successful measurement outcome) $\sum_{i^n}
\sqrt{p_{i^n}}\ket{i^n}^R\ket{i^n}^{X_A}$ will be 
coherently mapped to 
$\sum_{i^n}
\sqrt{p_{i^n}}\ket{i^n}^{R}\ket{i^n}^{X_A}\ket{\psi''_{i^n}}^{AB}$.
This achieves a simulation of $\<U_\cE : \cE^{X_A}\>$ using $I(X_A;B)$
cobits and $H(B)$ ebits.  According to Rule I, the coherent
communication returns $I(X_A;B)$ ebits at the end of the protocol,
bringing the net entanglement cost down to $H(B|X_A)$.  Thus we have
proven \eq{coherent-rsp}.
\qed\bigskip

\subsection{Proof of \thm{bidi-ccc}}
For ease of notation, we first consider the $E=0$ case, so our
 starting hypothesis is that $\<U\>\geq C_1\ctc + C_2\cbc$.
At the end of the proof we will return to the $E \neq 0$ case.

\subsubsection{The definition of $\cP_n$}  

Formally, \eq{cbit-toff} indicates the existence of sequences of
nonnegative real numbers $\{\epsilon_n\},\{\delta_n\}$ satisfying
$\epsilon_n, \delta_n {\; \ra \; } 0$ as $n{\; \ra\; }\infty$; a
sequence of protocols $\cP_n = (V_n \! \otimes \! W_n) \, U \, \cdots
\, U \, (V_1 \!  \otimes \! W_1) \, U \, (V_0 \!  \otimes \! W_0)$,
where $V_j,W_j$ are local isometries that may also act on extra local
ancilla systems, and sequences of integers $C_1^{(n)},C_2^{(n)}$
satisfying $nC_1 \geq C_1^{(n)} \geq n(C_1{-}\delta_n)$, $nC_2 \geq
C_2^{(n)} \geq n(C_2{-}\delta_n)$, such that the following success
criterion holds.

Let $a \in \{0,1\}^{C_1^{(n)}}$ and $b \in \{0,1\}^{C_2^{(n)}}$ be the
respective messages of Alice and Bob.  Let $\ket{\varphi_{ab}}:=\cP_n
(\ket{a}_{\A_1}\ket{b}_{\B_1})$.  Note that $\ket{\varphi_{ab}}$
generally occupies a space of larger dimension than $\A_1 \ot \B_1$
since $\cP_n$ may add local ancillas.
To say that $\cP_n$ can transmit classical messages, we require that
local measurements on 
$\ket{\varphi_{ab}}$ can generate messages $b'$ for Alice and $a'$ for
Bob according to a distribution $\Pr(a'b'|ab)$ such that
\be 
\forall_{a,b} \quad
\sum_{a',b'} \smfrac{1}{2}\l| \, 
\Pr(a'b'|ab) - \delta_{a,a'}\delta_{b,b'}\r| \leq \epsilon_n
\label{eq:cc-condition}
\ee
where $a', b'$ are summed over $\{0,1\}^{C_1^{(n)}}$ and
$\{0,1\}^{C_2^{(n)}}$ respectively.  
\eq{cc-condition} follows from applying our definition of a
protocol to classical
communication, taking the final state to be the distribution of
the output classical messages.
Since any measurement can be implemented as a joint unitary on the
system and an added ancilla, up to a redefinition of $V_n, W_n$, we
can assume
\be
 \ket{\varphi_{ab}} := \cP_n (\ket{a}_{\A_1}\ket{b}_{\B_1}) 
= \sum_{a'\!,b'} |b'\>_{\A_1} |a'\>_{\B_1} 
|\gamma_{a' \!, b'}^{a,b} \>_{\A_2 \B_2} 
\label{eq:coh-comm} \, \ee
where the dimensions of $\A_1$ and $\B_1$ are interchanged by $\cP_n$,
and $|\gamma_{a'\!,b'}^{a,b}\>$ are subnormalized states with 
$\Pr(a'b'|ab):=\<\gab|\gab\>$ satisfying \eq{cc-condition}.  Thus, for
each $a,b$ most of the weight of $\ket{\varphi_{ab}}$ is contained in
the $|\gamma_{a,b}^{a,b}\>$ term, corresponding to error-free
transmission of the messages.  See Fig.\ I(a).

\subsubsection{The three main ideas for turning classical communication
into coherent classical communication}

We first give an informal overview of the construction and the
intuition behind it.  For simplicity, consider the error-free term 
with $|\gamma_{a,b}^{a,b}\>$ in ${\A_2 \B_2}$.
To see why classical communication via unitary means should be
equivalent to coherent classical communication, consider the special
case when $|\gamma_{a,b}^{a,b}\>_{\A_2 \B_2}$ is independent of $a,b$.  
In this case, copying $a,b$ to local ancilla systems $\A_0,\B_0$
before $\cP_n$ and discarding $\A_2 \B_2$ after $\cP_n$ leaves a state
within trace distance $\eps_n$ of $\ket{b}_{\A_1} \ket{a}_{\A_0}
\ket{a}_{\B_1}\ket{b}_{\B_0}$---the desired coherent classical
communication. See Fig.\ I(b).
In general $|\gamma_{a,b}^{a,b}\>_{\A_2 \B_2}$ will carry information
about $a,b$, so tracing $\A_2 \B_2$ will break
the coherence of the classical communication.
Moreover, if the Schmidt coefficients of $|\gamma_{a,b}^{a,b}\>_{\A_2
\B_2}$ depend on $a,b$, then knowing $a,b$ is not sufficient to
coherently eliminate $|\gamma_{a,b}^{a,b}\>_{\A_2 \B_2}$ without
some additional communication.  The remainder of our proof is built
around the need to coherently eliminate this ancilla.

Our first strategy is to {\em encrypt} the classical messages $a,b$ by
a shared key, in a manner that preserves coherence (similar to that in
\mscite{Leung00}).  The coherent version of a shared key is a maximally
entangled state.  Thus Alice and Bob (1) again copy their messages to
$\A_0, \B_0$, then (2) encrypt, (3) apply $\cP_n$, and (4) decrypt.
Encrypting the message makes it possible to (5) almost decouple the
message from the combined ``key-and-ancilla'' system, which is
approximately in a state $|\Gamma_{00}\>$ independent of $a,b$ (exact
definitions will follow later).
(6) Tracing out $|{\Gamma}_{00}\>$ gives the desired coherent
communication.  Let $\cP_n'$ denote steps (1)-(5) (see Fig.\ I(c)).



\begin{figure}[h]
\centering \setlength{\unitlength}{0.56mm}
\begin{picture}(300,55)

\put(137,50){\makebox{\bf (c)}}

\put(153,27.5){\line(1,5){4}}
\put(153,27.5){\line(1,-5){4}}
\put(157,47.5){\line(1,0){45}}
\put(157,7.5){\line(1,0){45}}

\put(150,27.5){\line(1,5){4.5}}
\put(150,27.5){\line(1,-5){4.5}}
\put(154.5,50){\line(1,0){47.5}}
\put(154.5,5){\line(1,0){47.5}}

\put(175,21.5){\line(1,0){10}}
\put(175,33.5){\line(1,0){10}}

\put(178,40){\circle{3}}
\put(178,33.5){\line(0,1){8}}
\put(178,21.5){\line(0,-1){8}}
\put(178,15){\circle{3}}

\put(182,33.5){\circle{3}}
\put(182,47.5){\line(0,-1){15.5}}
\put(182,5){\line(0,1){18}}
\put(182,21.5){\circle{3}}

\put(198,33.5){\circle{3}}
\put(198,50){\line(0,-1){18}}
\put(198,7.5){\line(0,1){15.5}}
\put(198,21.5){\circle{3}}

\multiput(202,50)(2,0){16}{{\line(1,0){1}}}
\multiput(202,5)(2,0){16}{{\line(1,0){1}}}
\multiput(202,47.5)(2,0){16}{{\line(1,0){1}}}
\multiput(202,7.5)(2,0){16}{{\line(1,0){1}}}

\multiput(210,40)(2,0){6}{{\line(1,0){1}}}
\multiput(210,15)(2,0){6}{{\line(1,0){1}}}
\multiput(210,21.5)(2,0){6}{{\line(1,0){1}}}
\multiput(210,33.5)(2,0){6}{{\line(1,0){1}}}

\multiput(228,25.5)(2,0){3}{{\line(1,0){1}}}
\multiput(228,29.5)(2,0){3}{{\line(1,0){1}}}

\multiput(214,40)(0,2){5}{{\line(0,1){1}}}
\multiput(218,33.5)(0,2){9}{{\line(0,1){1}}}

\multiput(214,15)(0,-2){6}{{\line(0,-1){1}}}
\multiput(218,21.5)(0,-2){8}{{\line(0,-1){1}}}

\put(218,50){\circle{3}}
\put(214,47.5){\circle{3}}
\put(218,7.5){\circle{3}}
\put(214,5){\circle{3}}

\put(195,21.5){\line(1,0){7}}
\put(195,33.5){\line(1,0){7}}

\put(175,15){\line(1,0){27}}
\put(175,40){\line(1,0){27}}

\put(195,25.5){\line(1,0){7}}
\put(195,29.5){\line(1,0){7}}

\put(201,26.5){\makebox{
$\} | \hs \gamma $}\raisebox{0.3ex}{\tiny 
$
^{a \hs {\oplus} \hs x \hs, b \hs \oplus \hs y}
_{\hs a \! {'} \! \oplus \hs x \hs, b \hs ' \! \hs \oplus \! y}
$}{$\hs \>$}}


\put(185,20){\framebox(10,15){$\cP_n$}}

\put(158,52){\makebox{$\A_4$}}
\put(158,43){\makebox{$\A_3$}}
\put(161,38){\makebox{$\A_0$}}
\put(161,32){\makebox{$\A_1$}}

\put(161,20){\makebox{$\B_1$}}
\put(161,14){\makebox{$\B_0$}}
\put(158,9){\makebox{$\B_3$}}
\put(158,0){\makebox{$\B_4$}}

\put(168,38){\makebox{$|0\>$}}
\put(168,32){\makebox{$|a\>$}}
\put(168,20){\makebox{$|b\>$}}
\put(168,14){\makebox{$|0\>$}}

\put(203,38.5){\makebox{\small $|a\>$}}
\put(203,32.5){\makebox{\small $|b'\>$}}
\put(203,20){\makebox{\small $|a'\>$}}
\put(203,14){\makebox{\small $|b\>$}}

\put(188,2){\makebox{\footnotesize{$y$}}}
\put(188,8){\makebox{\footnotesize{$x$}}}

\put(188,52){\makebox{\footnotesize{$y$}}}
\put(188,45){\makebox{\footnotesize{$x$}}}

\put(230,2.5){\makebox(10,50){$\left. 
\begin{array}{c} {}\\{~}\\{~}\\{~}\\{~}\\{~}\\{~}\\{~} \end{array} \right\} $}}

\put(240,26){\makebox{$|\Gamma_{\!a \hs \oplus \hs a' \hs, b \hs \oplus 
\hs b' \hs }\>$}}

\put(70,50){\makebox{\bf (b)}}

\put(85,21.5){\line(1,0){10}}
\put(85,33.5){\line(1,0){10}}

\put(88,40){\circle{3}}
\put(88,33.5){\line(0,1){8}}
\put(88,21.5){\line(0,-1){8}}
\put(88,15){\circle{3}}

\put(105,21.5){\line(1,0){8}}
\put(105,33.5){\line(1,0){8}}

\put(85,15){\line(1,0){28}}
\put(85,40){\line(1,0){28}}

\put(95,20){\framebox(10,15){$\cP_n$}}

\put(71,38){\makebox{$\A_0$}}
\put(71,32){\makebox{$\A_1$}}

\put(71,20){\makebox{$\B_1$}}
\put(71,14){\makebox{$\B_0$}}

\put(78,38){\makebox{$|0\>$}}
\put(78,32){\makebox{$|a\>$}}
\put(78,20){\makebox{$|b\>$}}
\put(78,14){\makebox{$|0\>$}}

\put(115,39){\makebox{\small $|a\>$}}
\put(115,33){\makebox{\small $|b'\>$}}
\put(115,19){\makebox{\small $|a'\>$}}
\put(115,13){\makebox{\small $|b\>$}}

\put(105,25.5){\line(1,0){5}}
\put(105,29.5){\line(1,0){5}}
\put(111,26.25){\makebox{\small $\} | \! \gab \>$}}

\put(0,50){\makebox{\bf (a)}}

\put(15,21.5){\line(1,0){5}}
\put(15,33.5){\line(1,0){5}}

\put(30,21.5){\line(1,0){7}}
\put(30,33.5){\line(1,0){7}}

\put(30,25.5){\line(1,0){4}}
\put(30,29.5){\line(1,0){4}}
\put(39,33){\small {$|b'\>$}}
\put(39,19){\small {$|a'\>$}}
\put(35,26){\small {$\} \, |\!\gab \>$}}

\put(20,20){\framebox(10,15){$\cP_n$}}

\put(1,32){\makebox{$\A_1$}}
\put(1,20){\makebox{$\B_1$}}

\put(8,32){\makebox{$|a\>$}}
\put(8,20){\makebox{$|b\>$}}

\end{picture}
\label{fig:3protocols}
\caption{Schematic diagrams for $\cP_n$ and $\cP_n'$.
(a) A given protocol $\cP_n$ for two-way classical communication.  
The output is a superposition (over all $a',b'$) of the depicted states, 
with most of the weight in the $(a',b')=(a,b)$ term.  
The unlabeled output systems in the state $|\gab\>$ are $\A_2,\B_2$.
(b) The same protocol with the inputs copied to local ancillas $\A_0,
\B_0$ before $\cP_n$.  If $|\g_{a,b}^{a,b}\>$ is independent of
$a,b$, two-way coherent classical communication is achieved.
(c) The five steps of $\cP_n'$.  Steps (1)-(4) are shown in solid
lines.  Again, the inputs are copied to local ancillas, but $\cP_n$ is
used on messages encrypted by a coherent one-time-pad (the input
$|a\>_{\A_1}$ is encrypted by the coherent version of the key
$|x\>_{\A_3}$ and the output $|a' \hs \oplus x\>_{\B_1}$ is decrypted by
$|x\>_{\B_3}$; similarly, $|b\>_{\B_1}$ is encrypted by $|y\>_{\B_4}$
and $|b' \hs \oplus y\>_{\A_1}$ decrypted by $|y\>_{\A_4}$.  The
intermediate state is shown in the diagram.  Step (5), shown in dotted
lines, decouples the messages in $\A_{0,1},\B_{0,1}$ from
$\A_{2,3,4},\B_{2,3,4}$, which is in the joint state very close to
$|\Gamma_{00}\>$.
}
\end{figure}


If entanglement were free, then our proof of Theorem
\ref{thm:bidi-ccc} would be finished.  However, we have borrowed
$C_1^{(n)}{+}C_2^{(n)}$ ebits as the encryption key and replaced it
with $|\Gamma_{00}\>$.  Though the entropy of entanglement has not
decreased (by any significant amount), $|\Gamma_{00}\>$ is not
directly usable in subsequent runs of $\cP_n'$.  To address this
problem, we use a second strategy of running $k$ copies of $\cP_n'$ in
parallel and performing entanglement concentration of
$|\Gamma_{00}\>^{\otimes k}$.
For sufficiently large $k$, with high probability, we recover most of
the starting ebits.  The regenerated ebits can be used for more
iterations of $\cP_n'^{\otimes k}$ to offset the cost of making the
initial $k \lpm \! C_1^{(n)}{+}C_2^{(n)} \! \rpm$ ebits, without
the need of borrowing from anywhere.

However, a technical problem arises with simple repetition of
$\cP_n'$, which is that errors accumulate.  In particular, a na\"\i ve
application of the triangle inequality gives an error $k \eps_n$ but
$k$, $n$ are not independent.  In fact, the entanglement concentration
procedure of \mscite{BBPS96} requires $k\gg
\Sch(|\Gamma_{00}\>) = \exp(O(n))$ and we cannot guarantee that $k\epsilon_n
\rightarrow 0$ as $k,n \ra\infty$.  Our third strategy is to treat the
$k$ uses of $\cP_n'$ as $k$ uses of a slightly noisy channel, and
encode only $l$ messages (each having $C_1^{(n)}, C_2^{(n)}$ bits in
the two directions) using classical error correcting codes.  The error
rate then vanishes with a negligible reduction in the communication
rate and now making no assumption about how quickly $\epsilon_n$
approaches zero.  We will see how related errors in decoupling and
entanglement concentration are suppressed.   

We now describe the construction and analyze the error in detail.  

\subsubsection{The definition of $\cP_n'$}
\vspace*{-2ex}
\begin{enumerate}
\item[0.]  
Alice and Bob begin with inputs $\ket{a}_{\A_1}\ket{b}_{\B_1}$ and the
entangled states $\ket{\Phi}^{\! \ot C_1^{(n)}}_{\A_3 \B_3}$ and
$\ket{\Phi}^{\! \ot C_2^{(n)}}_{\A_4 \B_4}$.  (Systems $3$ and $4$
hold the two separate keys for the two messages $a$ and $b$.)  The
initial state can then be written as
\be
\frac{1}{\sqrt{N}}
\sum_x \ket{xx}_{\A_3 \B_3}
\sum_y \ket{yy}_{\A_4 \B_4}
~\ket{a}_{\A_1}\ket{b}_{\B_1} 
\ee
where $x$ and $y$ are summed over $\{0,1\}^{C_1^{(n)}}$ and
$\{0,1\}^{C_2^{(n)}}$, and $N = \exp \lpm \! C_1^{(n)} {+} C_2^{(n)}
\! \rpm$.

\item
They coherently copy the messages to $\A_0, \B_0$.

\item 
They encrypt the messages using the one-time-pad 
$\ket{a}_{\A_1} \ket{x}_{\A_3} \ra \ket{a\oplus x}_{\A_1}\ket{x}_{\A_3}$
and
$\ket{b}_{\B_1} \ket{y}_{\B_4} \ra \ket{b\oplus y}_{\B_1}\ket{y}_{\B_4}$
coherently to obtain
\be
\ket{a}_{\A_0} \ket{b}_{\B_0} \; 
\frac{1}{\sqrt{N}}\sum_{xy}
\ket{x}_{\A_3}\ket{y}_{\A_4} 
\ket{x}_{\B_3}\ket{y}_{\B_4} 
~\ket{a \hs \oplus \hs x}_{\A_1}
\ket{b \hs \oplus \hs y}_{\B_1}
\,.
\ee
\item
Using $U$ $n$ times, they apply $\cP_n$ to registers $\A_1$ and $\B_1$,
obtaining an output state 
\be
\ket{a}_{\A_0} \ket{b}_{\B_0}
\frac{1}{\sqrt{N}}\sum_{xy}
\ket{x}_{\A_3} \ket{y}_{\A_4} 
\ket{x}_{\B_3} \ket{y}_{\B_4} 
\sum_{a',b'} 
| b' \! \oplus y\>_{\A_1} |a' \! \oplus x\>_{\B_1} \gmany_{\A_2 \B_2}
\,.
\label{eq:ready-to-measure}
\ee

\item Alice decrypts her message in $\A_1$ using her key $\A_4$ and Bob
decrypts $\B_1$ using $\B_3$ coherently as 
$|b' \oplus y\>_{\A_1} |y\>_{\A_4} \ra |b'\>_{\A_1} |y\>_{\A_4}$
and
$|a' \oplus x\>_{\B_1} |x\>_{\B_3} \ra |a'\>_{\B_1} |x\>_{\B_3}$
producing a state 
\be
\ket{a}_{\A_0} \ket{b}_{\B_0}
\frac{1}{\sqrt{N}}\sum_{xy}
\ket{x}_{\A_3} \ket{y}_{\A_4} 
 \ket{x}_{\B_3} \ket{y}_{\B_4} 
\sum_{a',b'}
\ket{b'}_{\A_1} \ket{a'}_{\B_1} \gmany_{\A_2 \B_2}
\,. \ee
\item 
Further {\sc cnot}s $\A_1 \ra \A_4$, $\A_0 \ra \A_3$, $\B_1 \ra \B_3$
and $\B_0 \ra \B_4$ will leave $\A_{2,3,4}$ and $\B_{2,3,4}$ almost
decoupled from the classical messages.  To see this, the state has become 
\bea
%
& & \ket{a}_{\A_0}  \ket{b}_{\B_0}
\sum_{a',b'} \ket{b'}_{\A_1}\ket{a'}_{\B_1}
\frac{1}{\sqrt{N}}\sum_{xy}
\ket{a\oplus x}_{\A_3} 
\ket{a'\oplus x}_{\B_3} 
\ket{b'\oplus y}_{\A_4}
\ket{b\oplus y}_{\B_4} 
\gmany_{\A_2 \B_2}
\non \\ & = & 
 \ket{a}_{\A_0}  \ket{b}_{\B_0}
\sum_{a',b'} \ket{b'}_{\A_1}\ket{a'}_{\B_1}
\; \ket{\Gamma_{a\oplus a', b\oplus b'}}_{\A_{2,3,4} \B_{2,3,4}}
\label{eq:phiab}
\,, 
\eea
where 
\bea 
\ket{\Gamma_{\! a\oplus a' \!, b\oplus b' \hs}}_{\A_{2,3,4} \B_{2,3,4}} := 
\frac{1}{\sqrt{N}} \sum_{xy}
\ket{a\oplus x}_{\A_3} 
\ket{a'\oplus x}_{\B_3} 
\ket{b'\oplus y}_{\A_4}
\ket{b\oplus y}_{\B_4} 
\gmany_{\A_2 \B_2} \,.
\eea
The fact $\ket{\Gamma_{\! a\oplus a' \!, b\oplus b' \hs}}$ depends
only on $a \oplus a'$ and $b \oplus b'$, without any other dependence
on $a$ and $b$, can be easily seen by replacing $x,y$ with $a\oplus x,
b\oplus y$ in $\sum_{xy}$ in the RHS of the above.
Note that $\<\Gamma_{\! a\oplus a' \!, b\oplus b' \hs} | \Gamma_{\!
a\oplus a' \!, b\oplus b' \hs} \> = \smfrac{1}{N}\sum_{xy}
\Pr(a'\oplus x, b'\oplus y \, | \, a \oplus x , b \oplus y)$, so in
particular for the state corresponding to the error-free term, we have
$\<\Gamma_{00}|\Gamma_{00}\> = \smfrac{1}{N}\sum_{xy} \Pr(xy|xy) :=
1 - \bar{\eps}_n \geq 1-\epsilon_n$.\footnote{Thus it turns out that
\eq{cc-condition} was more than we 
needed; the {\em average} error (over all $a,b$) would have been
sufficient.  In general, this argument shows that using shared
entanglement (or randomness in the case of classical communication)
can convert an average error condition into a maximum error condition,
and will be further developed in \cite{DW05}.}

Suppose that Alice and Bob could project onto
the space where $a'=a$ and $b'=b$, and tell each other they
have succeeded (by using a little extra communication); then the
resulting ancilla state $\smfrac{1}{\sqrt{1{-}\bar{\eps}_n}}
|\Gamma_{00}\>$ has at least $C_1^{(n)} {+} \, C_2^{(n)} {+} \log
(1{-}\eps_n)$ ebits, since its largest Schmidt coefficient is 
$\leq \lbm \exp(C_1^{(n)}{+}C_2^{(n)} ) (1{-}\bar{\eps}_n) \rbm^{-1/2}$ and
$\bar{\eps}_n \leq \eps_n$ (cf.~\prop{U-CgE}).
Furthermore, $|\Gamma_{00}\>$ is manifestly independent of $a,b$.
We will see how to improve the probability of successful projection
onto the error free subspace by using block codes for error
correction, and how correct copies of $|\Gamma_{00}\>$ can be
identified if Alice and Bob can exchange a small amount of
information.
\end{enumerate}

\subsubsection{Main idea on how to perform error correction}

As discussed before, $|\Gamma_{00}\>$ cannot be used directly as an
encryption key -- our use of entanglement in $\cP_n'$ is not
catalytic.
Entanglement concentration of many copies of $|\Gamma_{00}\>$ obtained
from many runs of $\cP_n'$ will make the entanglement overhead for the
one-time-pad negligible, but errors will accumulate.
The idea is to suppress the errors in many uses of $\cP_n'$ by error
correction.
This has to be done with care, since we need to simultaneously ensure
low enough error rates in both the classical message and the state to
be concentrated, as well as sufficient decoupling of the classical
messages from other systems.

Our error-corrected scheme will have $k$ parallel uses of $\cP_n'$,
but the $k$ inputs are chosen to be a valid codeword of an error
correcting code.  Furthermore, for each use of $\cP_n'$, the state in
$\A_{2,3,4} \B_{2,3,4}$ will only be collected for entanglement
concentration if the error syndrome is trivial for that use of
$\cP_n'$.  We use the fact that errors occur rarely (at a rate of
$\epsilon_n$, which goes to zero as $n\ra\infty$) to show that (1) most
states are still used for concentration, and (2) communicating the
indices of the states with non trivial error syndrome requires a
negligible amount of communication.


\subsubsection{Definition of $\cP_{nk}''$: error corrected version of
$(\cP_n')^{\ot k}$ with entanglement concentration}

We construct two codes, one used by Alice to signal to Bob and one
from Bob to Alice.  We consider high distance codes.  The distance of
a code is the minimum Hamming distance between any two codewords,
i.e.\ the number of positions in which they are different.

First consider the code used by Alice.  Let $N_1=2^{C_1^{(n)}}$.
Alice is coding for a channel that takes input symbols from
$[N_1]:=\{1,\ldots,N_1\}$ and has probability $\leq \epsilon_n$ of
error on any input (the error rate depends on both $a$ and $b$).
We would like to encode $[N_1]^l$ in $[N_1]^k$ using a code with
distance $2k\alpha_n$, where $\alpha_n$ is a parameter that will be
chosen later.  Such a code can correct up to any $\lfloor k\alpha_n
{-} \smfrac{1}{2} \rfloor$ errors (without causing much problem, we
just say that the code corrects $k\alpha_n$ errors).  Using standard
arguments\footnote{We show the existence of a maximal code by
repeatedly adding new 
codewords that have distance $\geq 2k\alpha_n$ from all other chosen
codewords.  This gives at least $N^k / \vol(N,2k\alpha_n,k)$
codewords, where $\vol(N,k\delta,k)$ is the number of words in $[N]^k$
within a distance $k\delta $ of a fixed codeword.
But $\vol(N,k\delta,k) \leq \binom{k}{k\delta}N^{k\delta} \leq 2^{k
H_2(\delta)} N^{k\delta}$.  (See \cite{CT91} or \eq{binom-bounds}
later in this thesis for a derivation of $\binom{k}{k\delta} \leq 2^{k
H_2(\delta)}$.)
Altogether, the number of codewords $:= N^l \geq N^k / (2^{k H_2(2 \alpha_n)}
N^{2 k \alpha_n})$, thus $l \geq k \lbm 1-2\alpha_n -
\smfrac{H_2(2\alpha_n)}{\log N}\rbm$.}, we can construct such a code with $l \geq k
\lbm 1 {-} 2\alpha_n {-} H_2(2\alpha_n)/C_1^{(n)} \rbm $, where
$H_2(p)=-p\log p-(1{-}p)\log (1{-}p)$ is the binary entropy.
The code used by Bob is chosen similarly, with $N_2=2^{C_2^{(n)}}$
input symbols to each use of $\cP_n'$.  For simplicity, Alice's and
Bob's codes share the same values of $l$, $k$ and $\alpha_n$.   
We choose $\alpha_n \geq \max(1/C_1^{(n)}, 1/C_2^{(n)})$ so that $l
\geq k(1{-}3\alpha_n)$.  

Furthermore, we want the probability of having $\geq k\alpha_n$ errors to be
vanishingly small.  This probability is $\leq\exp(-k
D(\alpha_n\|\epsilon_n)) \leq \exp(k + k\alpha_n\log\epsilon_n)$
(using arguments from \cite{CT91}) $\leq \exp(-k)$ if
$\alpha_n \geq -2/\log\epsilon_n$.  

Using these codes, Alice and Bob construct $\cP_{nk}''$ as follows
(with steps 1-3 performed coherently).
\vspace*{-2ex} 
\begin{enumerate} 
\item[0.]  
Let $(a^{\rm o}_1, \cdots, a^{\rm o}_l)$ be a vector of $l$ messages
each of $C_1^{(n)}$ bits, and $(b^{\rm o}_1, \cdots, b^{\rm o}_l)$ be
$l$ messages each of $C_2^{(n)}$ bits.
\item 
Using her error correcting code, Alice encodes $(a^{\rm o}_1, \cdots,
a^{\rm o}_l)$ in a valid codeword $\vec{a} = (a_1, \cdots, a_k)$ which
is a $k$-vector.  Similarly, Bob generates a valid codeword $\vec{b} =
(b_1, \cdots, b_k)$ using his code.  
\item  
Let $\vec \A_1 := \A_1^{\otimes k}$ denote a tensor product of $k$ input
spaces each of $C_1^{(n)}$ qubits.  Similarly, $\vec \B_1:=
\B_1^{\otimes k}$.  (We will also denote $k$ copies of $\A_{0,2,3,4}$,
and $\B_{0,2,3,4}$ by adding the vector symbol.)
Alice and Bob apply $(\cP_n')^{\otimes k}$ to $\ket{\vec{a}}_{\vec
\A_1} |\vec{b}\>_{\vec \B_1}$; that is, in parallel, they apply
$\cP_n'$ to each pair of inputs $(a_j, b_j)$.  The resulting state is
a tensor product of states of the form given by \eq{phiab}:
\be
\bigotimes_{j=1}^k \lbL
\ket{a_j}_{\A_0}  \ket{b_j}_{\B_0}
\sum_{a_j',b_j'} |b_j'\>_{\A_1} |a_j'\>_{\B_1}
\; |\Gamma_{a_j \oplus a_j', b_j \oplus b_j'}\>_{\A_{2,3,4} \B_{2,3,4}} 
\rbL .
\label{eq:Pn-parallel}
\ee

Define $|\Gamma_{\va \oplus \va', \vb \oplus \vb'}\>_{\vec\A_{234}
\vec\B_{234}} := \bigotimes_{j=1}^k |\Gamma_{a_j \oplus a_j', b_j
\oplus b_j'}\>_{\A_{2,3,4} \B_{2,3,4}}$.  Then, \eq{Pn-parallel} can be
written more succinctly as
\be 
\ket{\va}_{\vec \A_0} |\vb\>_{\vec \B_0} 
 \sum_{\va',\vb'} |\vb'\>_{\vec \A_1} \ket{\va'}_{\vec \B_1}
|\Gamma_{\va \oplus \va', \vb \oplus \vb'}\>_{\vec\A_{234} \vec\B_{234}} \,.
\ee

\item
Alice performs the error correction step on $\vec \A_1$ and Bob
does the same on $\vec \B_1$.  According to our code
constructions, this (joint) step fails with probability $\pfail\leq
2\cdot 2^{-k}$.  (We will see below why $\pfail$ is
independent of $\va$ and $\vb$.)

In order to describe the residual state, we now introduce $\cG_{\hs
\A} = \{\vx \, {\in} \, [N_1]^k : |\vx| \, {\leq} \, k\alpha_n\}$ and
$\cG_{\B} = \{\vx \,{\in}\, [N_2]^k : |\vx| \, {\leq} \, k\alpha_n\}$,
where $|\vx|:=|\{j : x_j \,{\neq}\, 0\}|$ denotes the Hamming weight
of $\vx$.  Thus $\cG_{\hs \A,\B}$ are sets of correctable (good)
errors, in the sense that there exist local decoding isometries
$\cD_{\hs \A},\cD_{\B}$ such that for any code word $\va\in [N_1]^k$
we have $\forall \va'\in \va\oplus\cG_{\hs \A}, \cD_{\hs \A}\ket{\va'}
= \ket{\va}\ket{\va\oplus \va'}$ (and similarly, if $\vb\in[N_2]^k$ is
a codeword, then $\forall \vb'\in \vb\oplus\cG_{\B}, \cD_{\B} |\vb'\>
= |\vb\> |\vb\oplus \vb'\>$).  For concreteness, let the decoding maps
take $\vec \A_1$ to $\vec \A_1 \vec \A_5$ and $\vec \B_1$ to $\vec
\B_1 \vec \B_5$.

Conditioned on success, Alice and Bob are left with
\bea
& & \frac{1}{\sqrt{1{-}\pfail}} \, |\va,\vb\>_{\vec \A_{0,1}} 
		|\va,\vb\>_{\vec \B_{0,1}} 
\sum_{\va' \hs \in \va \oplus \cG_{\hs \A}} 
\sum_{~\vb' \hs \in \vb \oplus \cG_{\B}}
|\vb \oplus \vb'\>_{\vec \A_5} |\va \oplus \va'\>_{\vec \B_5}
|\Gamma_{\va\oplus \va', \vb\oplus \vb'}\>_{\vec \A_{234} \vec \B_{234}}
\\ &:=& \frac{1}{\sqrt{1{-}\pfail}} \, |\va,\vb\>_{\vec \A_{0,1}} 
		|\va,\vb\>_{\vec \B_{0,1}} 
 \sum_{\va'' \hs \in\cG_{\hs \A}} \sum_{~\vb'' \hs \in \cG_{\B}}
|\vb''\>_{\vec \A_5} |\va''\>_{\vec \B_5}
|\Gamma_{\va'',\vb''}\>_{\vec \A_{234}\vec \B_{234}},
\eea
\sloppypar{where we have defined $\va'' := \va\oplus \va'$ and $\vb''
 := \vb\oplus\vb'$.  Note that $2^{-k+1} \geq \pfail =
\sum_{(\va'',\vb'')\not\in \cG_{\hs \A} \times \cG_{\B}}
\dblbraket{\Gamma_{\va'',\vb''}}$, which is manifestly independent
of $\vec a,\vec b$.}
The ancilla is now {\em completely} decoupled from the message,
resulting in coherent classical communication.  The only remaining
issue is recovering entanglement from the ancilla, so for the
remainder of the protocol we ignore the now decoupled states $|\vec a,
\vec b\>_{\vec\A_{0,1}} |\vec a,\vec b\>_{\vec\B_{0,1}}$.

\item
For any $\vx$, define $S(\vx):=\{j : x_j \,{\neq}\, 0\}$ to be set of
positions where $\vx$ is nonzero.  If $\vx\in\cG_{\hs \A}$ (or
$\cG_{\B}$), then $|S(\vx)| \leq k\alpha_n$.  Thus, $S(\vx)$ can be
written using $\leq \log \sum_{j \leq k\alpha_n} \!\! \binom{k}{j}
\leq \log \binom{k}{k \alpha_n} 
+ \log (k\alpha_n) \leq kH_2(\alpha_n) + \log (k \alpha_n)$ bits.

The next step is for Alice to compute $|S(\vb'')\>$ from $|\vb''\>$
and communicate it to Bob using $\lpm \! kH_2(\alpha_n) + \log (k\alpha_n) 
\! \rpm \ctc$.  Similarly, Bob sends $\ket{S(\va'')}$ to Alice using
$\lpm \! kH_2(\alpha_n) + \log (k\alpha_n) \! \rpm \cbc$.  Here
we need to assume that some (possibly inefficient) protocol to send
$O(k)$ bits in either direction with error $\exp(-k-1)$ (chosen for
convenience) and with $Rk$ uses of $U$ for some constant $R$.  Such a
protocol was given by \prop{U-nonzero-CC} and the bound on the error can be
obtained from the HSW theorem\cite{Holevo98,SW97,Hayashi:02f}.

Alice and Bob now have the state
\be \frac{1}{\sqrt{1{-}\pfail}} \,
\sum_{\va''\in\cG_{\hs \A}} \sum_{\vb''\in\cG_{\B}}
|S(\va'') S(\vb'')\>_{\vec\A_6} \, |\vb''\>_{\vec\A_5} \, 
|S(\va'') S(\vb'')\>_{\vec\B_6} \, |\va''\>_{\vec\B_5} \,  
|\Gamma_{\va'',\vb''}\>_{\vec\A_{234} \vec \B_{234}}.
\ee
Conditioning on their knowledge of $S(\va''),S(\vb'')$, Alice and Bob
can now identify $k'\geq k(1-2\alpha_n)$ positions where
$a_j''=b_j''=0$, and extract $k'$ copies of
$\smfrac{1}{\sqrt{1{-}\pfail}}|\Gamma_{00}\>$.
Note that leaking $S(\va''), S(\vb'')$ to the environment will not
affect the extraction procedure, therefore, coherent computation and
communication of $S(\va''), S(\vb'')$ is unnecessary.  (We have not
explicitly included the environment's copy of $|S(\va'') S(\vb'')\>$ 
in the equations to minimize clutter.) 
After extracting $k'$ copies of
$\smfrac{1}{\sqrt{1{-}\pfail}}|\Gamma_{00}\>$, we can safely discard
the remainder of the state, which is now completely decoupled from both
$\lbm \!  \smfrac{1}{\sqrt{1{-}\pfail}}|\Gamma_{00}\> \! \rbm
^{\otimes k'}$ and the message $|\vec a\>_{\A_0} |\vec b\>_{\A_1}
|\vec b\>_{\B_0} |\vec a\>_{\B_1}$.

\item
Alice and Bob perform entanglement concentration ${\cal E}_{\rm conc}$
(using the techniques of \mscite{BBPS96}) on
$\lbm \! \smfrac{1}{\sqrt{1{-}\pfail}}|\Gamma_{00}\> \! \rbm^{\otimes k'}$.  
Note that since $\smfrac{1}{\sqrt{1{-}\pfail}}|\Gamma_{00}\>$ can be
created using $U$ $n$ times and then using classical communication and
postselection, it must have Schmidt rank $\leq {\rm Sch}(U)^n$, where
${\rm Sch}(U)$ is the Schmidt number of the gate $U$.
Also recall that $E \lbm \! \smfrac{1}{\sqrt{1{-}\pfail}}|\Gamma_{00}\> 
\!\rbm \geq C_1^{(n)} + C_2^{(n)} + \log (1{-}\eps_n)$.
According to \mscite{BBPS96}, ${\cal E}_{\rm conc}$ requires no
communication and with probability 
$\geq 1 - \exp \lbm{-}{\rm Sch}(U)^n \lpm\! \sqrt{k'}-\log(k'{+}1) \!\rpm 
\rbm$
produces at least $k' \lbm C_1^{(n)} {+} C_2^{(n)} {+} \log (1{-}\eps_n) \rbm 
- {\rm Sch}(U)^n \lbm \sqrt{k'} {-} \log(k'{+}1) \rbm$ ebits.

\end{enumerate}

\subsubsection{Error and resource accounting} 

$\cP_{nk}''$ consumes a total of \\[1ex]
$~~~$(0) $nk$ uses of $U$ (in the $k$ executions of $\cP_n'$) \\
$~~~$(1) $Rk$ uses of $U$
	(for communicating nontrivial syndrome locations) \\
$~~~$(2) $k \lbm \!\! C_1^{(n)} {+} C_2^{(n)} \!\! \rbm \qq$  
	(for the encryption of classical messages). \\[1ex]
$\cP_{nk}''$ produces, with probability and fidelity no less than 
$$1{-}2^{{-}(k-1)}{-}2^{{-}(k-1)}-\exp \lbm \!\!  {-}{\rm Sch}(U)^n \lpm \!\!
\sqrt{k'} {-} \log(k'{+}1) \!\!\rpm \!\!\rbm ,$$
 at least \\[1ex]
$~~~$(1) $l \, C_1^{(n)}\cof +l \, C_2^{(n)}\cob$  
\\
$~~~$(2) $k' \lpm \! C_1^{(n)}{+}C_2^{(n)} {+} \log (1{-}\eps_n)\! \rpm 
- {\rm Sch}(U)^n \lpm \! \sqrt{k'}{-}\log(k'{+}1) \! \rpm \qq$ . 

We restate the constraints on the above parameters: 
$\epsilon_n, \delta_n {\; \ra \; } 0$ as $n{\; \ra\; }\infty$;
$C_1^{(n)} \geq n(C_1{-}\delta_n)$, $C_2^{(n)} \geq
n(C_2{-}\delta_n)$;
$\alpha_n \geq \max(1/C_1^{(n)}, 1/C_2^{(n)}, -2/\log\epsilon_n)$; 
$k' \geq k(1{-}2\alpha_n)$; $l \geq k(1{-}3\alpha_n)$. 
 
We define ``error'' to include both infidelity and the probability of
failure.  To leading orders of $k,n$, this is equal to $2^{{-}(k-2)} +
\exp \lbm \!\!  {-} \sqrt{k} \; {\rm Sch}(U)^n \! \! \rbm $.  We
define ``inefficiency'' to include extra uses of $U$, net consumption of entanglement, and the amount by which
the coherent classical communication rates fall short of the classical
capacities.  To leading order of $k,n$, these are respectively
$Rk$, $2\alpha_nk(C_1^{(n)}{+}C_2^{(n)}) + \sqrt{k} \, {\rm Sch}(U)^n \approx 
 2\alpha_nk n(C_1{+}C_2) + \sqrt{k} \, {\rm Sch}(U)^n$,
and $nk(C_1{+}C_2) - l(C_1^{(n)} {+} C_2^{(n)}) \leq
nk(3\alpha_n(C_1 {+} C_2) + 2\delta_n)$.  We would like the error to
vanish, as well as the fractional inefficiency, defined as the
inefficiency divided by $kn$, the number of uses of $U$.
Equivalently, we can define $f(k,n)$ to be the {\em sum} of the error
and the fractional inefficiency, and require that $f(k,n)\ra 0$ as
$nk\ra \infty$.
By the above arguments,
\be f(k,n) \leq 2^{{-}(k-2)} + \exp(-\sqrt{k} \; {\rm Sch}(U)^n)
 + 2\alpha_n(C_1 {+} C_2) + \smfrac{1}{n \sqrt{k}} \; {\rm Sch}(U)^n
 + \frac{R}{n}
 + 3\alpha_n(C_1 {+} C_2) + 2 \delta_n \,.
\label{eq:frac-errors}\ee
Note that for any fixed value of $n$, $\lim_{k\ra\infty} f(k,n) =
5\alpha_n(C_1{+}C_2)+2\delta_n + R/n$.  (This requires $k$
to be sufficiently large and also $k \gg {\rm Sch}(U)^{2n}$.)  Now,
allowing $n$ to grow, we have
\be 
	\lim_{n\ra\infty}\lim_{k\ra\infty} f(k,n) = 0. 
\ee 
The order of limits in this equation is crucial due to the dependence 
of $k$ on $n$. 

The only remaining problem is our catalytic use of 
$O(nk)$ ebits.  In order to construct a protocol that uses only $U$,
we need to first use $U$ $O(nk)$ times to generate the starting
entanglement.  Then we repeat $\cP_n''$ $m$ times, reusing the same
entanglement.  The catalyst results in an additional fractional
inefficiency of $c/m$ (for some constant $c$ depending only on $U$)
and the errors and 
inefficiencies of $\cP_n''$ 
add up to no more than $mf(k,n)$.  Choosing $m = \lfloor
1/\sqrt{f(k,n)} \rfloor$ will cause all of these errors and
inefficiencies to simultaneously vanish.  (This technique is
essentially equivalent to using Lemmas~\ref{lemma:noo} and
\ref{lemma:cancel} and \thm{composability}.)  The actual error condition is
that 
\be
\lim_{m\ra\infty}\lim_{n\ra\infty}\lim_{k\ra\infty} \; mf(k,n) +
\frac{c}{m} \; = \; 0 \,.
\ee 
This proves the resource inequality
\be U \geq C_1\cof + C_2\cob.\ee

\subsubsection{The $E<0$ and $E>0$ cases} 

If $E<0$ then entanglement is consumed in $\cP_n$, so there exists a
sequence of integers $E^{(n)} \leq n(E+\delta_n)$ such that
\be \cP_n \! \left( \ket{a}_{\A_1}\ket{b}_{\B_1}
\ket{\Phi}^{E^{(n)}}_{\A_5 \B_5} \right)
= 
\sum_{a', b'} |b'\>_{\A_1} |a'\>_{\B_1} |\g_{a',b'}^{a,b}\>_{\A_2 \B_2}\,.
\ee
In this case, the analysis for $E^{(n)}=0$ goes through, only with
additional entanglement consumed.  Almost all equations are the same,
except now the Schmidt rank for $|\Gamma_{00}\>$ is upper-bounded by
$\l[\Sch(U)2^{E+\delta_n}\r]^n$ instead of $\Sch(U)^n$.  This is still
$\leq c^n$ for some constant $c$, so the same proof of correctness applies.

If instead $E>0$, entanglement is created, so for some
$E^{(n)}\geq n(E-\delta_n)$ we have
\be \cP_n \! \left( \ket{a}_{\A_1}\ket{b}_{\B_1} \right)
= 
\sum_{a', b'} |b'\>_{\A_1} |a'\>_{\B_1} |\g_{a',b'}^{a,b}\>_{\A_2 \B_2}\,.
\ee
for $E(|\g_{a,b}^{a,b}\>_{\A_2 \B_2}) \geq E^{(n)}$. 
Again, the previous construction and analysis go through, with an
extra $E^{(n)}$ ebits of entanglement of entropy in $|\Gamma_{00}\>$,
and thus an extra fractional efficiency of $\leq 2\alpha_nE$ in
\eq{frac-errors}.   The Schmidt rank of $|\Gamma_{00}\>$ is still
upper bounded by Sch$(U)^n$ in this case. 
\qed\bigskip

\begin{observation}
If $(C_1,C_2,E)\in\CCE(U)$, but $(C_1,C_2,E+\delta)\not\in\CCE(U)$ for
any $\delta>0$, then for any $\epsilon,\delta>0$ and for $n$
sufficiently large there is a protocol $\cP_n$ and a state
$\ket{\varphi}^{AB}$ on $\leq \kappa n\delta$ qubits (for a universal
constant $\kappa$), such that for any
$x\in\{0,1\}^{\lfloor n(C_1-\delta)\rfloor}, 
y\in\{0,1\}^{\lfloor n(C_2-\delta)\rfloor}$ we have either
$$ \cP_n \ket{x}^A\ket{y}^B \approx_\eps
\ket{xy}^A\ket{xy}^B\ket{\Phi}^{\lfloor n(E-\delta)\rfloor}
\ket{\varphi}$$
if $E>0$ or
$$ \cP_n \ket{x}^A\ket{y}^B \ket{\Phi}^{\lfloor -n(E-\delta)\rfloor}
\approx_\eps
\ket{xy}^A\ket{xy}^B\ket{\varphi}$$
if $E<0$.

The key point here is that if $E$ taken to be the maximum possible for
a given $C_1,C_2$, then the
above proof of \thm{bidi-ccc} in fact produces ancilla systems of a
sublinear size.
\end{observation}

\section{Discussion}\label{sec:ccc-discuss}

Quantum information, like quantum computing, has often been studied
under an implicit ``quantum co-processor'' model, in which quantum
resources are used by some controlling classical computer.  Thus, we
might use quantum computers or quantum channels to perform classical
tasks, like solving computational problems, encrypting or
authenticating a classical message, demonstrating nonlocal classical
correlations, synchronizing classical clocks and
so on.  On the other hand, since the quantum resources are manipulated
by a classical computer, it is natural to think of conditioning
quantum logical operations on classical information.

This framework has been quite useful for showing the strengths of
quantum information relative to classical information processing
techniques; e.g. we find that secure communication is possible,
distributed computations require less communication and so on.
However, in quantum Shannon theory, it is easy to be misled by the
central role of classical information in the quantum co-processor
model.  While classical communication may still be a useful {\em goal}
of quantum Shannon theory, it is often inappropriate as an
intermediate step.  Rather, we find in protocol after protocol that
coherently decoupled cbits are better thought of
as cobits.

Replacing cbits with cobits has significance beyond merely improving
the efficiency of quantum protocols.  In many cases, cobits give rise
to asymptotically reversible protocols, such as coherent teleportation
and super-dense coding, or more interestingly, remote state
preparation and HSW coding.  The resulting resource equalities go a
long way towards simplifying the landscape of quantum Shannon theory: (1)
The duality of teleportation and super-dense coding resolves a
long-standing open question about how the original forms of these
protocols could be individually optimal, but wasteful when composed;
we now know that all the irreversibility from composing teleportation
and super-dense coding is due to the map $\cof\geq\ctc$. (2) Coherent
RSP and HSW coding give a resource equality that allows us to easily derive
an expression for unitary gate capacity regions.  In the next chapter,
we will see more examples of how making classical communication
coherent leads to a wide variety of optimal coding theorems.

Although the implications of coherent classical communication are
wide-ranging, the fundamental insight is quite simple: when studying
quantum Shannon theory, we should set aside our intuition about the
central role of classical communication, and instead examine carefully
which systems are discarded and when communication can be coherently
decoupled.

\chapter{Optimal trade-offs in quantum Shannon theory}\label{chap:family}
The main purpose of quantum information theory, or more particularly
\emph{quantum Shannon theory}, is to characterize asymptotic resource
inter-conversion tasks in terms of quantum information theoretical
quantities such as von Neumann entropy, quantum mutual and coherent
informations. A particularly important class of problems involves a
noisy quantum channel or shared noisy entanglement between two parties
which is to be converted into qubits, ebits and/or cbits, possibly
assisted by limited use of qubits, ebits or cbits as an auxiliary
resource.  In this final chapter on quantum Shannon theory, we give a
full solution for this class of problems.

In \sect{family}, we will state two dual, purely quantum protocols:
for entanglement distillation assisted by quantum communication
(the ``mother'' protocol) and for entanglement assisted quantum
communication (the ``father'' protocol).  From these two, we can derive a
large class of ``children'' (including many previously known resource
inequalities) by direct application of teleportation or super-dense
coding.  The key ingredient to deriving the parents, and thus
obtaining the entire family, is coherent classical communication.
Specifically, we will show how the parents can be obtained by applying
Rules I and O to many of the previously known children.  In each
scenario, we will find that previous proofs of the children already use
coherently decoupled cbits (or can be trivially modified to do so),
so that the only missing ingredient is coherent classical communication.

Next, we address the question of optimality.  Most of the
protocols we involve one noisy resource (such as $\<\cN\>$) and two
noiseless standard ones (such as qubits and ebits), so instead of
capacities we need to work with two-dimensional capacity regions whose
boundaries determine trade-off curves.  We state and prove formulae
for each of these capacity regions in \sect{trade-off}.

Finally we give some ideas for improving these results in
\sect{fam-conclusion}.

{\em Bibliographical note:}  Most of the chapter is based on
\cite{DHW05}, though parts of \sect{family} appeared before in
\cite{DHW03}.  Both are joint work with Igor Devetak and Andreas
Winter.

\section{A family of quantum protocols.} 
\label{sec:family}
In this section, we consider a family of resource inequalities with
one noisy resource in the input and two noiseless resources in either
the input or the output.  The ``static'' members of the family involve
a noisy bipartite state $\rho^{AB}$, while the ``dynamic'' members
involve a general quantum channel $\cN: {A'}
\rightarrow B$.  In the former case one may define a class of
purifications $\proj{\psi}^{ABE} \ext \rho^{AB}$.  In the latter case
one may define a class of pure states $\ket{\psi}^{RBE}$, which
corresponds to the outcome of sending half of some $\ket{\phi}^{RA'}$
through the channel's isometric extension $U_\cN : {A'} \rightarrow
BE$, $U_\cN \ext \cN$.

Recall the identities, for a tripartite pure state $\ket{\psi}^{ABE}$,
\ben
  \frac{1}{2} I(A;B)_\psi + \frac{1}{2} I(A;E)_\psi &= & H(A)_\psi, \\
  \frac{1}{2} I(A;B)_\psi - \frac{1}{2} I(A;E)_\psi &= & I (A\,\rangle B)_\psi.
\een
Henceforth, all entropic quantities will be defined with respect
to $\ket{\psi}^{RBE}$ or $\ket{\psi}^{ABE}$, depending on the context,
so we shall drop the $\psi$ subscript.

We now introduce the ``parent'' resource inequalities, deferring their
construction until the end of the section.
The ``mother'' RI is a method for distillating entanglement from a
noisy state using quantum communication:
\begin{equation}
  \< \rho \>  + \frac{1}{2} I(A;E) \, [q \rightarrow q] 
                          \geq \frac{1}{2} I(A;B)\,[q \, q]. 
\tag{\female}\label{eq:mama}
\end{equation}
There exists a  dual ``father'' RI for entanglement-assisted quantum
communication, which is related to the mother by
interchanging dynamic and static resources, and the $A$ and $R$ systems:
\begin{equation}
  \frac{1}{2} I(R;E) \, [q \, q] + \< \cN \>
                          \geq \frac{1}{2} I(R;B)\,  [q  \rightarrow q]. 
\tag{\male}\label{eq:papa}
\end{equation}
We shall combine these parent RIs
 with the unit RIs corresponding
to teleportation, super-dense coding and entanglement distribution
($\qtq\geq \qq$)
to recover several previously known ``children'' protocols.

\par
Each parent has her or his own children (like the Brady
Bunch\footnote{\emph{The Brady Bunch}, running from 26 September 1969
till 8 March 1974, was a popular show of the American Broadcasting
Company about a couple with three children each from their previous
marriages.  For more information, see \cite{Moran95}.}).

Let us consider the mother first; she has three children.
The first is a variation of 
the  hashing inequality \eq{hashing},
which follows from the mother and teleportation.
\ben
\<\rho\> +  I(A;E) \, [c \rightarrow c] + \frac{1}{2} I(A;E) [q \, q]
&  \geq & \<\rho\> + \frac{1}{2} I(A;E) [q \rightarrow q]\\
&  \geq &  \frac{1}{2} I(A;B) [q \, q] \\
& = & I(A\,\rangle B) \,[q \, q] + \frac{1}{2} I(A;E) [q \, q].
\een
By the Cancellation Lemma (\ref{lemma:cancel}),
\be
\<\rho\> +   I(A;E) \, [c \rightarrow c] + o[q \, q] 
 \geq I(A\,\rangle B) \,[q \, q].
  \label{eq:hashing2}
\ee
This is slightly weaker than \eq{hashing}.
Further combining with teleportation gives a variation
on  noisy teleportation \eq{ntp}:
\be
\<\rho\> + I(A;B) \, [c \rightarrow c] + o[q \, q] 
 \geq  I(A\,\rangle B) \,[q \rightarrow q].
\label{eq:ntp2}
\ee 
The third child is noisy super-dense coding
(\eq{nsd}), obtained by combining the mother with super-dense coding:
\ben
 H(A) \, [q \rightarrow q] +  \< \rho \> 
& = & \frac{1}{2} I(A;B)  \, [q \rightarrow q] +
\frac{1}{2} I(A;E)  \, [q \rightarrow q] +
 \< \rho \> \\
& \geq & \frac{1}{2} I(A;B) [q \rightarrow q] + 
\frac{1}{2} I(A;B) [q \, q]\\
& \geq & I(A; B) \,[c \rightarrow c].
  \label{eq3}
\een
The father happens to have only two children.
One of them is the entanglement-assisted classical capacity
RI (\ref{eq:eac}), obtained by combining the father with (SD)
\ben
 H(R) \, [q \, q] +  \< \cN \> 
& = & \frac{1}{2} I(R;B)  \, [q \, q] +
\frac{1}{2} I(R;E)  \, [q \, q] +
 \< \cN \> \\
& \geq & \frac{1}{2} I(R;B) [q \, q] + 
\frac{1}{2} I(R;B) [q \rightarrow q]\\
& \geq & I(R; B) \,[c \rightarrow c].
  \label{eq13}
\een
The second is a variation on the quantum channel capacity result 
(\eq{lsd}). It is obtained by combining the father with entanglement
distribution. 
\ben
 \frac{1}{2} I(R;E) \, [q \, q] + \< \cN \>
   & \geq & \frac{1}{2} I(R;B)\,  [q  \rightarrow q] \\
 & =  & \frac{1}{2} I(R;E)\,  [q  \rightarrow q] 
+ \frac{1}{2} I(R \, \> B)\,  [q  \rightarrow q] \\
  & =  & \frac{1}{2} I(R;E)\,  [q \, q] 
+  \frac{1}{2} I(R \, \> B)\,  [q  \rightarrow q].
\een
Hence, by the Cancellation Lemma
\be
   \< \cN \> + o[q \, q]   \geq I(R\,\rangle B) \,[q  \rightarrow q].
  \label{eq5}
\ee

Alas, we do not know how to get rid of the $o$ term without invoking
further results. For instance, the original proof of the hashing
inequality and the HSW theorem allow us to get rid of the $o$ term, by
\lem{noo}.  Quite possibly the original proof
\cite{Lloyd96,Shor02,Devetak03} is needed.

{\bf Constructing the parent protocols using coherification rules.}
\label{sec:ccc}

Having demonstrated the power of the parent resource inequalities,
we now address the question of constructing protocols implementing them.

\begin{corollary}
The mother RI is obtained from the hashing inequality (\eq{hashing})
by applying rule I.  
\end{corollary}
It can be readily checked that the protocol from
\cite{DW03b,DW03c} implementing \eq{hashing} indeed satisfies the
conditions of rule I. The approximate uniformity condition is in fact
exact in this case.
\qed

\begin{corollary}
The father RI follows from the EAC protocol from \cite{BSST01}.
\end{corollary}
\begin{proof}
The main observation is that the protocol from \cite{BSST01} implementing 
\eq{eac} in fact outputs a \emph{private} classical channel as it is!
We shall analyze the protocol in the CP picture. 
Alice and Bob share a maximally entangled state $\ket{\Phi_D}^{AB'}$. 
Alice encodes her message 
$m$ via a unitary $U_m$: 
$$
m \mapsto (U_m^A\otimes \one^{B'}) \ket{\Phi_D}^{AB'} = 
(\one^A \ot (U^T_m)^{B'}) \ket{\Phi_D}^{A B'}.
$$
Applying the channel $(U_\cN^{A\ra BE})^{\otimes n}$ yields
$$
\ket{\Upsilon_m}^{BB'E} = ((U^T_m)^{B'} \otimes \one^{BE}) \ket{\Psi}^{BB'E},
$$
where $\ket{\Psi}^{BB'E} = (U_\cN^{A\ra BE})^{\otimes n}\ket{\Phi_D}^{A B'}$.
Bob's decoding operation consists of adding an ancilla system $\bar{B}$ 
in the state $\ket{0}^{\bar{B}}$, performing some unitary $U^{B B'\bar{B}}$ 
and von Neumann measuring the ancilla $\bar{B}$. Before the von Neumann
measurement the state of the total system is
$$
\ket{\Upsilon''_m}^{BB'\bar{B}E} = U^{BB'\bar{B}}
\ket{\Upsilon_m}^{BB'E} \ket{0}^{\bar{B}}. 
$$
After the measurement, the message $m$ is correctly decoded
with probability $1 - \epsilon$.  
By the gentle operator lemma\cite{Winter99}, $U^{BB'\bar{B}}$ could
have been chosen so that upon correct decoding,
 the post-measurement state $\ket{\Upsilon'_m}^{BB'E}$ satisfies 
$$
\| {\Upsilon'_m} -  \Upsilon_m\|_1 \leq \sqrt{8 \epsilon}.
$$
Assuming Bob correctly decodes $m$, he then applies $U_m^*$ to $B'$,
bringing the system $BB'E$ 
into the state $\ket{\Psi'_m} = ((U^*_m)^{B'} \otimes \one^{BE}) \ket{\Upsilon'_m}$,
for which
$$
\|\Psi'_m - \Psi \|_1 \leq  \sqrt{8 \epsilon},
$$
for all $m$.
Thus $m$ is coherently decoupled from $BB'E$, and we may apply Rule O.
\end{proof}

\begin{corollary}
The mother  RI follows from the NSD protocol from \cite{HHHLT01}.
\end{corollary}
\begin{proof}
The proof is almost the same as for the previous Corollary.
\end{proof}

\section{Two dimensional trade-offs for the family}
\label{sec:trade-off}

It is natural to ask about the optimality of our family of resource
inequalities.  In this section we show that they indeed give rise to
optimal two dimensional capacity regions, the boundaries of which are
referred to as trade-off curves.  To each family member corresponds a
theorem identifying the operationally defined capacity region
$C(\rho^{AB})$ ($C(\cN)$) with a formula $\tilde{C}(\rho^{AB})$
($\tilde{C}(\cN)$) given in terms of entropic quantities evaluated on
states associated with the given noisy resource $\rho^{AB}$ ($\cN$).
Each such theorem consists of two parts: the
\emph{direct coding theorem} which establishes $\tilde{C} \subseteq C$
and the \emph{converse} which establishes $C \subseteq \tilde{C} $.

\subsection{Grandmother protocol}

To prove the trade-offs involving static resources, we will first need
to extend the mother protocol \peq{mama} to a ``grandmother'' RI
by combining it with instrument compression \peq{ict2}.


\begin{theorem} [Grandmother]
Given a static resource $\rho^{AB}$, 
for any remote instrument ${\bf T}: A \rightarrow A' X_B$,
the following RI holds
\begin{equation}
  \frac{1}{2} I({A'};EE'|X_B)_\sigma  \, [q \rightarrow q] 
 + I(X_B; BE)_\sigma   [c \rightarrow c] +
\< \rho^{AB} \> \geq \frac{1}{2} I({A'};B|X_B)_\sigma   \,[q \, q]. 
\label{eq:granny}
\end{equation}
In the above,
the state $\sigma^{X_B A' B E E'}$ is defined by
$$
\sigma^{X_B A' B E E'} = {\tilde{\bf T}}^{A\ra A'E'X_B} (\psi^{ABE}),
$$
where $\proj{\psi}^{ABE} \ext \rho^{AB}$ and
$\tilde{\bf T}: A \rightarrow {A'} E' X_B$ is 
a QP extension of ${\bf T}$.
\end{theorem}
\begin{proof}
By the instrument compression RI (\ref{eq:ict2}),
\ben
\< \rho^{AB} \> + I(X_B; BE)_\sigma [c \rightarrow c] + 
H(X| BE)_\sigma [c \, c] 
& \geq  & \< \rho^{AB} \> + \< \bar{\Delta}^{X_B \rightarrow X_A X_B} \circ  
{\bf T} : \rho^{A} \>\\
& \geq  & \< \bar{\Delta}^{X_B \rightarrow X_A X_B} (\sigma^{X_B A}) \>. 
\label{parte1}
\een
On the other hand, by \thm{alfav} and the 
mother inequality (\ref{eq:mama}), 
$$ 
\< \bar{\Delta}^{X_B \rightarrow X_A X_B} (\sigma^{X_B A'}) \>
  + \frac{1}{2} I(A';EE'|X_B)_\sigma \, [q \rightarrow q] 
\geq \frac{1}{2} I(A'; B| X_B)_\sigma \,[q \, q].
$$
The grandmother RI is obtained by adding the above RIs,
followed by a derandomization via \cor{purim}. 
\end{proof}

\begin{corollary}
\label{cor:granny2}
In the above theorem, one may consider the special case
where ${\bf T}: A \rightarrow A' X_B$ corresponds to
some ensemble of operations $(p_x,  \cE_x)$, 
$\cE_x: A \rightarrow A'$, via the identification
$$
{\bf T}: \rho^{A} \mapsto 
\sum_x p_x \proj{x}^{X_B} \otimes \cE_x(\rho^{A}). 
$$
Then the $[c \rightarrow c]$ term from \eq{granny}
vanishes identically.
\qed
\end{corollary}


\subsection{Trade-off for noisy super-dense coding}
\label{sec:nSD-toff}

Now that we are comfortable with the
various formalisms, the formulae will reflect the
QP formalism, whereas the language  will be more 
in the CQ spirit.

\begin{figure}
\centerline{ {\scalebox{0.50}{\includegraphics{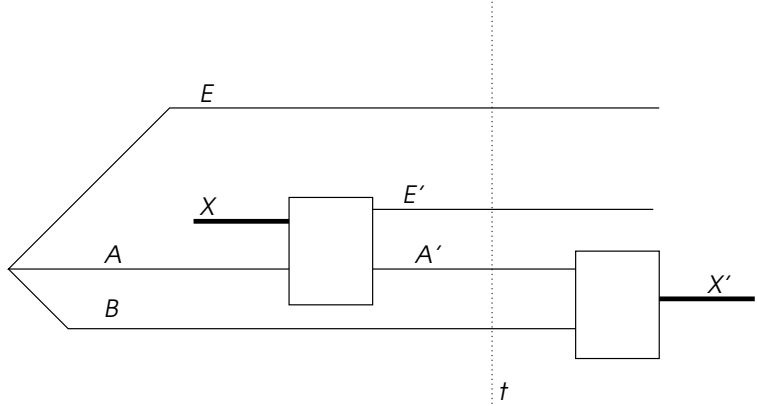}}}}
\caption{A general protocol for noisy super-dense coding.}
\label{fig:nsdfig}
\end{figure}

Given a bipartite state $\rho^{AB}$, 
the noisy super-dense coding capacity region 
$C_{\rm NSD}(\rho^{AB})$ is the two-dimensional region in the 
$(Q,R)$ plane with $Q \geq 0$ and $R \geq 0$ satisfying the RI
\begin{equation}
  \< \rho^{AB} \>  + Q \,[q \rightarrow q] \geq R \,[c \rightarrow c].
\label{nsddef}
\end{equation}

\begin{theorem}
The capacity region $C_{\rm NSD}(\rho^{AB})$ is given by
$$
C_{\rm NSD}(\rho^{AB}) = \tilde{C}_{\rm NSD} (\rho^{AB}):=
\bar{\bigcup_{l=1}^\infty \frac{1}{l}
\tilde{C}_{\rm NSD}^{(1)}( (\rho^{AB})^{\otimes l})},
$$
where the $\bar{S}$ means the closure of a set $S$ and
$\tilde{C}_{\rm NSD}^{(1)}(\rho^{AB})$
is the set of all $R \geq 0$, $Q \geq 0$ such that
$$
 R \leq  Q + \max_\sigma \left\{ I(A' \, \> BX)_\sigma : 
H(A'|X)_\sigma \leq Q \right\}.
$$
In the above, $\sigma$ is of the form
\be
\sigma^{XA'B} = \sum_x p_x \proj{x}^{X} \otimes \cE_x^{A\ra
A'}(\rho^{AB}). 
\label{eq:nsdsig} \ee
for some ensemble of operations $(p_x,  \cE_x)$, 
$\cE_x: A \rightarrow A'$.
\end{theorem}
\begin{proof}
We first prove the converse. Fix $n,R,Q, \delta, \epsilon$,
and use the Flattening Lemma (\ref{lemma:flattening}) so that we can
assume that $k = 1$.
The resources available are
\begin{itemize}
\item The state $(\rho^{AB})^{\otimes n}$
shared between Alice and Bob. Let it be
contained in the system $A^n B^n$, of total dimension
$d^n$, which we shall call $AB$ for short.
\item A perfect quantum channel $\id: A' \rightarrow A'$,
$ \dim A' = 2^{n Q}$,
from Alice to Bob
(after which $A'$ belongs to Bob despite the notation!).
 \end{itemize}
The resource to be simulated is the perfect
classical channel of size $D = 2^{n(R - \delta)}$
on any source, in particular on the
random variable $X$ corresponding to
the uniform distribution $\pi_{D}$.

In the protocol (see \fig{nsdfig}), 
Alice performs a $\{ c q \rightarrow q \}$ 
encoding $(\cE_x: A \rightarrow A')_x$, 
depending on the source random variable, 
and then sends the $A'$ system through the perfect quantum channel.
After time $t$ Bob performs a POVM 
$\Lambda: A'B \rightarrow X'$, 
on the system $A'B$, yielding the random variable $X'$.
The protocol ends at time $t_f$.
Unless otherwise stated, the entropic quantities below
refer to the state of the system at time $t$. 

Since at time $t_f$
the state of the system $XX'$ is supposed to be $\epsilon$-close to 
$\bar{\Phi}_{D}$,
\lem{fano} implies
$$
I(X;X')_{t_f} \geq n (R - \delta) - \eta'(\epsilon) - K \epsilon n R.
$$ 
By the Holevo bound \cite{Holevo73},
$$
I(X;X')_{t_f} \leq I(X; A'B).
$$
Recall from \eq{trip-entropy} the identity 
$$
I(X; A'B) = H(A') + I(A'\,\rangle BX) - I(A';B) + I(X;B). 
$$
Since $I(A';B) \geq 0$, and in our protocol $I(X;B) = 0$,
this becomes
$$
I(X; A'B) \leq H(A') + I(A'\,\rangle BX). 
$$
Observing that 
$$
nQ \geq H(A') \geq H(A'|X),
$$
these all add up to
$$
R \leq Q + \frac{1}{n} I(A'\,\rangle BX) + \delta + KR \epsilon
+ \frac{\eta'(\epsilon)}{n}.
$$
As these are true for any $\epsilon, \delta > 0$ and 
sufficiently large $n$, the converse holds.

Regarding the direct coding theorem, it suffices to
demonstrate the RI
$$
 \< \rho^{AB} \>  + H(A'|X)_\sigma \,[q \rightarrow q] 
\geq I(A'; B|X)_\sigma  \,[c \rightarrow c].
$$
This, in turn, follows from linearly combining
\cor{granny2} with super-dense coding \peq{sd}
much in the same way the noisy super-dense coding RI \peq{nsd}
follows from the mother \peq{mama}.
\end{proof}

\subsection{Trade-off for quantum communication assisted
            entanglement distillation}

Given a bipartite state $\rho^{AB}$, 
the quantum communication assisted entanglement distillation  
capacity region ( or ``mother'' capacity region for short) 
$C_{\rm M}(\rho^{AB})$ is the set of
$(Q,E)$ with $Q \geq 0$ and $E \geq 0$ satisfying the RI
\begin{equation}
  \< \rho^{AB} \>  + Q \,[q \rightarrow q] \geq E \,[q \, q].
\label{eq:momdef}
\end{equation}
(This RI is trivially false for $Q<0$ and trivially true for $Q\geq 0$ and
$E\geq 0$.)
\begin{theorem}
The capacity region $C_{\rm M}(\rho^{AB})$ is given by
$$
C_{\rm M}(\rho^{AB}) = \tilde{C}_{\rm M} (\rho^{AB}):=
\bar{\bigcup_{l=1}^\infty \frac{1}{l}
\tilde{C}_{\rm M}^{(1)}( (\rho^{AB})^{\otimes l})},
$$
where
$\tilde{C}_{\rm M}^{(1)}(\rho^{AB})$
is the set of all $Q \geq 0$, $E \geq 0$ such that
\be
E \leq  Q + \max_\sigma \left\{ I(A' \, \> BX)_\sigma: 
\frac{1}{2} I(A';EE'|X)_\sigma \leq Q \right\}.
\label{eq:mom-toff}\ee
In the above, $\sigma$ is the QP version of \eq{nsdsig},
namely
\be
\sigma^{XA'BEE'} = 
\sum_x p_x \proj{x}^{X} \otimes U_x^{A\ra A'E'}(\psi^{ABE}).
\label{eq:nsdsig-QP}
\ee
for some ensemble of isometries $(p_x, U_x)$, 
$U_x: A \rightarrow A'E'$, and purification
$\proj{\psi}^{ABE} \ext \rho^{AB}$.
\end{theorem}
\begin{proof}
We first prove the converse, which in this case follows
from the converse for the noisy super-dense coding trade-off.
The main observation is that super-dense coding (\eq{sd}) induces
an invertible linear map $f$ 
between the $(Q,E)$ and $(Q,R)$ planes corresponding
to the mother capacity region and that of noisy super-dense coding,
respectively, defined by
$$
f: (Q, E) \mapsto (Q + E, 2E).
$$
By adding superdense coding (i.e. $E\qq+E\qtq \geq 2E\ctc$) to the
mother \peq{momdef}, we find
\be
f( C_{\rm M}) \subseteq C_{\rm NSD}.
\ee
On the other hand, by inspecting the definitions of $\tilde{C}_{\rm
NSD}$ and $\tilde{C}_{\rm M}$, we can verify
\be 
\tilde{C}_{\rm NSD} =  f(\tilde{C}_{\rm M}).
\ee
The converse for the noisy super-dense coding trade-off 
is written as $ C_{\rm NSD} \subseteq \tilde{C}_{\rm NSD}$.
As $f$ is a bijection, putting everything together we have
$$
{C}_{\rm M} \subseteq f^{-1} (C_{\rm NSD})
\subseteq  f^{-1} (\tilde{C}_{\rm NSD}) = \tilde{C}_{\rm M},
$$
which is the converse for the mother trade-off.

The direct coding theorem follows immediately from 
\cor{granny2}.
\end{proof}

\subsection{Trade-off for noisy teleportation}
\label{ntp-toff}

Given a bipartite state $\rho^{AB}$, 
the noisy super-dense coding capacity region 
$C_{\rm NTP}(\rho^{AB})$ is a two-dimensional region in the 
$(R,Q)$ plane with $R \geq 0$ and $Q \geq 0$ satisfying the RI
\begin{equation}
  \< \rho^{AB} \>  + R \,[c \rightarrow c] \geq Q \,[q \rightarrow q].
\label{eq:ntpdef}
\end{equation}

\begin{theorem}
The capacity region $C_{\rm NTP}(\rho^{AB})$ is given by
$$
C_{\rm NTP}(\rho^{AB}) = \tilde{C}_{\rm NTP} (\rho^{AB}):=
\bar{\bigcup_{l=1}^\infty \frac{1}{l}
\tilde{C}_{\rm NTP}^{(1)}( (\rho^{AB})^{\otimes l})},
$$
where $\tilde{C}_{\rm NTP}^{(1)}(\rho^{AB})$
is the set of all $R \geq 0$, $Q \geq 0$ such that
\be
Q \leq  \max_\sigma \left\{ I(A' \, \> BX)_\sigma : 
 I(A'; B| X)_\sigma + I(X; BE)_\sigma \leq R \right\}.
\label{eq:ntp-toff}\ee
In the above, $\sigma$ is of the form
\be
\sigma^{XA'BE} =  \mathbf{T}(\psi^{ABE}),
\label{eq:ntpsig}
\ee
for some instrument $\mathbf{T}: A \rightarrow A'X$
and purification $\proj{\psi}^{ABE} \ext \rho^{AB}$.
\end{theorem}
\begin{proof}
We first prove the converse. Fix $n,Q,R, \delta, \epsilon$,
and use the Flattening Lemma so we can assume that the depth is one.
The resources available are
\begin{itemize}
\item The state $(\rho^{AB})^{\otimes n}$
shared between Alice and Bob. Let it be
contained in the system $A^n B^n$, which we shall call $AB$ for
short. 
\item A perfect classical channel of size $2^{n R}$.
 \end{itemize}
The resource to be simulated is
the perfect quantum channel
 $\id_D: A_1 \rightarrow B_1$,
$D = \dim A_1 = 2^{n (Q - \delta)}$,
from Alice to Bob,
on any source, in particular on the maximally entangled
state $\Phi^{A' A_1}$. 

\begin{figure}
\centerline{ {\scalebox{0.50}{\includegraphics{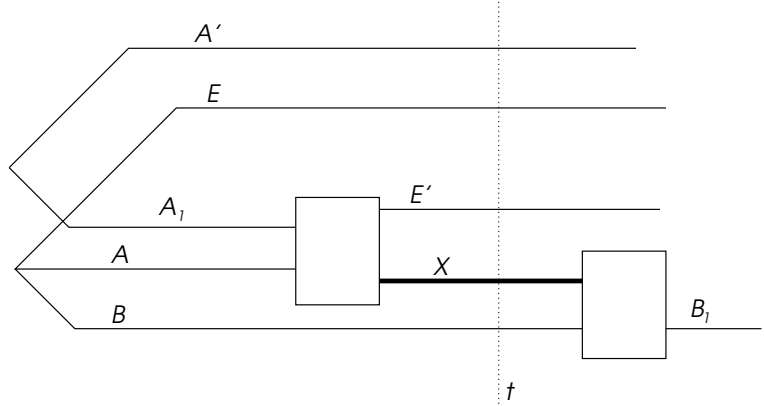}}}}
\caption{A general protocol for noisy teleportation.}
\label{fig:ntpfig}
\end{figure}

In the protocol (see \fig{ntpfig}), 
Alice performs a 
POVM $\Lambda: AA_1 \rightarrow  X$ on the system $AA_1$,
and  sends the outcome random variable $X$ 
through the classical channel.
After time $t$ Bob performs a $\{ cq \rightarrow q \}$ 
decoding quantum operation $\cD: XB \rightarrow B_1$.
The protocol ends at time $t_f$.
Unless otherwise stated, the entropic quantities below
refer to the time $t$. 

Our first observation is that 
performing the POVM $\Lambda$ induces 
an instrument 
${\bf T}: A \rightarrow A' X$,\footnote{  
Indeed, first a pure ancilla 
$A' A_1$ was appended, then another pure ancilla $X$ was appended,
the system  $A A'A_1 X$ was rotated to 
$A' E' X$, and finally $X$ was measured and $E'$ was traced out.}
so that the state of the system $XA'BE$ at time $t$ is indeed of the form
of \eq{ntpsig}.

Since at time $t_f$ the state of the system $A'B_1$ 
is supposed to be $\epsilon$-close to 
${\Phi}_{D}$,
\lem{fano} implies
$$
I(A' \> B_1)_{t_f} \geq n (Q - \delta) - \eta'(\epsilon) - K \epsilon n Q.
$$ 
By the data processing inequality,
$$
I(A' \> B_1)_{t_f} \leq I(A' \> BX).
$$
Thus
\be
Q \leq  \frac{1}{n} I(A' \,\rangle BX) + \delta + KQ \epsilon
+ \frac{\eta'(\epsilon)}{n}.
\label{eq:NTP-Q-bound}\ee
To bound $R$, start with the identity
$$
I(X; A'B E) = H(A')  + I(A'\,\rangle BEX) - I(A';BE) + I(X;BE). 
$$
Since $I(A';BE) = 0$, $H(A') \geq H(A'|X) $ and 
$I(A'\,\rangle BEX) \geq I(A'\,\rangle BX)$,
this becomes
$$
I(X; A'B E) \geq  I(A'; B|X) + I(X;BE).
$$
Combining this with 
$$
nR  \geq   H(X)  \geq   I(X; A'BE)
$$
gives the desired
\be
R \geq \frac{1}{n} [ I(A' ;B |X) + I(X;BE)].
\label{eq:NTP-R-bound}\ee
As \eqs{NTP-Q-bound}{NTP-R-bound} are true for any $\epsilon, \delta >
0$ and sufficiently large $n$, the converse holds.

Regarding the direct coding theorem, it suffices to
demonstrate the RI
\be
 \< \rho^{AB} \>  + (I(A'; B|X)_\sigma + I(X;BE)_\sigma)
 \,[c \rightarrow c]
\geq  I(A'\, \> BX)_\sigma \,[q \rightarrow q]. 
\label{dctntp}
\ee
Linearly combining
the grandmother RI (\eq{granny}) with teleportation (\eq{tp}),
much in the same way the variation on the noisy teleportation
RI (\eq{ntp2}) was obtained  from the mother (\eq{mama}), we have
$$
 \< \rho^{AB} \>  + (I(A'; B|X)_\sigma + I(X;BE)_\sigma)
 \,[c \rightarrow c] + o [q \, q]
\geq  I(A'\, \> BX)_\sigma \,[q \rightarrow q]. 
$$
Equation (\ref{dctntp}) follows by invoking \lem{noo}
and \eq{hashing}.
\end{proof}

\subsection{Trade-off for classical communication assisted
            entanglement distillation}
\label{sec:distill-toff}

Given a bipartite state $\rho^{AB}$, 
the classical communication assisted entanglement distillation  
capacity region (or ``entanglement distillation'' capacity region for short) 
$C_{\rm ED}(\rho^{AB})$ is the two-dimensional region in the 
$(R,E)$ plane with $R \geq 0$ and $E \geq 0$ satisfying the RI
\begin{equation}
  \< \rho^{AB} \>  + R \,[c \rightarrow c] \geq E \,[q \, q].
  \label{eq:eddef}
\end{equation}

\begin{theorem}
The capacity region $C_{\rm ED}(\rho^{AB})$ is given by
$$
C_{\rm ED}(\rho^{AB}) = \tilde{C}_{\rm ED} (\rho^{AB}):=
\bar{\bigcup_{l=1}^\infty \frac{1}{l}
\tilde{C}_{\rm ED}^{(1)}( (\rho^{AB})^{\otimes l})},
$$
where
$\tilde{C}_{\rm ED}^{(1)}(\rho^{AB})$
is the set of all $R \geq 0$, $E \geq 0$ such that
\be
E \leq  \max_\sigma \left\{ I(A' \, \> BX)_\sigma: 
I(A'; EE'|X)_\sigma + I(X; BE)_\sigma \leq R \right\},
\label{eq:ed-toff}\ee
In the above, $\sigma$ is the
fully QP version of \eq{ntpsig},
namely
\be
\sigma^{XA'BEE'} =  \mathbf{T}'(\psi^{ABE}),
\label{edsig}
\ee
for some instrument $\mathbf{T}: A \rightarrow A'E'X$
with pure quantum output
and purification $\proj{\psi}^{ABE} \ext \rho^{AB}$.
\end{theorem}
\begin{proof}
We first prove the converse, which in this case follows
from the converse for the noisy teleportation trade-off.
The argument very much parallels that of the converse for the 
mother trade-off.
The main observation is that teleportation (\eq{tp}) induces
an invertible linear map $g$ 
between the $(R,E)$ and $(R,Q)$ planes corresponding
to the entanglement distillation capacity region
and that of noisy teleportation,
respectively, defined by
$$
g: (R, E) \mapsto (R + 2E, E).
$$
By applying TP to \eq{eddef}, we find
\be
g( C_{\rm ED}) \subseteq C_{\rm NTP}.
\ee
On the other hand, from the definitions of $\tilde{C}_{\rm ED}$ and
$\tilde{C}_{\rm NTP}$ (\eqs{ed-toff}{ntp-toff}), we have
\be 
\tilde{C}_{\rm ED} =  g(\tilde{C}_{\rm NTP}).
\ee
The converse for the noisy teleportation trade-off 
is written as $ C_{\rm NTP} \subseteq \tilde{C}_{\rm NTP}$.
As $g$ is a bijection, putting everything together we have
$$
{C}_{\rm ED} \subseteq g^{-1} (C_{\rm NTP})
\subseteq  g^{-1} (\tilde{C}_{\rm NTP}) = \tilde{C}_{\rm ED},
$$
which is the converse for the entanglement distillation trade-off.

Regarding the direct coding theorem, it suffices to
demonstrate the RI
\be
 \< \rho^{AB} \>  + (I(A'; EE'|X)_\sigma + I(X;BE)_\sigma)
 \,[c \rightarrow c]
\geq  I(A'\, \> BX)_\sigma \,[q \, q]. 
\label{eq:dcted}
\ee
Linearly combining
the grandmother RI (\eq{granny}) with teleportation (\ref{eq:tp}),
much in the same way the variation on the hashing
RI (\eq{hashing2}) was obtained  from the mother (\eq{mama}), we have
$$
 \< \rho^{AB} \>  + (I(A'; EE'|X)_\sigma + I(X;BE)_\sigma)
 \,[c \rightarrow c] + o [q \, q]
\geq  I(A'\, \> BX)_\sigma \,[q \rightarrow q]. 
$$
\eq{dcted} follows by invoking \lem{noo}
and \eq{hashing}.
\end{proof}

\subsection{Trade-off for entanglement assisted
            quantum communication}
\label{sec:father-toff}

Given a noisy quantum channel $\cN: A' \rightarrow B$, 
the entanglement  assisted quantum communication
capacity region ( or ``father'' capacity region for short) 
$C_{\rm F}(\cN)$ is the region of
$(E,Q)$ plane with $E \geq 0$ and $Q \geq 0$ satisfying the RI
\begin{equation}
  \< \cN \>  + E \,[q \, q] \geq Q \,[q \rightarrow  q].
  \label{daddef}
\end{equation}

\begin{theorem}\label{thm:father-toff}
The capacity region $C_{\rm F}(\cN)$ is given by
$$
C_{\rm F}(\cN) = \tilde{C}_{\rm F} (\cN):=
\bar{\bigcup_{l=1}^\infty \frac{1}{l}
\tilde{C}_{\rm F}^{(1)}( \cN^{\otimes l})},
$$
where
$\tilde{C}_{\rm F}^{(1)}(\cN)$
is the set of all $E \geq 0$, $Q \geq 0$ such that
\ben
Q & \leq &  E + I(A \, \> B)_\sigma \\
Q & \leq &  \frac{1}{2} I(A;B)_\sigma.
\een
In the above, $\sigma$ is of the form 
$$
\sigma^{ABE} =  U_\cN \circ  \cE (\phi^{A A''}),
$$
for some pure input state $\ket{\phi^{A A''}}$,
encoding operation $\cE: A'' \rightarrow A'$,
and where $U_\cN: A' \rightarrow BE$ is an isometric extension of $\cN$.


\end{theorem}

This tradeoff region includes two well-known limit points.  When
$E=0$, the quantum capacity of $\cN$ is
$I(A\>B)$\cite{Lloyd96,Shor02,Devetak03}, and for $E>0$, entanglement
distribution ($\qtq\geq\qq$) means it should still be bounded by
$I(A\>B) + E$.  On the other hand, when given unlimited entanglement,
the classical capacity is $I(A;B)$\cite{BSST01} and thus the quantum
capacity is never greater than $\half I(A;B)$ no matter how much
entanglement is available.  These bounds meet when $E=\half I(A;E)$
and $Q=\half I(A;E)$, the point corresponding to the father protocol.
Thus, the goal of our proof is to show that the father protocol is
optimal.

\begin{proof}
We first prove the converse. Fix $n,E,Q, \delta, \epsilon$,
and use the Flattening Lemma to reduce the depth to one.
The resources available are
\begin{itemize}
\item The channel  $\cN^{\otimes n}: {A'}^n \rightarrow {B^n}$ 
from Alice to Bob. We shall shorten ${A'}^n$ to $A'$ and 
${B}^n$ to $B$.
\item  The maximally entangled state $\Phi^{T_A T_B}$, 
$\dim T_A = \dim T_B = 2^{n E }$,
shared between Alice and Bob.
 \end{itemize}
The resource to be simulated is
the perfect quantum channel
 $\id_D: A_1 \rightarrow B_1$,
$D = \dim A_1 = 2^{n (Q - \delta)}$,
from Alice to Bob,
on any source, in particular on the maximally entangled
state $\Phi^{R A_1}$. 

\begin{figure}
\centerline{ {\scalebox{0.50}{\includegraphics{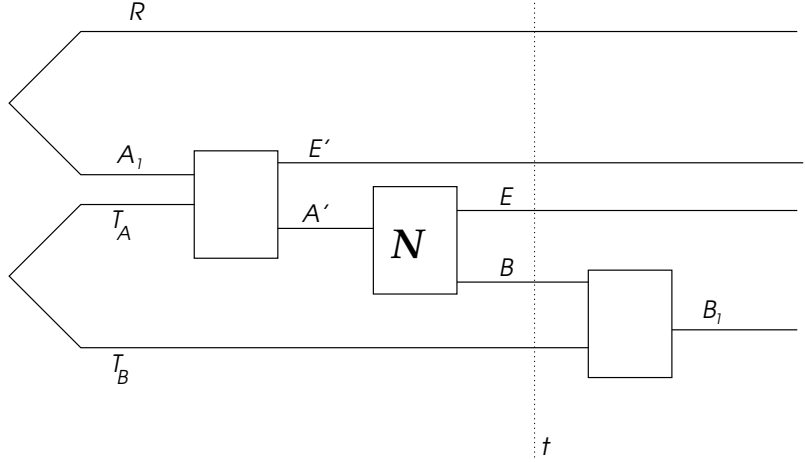}}}}
\caption{A general protocol for entanglement assisted
quantum communication.}
\label{fig:dadfig}
\end{figure}

In the protocol (see \fig{dadfig}), 
Alice performs a general
encoding map $\cE: A_1 T_A \rightarrow A' E'$
and sends the system $A'$ through the noisy channel
$\cN:{A' \rightarrow B}$.
After time $t$ Bob performs a decoding 
operation $\cD: B T_B \rightarrow B_1$.
The protocol ends at time $t_f$.
Unless otherwise stated, the entropic quantities below
refer to the time $t$.

Define $A := R T_B$ and $A'' := A_1 T_A$.
Since at time $t_f$ the state of the system $RB_1$ is supposed 
to be $\epsilon$-close to ${\Phi}_{D}$,
\lem{fano} implies
$$
I(R \, \> B_1)_{t_f} \geq n (Q - \delta) - \eta' (\epsilon) - 
K \epsilon n Q.
$$ 
By the data processing inequality,
$$
I(R \,\> B_1)_{t_f} \leq I(R \, \> BT_B).
$$
Together with the inequality 
$$ 
I(R \,\rangle B T_B) \leq
I(R T_B\,\rangle B) + H(T_B),
$$
since $E = H(T_B)$, the above implies
$$
Q  \leq  E + \frac{1}{n} I(A \> B) 
+ \delta + KQ \epsilon + \frac{\eta'(\epsilon)}{n}. 
$$
Combining this  with
$$
H(A) = H(R) + H(T_B) = nQ + nE.
\label{sumy}
$$
gives
$$
Q \leq \frac{1}{2n} I(A;B) + 
\delta/2 + KQ \epsilon/2 + \frac{\eta'(\epsilon)}{2n}.
$$
As these are true for any $\epsilon, \delta > 0$ and 
sufficiently large $n$, the converse holds.

Regarding the direct coding theorem, it follows 
directly form the father RI
$$
 \< \cN \>  + \half I(A;E)_\sigma \,[q \, q] \geq 
\half I(A;B)_\sigma \,[q \rightarrow  q].
$$

\end{proof}

\subsection{Trade-off for entanglement assisted
            classical communication}
\label{sec:NCE-toff}

The result of this subsection was first proved by Shor 
in \cite{Shor04}. Here we state it for completeness, and 
give an independent proof of the converse.
An alternative proof of the direct coding theorem was sketched 
in \cite{DS03} and is pursued in \cite{DHLS05} to
unify this result with the father trade-off.

Given a noisy quantum channel $\cN: A' \rightarrow B$, 
the entanglement assisted classical communication
capacity region (or ``entanglement assisted'' capacity region 
for short)
$C_{\rm EA}(\cN)$ is the set of all points
$(E,R)$ with $E \geq 0$ and $R \geq 0$ satisfying the RI
\begin{equation}
  \< \cN \>  + E \,[q \, q] \geq R \,[c \rightarrow  c].
  \label{cl-daddef}
\end{equation}

\begin{theorem}
The capacity region $C_{\rm EA}(\cN)$ is given by
$$
C_{\rm EA}(\cN) = \tilde{C}_{\rm EA} (\cN):=
\bar{\bigcup_{l=1}^\infty \frac{1}{l}
\tilde{C}_{\rm EA}^{(1)}( \cN^{\otimes l})},
$$
where
$\tilde{C}_{\rm EA}^{(1)}(\cN)$
is the set of all $E \geq 0$, $R \geq 0$ such that
\be
R \leq  \max_\sigma \left\{ I(AX;B)_\sigma
    : E \geq H(A|X)_\sigma  \right\}.
\label{eq:eac-toff}
\ee
In the above, $\sigma$ is of the form 
\be
\sigma^{XAB} = \sum_x p_x \proj{x}^X \otimes \cN(\phi_x^{AA'}),
\label{sigea}
\ee
for some pure input ensemble $( p_x, \ket{\phi_x}^{AA'} )_x$.
\end{theorem}
\begin{proof}
We first prove the converse. Fix $n,E,Q, \delta, \epsilon$,
and again use the flattening lemma to reduce depth to one.
The resources available are
\begin{itemize}
\item The channel  $\cN^{\otimes n}: {A'}^n \rightarrow {B^n}$ 
from Alice to Bob. We shall shorten ${A'}^n$ to $A'$ and 
${B}^n$ to $B$.
\item  The maximally entangled state $\Phi^{T_A T_B}$, 
$\dim T_A = \dim T_B = 2^{n E }$,
shared between Alice and Bob.
 \end{itemize}
The resource to be simulated is the perfect
classical channel of size $D = 2^{n(R - \delta)}$
on any source, in particular on the
random variable $X$ corresponding to
the uniform distribution $\pi_{D}$.

\begin{figure}
\centerline{ {\scalebox{0.50}{\includegraphics{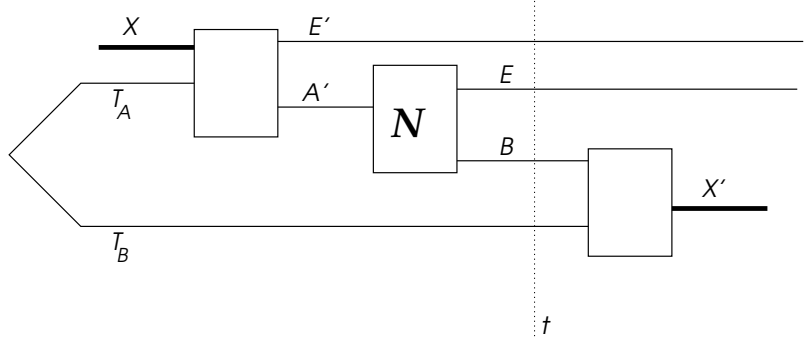}}}}
\caption{A general protocol for entanglement assisted
classical communication.}
\label{fig:eafig}
\end{figure}

In the protocol (see \fig{eafig}), 
Alice performs a $\{ c q \rightarrow q \}$ 
encoding $(\cE_x: T_A \rightarrow A')_x$, 
depending on the source random variable, 
and then sends the $T_A$ system through the noisy channel
$\cN:{A' \rightarrow BE}$.
After time $t$ Bob performs a POVM 
$\Lambda: T_B B \rightarrow X'$, 
on the system $T_B B$, yielding the random variable $X'$.
The protocol ends at time $t_f$.
Unless otherwise stated, the entropic quantities below
refer to the state of the system at time $t$. 

\par
Since at time $t_f$
the state of the system $XX'$ is supposed to be $\epsilon$-close to 
$\bar{\Phi}_{D}$,
\lem{fano} implies
$$
I(X;X')_{t_f} \geq n (R - \delta) - \eta'(\epsilon) - K \epsilon n R.
$$ 
By the Holevo bound
$$
I(X;X')_{t_f} \leq I(X; T_BB).
$$
Using the chain rule twice, we find
\ben
I(X; T_BB) &=& I(X;B|T_B) + I(X;T_B)
\\ &=& I(XT_B;B) + I(X;T_B) - I(T_B;B)
\een
Since  $I(T_B;B) \geq 0$  
and in this protocol $I(X;T_B) = 0$, this becomes
$$
I(X; T_BB) \geq  I(XT_B;B).
$$
These all add up to
$$
R \leq \frac{1}{n}I(XT_B;B) + \delta + Kd \epsilon
+ \frac{\eta'{\epsilon}}{n},
$$
while on the other hand,
$$
nE \geq H(T_B|X).
$$
As these are true for any $\epsilon, \delta > 0$ and 
sufficiently large $n$, we have thus shown a variation
on the converse with the state $\sigma$ from
(\ref{sigea}) replaced by $\tilde{\sigma}$,
$$
\tilde{\sigma}^{XABE'} = \sum_x p_x \proj{x}^X \otimes 
\cN\circ U_x^{A'' \rightarrow 
A' E'}(\phi^{AA''}),
$$
defining $A := T_B$  and letting $U_x: T_A \rightarrow 
A' E'$ be the isometric extension of $\cE_x$.

However, this is a weaker result than we would like; the converse we
have proved allows arbitrary noisy encodings and we would like to show
that isometric encodings are optimal, or equivalently that the $E'$
register is unnecessary.  We will accomplish this, following Shor
\cite{private-shor-04}, by using a standard trick
of measuring $E'$ and showing that the protocol can only improve.
If we apply the dephasing map $\bar{\id}: E' \rightarrow Y$ 
to $\tilde{\sigma}^{ABE'}$, we obtain a state of the form
$$
{\sigma}^{XYAB} = \sum_{xy} p_{xy} \proj{x}^X \otimes \proj{y}^{Y}
 \otimes \cN (\psi_{xy}^{AA'}).
$$
The converse now follows from
\ben
I(B;AX)_{\tilde{\sigma}} & \leq & I(B;AXY)_{\sigma} \\
H(A|X)_{\tilde{\sigma}} & \geq & H(A|XY)_{\sigma}.
\een
\end{proof}

\section{Conclusion}
\label{sec:fam-conclusion}
The goal of quantum Shannon theory is to give information-theoretic
formulae for the rates at which noisy quantum resources can be
converted into noiseless ones.  This chapter has taken a major step
towards that goal by finding the trade-off curves for most one-way
communication scenarios involving a noisy state or channel and two
of the three basic noiseless resources (cbits, ebits and qubits).
The main tools required for this were the resource formalism of
\chap{shannon}, coherent classical communication (from \chap{ccc}),
derandomization and basic protocols like HSW coding.

However, our expressions for trade-off curves also should be seen more
as first steps rather than final answers.  For one thing, we would
ultimately like to have formulae for the capacity that can be
efficiently computed, which will probably require replacing our
current regularized expressions with single-letter ones.  This is
related to the additivity conjectures, which are equivalent for some
channel capacities\cite{Shor03}, but are false for others\cite{DSS97}.

A more reasonable first goal is to strengthen some of the
converse theorems, so that they do not require maximizing over as many
different quantum operations.  As inspiration, note that \cite{BKN98}
showed that isometric encodings suffice to achieve the optimal rate of
quantum communication through a quantum channel.
However, the analogous result for entanglement-assisted quantum
communication is not known.  Specifically, in \fig{dadfig}, I suspect
that the $E'$ register (used to discard some of the inputs) is only
necessary when Alice and Bob share more entanglement than the protocol
can use.  Similarly, it seems plausible to assume that the optimal
form of protocols for noisy teleportation (\fig{ntpfig}) is to perform
a general TPCP preprocessing operation on the shared entanglement,
followed by a unitary interaction between the quantum data  and Alice's
part of the entangled state.  These are only two of the more obvious
examples and there ought to be many possible ways of improving
our formulae.
\chapter{The Schur transform}\label{chap:schur}
\section{Overview}
The final four chapters will explore the uses of Schur duality in
quantum computing and information theory.  Schur duality is a natural
way to decompose $\nqudits$ in terms of representations of the
symmetric group $\cS_n$ and the unitary group $\cU_d$.  In this
chapter, we will describe Schur duality and develop its
representation-theoretic background within the framework of quantum
information.  The primary connection between these fields is that a
vector space can be interpreted either as a representation of a group
or as state-space of a quantum system.  Thus, Schur duality can be
interpreted both as a mathematical fact about representations and
operationally as a fact about the transformations possible on a
quantum system.

\chap{sch-qit} will describe how Schur duality is useful in
quantum information theory.  We will see that Schur duality is a
quantum analogue of the classical method of types, in which strings
are described in terms of their empirical distributions.  This has a
number of applications in information theory, which we will survey
while highlighting the role of Schur duality.  
The chapter concludes with new work describing how i.i.d. quantum channels can
be decomposed in the Schur basis.

We then turn to computational issues in \chaps{sch-algo}{sch-Sn}.  The
unitary transform that relates the Schur basis to the 
computational basis is known as the {\em Schur transform} and
presenting efficient circuits for the Schur transform is the
main goal of \chap{sch-algo}.  These circuits mean that the
information-theoretic tasks described in \chap{sch-qit} can now all be
implemented efficiently on a quantum computer; even though
computational efficiency is not often considered in quantum
information theory, it will be necessary if we ever expect to
implement many of the coding schemes that exist.

Finally, \chap{sch-Sn} discusses algorithmic
connections between the Schur transform and related efficient
representation-theoretic transforms, such as the quantum Fourier
transform on $\cS_n$.  Ultimately the goal of this work is to find
quantum speedups that use either the Schur transform or the $\cS_n$
Fourier transform.

Most of the original work in this chapter has not yet been published.
The next two chapters are mostly review, although there are
several places where the material is assembled and presented in ways
that have not seen before in the literature.  The exception is the
last section of \chap{sch-qit} on decomposing i.i.d. quantum channels,
which is a new contribution.  The last two chapters are joint work
with Dave Bacon and Isaac Chuang.  Parts of \chap{sch-algo} appeared
in \cite{BCH04} and the rest of the chapter will be presented in
\cite{BCH05a}.  \chap{sch-Sn} will become \cite{BCH05b}.

\section{Representation theory and quantum computing}
\subsection{Basics of representation theory}
In this section, we review aspects of representation theory that will
be used in the second half of the thesis.  For a more detailed description of
representation theory, the reader should consult \cite{Artin95} for
general facts about group theory and representation theory or
\cite{GW98} for representations of Lie groups.  See also
\cite{Fulton91} for a more introductory and informal approach to Lie
groups and their representations.

{\em Representations:} For a complex vector space $V$, define
$\End(V)$ to be set of linear maps from $V$ to itself (endomorphisms).
 A representation of a group $G$ is a vector
space $V$ together with a homomorphism from $\ccG$ to $\End(V)$, i.e. a
function ${\bf R}:\ccG\ra
\End(V)$ such that ${\bf R}(g_1) {\bf R}(g_2)= {\bf R}(g_1 g_2)$.  If
${\bf R}(g)$ is a unitary operator for all $g$, then we say ${\bf R}$
is a unitary representation.  Furthermore, we say a representation
$(\bR,V)$ is finite dimensional if $V$ is a finite dimensional vector
space.  In this thesis, we will always consider complex finite
dimensional, unitary representations and use the generic term
`representation' to refer to complex, finite dimensional, unitary
representations.  Also, when clear from the context, we will denote a
representation $(\bR,V)$ simply by the representation space $V$.

The reason we consider only complex, finite dimensional, unitary
representations is so that we can use them in quantum computing.  If
$d=\dim V$, then a $d$-dimensional quantum system can hold a unit
vector in a representation $V$.  A group element $g\in\ccG$ corresponds
to a unitary rotation $\bR(g)$, which can in principle be performed by
a quantum computer.  

{\em Homomorphisms:} For any two vector spaces $V_1$ and $V_2$, define
$\Hom(V_1,V_2)$ to be the set of linear transformations from $V_1$ to
$V_2$.  If $\ccG$ acts on $V_1$ and $V_2$ with representation matrices
$\bR_1$ and $\bR_2$ then the canonical action of $\ccG$ on
$\Hom(V_1,V_2)$ is given by the map from $M$ to
$\bR_2(g)M\bR_1(g)^{-1}$ for any $M\in\Hom(V_1,V_2)$.  For any
representation $(\bR,V)$ define $V^\ccG$ to be the space of
$\ccG$-invariant vectors of $V$: i.e. $V^{\ccG}:=\{\ket{v}\in V :
\bR(g)\ket{v}=\ket{v} \,\forall g\in\ccG\}$.  Of particular interest is
the space $\Hom(V_1,V_2)^\ccG$, which can be thought of as the linear
maps from $V_1$ to $V_2$ which commute with the action of $\ccG$.
If $\Hom(V_1,V_2)^\ccG$ contains any invertible maps (or equivalently,
any unitary maps) then we say that
$(\bR_1,V_1)$ and $(\bR_2,V_2)$ are {\em equivalent} representations
and write 
$$V_1 \stackrel{\ccG}{\cong} V_2.$$
This means that there exists a unitary change of basis $U:V_1\ra V_2$ such
that for any $g\in\ccG$, $U\bR_1(g)U^\dag = \bR_2(g)$.

{\em Dual representations:} Recall that the {\em dual} of a vector
space $V$ is the set of linear maps from $V$ to $\bbC$ and is denoted
$V^*$.  Usually if vectors in $V$ are denoted by kets (e.g. $\ket{v}$)
then vectors in $V^*$ are denoted by bras (e.g. $\bra{v}$).  If we fix
a basis $\{\ket{v_1},\ket{v_2},\ldots\}$ for $V$ then the transpose is
a linear map from $V$ to $V^*$ given by $\ket{v_i}\ra \bra{v_i}$.
Now, for a representation $(\bR,V)$ we can define the {\em dual
representation} $(\bR^*,V^*)$ by $\bR^*(g)\bra{v^*} := \bra{v^*}
\bR(g^{-1})$.  If we think of $\bR^*$ as a representation on $V$
(using the transpose map to relate $V$ and $V^*$), then
it is given by $\bR^*(g)=(\bR(g^{-1}))^T$.  When $\bR$ is a unitary
representation, this is the same as the {\em conjugate representation}
$\bR(g)^*$, where here $^*$ denotes the entrywise complex conjugate.
One can readily verify that the dual and conjugate representations are
indeed representations and that
$\Hom(V_1,V_2)\stackrel{\ccG}{\cong}V_1^* \ot V_2$.

{\em Irreducible representations:} Generically the unitary operators
of a representation may be specified (and manipulated on a quantum
computer) in an arbitrary orthonormal basis.  The added structure of
being a representation, however, implies that there are particular
bases which are more fundamental to expressing the action of the
group.  We say a representation $(\bR,V)$ is irreducible (and call it
an irreducible representaiton, or {\em irrep}) if the only subspaces
of $V$ which are invariant under $\bR$ are the empty subspace $\{0\}$
and the entire space $V$.  For finite groups, any finite-dimensional
complex representation is reducible; meaning it is decomposable into a
direct sum of irreps.  For Lie groups, we need additional conditions,
such as demanding that the representation $\bR(g)$ be {\em rational};
i.e. its matrix elements are polynomial functions of the matrix
elements $g_{ij}$ and $(\det g)^{-1}$.  We say a representation of a
Lie group is {\em polynomial} if its matrix elements are polynomial
functions only of the $g_{ij}$.

{\em Isotypic decomposition:} Let $\hat{\ccG}$ be a complete set of
inequivalent irreps of $\ccG$.  Then for any reducible representation
$({\bf R},V)$ there is a basis under which the action of ${\bf R}(g)$
can be expressed as
\begin{equation}
{\bf R}(g)\cong \bigoplus_{\lambda\in\hat{\ccG}} 
\bigoplus_{j=1}^{n_\lambda}{\bf r}_\lambda(g)= 
\bigoplus_{\lambda\in\hat{G}}  {\br}_\lambda(g)
\ot { I}_{n_\lambda}  \label{eq:directsum}
\end{equation}
where $\lambda\in \hat{\ccG}$ labels an irrep $({\bf
r}_\lambda,V_\lambda)$ and $n_\lambda$ is the multiplicity of the
irrep $\lambda$ in the representation $V$.  Here we use $\cong$ to
indicate that there exists a unitary change of basis relating the
left-hand size to the right-hand side.\footnote{We only need to use
$\iso{G}$ when relating representation spaces.  In \eq{directsum} and
other similar isomorphisms, we instead explicitly specify the
dependence of both sides on $g\in G$.}  Under this change of basis we
obtain a similar decomposition of the representation space $V$ (known
as the {\em isotypic decomposition}):
\be V \stackrel{\ccG}{\cong} \bigoplus_{\lambda\in\hat{\ccG}}
 V_\lambda \ot \bbC^{n_\lambda}. 
\label{eq:rep-space-decomp}\ee
Thus while generically we may be given a representation in some
arbitrary basis, the structure of being a representation picks out a
particular basis under which the action of the representation is not
just block diagonal but also maximally block diagonal: a direct sum of
irreps.

Moreover, the multiplicity space $\bbC^{n_\lambda}$
in \eq{rep-space-decomp} has the structure of $\Hom(V_\lambda,V)^\ccG$.
This means that for any representation $(\bR,V)$,
\eq{rep-space-decomp} can be restated as 
\be V \stackrel{\ccG}{\cong}
 \bigoplus_{\lambda\in\hat{\ccG}} V_\lambda \ot \Hom(V_\lambda, V)^{\ccG}.
 \label{eq:hom-decomp}\ee
Since $\ccG$ acts trivially on $\Hom(V_\lambda, V)^{\ccG}$,
\eq{directsum} remains the same.  As with the other results in this
chapter, a proof of \eq{hom-decomp} can be found in
\cite{GW98}, or other standard texts on representation theory.

The value of \eq{hom-decomp} is that the unitary mapping from the
right-hand side (RHS)
to the left-hand side (LHS) has a simple explicit expression: it
corresponds to the canonical map $\varphi:A\otimes \Hom(A,B)\ra B$ given by
$\varphi(a\otimes f)=f(a)$.  Of course, this doesn't tell us how to
describe $\Hom(V_\lambda,V)^{\ccG}$, or how to specify an orthonormal
basis for the space, but we will later find this form of the
decomposition useful.

\subsection{The Clebsch-Gordan transform}
\label{sec:CG-general}

If $(\bR_\mu,V_\mu)$ and $(\bR_\nu,V_\nu)$ are representations of $\ccG$, their
tensor product $(\bR_\mu\ot\bR_\nu, V_\mu \ot V_\nu)$ is another
representation of $\ccG$.  In general if $V_\mu$ and $V_\nu$ are
irreducible, their tensor product will not necessarily be.  According
to \eq{hom-decomp}, the tensor product decomposes as 
\be V_\mu \ot V_\nu \iso{\ccG} 
 \bigoplus_{\lambda\in\hat{\ccG}} V_\lambda \ot 
\Hom(V_\lambda, V_\mu \ot V_\nu)^{\ccG}
\iso{\ccG} 
 \bigoplus_{\lambda\in\hat{\ccG}} V_\lambda \ot 
\bbC^{M_{\mu\nu}^{\lambda}},
 \label{eq:CG-hom-decomp}\ee
where we have defined the multiplicity $M_{\mu,\nu}^\lambda := \dim
\Hom(V_\lambda, V_\mu \ot V_\nu)^{\ccG}$.  When $\ccG=\cU_d$, the
$M_{\mu\nu}^\lambda$ are known as {\em Littlewood-Richardson}
coefficients.

The decomposition in \eq{CG-hom-decomp} is known as the Clebsch-Gordan
(CG) decomposition and the corresponding unitary map $\Ucg^{\mu,\nu}$
is called the CG transform.  On a quantum
computer, we can think of $\Ucg^{\mu,\nu}$ as a map from states of the form
$\ket{v_\mu}\ket{v_\nu}$ to superpositions of states
$\ket{\lambda}\ket{v_\lambda}\ket{\alpha}$, where $\lambda\in\hat{\ccG}$
labels an irrep, $\ket{v_\lambda}$ is a basis state for $V_\lambda$ and
$\alpha\in\Hom(V_\lambda, V_\mu \ot V_\nu)^{\ccG}$.  
Using the
isomorphism  $\Hom(A,B)\iso{\ccG} A^*\ot B$ we could also write that
$\ket{\alpha}\in(V_\lambda^* \ot V_\mu \ot V_\nu)^{\ccG}$; an
interpretation which makes it more obvious how to normalize $\alpha$.

There are a few issues that arise when implementing the map $\Ucg^{\mu,\nu}$.
For example, since different $V_\lambda$ (and different multiplicity
spaces) have different dimensions, the register for $\ket{v_\lambda}$
will need to be padded to at least $\lceil \log\max_\lambda\dim
V_\lambda\rceil$ qubits.  This means that the overall transformation
will be an isometry that slightly enlarges the Hilbert space, or
equivalently, will be a unitary that requires the input of a small
number of ancilla qubits initialized to $\ket{0}$.  Also, when $\ccG$
has an infinite number of inequivalent irreps (e.g. when $\ccG$ is a
Lie group) then in order to store $\lambda$, we need to consider only
some finite subset of $\hat{\ccG}$.  Fortunately, there is usually a
natural way to perform this restriction.

Returning to \eq{CG-hom-decomp} for a moment, 
note that all the complexity of the CG transform is pushed into
the multiplicity space $\Hom(V_\lambda, V_\mu \ot V_\nu)^{\ccG}$.   For
example, the fact that some values of $\lambda$ don't appear on the
RHS means that some of the  multiplicity spaces may be zero.  Also,
the inverse transform $(\Ucg^{\mu,\nu})^\dag$ is given simply by the map 
\be (\Ucg^{\mu,\nu})^\dag \ket{\lambda}\ket{v_\lambda}\ket{\alpha} =
\alpha\ket{v_\lambda}.\label{eq:Ucg-dag}\ee
  We will use these properties of the CG
transform when decomposing i.i.d. channels in \sect{normal-form} and
in giving an efficient construction of the CG transform in
\sect{cg-construct}.

\subsection{The quantum Fourier transform}\label{sec:QFT}
Let $\ccG$ be a finite group (we will return to Lie groups later).  
A useful representation is given by letting each $g\in\ccG$ define an
orthonormal basis vector $\ket{g}$.  The resulting space
$\Span\{\ket{g} : g\in \ccG\}$ is denoted $\bbC[\ccG]$ and is called
the {\em regular representation}.  $\ccG$ can act on $\bbC[\ccG]$ in two
different ways: left multiplication $\bL(g)\ket{h} :=\ket{gh}$, and
right multiplication $\bR(g)\ket{h} := \ket{hg^{-1}}$.  This means
that there are really two different regular representations: the left
regular representation $(\bL,\bbC[\ccG])$ and the right regular
representation $(\bR,\bbC[\ccG])$.  Since these representations
commute, we could think of $\bL(g_1)\bR(g_2)$ as a representation of
$\ccG\times \ccG$.  Under this action, it can be shown that $\bbC[\ccG]$
decomposes as
\be \bbC[\ccG] \iso{\ccG\times\ccG}
\bigoplus_{\lambda\in\hat{\ccG}} V_\lambda \hat{\ot} V_\lambda^*.
\label{eq:fourier-decomp}\ee
Here the $V_\lambda$ correspond to $\bL$ and $V_\lambda^*$ corresponds
to $\bR$, and $\hat{\ot}$ is used to emphasize that we are not
considering the tensor product action of a single group, but rather
are taking the tensor product of two irreps from two different copies
of the group $G$.
This means that if we decompose only one of the regular
representations, e.g. $(\bL,\bbC[\ccG])$, the $V_\lambda^*$ in
\eq{fourier-decomp} becomes the multiplicity space for $V_\lambda$ as
follows:
\be \bbC[\ccG] \iso{\ccG}
\bigoplus_{\lambda\in\hat{\ccG}} V_\lambda \ot \bbC^{\dim V_\lambda}.
\label{eq:left-reg-decomp}\ee
A similar expression holds for $(\bR,\bbC[\ccG])$ with $V_\lambda^*$
appearing instead of $V_\lambda$.

The unitary matrix corresponding to the isomorphism
in \eq{fourier-decomp} is called the Fourier transform, or when it
acts on quantum registers, the quantum Fourier transform (QFT).
Denote this matrix by $\Uqft$.  For any $g_1,g_2\in\ccG$ we have
\be \hat{\bL}(g_1)\hat{\bR}(g_2):= 
\Uqft \bL(g_1)\bR(g_2) \Uqft^\dag = 
\sum_{\lambda\in\hat{\ccG}} \oprod{\lambda}
\ot \br_\lambda(g_1) \ot \br_\lambda(g_2)^*,
\label{eq:uqft-rep-matrices}\ee
where $\hat{\bL}$ and $\hat{\bR}$ are the Fourier transformed versions
of $\bL$ and $\bR$; $\hat{\bL}(g) := \Uqft\bL(g)\Uqft^\dag$ and
$\hat{\bR}(g) := \Uqft\bR(g)\Uqft^\dag$.

Unlike the CG transform, the Fourier transform has a simple explicit
expression. 
\be \Uqft = \sum_{g\in\ccG}\sum_{\lambda\in\hat{\ccG}}
\sum_{i,j=1}^{\dim V_\lambda} \sqrt{\frac{\dim V_\lambda}{|\ccG|}}
\br_\lambda(g)_{ij} \ket{\lambda,i,j}\bra{g}
\label{eq:uqft-explicit}\ee
The best-known quantum Fourier transform is over the cyclic group
$\ccG=\bbZ_N$.  Here the form is particularly simple, since all irreps
are one-dimensional and the set of irreps $\hat{\ccG}$ is equivalent to
$\bbZ_N$.  Thus the $\ket{i,j}$ register can be neglected and we obtain
the familiar expression $\sum_{x,y\in\bbZ_N}N^{-1/2} e^{2\pi
ixy/N}\ket{y}\bra{x}$.  The ability of a quantum computer to
efficiently implement this Fourier transform is at the heart of
quantum computing's most famous advantages over classical
computation\cite{Shor94}.

Quantum Fourier transforms can also be efficiently implemented for
many other groups.  Beals\cite{Beals97} has shown how to implement
the $\cS_n$ QFT on a quantum computer in $\poly(n)$ time, P\"{u}schel,
R\"{o}tteler and Beth\cite{PRB99} have given efficient QFTs for other nonabelian
groups and Moore,
Rockmore and Russell\cite{MRR03} have generalized these approaches to
many other finite groups.  Fourier transforms on Lie groups are also
possible, though the infinite-dimensional spaces involved lead to
additional complications that we will not discuss here.  Later
(\sect{cg-construct}) we will give an efficient algorithm for a $\cU_d$
CG transform.  However, if some sort of $\cU_d$ QFT could be
efficiently constructed on a quantum computer, then it would yield an
alternate algorithm for the $\cU_d$ CG transform.  We will discuss
this possibility further in \sect{GPE-CG} (see also Prop~9.1 of
\cite{Kup03}) and will discuss the $\cS_n$ QFT more broadly in 
\chap{sch-Sn}.

\section{Schur duality}\label{sec:schur-def}
We now turn to the two representations relevant to the Schur transform.  Recall
that the symmetric group of degree $n$, ${\mathcal S}_n$, is the group of all
permutations of $n$ objects. Then we have the following natural representation
of the symmetric group on the space $(\bbC^d)^{\ot n}$:
\begin{equation}
\bP(s) |i_1\rangle \otimes |i_2\rangle \otimes \cdots
\otimes|i_n\rangle = |i_{s^{-1}(1)}\rangle\otimes
|i_{s^{-1}(2)}\rangle\otimes \cdots
\otimes|i_{s^{-1}(n)}\rangle
\end{equation}
where $s \in {\mathcal S}_n$ is a permutation and $s(i)$ is the label
describing the action of $s$ on label $i$.  For example, consider the
transposition $s=(12)$ belonging to the group $\cS_3$.  Then $\bP(s)
|i_1,i_2,i_3\rangle=|i_2,i_1,i_3\rangle$.  $(\bP,\nqudits)$ is the
representation 
of the symmetric group which will be relevant to the Schur transform.
Note that $\bP$ obviously depends on $n$, but also has an implicit
dependence on $d$.

Now we
turn to the representation of the unitary group.  Let ${\mathcal U}_d$ denote
the group of $d\times d$ unitary operators. Then there is a representation of
$\cU_d$ given by the $n$-fold product action as
\begin{equation}
{\bf Q}(U)|i_1\rangle \otimes|i_2\rangle \otimes\cdots\otimes |i_n\rangle =
U |i_1\rangle\otimes U|i_2\rangle \otimes\cdots \otimes U|i_n\rangle
\label{eq:bQ-def}
\end{equation}
for any $U \in {\mathcal U}_d$. More compactly, we could write that
$\bQ(U)=U^{\ot n}$.  $({\bf Q}, \nqudits)$ is the representation of the
unitary group 
which will be relevant to the Schur transform.

Since both $\bP(s)$ and ${\bf Q}(U)$  meet our above criteria for
reducibility, they can each be decomposed into a direct sum of irreps
as in \eq{directsum}, 
\begin{eqnarray}
\bP(s)&\stackrel{\cS_n}{\cong}
&\bigoplus_\alpha I_{n_\alpha} \otimes
{\bf p}_\alpha(s)
\nonumber \\
{\bf Q}(U)& \stackrel{\cU_d}{\cong}
&\bigoplus_\beta I_{m_\beta} \otimes {\bf
q}_\beta(U) \label{eq:fistdecomp}
\end{eqnarray}
where $n_\alpha$ ($m_\beta$) is the multiplicity of the $\alpha$th
($\beta$th) irrep ${\bf p}_\alpha(s)$ (${\bf q}_\beta(U)$) in
the representation $\bP(s)$ (${\bf Q}(U)$).  At this point there is
not necessarily any relation between the two different unitary
transforms implementing the isomorphisms in \eq{fistdecomp}.  However,
further structure in this
decomposition follows from the fact that $\bP(s)$
commutes with ${\bf Q}(U)$: $\bP(s) {\bf Q}(U)={\bf Q}(U)
\bP(s)$.  This implies, via Schur's Lemma, that the action
of the irreps of $\bP(s)$ must act on the multiplicity
labels of the irreps ${\bf Q}(U)$ and vice versa.  Thus, the
simultaneous action of $\bP$ and $\bQ$ on $(\bbC^d)^{\ot n}$
decomposes as 
\begin{equation}
\bQ(U)\bP(s)  \stackrel{\cU_d\times\cS_n}{\cong}
 \bigoplus_\alpha\bigoplus_\beta
 I_{m_{\alpha,\beta}} \otimes
\bq_\beta(U) \ot {\bf p}_\alpha(s)
\label{eq:pqdecomp}\end{equation}
where $m_{\alpha,\beta}$ can be thought of as the multiplicity of the
irrep $\bq_\beta(U) \hat{\ot} {\bf p}_\alpha(s)$ of the group $\cU_d
\times \cS_n$. 

Not only do $\bP$ and $\bQ$ commute, but the algebras they generate
(i.e. ${\cal A} := \bP(\bbC[\cS_n]) = \Span\{\bP(s) : s\in\cS_n\}$
and ${\cal B} := \bQ(\bbC[\cU_d]) = \Span\{\bQ(U) : U\in\cU_d\}$) {\em
centralize} each other\cite{GW98}, meaning that $\cB$ is the
set of operators in 
$\End((\bbC^d)^{\ot n})$ commuting with $\cA$ and vice versa, $\cA$ is
the set of operators in 
$\End((\bbC^d)^{\ot n})$ commuting with $\cB$.  This means that the
multiplicities $m_{\alpha,\beta}$ are either zero or one, and that
each $\alpha$ and $\beta$ appears at most once.  Thus \eq{pqdecomp}
can be further simplified to
\be \bQ(U)\bP(s) \stackrel{\cS_n\times\cU_d}{\cong}
 \bigoplus_\lambda   {\bf q}_\lambda(U)
\ot \bp_\lambda(s)\label{eq:preschur-decomp}\ee
where $\lambda$ runs over some unspecified set.

Finally, Schur duality (or Schur-Weyl
duality)\cite{GW98} provides a simple 
characterization of the range of $\lambda$ in \eq{preschur-decomp} and
shows how the decompositions are related for different values of $n$
and $d$.  To define Schur duality, we will need to somehow specify the
irreps of $\cS_n$ and $\cU_d$.

Let $\mathcal{I}_{d,n}=\{\lambda=(\lambda_1,
\lambda_2,\dots,\lambda_d) |\lambda_1 \geq \lambda _2 \geq \cdots
\geq \lambda_d \geq 0$ and $\sum_{i=1}^d \lambda_i = n\}$ denote
partitions of $n$ into $\leq d$ parts.  We consider two partitions
$(\lambda_1,\ldots,\lambda_d)$ and
$(\lambda_1,\ldots,\lambda_d,0,\ldots,0)$ 
equivalent if they differ only by trailing zeroes; according to this
principle, $\cI_n := \cI_{n,n}$ contains all the partitions of $n$.
Partitions label irreps of $\cS_n$ and $\cU_d$ as follows: if we let $d$
vary, then $\cI_{d,n}$ labels irreps of $\cS_n$, and if we let $n$
vary, then $\cI_{d,n}$ labels polynomial irreps of $\cU_d$. Call these
$({\bf p}_\lambda,\cP_\lambda)$ and $({\bf q}_\lambda^d,
\cQ_\lambda^d)$ respectively, for $\lambda\in\cI_{d,n}$.
We need the superscript $d$ because the same partition $\lambda$ can
label different irreps for different $\cU_d$; on the other hand the
$\cS_n$-irrep 
$\cP_\lambda$ is uniquely labeled by $\lambda$ since $n=\sum_i \lambda_i$.

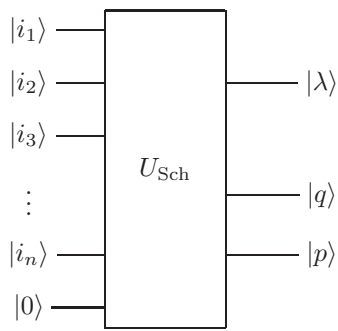
\begin{figure}[ht]
\begin{centering}
\leavevmode\xymatrix@R=4pt@C=4pt{
{\ket{i_1}} & *{~~~} & \gnqubit{{~~~\Usch~~~}}{ddddd}\ar@{-}[ll]\\
{\ket{i_2}} & \nw & \gspace{~~~\Usch~~~}\w && |\lambda\>\ar@{-}[ll] \\
{\ket{i_3}} & \nw &\gspace{~~~\Usch~~~}\w & ~~~~\\
{\vdots} &  & \gspace{~~~\Usch~~~} && |q\>\ar@{-}[ll]\\
{\ket{i_n}} & \nw & \gspace{~~~\Usch~~~}\w && |p\>\ar@{-}[ll]\\
{\ket{0}}& \nw & \gspace{~~~\Usch~~~}\w
}

\caption{The Schur transform.  Notice how the direct sum over
  $\lambda$ in \eq{schur-decomp} becomes a tensor product between
the $\ket{\lambda}$ register and the $\ket{q}$ and $\ket{p}$
registers.  Since the number of qubits needed for $\ket{q}$ and
$\ket{p}$ vary with $\lambda$, we need slightly more spatial
resources, which are here denoted by the ancilla input $|0\rangle$.}
\label{fig:schur}
\end{centering}
\end{figure}

For the case of $n$ qudits, Schur
duality states that there exists a basis (which we label
$\ket{\lambda}\ket{q_\lambda}\ket{p_\lambda}_{\rm Sch}$ and call the
{\em Schur basis}) which simultaneously decomposes the action of
$\bP(s)$ and $\bQ(U)$ into irreps:
\begin{eqnarray}
{\bf Q}(U)\ket{\lambda}\ket{q_\lambda}\ket{p_\lambda}_{\rm Sch}&=&
\ket{\lambda}({\bf q}_\lambda^d(U) \ket{q_\lambda})
\ket{p_\lambda}_{\rm Sch}
\nonumber \\
\bP(s) \ket{\lambda}\ket{q_\lambda}\ket{p_\lambda}_{\rm Sch}
&=& \ket{\lambda} \ket{q_\lambda} ({\bf
p}_\lambda(s)\ket{p_\lambda})_{\rm Sch} \label{eq:schur}
\end{eqnarray}
and that the common representation space $(\bbC^d)^{\ot n}$ decomposes as
\be (\bbC^d)^{\ot n} \iso{\cU_d \times \cS_n}
\bigoplus_{\lambda\in\cI_{d,n}} \cQ_\lambda^d \,\hat{\ot}\,\cP_\lambda.
\label{eq:schur-decomp}\ee
The Schur basis can be expressed as superpositions over the
standard computational basis states $|i_1,i_2,\dots,i_n\rangle$ as
\begin{equation}
|\lambda,q_\lambda,p_\lambda\rangle_{\rm
Sch}=\sum_{i_1,i_2,\dots,i_n} \left[ {\bf U}_{\rm Sch}
\right]^{\lambda,q_\lambda,p_\lambda}_{i_1,i_2,\dots,i_n}
|i_1i_2\dots i_n \rangle,
\label{eq:Usch-def}\end{equation}
where ${\bf U}_{\rm Sch}$ is the unitary transformation implementing
the isomorphism in \eq{schur-decomp}.  Thus, for any $U\in\cU_d$ and
any $s\in\cS_n$, 
\be \Usch \bQ(U)\bP(s) \Usch^\dag =
\sum_{\lambda\in\cI_{d,n}}
\oprod{\lambda} \ot \bq_\lambda^d(U) \ot \bp_\lambda(s).
\label{eq:schur-decomp2}\ee
If we now think of $\Usch$ as a quantum circuit, it will
map the Schur basis state $\ket{\lambda,q_\lambda,p_\lambda}_{\rm
Sch}$ to the computational basis state
$\ket{\lambda,q_\lambda,p_\lambda}$ with $\lambda$, $q_\lambda$, and
$p_\lambda$ expressed as bit strings.  The dimensions of the irreps
${\bf p}_\lambda$ and ${\bf q}^d_\lambda$ vary with $\lambda$, so we
will need to pad the $|q_\lambda,p_\lambda\rangle$ registers when they
are expressed as bitstrings.  We will label the padded basis as
$\ket{\lambda}\ket{q}\ket{p}$, explicitly dropping the $\lambda$
dependence.  In \chap{sch-algo} we will show how to do this padding
efficiently with only a logarithmic spatial overhead.  We will refer to
the transform from the computational basis $|i_1,i_2,\dots,i_n\rangle$
to the basis of three bitstrings $|\lambda\rangle|q\rangle|p\rangle$
as the Schur transform. The Schur transform is shown schematically in
Fig.~\ref{fig:schur}.  Notice that just as the standard computational
basis $|i\rangle$ is arbitrary up to a unitary transform, the bases
for $\cQ_\lambda^d$ and $\cP_\lambda$ are also both arbitrary up to a
unitary transform, though we will later choose particular bases for
$\cQ_\lambda^d$ and $\cP_\lambda$.

{\em Example of the Schur transform:}Let $d=2$.  Then for $n=2$
there are two valid partitions, $\lambda_1=2,\lambda_2=0$ and
$\lambda_1=\lambda_2=1$.  Here the Schur transform corresponds to
the change of basis from the standard basis to the singlet and
triplet basis: $|\lambda=(1,1),q_\lambda=0,p_\lambda=0\rangle_{\rm
Sch}=\frac{1}{\sqrt{2}}(|01\rangle-|10\rangle)$,
$|\lambda=(2,0),q_\lambda=+1,p_\lambda=0\rangle_{\rm
Sch}=|00\rangle$,
$|\lambda=(2,0),q_\lambda=0,p_\lambda=0\rangle_{\rm Sch}=\frac{1}
{\sqrt{2}} (|01\rangle+|10\rangle)$, and
$|\lambda=(2,0),q_\lambda=-1,p_\lambda=0\rangle_{\rm
Sch}=|11\rangle$. Abstractly, then, the Schur transform then
corresponds to a transformation
\begin{equation}
{\bf U}_{\rm Sch}=\left.\begin{array}{c} |\lambda=(1,1),q_\lambda=0,p_\lambda=0\rangle_{\rm Sch} \\
|\lambda=(2,0),q_\lambda=+1,p_\lambda=0\rangle_{\rm Sch}
\\ |\lambda=(2,0),q_\lambda=0,p_\lambda=0\rangle_{\rm Sch}\\ |\lambda=(2,0),q_\lambda=-1,p_\lambda=0\rangle_{\rm
Sch}\end{array}\right \{ \overbrace{\left[\begin{array}{cccc} 0 & 
\frac{1}{\sqrt{2}} & - \frac{1}{\sqrt{2}} & 0 \\ 1 & 0 & 0 & 0 \\ 0 &
\frac{1}{\sqrt{2}} &  \frac{1}{\sqrt{2}} & 0 \\ 0 & 0 & 0 & 1
\end{array}\right]}^{|00\rangle~|01\rangle~|10\rangle~|11\rangle}
\end{equation}
It is easy to verify that the $\lambda=(1,1)$ subspace transforms as a
one dimensional irrep of ${\mathcal U}_2$ and as the alternating sign
irrep of ${\mathcal S}_2$ while the $\lambda=(2,0)$ subspace
transforms as a three dimensional irrep of ${\mathcal U}_2$ and as the
trivial irrep of ${\mathcal S}_2$.  Notice that the labeling scheme
for the standard computational basis uses $2$ qubits while the
labeling scheme for the Schur basis uses more qubits (one such
labeling assigns one qubit to $|\lambda\rangle$, none to $|p\rangle$
and two qubits to $|q\rangle$).  Thus we see how padding will be
necessary to directly implement the Schur transform.

To see a more complicated example of the Schur basis, let $d=2$
and $n=3$.  There are again two valid partitions, $\lambda=(3,0)$
and $\lambda=(2,1)$.  The first of these partitions labels to
the trivial irrep of ${\mathcal S}_3$ and a $4$ dimensional irrep
of ${\mathcal U}_3$.  The corresponding Schur basis vectors can be
expressed as 
\be\begin{split}
|\lambda=(3,0),q_\lambda=+3/2,p_\lambda=0\rangle_{\rm
Sch} & = |000\rangle \\
|\lambda=(3,0),q_\lambda=+1/2,p_\lambda=0\rangle_{\rm Sch}
&= 
\frac{1}{\sqrt{3}}\left(|001\rangle+|010\rangle+|100\rangle\right)\\
|\lambda=(3,0),q_\lambda=-1/2,p_\lambda=0\rangle_{\rm Sch}
&=
\frac{1}{\sqrt{3}}\left(|011\rangle+|101\rangle+|110\rangle\right)\\
|\lambda=(3,0),q_\lambda=-3/2,p_\lambda=0\rangle_{\rm
Sch} &= |111\rangle.\end{split}\label{eq:3qubit-sym}\ee

The second of these partitions labels
a two dimensional irrep of ${\mathcal S}_3$ and a two dimensional
irrep of ${\mathcal U}_2$.  Its Schur basis states can be expressed
as
\be\begin{split}
|\lambda=(2,1),q_\lambda=+1/2,p_\lambda=0\rangle_{\rm Sch}&=
\frac{1}{\sqrt{2}} \left(|100\rangle - |010\rangle\r)\\
|\lambda=(2,1),q_\lambda=-1/2,p_\lambda=0\rangle_{\rm Sch}&=
\frac{1}{\sqrt{2}} \left(|101\rangle - |011\rangle\r)\\
|\lambda=(2,1),q_\lambda=+1/2,p_\lambda=1\rangle_{\rm Sch}&=
\sqrt{\frac{2}{3}}|001\> - \frac{|010\> + |100\>}{\sqrt{6}}\\
|\lambda=(2,1),q_\lambda=-1/2,p_\lambda=1\rangle_{\rm Sch}&=
\sqrt{\frac{2}{3}}|110\> - \frac{|101\> + |011\>}{\sqrt{6}}.
\end{split}\label{eq:3qubit-mixed}\ee
We can easily verify that \eqs{3qubit-sym}{3qubit-mixed} indeed
transform under $\cU_2$ and $\cS_3$ the way we expect; not so easy
however is coming up with a circuit that relates this basis to the
computational basis and generalizes naturally to other values of $n$
and $d$.  However, note that $p_\lambda$ determines whether the first
two qubits are in a singlet or a triplet state.  This gives a hint of
a recursive structure that we will exploit in \chap{sch-algo} to
construct an efficient general algorithm for the Schur transform.

\subsection{Constructing $\cQ_\lambda^d$ and $\cP_\lambda$ using
Schur duality}\label{sec:weyl-symmetrizer}
So far we have said little about the form of $\cQ_\lambda^d$
and $\cP_\lambda$, other than that they are indexed by partitions.  It
turns out that Schur duality gives a straightforward description of
the irreps of $\cU_d$ and $\cS_n$.  We will not use this explicit
description to construct the Schur transform, but it is
still helpful for understanding the irreps $\cQ_\lambda^d$
and $\cP_\lambda$.  As with the rest of this chapter, proofs and
further details can be found in \cite{GW98}.

We begin by expressing $\lambda\in\cI_{d,n}$ as a Young diagram in
which there are up to $d$ rows with $\lambda_i$ boxes in row $i$.  For
example, to the partition $(4,3,1,1)$ we associate the diagram
\be \yng(4,3,1,1). \label{eq:ferrers-example}\ee
Now we define a Young tableau $T$ of shape $\lambda$ to be a way of
filling the $n$ boxes of $\lambda$ with the integers $1,\ldots,n$,
using each number once and so that integers increase from left to
right and from top to bottom.  For example, one valid Young tableau
with shape $(4,3,1,1)$ is
$$\young(1467,258,3,9).$$
For any Young tableau $T$, define
$\operatorname{Row}(T)$ to be set of permutations obtained by
permuting the integers within each row of $T$; similarly define
$\operatorname{Col}(T)$ to be the permutations that leave each integer
in the same column of $T$.  Now we define the {\em Young symmetrizer}
$\Pi_{\lambda:T}$ to be an operator acting on $(\bbC^d)^{\ot n}$ as follows:
\be \Pi_{\lambda:T} := \frac{\dim\cP_\lambda}{n!}
\l(\sum_{c\in\operatorname{Col}(T)} \sgn(c)\bP(c)\r)
\l(\sum_{r\in\operatorname{Row}(T)} \bP(r)\r).\ee

It can be shown that the Young symmetrizer $\Pi_{\lambda:T}$ is a projection
operator whose support is a
subspace isomorphic to $\cQ_\lambda^d$.  In particular $\Usch\Pi_{\lambda:T}
\Usch^\dag = \oprod{\lambda} \ot \oprod{y(T)} \ot I_{\cQ_\lambda^d}$
for some unit vector $\ket{y(T)}\in\cP_\lambda$.  Moreover, these vectors
$\ket{y(T)}$ form a basis known as
Young's natural basis, though the $\ket{y(T)}$ are not orthogonal, so
we will usually not work with them in quantum circuits.

Using Young symmetrizers, we can now explore some more general
examples of $\cQ_\lambda^d$ and $\cP_\lambda$.  If $\lambda=(n)$, then
the only valid tableau is
$$ \young(12)\cdots\young(n).$$
The corresponding $\cS_n$-irrep $\cP_{(n)}$ is trivial and the
$\cU_d$-irrep is given by the action of $\bQ$ on the totally symmetric
subspace of $(\bbC^d)^{\ot n}$, i.e. $\{\ket{v} : \bP(s)\ket{v}=\ket{v}
\forall s\in\cS_n\}$.  On the other hand, if $\lambda=(1^n)$, meaning
$(1,1,\ldots, 1)$ ($n$ times), then the only valid tableau is
$$ \begin{array}{c}\young(1,2)\\\vdots\\\young(n)\end{array}.$$
The $\cS_n$-irrep $\cP_{(1^n)}$ is still one-dimensional, but now
corresponds to the sign irrep of $\cS_n$, mapping $s$ to $\sgn(s)$.
The $\cU_d$-irrep $\cQ_{(1^n)}^d$ is equivalent to the totally
antisymmetric subspace of $(\bbC^d)^{\ot n}$, i.e. $\{\ket{v} :
\bP(s)\ket{v}=\sgn(s)\ket{v} \forall s\in\cS_n\}$.  Note that if
$d>n$, then this subspace is zero-dimensional, corresponding to the
restriction that irreps of $\cU_d$ are indexed only by partitions with
$\leq d$ rows.

Other explicit examples of $\cU_d$ and $\cS_n$ irreps are presented from a
particle physics perspective in \cite{Georgi99}.  We also give more
examples in \sect{gz-yy}, when we introduce explicit bases for
$\cQ_\lambda^d$ and $\cP_\lambda$.

\section{Dual reductive pairs}\label{sec:dual-reductive}

Schur duality can be generalized to groups other than $\cU_d$ and
$\cS_n$.  The groups for which this is possible are known as dual
reductive pairs, and in this section we give an overview of their
definition and properties (following Sec 9.2 of \cite{GW98}).  The
next two chapters will focus primarily on Schur duality, but here we
give some ideas about how the techniques used in those chapters could
be applied 
to other groups and other representations.

Suppose $G$ and $K$ are groups with irreps
$(\rho_\mu,U_\mu)_{\mu\in\hat{G}}$ and
$(\sigma_\nu,V_\nu)_{\nu\in\hat{K}}$ respectively. 
Then the irreps of $G\times K$ are given by $(\rho_\mu \ot
\sigma_\nu, U_\mu \hat{\ot} V_\nu)$.  Now suppose $(\gamma,Y)$ is a
representation of $G\times K$.  Its isotypic decomposition
(cf.~\eq{rep-space-decomp}) is of the form
\be Y \iso{G\times K} \bigoplus_{\mu\in\hat{G}}
\bigoplus_{\nu\in\hat{K}} U_\mu \hat{\ot} V_\nu \ot \bbC^{m_{\mu,\nu}},
\label{eq:GK-isotypic}\ee
where the $m_{\mu,\nu}$ are multiplicity factors.
Define the algebras $\cA = \gamma(\bbC[G\times\{e\}])$ and 
$\cB = \gamma(\bbC[\{e\}\times K])$.  Then \cite{GW98} proves the
following generalization of Schur duality:
\begin{proposition}\label{prop:dual-reductive}
The following are equivalent:
\bi\item[(1)] Each $m_{\mu,\nu}$ is either 0 or 1, and at most one 
$m_{\mu,\nu}$ is nonzero for each $\mu$ and each
$\nu$.  In other words,
\eq{GK-isotypic} has the form 
\be W \iso{G\times K} \bigoplus_{\lambda\in S}
U_{\varphi_G(\lambda)} \hat{\ot}V_{\varphi_K(\lambda)}
\label{eq:dual-isotypic}\ee 
where $S$ is some set and $\varphi_G:S\ra\hat{G},
\varphi_K:S\ra\hat{K}$ are injective maps.
\item[(2)] $\cB$ is the commutant of $\cA$ in $\End(V)$
(i.e. $\cB=\{x\in End(V) : [x,a]=0\, \forall a\in\cA\}$) and $\cA$ is
the commutant of $\cB$.  When this holds we say that $\cA$ and $\cB$
are {\em double commutants}.
\ei
\end{proposition}
When these conditions hold we say that the groups $\gamma(G \times
\{e\})$ and $\gamma(\{e\} \times K)$ form a {\em dual reductive
pair}.  In this case, \eq{dual-isotypic} gives us a one-to-one
correspondence between the subsets of $\hat{G}$ and $\hat{K}$ that
appear in $W$.  This redundancy can often be useful.   For example,
measuring the $G$-irrep automatically also measures the $K$-irrep.
In fact, the key idea behind the algorithms we will encounter in
\chaps{sch-algo}{sch-Sn} is that the Schur transform can be approached by
working only with $\cU_d$-irreps or only with $\cS_n$ irreps.

Many of the examples of dual reductive pairs that are known relate to
the orthogonal and symplectic Lie groups\cite{Howe89}, and so are not
immediately applicable to quantum information.  However, in this
section we will point out one example of a dual reductive pair that
could arise naturally when working with quantum states.

Let $G=\cU_{d_A}$ and $K=\cU_{d_B}$ and define $W$ to be the
$n^{\text{th}}$ symmetric product of $\bbC^{d_A}\ot\bbC^{d_B}$; i.e.
\be W := \l((\bbC^{d_A}\ot\bbC^{d_B})^{\ot n}\r)^{\cS_n}
= \l\{\ket{v}\in\l(\bbC^{d_A}\ot\bbC^{d_B}\r)^{\ot n}
: \bP(s)\ket{v} = \ket{v} \forall s\in\cS_n\r\}.\ee
We have seen in \sect{weyl-symmetrizer} that $W\iso{\cU_{d_Ad_B}}
\cQ_{(1^n)}^{d_Ad_B}$.  However, here we are interested in the action
of $\cU_{d_A}\times \cU_{d_B}$ on $W$, which we define in the natural
way; i.e. $(U_A,U_B)$ is mapped to $(U_A\ot U_B)^{\ot n}$.  It is
straightforward to show that $\cU_{d_A}$ and $\cU_{d_B}$ generate
algebras that are double commutants.  This means that $W$ decomposes
under $\cU_{d_A}\times \cU_{d_B}$ as
\be W \iso{\cU_{d_A}\times \cU_{d_B}}
\bigoplus_{\lambda\in\cI_{d,n}} \cQ_\lambda^{d_A} \hat{\ot}
\cQ_\lambda^{d_B},\label{eq:dual-unitary-decomp}\ee
where $d=\min(d_A,d_B)$.
This yields several nontrivial conclusions.  For example, if the
system were shared between two parties, then this would mean 
that the states of both parties would have the same Young frame.
Also, it turns out that applying the Schur transform circuit in
\sect{schur-circuit} to either $A$ or $B$ gives an efficient method
for performing the isomorphism in \eq{dual-unitary-decomp}.

The implications of other dual reductive pairs for quantum information
are largely unknown.  However, in principle they offer far-ranging
generalizations of Schur duality that remain amenable to manipulation
by the same sorts of algorithms.

\chapter{Applications of the Schur transform to quantum information
theory}\label{chap:sch-qit}

In physics, the Schur basis is a natural way to study systems with
permutation symmetry.  In quantum information theory, the Schur basis
is well suited to i.i.d. states and channels, such as $\rho^{\ot n}$
and $\cN^{\ot n}$.  For example, if $\rho$ is a $d\times d$ density
matrix, then $\rho^{\ot n}$ decomposes under the
Schur transform as
\be \Usch \rho^{\ot n} \Usch^\dag
= \sum_{\lambda\in\cI_{d,n}}
\oprod{\lambda} \ot \bq_\lambda^d(\rho) \ot I_{\cP_\lambda}.
\label{eq:rho-decomp}\ee
To prove this, and to interpret the $\bq_\lambda^d(\rho)$ term, we
note that irreps of $\cU_d$ can also be interpreted as irreps of
$\GL_d$ (the group of $d\times d$ complex invertible
matrices)\footnote{This is because $\GL_d$ is the complexification of
$\cU_d$, meaning that its Lie algebra (the set of all $d\times d$
complex matrices) is equal to the tensor product of $\bbC$ with the
Lie algebra of $\cU_d$ (the set of $d\times d$ Hermitian matrices).
See \cite{GW98,Fulton91} for more details.   For this reason,
mathematicians usually discuss the representation theory of $\GL_d$
instead of $\cU_d$.}.  If $\rho$ is not an invertible matrix then we
can still express $\rho$ as a limit of elements of $\GL_d$ and can use
the continuity of 
$\bq_\lambda^d$ to define $\bq_\lambda^d(\rho)$.  Then \eq{rho-decomp}
follows from \eq{schur-decomp2}.

The rest of the chapter will explore the implications of
\eq{rho-decomp} and related equations.  We will see that the first
register, $\ket{\lambda}$, corresponds to the spectrum of $\rho$, and indeed
that a good estimate for the spectrum of $\rho$ is given by measuring
$\ket{\lambda}$ and guessing $(\lambda_1/n,\ldots,\lambda_d/n)$ for the
spectrum.  The $\cQ_\lambda^d$ register depends on the spectrum
($\lambda$) for its structure, but itself contains information only
about the eigenbasis of $\rho$.  Both of these registers are
vanishingly small---on the order of $d^2 \log n$ qubits---but contain
all the features of $\rho$.  The $\cP_\lambda$ register contains
almost all the entropy, but always carries a uniform distribution that
is independent of $\rho$ once we condition on $\lambda$.

This situation can be thought of as generalization of the classical
method of types, a technique in information theory in which strings
drawn from i.i.d. distributions are classified by their empirical
distributions.  We give a brief review of this method in
\sect{classical-types} so that the reader will be able to appreciate
the similarities with the quantum case.  In \sect{quantum-types}, we
show how \eq{rho-decomp} leads to a quantum method of types, and
give quantitative bounds to make the theory useful.  We survey
known applications of Schur duality to quantum information theory in
\sect{schur-qit-apps}, using our formulae from \sect{quantum-types} to
give concise proofs of the some of the main results from the
literature.  Finally, we show how Schur duality may be used to
decompose i.i.d. quantum channels in
\sect{normal-form}.

Only the last section represents completely new work.
The idea of Schur duality as a quantum method of
types has been known for years, beginning with applications to quantum
hypothesis testing\cite{Hayashi:01a} and spectrum
estimation\cite{Keyl:01}, further developed in a series of papers by
Hayashi and Matsumoto\cite{Hayashi:01b, Hayashi:02a, Hayashi:02b,
Hayashi:02c, Hayashi:02d,Hayashi:02e}, extended to other applications in
\cite{Bacon:01a, Kempe:01a, Bartlett03, Bartlett04, Korff04,
Hayashi:04a}, and recently applied to information theory in
\cite{CM04}.  The contribution of the first three sections is to
present these results together as applications of the same general
method.

\section{The classical method of types}\label{sec:classical-types}

The method of types is a powerful tool in classical information
theory.  Here we briefly review the method of types (following
\cite{CT91,CK81}) to give an idea of how the Schur basis will later be
used for the quantum generalization.

Consider a string $x^n=(x_1,\ldots,x_n)\in [d]^n$, where
$[d]:=\{1,\ldots, d\}$.  Define the type of $x^n$ to be the $d$-tuple
of integers $t(x^n) := \sum_{j=1}^n e_{x_j}$, where $e_i\in\bbZ^d$ is
the unit vector with a one in i$^{\text{th}}$ position.  Thus $t(x^n)$
counts the frequency of each symbol $1,\ldots,d$ in $x^n$.  Let
$\cT_d^n:=\{(n_1,\ldots,n_d):n_1+\ldots+n_d=n, n_i\geq 0\}$ denote the
set of all possible types of strings in $[d]^n$ (also known as the
weak $d$-compositions of $n$).  Since an element of $\cT_d^n$ can be
written as $d$ numbers ranging from $0,\ldots,n$ we obtain the simple
bound $|\cT_d^n|\leq (n+1)^d$.  In fact, $|\cT_d^n| =
\binom{n+d-1}{d-1}$, but knowing the exact number is rarely necessary.
For a type $t$, let the normalized probability distribution
$\bar{t}:=t/n$ denote its empirical distribution.

The set of types $\cT_d^n$ is larger than the set of partitions
$\cI_{d,n}$ because symbol frequencies in types do not have to occur in
decreasing order.  In principle, we could separate a type
$t\in\cT_d^n$ into a 
partition $\lambda\in\cI_{d,n}$ (with nonincreasing parts) and a
mapping of the parts of $\lambda$ onto $[d]$, which we call
$q_\lambda$.  The map $q_\lambda$ corresponds to some
$(a_1,\ldots,a_d)\in \cS_d$ for which there are $\lambda_i$ symbols
equal to $a_i$ for each $i\in\{1,\ldots,d\}$.  However, if not all the
$\lambda_i$ are distinct, then this is more information than we need.
In particular, if $\lambda_{i}=\lambda_{i+1}=\ldots = \lambda_{j}$,
then we don't care about the ordering of $a_i,\ldots,a_j$.  Define
$m_i(\lambda)$ to be the number of parts of $\lambda$ equal to $i$,
i.e. $|\{j:\lambda_j=i\}|$.  Then the number of distinct $q_\lambda$
is $d! / m_1!\ldots m_n! =: \binom{d}{m}$.  This separation is not
usually used for classical information theory, but helps show how the
quantum analogue of type is split among the $\ket{\lambda}$ and
$\ket{q}$ registers.

For a particular type $t\in\cT_d^n$, denote the set of all strings in
$[d]^n$ with type $t$ by $T_t=\{x^n\in[d^n] : t(x^n)=t\}$.  There are
two useful facts about $T_t$.  First, $|T_t| = \binom{n}{t} := n! /
t_1!\ldots t_d!$ (or equivalently $|T_t|=\binom{n}{\lambda}$, where 
$\lambda$ is a sorted version of $t$).  Second, let $P$ be a
probability distribution on $[d]$ and $P^{\ot n}$ the probability
distribution on $[d]^n$ given by $n$ i.i.d. copies of $P$,
i.e. $P^{\ot n}(x^n) := P(x_1)\cdots P(x_n)$.  Then for any $x^n\in T_t$ we
have $P^{\ot n}(x^n) = P(1)^{t_1} \cdots P(d)^{t_d} = \exp(\sum_{j=1}^d
t_j\log P(j))$. This has a natural expression in terms of the entropic
quantities $H(\bar{t}):=-\sum_j \bar{t}_j\log \bar{t}_j$ and
$D(\bar{t}\| P) := \sum_j \bar{t}_j \log \bar{t}_j/P(j)$ as
\be P^{\ot n}(x^n) = \exp\l(-n\l(H(\bar{t}) + D(\bar{t}\|P)\r)\r).
\label{eq:Px-type}\ee

These basic facts can be combined with simple probabilistic arguments
to prove many results in classical information theory.  For example,
if we define $P^{\ot n}(T_t) := \sum_{x^n\in T_t} P^{\ot n}(x^n)$, then 
\be \bar{t}^{\ot n}(T_t) = \binom{n}{t}\exp(-nH(\bar{t})).\ee
Since $\bar{t}^{\ot n}(T_t)\leq 1$, we get the bound $\binom{n}{t} \leq
\exp(nH(\bar{t}))$.  On the other hand, by doing a bit of
algebra\cite{CT91} one can show that $\bar{t}^{\ot n}(T_t)\geq
\bar{t}^{\ot n}(T_{t'}$) for any $t'\in\cT_d^n$; i.e. under the probability
distribution $\bar{t}^{\ot n}$, the most likely type is $t$.  This
allows us to lower bound $\binom{n}{t}$ by $\exp(nH(\bar{t})) /
|\cT_{d,n}|$.  Together these bounds are
\be (n+1)^{-d} \exp(nH(\bar{t})) \leq |T_t| = \binom{n}{t} \leq
\exp(nH(\bar{t})). \label{eq:binom-bounds}\ee
Combining \eqs{binom-bounds}{Px-type} for an arbitrary distribution
$P$ then gives
\be (n+1)^{-d}\exp\l(-n D(\bar{t}\|P)\r) \leq 
 P^{\ot n}(T_t) \leq \exp\l(-n D(\bar{t}\|P)\r),\ee

Thus, as $n$ grows large, we are likely to observe an empirical
distribution $\bar{t}$ that is close to the actual distribution $P$.
To formalize this, define the set of {\em typical sequences}
$T_{P,\delta}^n$ by
\be T_{P,\delta}^n :=
\bigcup_{\substack{t\in\cT_{d,n}\\\|\bar{t}-P\|_1\leq\delta}}
 T_t.\ee
To bound $P^{\ot n}(T_{P,\delta}^n)$, we apply Pinsker's
inequality\cite{Pinsker64}:
\be D(Q\|P) \geq \half\|P-Q\|_1^2.\ee
Denote the complement of $T_{P,\delta}^n$ by $[d]^n - {T_{P,\delta}^n}$.
Then
\be P^{\ot n}([d]^n - {T_{P,\delta}^n}) 
= \sum_{\substack{t\in\cT_{d,n}\\\|\overline{t}-P\|_1 > \delta}}
P^{\ot n}(T_t)
\leq \sum_{\substack{t\in\cT_{d,n}\\\|\overline{t}-P\|_1 > \delta}}
 \exp\l(-nD(\bar{t}\| P)\r)
\leq (n+1)^d \exp\l(-\frac{n\delta^2}{2}\r)
\label{eq:P-atypical}\ee
and therefore
\be P^{\ot n}(T_{P,\delta}^n) 
\geq 1- (n+1)^d \exp\l(-\frac{n\delta^2}{2}\r).
\label{eq:P-typical}\ee

This has several useful consequences:
\begin{itemize}
\item {\em Estimating the probability distribution $P$:}  If the true
probability distribution of an i.i.d. process is $P$ and we
observe empirical distribution $\bar{t}$ on $n$ samples, the
probability that $\|t-P\|_1 > \delta$ is $\leq (n+1)^d
\exp\l(-\frac{n\delta^2}{2}\r)$, which decreases exponentially with
$n$ for any constant value of $\delta$.
\item {\em Data compression (cf.~\eq{shannon-compression}):} We can
compress $n$ letters from an 
i.i.d. source with distribution $P$ by transmitting only strings in
$T_{P,\delta}^n$.  Asymptotically, the probability of error is $\leq (n+1)^d
\exp\l(-\frac{n\delta^2}{2}\r)$, which goes to zero as $n\ra\infty$.
The number of bits required is $\lceil \log |T_{P,\delta}^n|\rceil$.
To estimate this quantity, use Fannes' inequality (\lem{fannes}) to bound
\be |H(\bar{t}) - H(P)| \leq \eta(\delta) + \delta \log d
\label{eq:classical-fannes}\ee
whenever $\|\bar{t}-P\|_1\leq\delta$.  Thus
\be \log |T_{P,\delta}^n| \leq
\log|\cT_{d,n}| + n\l[H(P) + \eta(\delta) + \delta\log d\r]
\leq n\l[H(P) + \eta(\delta) + \delta\log d +
\frac{d}{n}\log(n+1)\r],\ee
which asymptotically approaches $H(P)$ bits per symbol.
\item {\em Randomness concentration (cf.~\eq{crc}):} Suppose we are
given a random 
variable $x^n$ distributed according to $P^{\ot n}$ and wish to produce from
it some uniformly distributed random bits.  Then since all $x^n$ with
the same type have the same probability, conditioning on the type
$t=t(x^n)$ is sufficient to give a uniformly distributed random
variable.  According to \eqs{classical-fannes}{P-typical}, this yields
$\geq n(H(P)-\eta(\delta)-\delta\log d) = n(H(P)-o(1))$ bits with
probability that asymptotically approaches one.
\end{itemize}

If we have two random variables $X$ and $Y$ with a
joint probability distribution $P(X,Y)$, then we can define joint
types and jointly typical sequences.  These can be used to prove
more sophisticated results, such as Shannon's noisy coding
theorem\cite{Shannon48} and the Classical Reverse Shannon
Theorem\cite{BSST01, Winter:02a}.  Reviewing classical joint types
would take us too far afield, but \sect{normal-form} will develop a
quantum analogue of joint types which can be applied to channels or
noisy bipartite states.

Let us now summarize in a manner that shows the parallels with the
quantum case.  A string
$x^n\in[d]^n$ can be expressed as a triple
$(\lambda,q_\lambda,p_\lambda)$ where $\lambda\in\cI_{d,n}$,
$q_\lambda\in Q_\lambda$ and $p_\lambda\in P_\lambda$ for sets
$Q_\lambda$ and $P_\lambda$ satisfying $|Q_\lambda| \leq \poly(n)$ and
$\exp(nH(\bar{\lambda})) / \poly(n) \leq |P_\lambda| \leq
\exp(nH(\bar{\lambda}))$, if we think of $d$ as a constant.
Furthermore, permuting $x^n$ with an 
element of $\cS_n$ affects only the $p_\lambda$ register and for $f\in
\cS_d$, the map $x^n\ra (f(x_1),\ldots, f(x_n))$ affects only the
$q_\lambda$ register.  This corresponds closely with the quantum
situation in \eq{rho-decomp}.  We now show how dimension counting in
the quantum case resembles the combinatorics of the classical method
of types.

\section{Schur duality as a quantum method of types}
\label{sec:quantum-types}
In this section, we generalize the classical method of types to
quantum states.  Our goal is to give asymptotically tight bounds on
$\cQ_\lambda^d$ and $\cP_\lambda$ and the other quantities appearing in
\eq{rho-decomp}(following \cite{GW98,Hayashi:02e,CM04}).

First recall
that $|\cI_{d,n}| \leq |\cT_{d,n}| = (n+1)^d = \poly(n)$.  
For $\lambda\in\cI_{d,n}$, define $\tilde{\lambda} := \lambda +
(d-1,d-2,\ldots,1,0)$.  
Then the dimensions of $\cQ_\lambda^d$ and $\cP_\lambda$ are given
by\cite{GW98}
\bea \dim \cQ_\lambda^d &=&\frac{\prod_{1\leq i<j\leq d}
(\tilde{\lambda}_i - \tilde{\lambda}_j)}{\prod_{m=1}^d m!} 
\label{eq:cQ-dim}\\
\dim \cP_\lambda&=& \frac{n!}{\tilde{\lambda}_1! \tilde{\lambda}_2!
\cdots \tilde{\lambda}_d!}
\prod_{1\leq i<j\leq d}
(\tilde{\lambda}_i - \tilde{\lambda}_j)
\label{eq:cP-dim}\eea
It is straightforward to bound these by\cite{Hayashi:02e,CM04}
\begin{align}  & \dim \cQ_\lambda^d \leq (n+d)^{d(d-1)/2} 
\label{eq:cQ-bound}\\
\binom{n}{\lambda} (n+d)^{-d(d-1)/2}  \leq &
\dim \cP_\lambda \leq  \binom{n}{\lambda}.
\label{eq:cP-bound1}\end{align}
Applying \eq{binom-bounds} to \eq{cP-bound1} yields the more useful
\bea \exp\l(nH(\bl)\r)(n+d)^{-d(d+1)/2} \leq \dim\cP_\lambda
\leq \exp\l(nH(\bl)\r).\label{eq:cP-bound2}\eea

We can use \eq{rho-decomp} to derive a quantum analogue of
\eq{Px-type}.  To do so, we will need to better describe the structure
of $\cQ_\lambda^d$.  Define the torus $\cU_1^{\times d} = \cU_1 \times
\ldots \cU_1 \subset \cU_d$ as the
subgroup of diagonal matrices (in some fixed basis of $\bbC^d$).  For
$x\in\bbC^d$ let $\diag(x)$ denote the diagonal matrix with entries
$x_1,\ldots,x_d$.  The (one-dimensional) irreps of $\cU_1^{\times d}$
are labeled by $\mu\in\bbZ^d$ and are 
given by $x^\mu := x_1^{\mu_1}\cdots x_d^{\mu_d}$.  We will be
interested only in $\mu$ with nonnegative entries, and we write
$\bbZ_+^d$ to denote this set (note that this is different from $\dom$
because the components of $\mu$ can be in any order).

If $(\bq,\cQ)$ is a polynomial representation of $\cU_d$, then upon
restriction to $\cU_1^{\times d}$ one can show that it breaks up into
orthogonal 
subspaces labeled by different $\mu\in\bbZ_+^d$. The subspace
corresponding to the $\cU_1^{\times d}$-representation $\mu$ is called
the $\mu$-weight space of $\cQ$ and is denoted $\cQ(\mu)$.  Formally,
we can define $\cQ(\mu)\subset\cQ$ by $\cQ(\mu):=\{\ket{q}\in\cQ :
\bq(\diag(x_1,\ldots,x_d))\ket{q} = 
x_1^{\mu_1} \cdots x_d^{\mu_d}\ket{q}\;\forall x_1,\ldots,x_d\in
\bbC\backslash\{0\}\}$.  For example $(\bbC^d)^{\ot n}(\mu) = \Span\{\ket{x^n}
: x^n\in T_\mu\}$.

To describe the weight spaces of $\cQ_\lambda^d$ we define the Kostka
coefficient $K_{\lambda\mu}:=\dim\cQ_\lambda^d(\mu)$ (as can be easily
checked, $K_{\lambda\mu}$ depends on $d$ only through $\lambda$ and
$\mu$)\cite{GW98}.  While no useful formula is known for $K_{\lambda\mu}$, they
do satisfy
\begin{itemize}
\item $\sum_\mu K_{\lambda\mu} =\dim\cQ_\lambda^d$
\item  $K_{\lambda\mu}\neq 0$ if and only if $\mu\prec\lambda$,
meaning that $|\mu|=|\lambda|$ and $\sum_{i=1}^c \mu_i \leq
\sum_{i=1}^c \lambda_i$ for $c=1,\ldots,d-1$.
\item $K_{\lambda\lambda}=1$
\end{itemize}
If we order weights according to the majorization relation $\prec$,
then there exists a {\em highest-weight vector} spanning the
one-dimensional space $\cQ_\lambda^d(\lambda)$.  At the risk of some
ambiguity, we call this vector $\ket{\lambda}$.  We will also define
an orthonormal basis for $\cQ_\lambda^d$, denoted $Q_\lambda^d$, in
which each basis vector lies in a single weight space.  This is
clearly possible in general, and also turns out to be consistent with
the basis we will introduce in \sect{gz-yy} for use in quantum
algorithms. To simplify notation later on, whenever we work with a
particular density matrix $\rho$, we will choose the torus $\cU_1^{\times d}$ to
be diagonal with respect to the same basis as $\rho$.  This means that
$\bq_\lambda^d(\rho)$ is diagonalized by $Q_\lambda^d$, the induced
weight basis for $\cQ_\lambda^d$.

We now have all the tools we need to find the spectrum of
$\bq_\lambda^d(\rho)$.  Let the eigenvalues of $\rho$ be given by $r_1
\geq \cdots \geq r_d$ (we sometimes write $r=\spec \rho$).  Then for
all $\mu\in\cT_d^n$, $\bq_\lambda^d(\rho)$ has eigenvector
$r^\mu=r_1^{\mu_1} \cdots r_d^{\mu_d}$ with multiplicity
$K_{\lambda\mu}$.  The highest eigenvalue is 
$r^\lambda = \exp[-n(H(\bl) + D(\bl\|r))]$ (since $r$ is nonincreasing
and $\mu\prec\lambda$ for any $\mu$ with $K_{\lambda\mu}\neq 0$).
Thus we obtain the 
following bounds  on $\tr\bq_\lambda^d(\rho)$:
\be r^\lambda \leq
\tr\bq_\lambda^d(\rho) = \sum_\mu K_{\lambda\mu}r^\mu \leq
r^\lambda\dim\cQ_\lambda^d.\ee

To relate this to quantum states, let $\Pi_\lambda$ denote the
projector onto $\cQ_\lambda^d\ot \cP_\lambda \subset (\bbC^d)^{\ot
n}$.  Explicitly $\Pi_\lambda$ is given by
\be \Pi_\lambda = \Usch^\dag\l(\oprod{\lambda}\ot I_{\cQ_\lambda^d} \ot
I_{\cP_\lambda}\r) \Usch.
\label{eq:Pi-lambda-def}\ee
From the bounds on $\dim\cQ_\lambda^d$ and
$\dim\cP_\lambda$ in \eqs{cQ-bound}{cP-bound2}, we obtain
\be \exp\l(nH(\bl)\r)(n+d)^{-d(d+1)/2} \leq \tr\Pi_\lambda
\leq \exp\l(nH(\bl)\r)(n+d)^{d(d-1)/2} 
\label{eq:Pi-lambda-dim}\ee
Also $\tr\Pi_\lambda \rho^{\ot n}\Pi_\lambda = \tr \bq_\lambda^d(\rho)\cdot
\dim\cP_\lambda$, which can be bounded by
\be \exp\l(-nD(\bl\|r)\r)(n+d)^{-d(d+1)/2} \leq 
\tr\Pi_\lambda \rho^{\ot n} \Pi_\lambda\leq 
\exp\l(-nD(\bl\|r)\r)(n+d)^{d(d-1)/2}
\label{eq:schur-proj}\ee
Similarly, we have
\be \Pi_\lambda\rho^{\ot n} = \rho^{\ot n}\Pi_\lambda
= \Pi_\lambda \rho^{\ot n}\Pi_\lambda \leq r^\lambda \Pi_\lambda
= \exp[-n(H(\bl) + D(\bl\|r))] \Pi_\lambda.\ee
 For some values of $\mu$, $r^\mu$ can be much smaller, so we cannot
express any useful lower bound on the eigenvalues of $\Pi_\lambda \rho^{\ot
n}\Pi_\lambda$, like we can with classical types. Of course, tracing
out $\cQ_\lambda^d$ gives us a maximally mixed state in $\cP_\lambda$,
and this is the quantum analogue of the fact that $P^{\ot n}(\cdot | t)$
is uniformly distributed over $T_t$.

We can also define the typical projector
\be \Pi_{r,\delta}^n = 
\sum_{\lambda : \bl\in\cB_{\delta}(r)} \Pi_\lambda = 
\Usch^\dag \l[
\sum_{\lambda : \bl\in\cB_{\delta}(r)}
\oprod{\lambda} \ot I_{\cQ_\lambda^d} \ot I_{\cP_\lambda}\r] \Usch,\ee
where $\cB_{\delta}(r) := \{\bl : \|\bl-r\|_1\leq
\delta\}$.  Using Pinsker's inequality, we find that
\be \tr \Pi_{r,\delta}^n \rho^{\ot n} \geq
1-\exp\l(-\frac{n\delta^2}{2}\r)(n+d)^{d(d+1)/2},
\label{eq:typ-proj}\ee
similar to the classical case.
The typical subspace is defined to be the support of the typical
projector.  Its dimension can be bounded (using
\eqs{typ-proj}{classical-fannes}) by 
\be \tr \Pi_{r,\delta}^n
\leq |\cI_{d,n}| \max_{\bl\in\cB_{\delta}(r)}\tr\Pi_\lambda
\leq (n+d)^{d(d+1)/2}
\exp(nH(r) + \eta(\delta) + \delta\log d),\ee
which is sufficient to derive Schumacher compression (cf.~\eq{schu}).

The bounds described in this section are fairly simple, but are
already powerful enough to derive many results in quantum information
theory.  Before discussing those applications, we will describe 
a variation of the decomposition of $\rho^{\ot n}$ given in
\eq{rho-decomp}.  Suppose we are given $n$ copies of a pure state
$\ket{\psi}^{AB}$ where $\rho^A = \tr_B \psi^{AB}$.  This situation
also arises when we work in the CP formalism
(see \sect{formalism}). Purifying both sides of \eq{rho-decomp}
then gives us the alternate decomposition
\be (\Usch^A \ot \Usch^B)(\ket{\psi}^{AB})^{\ot n} 
= \sum_{\lambda\in\cI_{d,n}}
c_\lambda \ket{\lambda}^{A_1}\ket{\lambda}^{B_1}
\ot \ket{q_\lambda}^{A_2B_2} \ot \ket{\Phi_{\cP_\lambda}}^{A_3B_3}
\label{eq:bipartite-decomp}\ee
Here $c_\lambda$ are coefficients satisfying $|c_\lambda|^2 = \tr
\Pi_\lambda \rho^{\ot n}$, $\ket{q_\lambda}$ are arbitrary states in
$\cQ_\lambda^d$ and $\ket{\Phi_{\cP_\lambda}}$ is a maximally
entangled state\footnote{In fact, we will see in \sect{nf-CG} that
$\ket{\Phi_{\cP_\lambda}}$ is uniquely determined.} on $\cP_\lambda
\ot \cP_\lambda$.

\section{Applications of Schur duality}
\label{sec:schur-qit-apps}

The Schur transform is useful in a surprisingly large number of
quantum information protocols.  Here we will review these applications
using the formulae from the last section to rederive the main results.
It is worth noting that an efficient implementation of the Schur
transform is the only nontrivial step necessary to perform these
protocols.  Thus our construction of the Schur transform in the next
chapter will simultaneously make all of these tasks computationally
efficient.

\subsubsection{Spectrum and state estimation}

Suppose we are given many copies of an unknown mixed quantum
state, $\rho^{\otimes n}$.  An important task is to obtain
an estimate for the spectrum of ${\rho}$ from these $n$
copies.  An asymptotically good estimate (in the sense of large
deviation rate) for the spectrum of ${\rho}$ can be obtained
by applying the Schur transform, measuring $\lambda$ and taking
the spectrum estimate to be $(\lambda_1/n,\ldots,
\lambda_d/n)$\cite{Keyl:01,Vidal:99a}.  Indeed the probability that
$\|\lambda - \spec\rho\|_1\leq \delta$ for any $\delta>0$ is bounded
by \eq{typ-proj}.  Thus an efficient implementation of the Schur transform will
efficiently implement the spectrum estimating protocol (note that it
is efficient in $d$, not in $\log(d)$).

The more general problem of estimating ${\rho}$ reduces
to measuring $|\lambda\rangle$ and $\cQ_\lambda^d$, but optimal estimators
have only been explicitly constructed for the case of
$d=2$\cite{Gill:02a}.  One natural estimation scheme is given by first
measuring $\lambda$ and then performing a covariant POVM on
$\cQ_\lambda^d$ with POVM elements
\be \bq_\lambda^d(U) \oprod{\lambda} \bq_\lambda^d(U)^\dag 
\dim\cQ_\lambda^d \,dU,\ee
where $\ket{\lambda}$ is the highest weight vector in $\cQ_\lambda^d$
and $dU$ is a Haar 
measure for $\cU_d$.  The corresponding state estimate is then
$\hat{\rho} = U\l(\sum_{i=1}^d \lambda_i \oprod{i}\r) U^\dag$.
In this estimation scheme, as $n\ra
\infty$ the probability
that $\|\rho-\hat{\rho}\|_1 > \delta$ scales as $\exp(-nf(\delta))$
with $f(\delta)>0$ whenever $\delta>0$; 
\mscite{Keyl04} proves this and derives the function $f(\delta)$.
However, it is not known whether the $f(\delta)$ obtained for this
measurement scheme is the best possible.

A related problem is quantum hypothesis testing (determining whether
one has been given the state $\rho^{\ot n}$ or some other state).  An
optimal solution to quantum hypothesis testing can
be obtained by a similar protocol\cite{Hayashi:02d}.

\subsubsection{Universal distortion-free entanglement concentration}

Let $|\psi\rangle_{AB}$ be a bipartite partially entangled state
shared between two parties, $A$ and $B$. Suppose we are given many
copies of $|\psi\rangle_{AB}$ and we want to transform these
states into copies of a maximally entangled state using only local
operations and classical communication.  Further, suppose that we
wish this protocol to work when neither $A$ nor $B$ know the state
$|\psi \rangle^{AB}$.  Such a scheme is called a universal
(meaning it works with unknown states $|\psi\rangle^{AB}$)
entanglement concentration protocol, as opposed to the original
entanglement concentration protocol described by \cite{BBPS96}.
Further we also would like the 
scheme to produce perfect maximally entangled states, i.e. to be
distortion free.  Universal distortion-free entanglement
concentration can be performed\cite{Hayashi:02a} by both parties
performing Schur transforms on their $n$ halves of
$|\psi\rangle^{AB}$, measuring their $|\lambda\rangle$, discarding
$\cQ_\lambda^d$ and retaining $\cP_\lambda$.  According to
\eq{bipartite-decomp}, the two parties will now
share a maximally entangled state of dimension $\dim\cP_\lambda$,
where $\lambda$ is observed with probability
$\dim\cP_\lambda \cdot \tr\bq_\lambda^d(\tr_B\oprod{\psi})$.

According to Eqns.~(\ref{eq:typ-proj}),
(\ref{eq:classical-fannes}) and (\ref{eq:Pi-lambda-dim}), this
produces at least $n(S(\rho)-\eta(\delta) - \delta\log d)-\half
d(d+1)\log(n+d)$ ebits with probability 
$\geq 1-\exp\l(-\frac{n\delta^2}{2}\r)(n+d)^{d(d+1)/2}$.  The rate at
which this error probability vanishes for any fixed $\delta$ can be
shown to be optimal among protocols of this form\cite{Hayashi:02a}.

\subsubsection{Universal Compression with Optimal Overflow
Exponent}

Measuring $|\lambda\rangle$ weakly so as to cause little disturbance, together
with appropriate relabeling, comprises a universal compression algorithm with
optimal overflow exponent (rate of decrease of the probability that the
algorithm will output a state that is much too
large)\cite{Hayashi:02b,Hayashi:02c}.

Alternatively, suppose we are given $R$ s.t. $H(\rho)<R$ and we want
to compress $\rho^{\ot n}$ into $nR$ qubits.  Define the projector
$\Pi_R^n$ by 
\be \Pi_R^n := \sum_{\substack{\lambda\in\cI_{d,n}\\
H(\bl)\leq R_n}} \Pi_\lambda, \ee
where $R_n := R - \half d(d+1)\log(n+d)$.
Since $\tr\Pi_R^n \leq \exp(nR)$, projecting onto $\Pi_R^n$ allows the
residual state to be compressed to $nR$ qubits.  The error can be
shown to be bounded by
\be \leq (n+d)^{d(d+1)/2} \exp\l[-n\min_{P:H(P)>R_n}
D(P\|\spec\rho)\r],\ee
which decreases exponentially with $n$ as long as $R>H(\rho)$.

\subsubsection{Encoding and decoding into decoherence-free subsystems}

Further applications of the Schur transform include encoding into
decoherence-free subsystems\cite{Zanardi:97a, Knill:00a, Kempe:01a,
Bacon:01a}.  Decoherence-free subsystems are subspaces of a system's
Hilbert space which are immune to decoherence due to a symmetry of the
system-environment interaction.  For the case where the environment
couples identically to all systems, information can be protected from
decoherence by encoding into the $|p_\lambda
\rangle$ basis.  We can use the inverse Schur transform (which, as
a circuit can be implemented by reversing the order of all gate elements and
replacing them with their inverses) to perform this encoding: simply feed in
the appropriate $|\lambda\rangle$ with the state to be encoded
into the $\cP_\lambda$ register and any state into the $\cQ_\lambda^d$
register into the inverse Schur transform.  Decoding can similarly
be performed using the Schur transform.

This encoding has no error and asymptotically unit efficiency, since
$\log \max_\lambda \dim\cP_\lambda$ qubits can be sent and
$\max_\lambda\dim\cP_\lambda \geq d^n / (|\cI_{d,n}|
\max_\lambda\dim\cQ_\lambda^d) \geq d^n (n+d)^{-d(d+1)/2}$.

\subsubsection{Communication without a shared reference frame}

An application of the concepts of decoherence-free subsystems comes
about when two parties wish to communicate (in either a classical or
quantum manner) but do not share a reference frame.  The
effect of not sharing a reference frame is the same as the effect of
collective decoherence: the same random unitary rotation is
applied to each subsystem.  Thus encoding information into the
$\cP_\lambda$ register will allow this information to be communicated
in spite of the fact that the two parties do not share a reference
frame\cite{Bartlett03}.  Just as with decoherence-free subsystems,
this encoding and decoding can be done with the Schur transform. 

\section{Normal form of memoryless channels}
\label{sec:normal-form}

So far we have has only discussed the decomposition of $\rho^{\ot n}$,
or equivalently, of pure bipartite entangled states.  However many
interesting problems in quantum information theory involve what are
effectively tripartite states.  Not only are tripartite states
$\ket{\psi}^{ABC}$ interesting in themselves\cite{Thap99}, they also
appear when a noisy bipartite state $\rho^{AB}$ is replaced by its
purification $\ket{\psi}^{ABE}$ and when a
noisy quantum channel $\cN^{A\ra B}$ is replaced by its purification
$U_\cN^{A\ra BE}$.  When considering $n$ copies of these resources,
much of their structure can be understood in terms of the vector spaces
$(\cP_{\lambda_A} \ot \cP_{\lambda_B} \ot \cP_{\lambda_E})^{\cS_n}$.
We explain how this follows from the $\cS_n$ CG transform in \sect{nf-CG}
and then apply this to quantum channels in \sect{nf-channel}.
Finally, we generalize the bounds from \sect{quantum-types} to a quantum
analogue of joint typicality in \sect{nf-typical}.

\subsection{The $\cS_n$ Clebsch-Gordan transformation}
\label{sec:nf-CG}
We begin by describing how the CG transform (cf.~\sect{CG-general})
specializes to $\cS_n$.  For $\lambda_A,\lambda_B\in\cI_n$,
\eq{CG-hom-decomp} implies
\be \cP_{\lambda_A} \ot \cP_{\lambda_B} \iso{\cS_n}
\bigoplus_{\lambda_C\in\cI_n} \cP_{\lambda_C} \ot
\Hom(\cP_{\lambda_C}, \cP_{\lambda_A} \ot \cP_{\lambda_B})^{\cS_n}
\iso{\cS_n} \bigoplus_{\lambda_C\in\cI_n} \cP_{\lambda_C} \ot
\bbC^{g_{\lambda_A\lambda_B\lambda_C}}
\label{eq:CG}\ee
Here we have defined the Kronecker coefficient
$g_{\lambda_A\lambda_B\lambda_C}:=\dim 
\Hom(\cP_{\lambda_C}, \cP_{\lambda_A} \ot \cP_{\lambda_B})^{\cS_n}$.

It can be shown that there is an orthonormal basis for $\cP_\lambda$, which we
call $P_\lambda$, in which $\bp_\lambda(s)$ are real and
orthogonal.\footnote{One way to prove this is to consider Young's
natural basis, which was introduced in \sect{weyl-symmetrizer}.  Since
the $\Pi_{\lambda:T}$ produce real linear combinations of the states
$\ket{i_1,\ldots,i_n}$, the matrices $\bp_\lambda(s)$ are also real
when written in Young's natural basis.  If we generate an orthonormal
basis by applying Gram-Schmidt to Young's natural basis, the matrices
$\bp_\lambda(s)$ remain real.

In \sect{gz-yy} we will introduce a different orthonormal basis for
$\cS_n$, known as Young's orthogonal basis, or as the Young-Yamanouchi
basis.  \mscite{JK81} gives an explicit formula in this basis for
$\bp_\lambda(s)$ in which the matrices are manifestly real.}  This
means that $\cP_\lambda\iso{\cS_n}\cP_\lambda^*$.  Since
$\Hom(A,B)\cong A^*\ot B$, it follows that
\be \Hom(\cP_{\lambda_C}, \cP_{\lambda_A} \ot \cP_{\lambda_B})^{\cS_n}
\iso{\cS_n} (\cP_{\lambda_A}\ot \cP_{\lambda_B}\ot
\cP_{\lambda_C})^{\cS_n}. \label{eq:CG-mult-state}\ee
As a corollary, $g_{\lambda_A\lambda_B\lambda_C}$ is unchanged by
permuting $\lambda_A,\lambda_B,\lambda_C$.  Unfortunately, no
efficient method of calculating $g_{\lambda_A\lambda_B\lambda_C}$ is
known, though asymptotically they have some connections to the quantum
mutual information that will be investigated in future work.  The
permutation symmetry of $g_{\lambda_A\lambda_B\lambda_C}$ also means
that we can consider CG transformations from $AB\ra C$, $AC\ra B$ or
$BC\ra A$, with the only difference being a normalization factor which
we will explain below.

According to \eq{CG-mult-state}, the CG transformation can be
understood in terms of tripartite $\cS_n$-invariant vectors.  Let
$\ket{\alpha}$ be a unit vector in $(\cP_{\lambda_A}\ot
\cP_{\lambda_B}\ot \cP_{\lambda_C})^{\cS_n}$, with corresponding
density matrix $\alpha=\oprod{\alpha}$.  Since $\alpha^A :=
\tr_{BC}\alpha$ is 
invariant under permutations and $\tr\alpha=1$, Schur's Lemma
 implies that $\alpha^A = I_{\cP_{\lambda_A}}/D_A$, with
$D_A:=\dim\cP_{\lambda_A}$.  This means we can Schmidt decompose
$\ket{\alpha}$ as
\be \ket{\alpha}^{ABC} = \frac{1}{\sqrt{D_A}} \sum_{p_A\in P_{\lambda_A}}
\ket{p_A}^A W_\alpha^{A'\ra BC}\ket{p_A}^{A'}
\label{eq:alpha-decomp}\ee
where $W_\alpha\in\Hom(\cP_{\lambda_A},\cP_{\lambda_B}\ot
\cP_{\lambda_C})^{\cS_n}$ is an isometry.   We can express
$\Ucg^{\lambda_B,\lambda_C}$ in terms of 
$W_\alpha$ according to 
\be(\Ucg^{\lambda_B,\lambda_C})^\dag\ket{\alpha}\ket{p_A} = W_\alpha\ket{p_A}.
\label{eq:CG-W}\ee

The simple form of \eq{alpha-decomp} suggests that the CG
transformation can also be implemented by teleportation.  For any
$\lambda\in\cI_n$, let $D_\lambda := \dim \cP_\lambda$ and define
$\ket{\Phi_\lambda} = D_\lambda^{-\frac{1}{2}}\sum_{p\in P_\lambda}
\ket{p}\ket{p}$.  Note that up to a phase $\ket{\Phi_\lambda}$ is the unique
invariant vector in $\cP_\lambda\ot\cP_\lambda$.  To see that it is
invariant, use the fact that $(A\ot I)\ket{\Phi_\lambda}=(I\ot
A^T)\ket{\Phi_\lambda}$ for any operator $A$ and the fact that
$\bp_\lambda$ are orthogonal matrices, so $\bp_\lambda(s)^T =
\bp_\lambda(s)^{-1}$.  Uniqueness follows from
\be 
\dim(\cP_\lambda \ot \cP_\lambda)^{\cS_n}
=\dim\Hom(\cP_\lambda,\cP_\lambda)^{\cS_n} = 1,
\label{eq:unique-invariant-bipartite}\ee
where the first equality is because
$\cP_\lambda\iso{\cS_n}\cP_\lambda^*$ and the second equality is
due to Schur's Lemma.

We use $\ket{\Phi_\lambda}$ for teleportation as follows.  If
$\ket{p_A}\in\cP_{\lambda_A}$ is a basis vector (i.e. $\ket{p_A}\in
P_{\lambda_A}$), 
then $\bra{\Phi_\lambda}^{AA'}\ket{p_A}^A = \frac{1}{\sqrt{D_A}}
\bra{p_A}$.  Also \eq{alpha-decomp} can be written more simply as
\be \ket{\alpha}^{ABC} = 
\l(I^A \ot W_\alpha^{A'\ra BC}\r) \ket{\Phi_{\lambda_A}}^{AA'}
\label{eq:alpha-decomp2}.
\ee
Combining \eqs{alpha-decomp2}{CG-W} then yields
\be \bra{\Phi_{\lambda_A}}^{AA'}\ket{\alpha}^{A'BE}\ket{p_A}^A =
\frac{1}{D_A} W_\alpha^{A\ra BE}\ket{p_A}^A= 
\frac{1}{D_A}(\Ucg^{\lambda_B,\lambda_C})^\dag\ket{\alpha}\ket{p_A}
\label{eq:CG-via-TP}\ee

This connection between the quantum CG transform and
 $\cS_n$-invariant tripartite states will now be used to decompose
 i.i.d. quantum channels.

\subsection{Decomposition of memoryless quantum channels}
\label{sec:nf-channel}
Let $\cN:A'\ra B$ be a quantum channel and $U_\cN:A'\ra BE$ its
isometric extension.  Let $d_A = \dim A', d_B=\dim B, d_E=\dim E$ and
$d:=\max(d_A,d_B,d_E)$.  We want to consider $n$ uses of $U_\cN$ in the
Schur basis.  In general this has the form
\be \Usch U_\cN^{\ot n} \Usch^\dag =
\!\!\!\sum_{\lambda_A,\lambda_B,\lambda_E\in\cI_N}\!\!\!\!\!
\ket{\lambda_B\lambda_E}\bra{\lambda_A}
\!\!\!\sum_{\substack{q_A\in Q_{\lambda_A},
q_B\in Q_{\lambda_B}\\q_E\in Q_{\lambda_E}}}\!\!\!\!\!
\ket{q_Bq_E}\bra{q_A}
\!\!\!\!\sum_{\substack{p_A\in P_{\lambda_A},
p_B\in P_{\lambda_B}\\p_E\in P_{\lambda_E}}}\!\!\!\!\!
\ket{p_Bp_E}\bra{p_A}\,
C^{\lambda_Aq_Ap_A}_{\lambda_B\lambda_Eq_Bq_Ep_Bp_E},\ee
for some coefficients
$C^{\lambda_Aq_Ap_A}_{\lambda_B\lambda_Eq_Bq_Ep_Bp_E}$.
So far this tells us nothing at all!
But we know that $U_\cN^{\ot n}$ is invariant under permutations;
i.e. $\l[\bP(s^{-1})^B\ot\bP(s^{{-1}})^E\r] U_\cN^{\ot n} \bP(s)^A =
U_\cN^{\ot n}$ for all 
$s\in\cS_n$.
Thus
\be U_\cN^{\ot n}\in \Hom(\bbC^{d_A^n}, \bbC^{d_B^n}\ot
\bbC^{d_E^n})^{\cS_n} \iso{\cU_d\times\cS_n}
\bigoplus_{\lambda_A,\lambda_B,\lambda_E\in\cI_n}
\Hom(\cQ_{\lambda_A}^{d_A},\cQ_{\lambda_B}^{d_B}\ot\cQ_{\lambda_E}^{d_E})
\hat{\ot} 
\Hom(\cP_{\lambda_A},\cP_{\lambda_B}\ot\cP_{\lambda_E})^{\cS_n}
\ee
Let $P[\lambda_A;\lambda_B,\lambda_E]$ be an orthonormal basis for 
$\Hom(\cP_{\lambda_A},\cP_{\lambda_B}\ot\cP_{\lambda_E})^{\cS_n}$.
Then we can expand 
$\Usch U_\cN^{\ot n} \Usch^\dag$ as
\be \Usch U_\cN^{\ot n} \Usch^\dag =
\sum_{\substack{\lambda_A,\lambda_B,\lambda_E\in\cI_n,
\alpha\in P[\lambda_A;\lambda_B,\lambda_E]\\
q_A\in Q_{\lambda_A}^{d_A},q_B\in Q_{\lambda_B}^{d_B},q_E\in
Q_{\lambda_E}^{d_E} 
}}
[V_\cN^n]_{\lambda_B\lambda_Eq_Bq_E\alpha}^{\lambda_Aq_A}
\ket{\lambda_B\lambda_E}\bra{\lambda_A}
\ot \ket{q_Bq_E}\bra{q_A} \ot W_\alpha,
\label{eq:normal-form}\ee
where the coefficients
$[V_\cN^n]_{\lambda_B\lambda_Eq_Bq_E\alpha}^{\lambda_Aq_A}$ correspond to an
isometry; i.e. 
\be \sum_{\lambda_B,\lambda_E,q_B,q_E,\alpha}
([V_\cN^n]_{\lambda_B\lambda_Eq_Bq_E\alpha}^{\lambda_Aq_A})^*
[V_\cN^n]_{\lambda_B\lambda_Eq_Bq_E\alpha}^{\lambda_A'q_A'} = 
\delta_{\lambda_A,\lambda_A'}\delta_{q_A,q_A'}. \ee
This is depicted as a quantum circuit in \fig{normal-form}.

\begin{figure}[ht]
\begin{centering}
\setlength{\unitlength}{3947sp}%
\begingroup\makeatletter\ifx\SetFigFont\undefined%
\gdef\SetFigFont#1#2#3#4#5{%
  \reset@font\fontsize{#1}{#2pt}%
  \fontfamily{#3}\fontseries{#4}\fontshape{#5}%
  \selectfont}%
\fi\endgroup%
\begin{picture}(4275,2724)(1726,-2323)
\thinlines
{\put(2881,-1531){\oval(210,210)[bl]}
\put(2881,284){\oval(210,210)[tl]}
\put(3496,-1531){\oval(210,210)[br]}
\put(3496,284){\oval(210,210)[tr]}
\put(2881,-1636){\line( 1, 0){615}}
\put(2881,389){\line( 1, 0){615}}
\put(2776,-1531){\line( 0, 1){1815}}
\put(3601,-1531){\line( 0, 1){1815}}}
{\put(3601,-1411){\line( 1, 0){225}}}
{\put(4276,-1411){\line( 1, 0){225}}}
{\put(4606,-2206){\oval(210,210)[bl]}
\put(4606,-1291){\oval(210,210)[tl]}
\put(5146,-2206){\oval(210,210)[br]}
\put(5146,-1291){\oval(210,210)[tr]}
\put(4606,-2311){\line( 1, 0){540}}
\put(4606,-1186){\line( 1, 0){540}}
\put(4501,-2206){\line( 0, 1){915}}
\put(5251,-2206){\line( 0, 1){915}}}
{\put(2176,-1936){\line( 1, 0){2325}}}
{\put(3601,-1036){\line( 1, 0){2250}}}
{\put(3601,-661){\line( 1, 0){2250}}}
{\put(3601,-286){\line( 1, 0){2250}}}
{\put(3601, 89){\line( 1, 0){2250}}}
{\put(5251,-1411){\line( 1, 0){600}}}
{\put(5251,-1936){\line( 1, 0){600}}}
{\put(2176, 14){\line( 1, 0){600}}}
{\put(2176,-736){\line( 1, 0){600}}}

\put(4876,-286){\circle*{100}}
\put(4876, 89){\circle*{100}}
\put(4876, 89){\line( 0,-1){1275}}

\put(3076,-736){\makebox(0,0)[lb]{$[V_\cN^n]$}}
\put(3901,-1486){\makebox(0,0)[lb]{$\ket{\alpha}$}}

\put(6001,-2011){\makebox(0,0)[lb]{$\ket{p_E}$}}
\put(6001,-1486){\makebox(0,0)[lb]{$\ket{p_B}$}}
\put(6001,-1111){\makebox(0,0)[lb]{$\ket{q_E}$}}
\put(6001,-736){\makebox(0,0)[lb]{$\ket{q_B}$}}
\put(6001,-361){\makebox(0,0)[lb]{$\ket{\lambda_E}$}}
\put(6001, 14){\makebox(0,0)[lb]{$\ket{\lambda_B}$}}

\put(1726,-61){\makebox(0,0)[lb]{$\ket{\lambda_A}$}}
\put(1726,-811){\makebox(0,0)[lb]{$\ket{q_A}$}}
\put(1726,-2011){\makebox(0,0)[lb]{$\ket{p_A}$}}
\put(4726,-1786){\makebox(0,0)[lb]{$\Ucg^\dag$}}
\end{picture}
\caption{The quantum channel $U_\cN^{\ot n}$ is decomposed in the
Schur basis as in \eq{normal-form}. Alice inputs an $n$ qudit state of
the form $\ket{\lambda_A}\ket{q_A}\ket{p_A}$ and the channel outputs
superpositions of $\ket{\lambda_B}\ket{q_B}\ket{p_B}$ for Bob and
$\ket{\lambda_E}\ket{q_E}\ket{p_E}$ for Eve.  The intermediate state
$\ket{\alpha}$ belongs to
$\Hom(\cP_{\lambda_A},\cP_{\lambda_B}\ot\cP_{\lambda_E})^{\cS_n}$.}
\label{fig:normal-form}
\end{centering}
\end{figure}

Using \eqs{CG-W}{CG-via-TP}, we can replace the CG transform in
\fig{normal-form} with a teleportation-like circuit.  Instead of
interpreting $\alpha$ as a member of
$\Hom(\cP_{\lambda_A},\cP_{\lambda_B}\ot\cP_{\lambda_E})^{\cS_n}$, we
say that $\ket{\alpha}\in
(\cP_{\lambda_A}\ot\cP_{\lambda_B}\ot\cP_{\lambda_E})^{\cS_n}$.  This
has the advantage of making its normalization more straightforward and
of enhancing the symmetry between $A$, $B$ and $E$.  The $\Ucg^\dag$
then becomes replaced with a projection onto
$\ket{\Phi_{\lambda_A}}$.  Since this only succeeds with probability
$1/D_A$, the resulting state needs to be normalized by multiplying by
$\sqrt{D_A}$.  The resulting circuit is given in \fig{normal-form2}.

\begin{figure}[ht]
\begin{centering}

\setlength{\unitlength}{3947sp}%
\begingroup\makeatletter\ifx\SetFigFont\undefined%
\gdef\SetFigFont#1#2#3#4#5{%
  \reset@font\fontsize{#1}{#2pt}%
  \fontfamily{#3}\fontseries{#4}\fontshape{#5}%
  \selectfont}%
\fi\endgroup%
\begin{picture}(4275,3174)(1726,-2773)
\thinlines
{\put(3601,-1036){\line( 1, 0){2250}}}
{\put(3601,-661){\line( 1, 0){2250}}}
{\put(3601,-286){\line( 1, 0){2250}}}
{\put(3601, 89){\line( 1, 0){2250}}}
{\put(2176, 14){\line( 1, 0){600}}}
{\put(2176,-736){\line( 1, 0){600}}}
{\put(2881,-2131){\oval(210,210)[bl]}
\put(2881,284){\oval(210,210)[tl]}
\put(3496,-2131){\oval(210,210)[br]}
\put(3496,284){\oval(210,210)[tr]}
\put(2881,-2236){\line( 1, 0){615}}
\put(2881,389){\line( 1, 0){615}}
\put(2776,-2131){\line( 0, 1){2415}}
\put(3601,-2131){\line( 0, 1){2415}}}
{\put(3601,-1711){\line( 1, 0){225}}}
{\put(4276,-1711){\line( 1, 0){1575}}}
{\put(4501,-1411){\line( 1, 0){1350}}}
{\put(2176,-2536){\line( 1, 0){2700}}}
{\put(4981,-2656){\oval(210,210)[bl]}
\put(4981,-2041){\oval(210,210)[tl]}
\put(5746,-2656){\oval(210,210)[br]}
\put(5746,-2041){\oval(210,210)[tr]}
\put(4981,-2761){\line( 1, 0){765}}
\put(4981,-1936){\line( 1, 0){765}}
\put(4876,-2656){\line( 0, 1){615}}
\put(5851,-2656){\line( 0, 1){615}}}
{\put(4276,-1711){\line( 3, 4){225}}}
{\put(4276,-1711){\line( 1,-2){225}}
\put(4501,-2161){\line( 1, 0){375}}}

\put(3076,-1036){\makebox(0,0)[lb]{$[V_\cN^n]$}}
\put(1726,-61){\makebox(0,0)[lb]{$\ket{\lambda_A}$}}
\put(1726,-811){\makebox(0,0)[lb]{$\ket{q_A}$}}
\put(6001, 14){\makebox(0,0)[lb]{$\ket{\lambda_B}$}}
\put(6001,-361){\makebox(0,0)[lb]{$\ket{\lambda_E}$}}
\put(6001,-736){\makebox(0,0)[lb]{$\ket{q_B}$}}
\put(6001,-1111){\makebox(0,0)[lb]{$\ket{q_E}$}}
\put(3951,-1786){\makebox(0,0)[lb]{$\ket{\alpha}$}}
\put(1726,-2611){\makebox(0,0)[lb]{$\ket{p_A}$}}
\put(5001,-2386){\makebox(0,0)[lb]{$\sqrt{D_A}\bra{\Phi_{\lambda_A}}$}}
\put(6001,-1486){\makebox(0,0)[lb]{$\ket{p_B}$}}
\put(6001,-1786){\makebox(0,0)[lb]{$\ket{p_E}$}}
\end{picture}

\caption{The quantum channel $U_\cN^{\ot n}$ is decomposed in the
Schur basis with teleportation replacing the $\cS_n$ CG transform.
Here the intermediate state
$\ket{\alpha}$ belongs to
$(\cP_{\lambda_A}\ot\cP_{\lambda_B}\ot\cP_{\lambda_E})^{\cS_n}$,
the box labeled $\sqrt{D_A}\bra{\Phi_{\lambda_A}}$ represents
projecting onto the 
maximally entangled state $\ket{\Phi_{\lambda_A}}$ and normalization
requires multiplying
the residual state by $\sqrt{D_A}$, where $D_A := \dim \cP_{\lambda_A}$.
}
\label{fig:normal-form2}
\end{centering}
\end{figure}

\subsection{Jointly typical projectors in the Schur basis}
\label{sec:nf-typical}
The channel decomposition in the last section is still extremely
general.  In particular, the structure of the map is given by the
$\lambda_A$, $\lambda_B$ and $\lambda_E$ which appear in
\eq{normal-form}, but generically all of the coefficients will be
nonzero.  However, for large values of $n$, almost all of the weight
will be contained in a small set of {\em typical} triples of
$(\lambda_A,\lambda_B,\lambda_E)$.  These triples are the quantum
analogue of joint types from classical information theory.

In this section we show the existence of typical sets of
$(\bl_A,\bl_B,\bl_E)$ onto which a channel's input and output can be
projected with little disturbance.
In fact, we will define three versions of the typical set $\cT_\cN^n$
and show that they 
are in a certain sense asymptotically equivalent.  For each version,
let $\rho^A$ be an arbitrary channel input, and
$\ket{\psi}^{ABE}=(I^A \ot U_\cN^{A'\ra BE})\ket{\Phi_\rho}^{AA'}$ the
purified channel output (following the CP formalism).  Now define
$R(\cN)$ to be set of $\psi^{ABE}$ that can be generated in this
manner. 

\begin{itemize}
\item Define $\cT_\cN^* :=\{(r_A,r_B,r_E) : \exists \psi^{ABE}\in
R(\cN) \st\! r_A\!=\spec(\psi^A), r_B\!=\spec(\psi^B),
r_E\!=\spec(\psi^E)\}$.  This set is simply the set of triples of
spectra that can arise from one use of the channel.  It has the
advantage of being easy to compute and to optimize over, but it
doesn't give us direct information about which values of
$(\lambda_A,\lambda_B,\lambda_E)$ we need to consider.


\item Define $\tilde{\cT}_\cN^n(\eps) := \{(\bl_A,\bl_B,\bl_E) :
\exists 
\psi^{ABE}\in R(\cN) \st \tr (\Pi_{\lambda_A}^A \ot \Pi_{\lambda_B}^B
\ot \Pi_{\lambda_E}^E) \psi^{\ot n} \geq \eps\}$.

This set tells us which $(\lambda_A,\lambda_B,\lambda_E)$ we need to
consider when working with purified outputs of $U_\cN^{\ot n}$.  To
see this note that if $\psi\in R(\cN)$, then projecting $\psi^{\ot n}$
onto $\tilde{\cT}_\cN^n(\eps)$ will succeed with probability $\geq
1-\eps(n+1)^{3d}$ since there are $\leq (n+1)^{3d}$ possible triples
$(\lambda_A,\lambda_B,\lambda_E)$.

\item Define $\cT_\cN^n(\eps)$ to be the set of $(\bl_A,\bl_B,\bl_E)$
s.t. there exists a subnormalized density matrix $\omega_A$ on 
 $\cQ_{\lambda_A}^{d_A}$ (i.e. $\tr\omega_A \leq 1$) s.t. 
\be \tr (\oprod{\lambda_B}\ot
\oprod{\lambda_E} \ot I_{\cQ_{\lambda_B}} \ot I_{\cQ_{\lambda_E}} \ot
 I_\alpha) V_\cN^n (\oprod{\lambda_A} \ot \omega_A) \geq \eps.\ee 
Since that $V_\cN^n$ completely determines the map from
$\lambda_A\mapsto (\lambda_B,\lambda_E)$, we don't need to consider
different values of the $\ket{p_A}$ register.

This set is useful when considering channel outputs in the CQ
formalism.  It says that if the input is encoded in
$\cQ_{\lambda_A}^{d_A}\ot \cP_{\lambda_A}$ then only certain output
states need be considered.  
\end{itemize}

All of these sets could also be generalized to include possible
$\cQ_\lambda^d$ states as well.  However, we focus attention on the
$(\lambda_A,\lambda_B,\lambda_E)$ since those determine the dimensions
of $\cP_\lambda$ and hence the possible communication rates.

We claim that the three typical sets described above are close to one
another. In other words, for any element in one typical set, the other
sets have nearby elements, although we may have to decrease $\eps$.
Here ``nearby'' means that the distance goes to zero for any fixed or
slowly-decreasing value of $\eps$ as $n\ra \infty$.

In the following proofs we will frequently omit mentioning $\Usch$,
implicitly identifying $\rho^{\ot n}$ with $\Usch\rho^{\ot
n}\Usch^\dag$ and $U_\cN^{\ot n}$ with $\Usch U_\cN^{\ot
n}\Usch^\dag$.
\begin{itemize}
\item $\cT_\cN^* \Rightarrow \tilde{\cT}_\cN^n(\eps)$ (i.e. for any
triple in $\cT_\cN^*$ there is a nearby triple in
$\tilde{\cT}_\cN^n(\eps)$)
\begin{proof}
Suppose $(r_A,r_B,r_E)\in \cT_\cN^*$ and let $\psi^{ABE}$ be the
corresponding state in $R(\cN)$ whose reduced states have spectra
$r_A$, $r_B$ and 
$r_E$.  Define the probability distribution
$\Pr(\lambda_A,\lambda_B,\lambda_E) := \tr (\Pi_{\lambda_A}^A \ot
\Pi_{\lambda_B}^B \ot \Pi_{\lambda_E}^E) \psi^{\ot n}$.  Then by
\eq{typ-proj}, $\Pr(\frac{1}{2}\|r_A-\bl_A\|_1 > \delta) \leq
(n+d)^{d(d+1)/2}\exp(-n\delta^2)$ for any $\delta>0$.  Repeating this
for $\bl_B$ and 
$\bl_E$, we find that
$$ \Pr\l[\!\l(\frac{1}{2}\|r_A-\bl_A\|_1 \!>\! \delta\r)
\lor\! \l(\frac{1}{2}\|r_B-\bl_B\|_1 \!>\! \delta\r) \lor\!
\l(\frac{1}{2}\|r_E-\bl_E\|_1 \!>\! \delta\r)\!\r] \!\leq
3(n+d)^{\frac{d(d+1)}{2}}\!\exp(\!-n\delta^2).$$
  Since the number of triples
$(\bl_A,\bl_B,\bl_E)$ is $\leq (n+1)^{3d}$, this means there exists a
triple $(\bl_A,\bl_B,\bl_E)$ with $\Pr(\bl_A,\bl_B,\bl_E)\geq
(n+1)^{-3d}(1 - 3(n+d)^{d(d+1)/2}\exp(-n\delta^2)) =: \eps$ (and so
$(\bl_A,\bl_B,\bl_E)\in \tilde{\cT}_\cN^n(\eps)$), satisfying
$\frac{1}{2}\|r_A-\bl_A\|_1\leq \delta$,
$\frac{1}{2}\|r_B-\bl_B\|_1\leq \delta$ and
$\frac{1}{2}\|r_E-\bl_E\|_1\leq \delta$.  One natural choice is to
take $\delta=(\log n)/\sqrt{n}$ and $\eps=1/\poly(n)$.
\end{proof}

\item $\tilde{\cT}_\cN^n(\eps) \subseteq {\cT}_\cN^n(\eps)$
\begin{proof}
Suppose $(\bl_A,\bl_B,\bl_E)\in\tilde{\cT}_\cN^n(\eps)$, meaning that there
exists $\psi^{ABE}\in R(\cN)$ s.t. $\tr (\Pi_{\lambda_A}^A \ot
\Pi_{\lambda_B}^B \ot \Pi_{\lambda_E}^E) \psi^{\ot n} \geq \eps$.  Thus
if we set $\rho^A=\tr_{BE}\psi^{\ot n}$ then 
\bea \eps &\leq& \tr (\Pi_{\lambda_A}^A \ot \Pi_{\lambda_B}^B \ot
\Pi_{\lambda_E}^E)
\l[(I^A \ot U_\cN^{A'\ra BE})\ket{\Phi_\rho}^{AA'}\r]^{\ot n}\\
&=& \tr (\Pi_{\lambda_B} \ot \Pi_{\lambda_E})
U_\cN^{\ot n}(\Pi_{\lambda_A}\rho^{\ot n}\Pi_{\lambda_A})\\
&=& \tr (\Pi_{\lambda_B} \ot \Pi_{\lambda_E})
U_\cN^{\ot n}(\oprod{\lambda_A} \ot \bq_{\lambda_A}(\rho)
\ot I_{\cP_{\lambda_A}}) \\
&=& \tr (\oprod{\lambda_B}\ot I_{\cQ_{\lambda_B}} \ot
\oprod{\lambda_E}\ot I_{\cQ_{\lambda_E}})
V_\cN^n(\oprod{\lambda_A} \ot \bq_{\lambda_A}(\rho)
\cdot \dim\cP_{\lambda_A})\\
&=& \tr \l(\oprod{\lambda_B}\ot I_{\cQ_{\lambda_B}} \ot
\oprod{\lambda_E}\ot I_{\cQ_{\lambda_E}}\r)
V_\cN^n(\oprod{\lambda_A} \ot \omega_A)
\eea

In the last step we have defined the (subnormalized) density matrix
$\omega_A := \bq_{\lambda_A}(\rho) \cdot \dim\cP_{\lambda_A}$. (It is
subnormalized because $\tr \omega_A = \tr \Pi_{\lambda_A}\rho^{\ot n}
\leq 1$.)  Thus $(\bl_A,\bl_B,\bl_E)\in \cT_\cN^n(\eps)$.
\end{proof}

\item ${\cT}_\cN^n(\eps) \subseteq \tilde{\cT}_\cN^n(\eps')$,
$\eps'=\eps(n+d)^{-d^2}$
\begin{proof}
If $(\bl_A,\bl_B,\bl_E)\in {\cT}_\cN^n(\eps)$ then there exists
a density matrix $\omega_A$ on $\cQ_{\lambda_A}^{d_A}$ s.t.
\be \tr (\Pi_{\lambda_B}\ot \Pi_{\lambda_E})
\cN^{\ot n}\l(\oprod{\lambda_A} \ot \omega_A \ot
\frac{I_{\cP_{\lambda_A}}}{\dim \cP_{\lambda_A}}\r) \geq \eps.\ee
In fact, this would remain true if we replaced $I_{\cP_{\lambda_A}} /
\dim \cP_{\lambda_A}$ with any normalized state.

Define $\rho_0 = \sum_{i=1}^{d_A}\bl_{A,i} \ket{i}\bra{i}$ and let $dU$
denote a Haar measure on $\cU_{d_A}$.  By Schur's Lemma, averaging
$\bq_{\lambda_A}^{d_A}(U\rho_0 U^\dag)$ over $dU$ gives a matrix
proportional to the identity. 
To obtain the proportionality factor, we use \eq{schur-proj} to bound
\be\beta := \tr \Pi_{\lambda_A} \rho_0^{\ot n} \Pi_{\lambda_A} =
\tr\bq_{\lambda_A}^{d_A}(\rho_0) \cdot \dim\cP_{\lambda_A} \geq
(n+d)^{-d(d+1)/2}.\ee
Upon averaging, we then find that
\be \beta\frac{I_{\cQ_{\lambda_A}^{d_A}}}{\dim \cQ_{\lambda_A}^{d_A}} =
\int dU \bq_{\lambda_A}^{d_A}(U\rho_0 U^\dag) \cdot \dim\cP_{\lambda_A}\ee
Now $\omega_A \leq I_{\cQ_{\lambda_A}^{d_A}}$, so $\cN^{\ot n}\l(\oprod{\lambda_A} \ot \omega_A \ot
\frac{I_{\cP_{\lambda_A}}}{\dim \cP_{\lambda_A}}\r) \leq
\cN^{\ot n}\l(\oprod{\lambda_A} \ot I_{\cQ_{\lambda_A}^{d_A}} \ot
\frac{I_{\cP_{\lambda_A}}}{\dim \cP_{\lambda_A}}\r)$ and
\bea \eps &\leq &
\tr \cN^{\ot n}\l(\oprod{\lambda_A} \ot \omega_A \ot
\frac{I_{\cP_{\lambda_A}}}{\dim \cP_{\lambda_A}}\r)
(\Pi_{\lambda_B}\ot \Pi_{\lambda_E}) \\
&\leq &
\tr \cN^{\ot n}\l(\oprod{\lambda_A} \ot I_{\cQ_{\lambda_A}^{d_A}} \ot
\frac{I_{\cP_{\lambda_A}}}{\dim \cP_{\lambda_A}}\r)
(\Pi_{\lambda_B}\ot \Pi_{\lambda_E}) \\
& = & \frac{\dim \cQ_{\lambda_A}^{d_A}}{\beta}
\int dU \tr (\cN^{\ot n}
(\Pi_{\lambda_A}(U\rho_0U^\dag)^{\ot n}\Pi_{\lambda_A}))
(\Pi_{\lambda_B}\ot \Pi_{\lambda_E})\\
&\leq & \max_U \frac{\dim \cQ_{\lambda_A}^{d_A}}{\beta}
 \tr (\cN^{\ot n}
(\Pi_{\lambda_A}(U\rho_0U^\dag)^{\ot n}\Pi_{\lambda_A}))
(\Pi_{\lambda_B}\ot \Pi_{\lambda_E}).
\eea
In the last step we have used the fact that $\int dU=1$ so that $\int
dU f(U) \leq \max_U f(U)$ for any function on $\cU_{d_A}$.
Therefore $\exists \rho=U\rho_0U^\dag$ with $\psi^{ABE}=(I^A \ot
U_\cN^{A'\ra BE})\ket{\Phi_\rho}^{AA'}$ such that $\tr
(\Pi_{\lambda_A}^A \ot \Pi_{\lambda_B}^B \ot \Pi_{\lambda_E}^E)
\psi^{\ot n} \geq \eps\beta/\dim\cQ_{\lambda_A}^{d_A} \geq \eps(n+d)^{-d^2}
=: \eps'$.

This means that $(\bl_A,\bl_B,\bl_E)\in\tilde{\cT}_\cN^n(\eps')$.
\end{proof}

\item $\tilde{\cT}_\cN^n(\eps) \Rightarrow {\cT}_\cN^*$
\begin{proof}
Again, we are given $\psi^{ABE}\in R(\cN)$ and a triple
$(\lambda_A,\lambda_B,\lambda_E)$ s.t. $\tr
(\Pi_{\lambda_A}^A \ot \Pi_{\lambda_B}^B \ot \Pi_{\lambda_E}^E)
\psi^{\ot n} \geq \eps$.  And again we define
$\Pr(\bl_A,\bl_B,\bl_E) := \tr (\Pi_{\lambda_A}^A \ot
\Pi_{\lambda_B}^B \ot \Pi_{\lambda_E}^E) \psi^{\ot n}$.
Now let $\delta := \max_{X\in\{A,B,E\}} \frac{1}{2}\| \bl_X - \spec
\psi^X \|_1$ and use \eq{typ-proj} to bound
\be \eps \leq \Pr(\bl_A,\bl_B,\bl_E)
\leq (n+d)^{d(d-1)/2} \exp(-n\delta^2).\ee
Thus $(r_A,r_B,r_E)=(\spec \psi^A,\spec \psi^B, \spec
\psi^E)\in \cT_\cN^*$ and satisfies
$\frac{1}{2}\|r_A-\bl_A\|_1\leq \delta$,
$\frac{1}{2}\|r_B-\bl_B\|_1\leq \delta$ and
$\frac{1}{2}\|r_E-\bl_E\|_1\leq \delta$ for $\delta$ s.t.
\be \delta^2 \leq
\frac{\binom{d}{2}\log(n+d) + \log 1/\eps}{n}.\ee
\end{proof}
\end{itemize}

The preceding set of proofs establishes more than will usually be
necessary.  The main conclusion to draw from this section is that one
can project onto triples $(\bl_A,\bl_B,\bl_E)$ that are
all within $\delta$ of triples in $\cT_\cN^*$ while disturbing the
state by no more than $\poly(n)\exp(-n\delta^2)$.

\subsection{Conclusions}
The results of this chapter should be thought of laying the groundwork
for a quantum analogue of joint types.  Although many coding theorems
have been proved for noisy states and channels without using this
formalism, hopefully joint quantum types will give proofs
that are simpler, more powerful, or not feasible by other means.  One
problem for which the technique seems promising is the Quantum Reverse
Shannon Theorem\cite{BDHSW05}, in which it gives a relatively simple
method for efficiently simulating a noisy quantum channel on arbitrary
sources. It remains to be seen where else the techniques will be useful.

\chapter{Efficient circuits for the Schur transform}
\label{chap:sch-algo}

The previous chapter showed how the Schur transform is a vital
ingredient in a wide variety of coding theorems of quantum information
theory.  However, for these protocols to be of practical value, an
efficient (i.e. polynomial time) implementation of the Schur transform
will be necessary. 

The goal of performing classical coding tasks in polynomial or even
linear time has long been studied, but quantum information theory has
typically ignored questions of efficiency.  For example, random coding
results (such as \cite{Holevo98,SW97,BHLSW03,DW03c}) require an
exponential number of bits to describe, and like classical random
coding techniques, do not yield efficient algorithms.
There are a few important exceptions.  Some quantum coding tasks, such
as Schumacher compression\cite{Sch95,JS94}, are essentially equivalent
to classical circuits, and as such can be performed
efficiently on a quantum computer by carefully modifying an efficient
classical algorithm to run reversibly and deal properly with ancilla
systems\cite{Cleve:96a}.  Another example, which illustrates some of
the challenges involved, is \mscite{Kaye01}'s efficient implementation
of entanglement concentration\cite{BBPS96}.  Quantum key
distribution\cite{BB84} not only runs efficiently, but can be
implemented with entirely, or almost entirely, single-qubit operations
and classical computation.  Fault tolerance\cite{Shor96} usually seeks
to perform error correction with as few gates as possible, although
using teleportation-based techniques\cite{Got99,Knill:04a}
computational efficiency may not be quite as critical to the threshold
rate.  Finally, some randomized quantum code
constructions have been given efficient constructions using classical
derandomization techniques in 
\cite{AS04}.  Our efficient construction of the Schur transform
adds to this list a powerful new tool for finding algorithms that
implement quantum
communication tasks.

From a broader perspective, the transforms involved in
quantum information protocols are important because they show a
connection between a quantum problem with structure and transforms
of quantum information which exploit this structure.  The theory
of quantum algorithms has languished relative to the tremendous
progress in quantum information theory due in large part to a lack
of exactly this type of construction: transforms with
interpretations. When we say a quantum algorithm is simply a
change of basis, we are doing a disservice to the fact that
efficient quantum algorithms must have efficient quantum circuits.
In the nonabelian hidden subgroup problem, for example, it is
known that there is a transform which solves the problem, but
there is no known efficient quantum circuit for this
transform\cite{Ettinger:99a}. There is great impetus, therefore,
to construct efficient quantum circuits for transforms of quantum
information where the transform exploits some structure of the
problem.

We begin in \sect{gz} by describing explicit bases (known as {\em
subgroup-adapted bases}) for the irreps of
the unitary and symmetric groups.  In \sect{schur-circuit}, we show
how these bases allow the Schur transform to be decomposed into a
series of CG transforms and in \sect{cg-construct} we give an
efficient construction of a CG transform.  Together these three
sections comprise an efficient (i.e. running time polynomial in $n$,
$d$ and $\log 1/\epsilon$ for error $\epsilon$) algorithm for the
Schur transform.

\section{Subgroup-adapted bases for $\cQ^d_\lambda$ and $\cP_\lambda$}
\label{sec:gz}
To construct a quantum circuit for
the Schur transform, we will need to explicitly specify the Schur
basis.  Since we want the Schur basis to be of the form
$\ket{\lambda,q,p}$, our task reduces to specifying orthonormal
bases for $\cQ_\lambda^d$ and $\cP_\lambda$.   We will call these bases
$Q_\lambda^d$ and $P_\lambda$, respectively.

We will choose $Q_\lambda^d$ and $P_\lambda$ to both be a type of
basis known as a {\em subgroup-adapted basis}.  In \sect{subgp-adapted}
we describe the general theory of subgroup-adapted bases, and in
\sect{gz-yy}, we will describe subgroup-adapted bases for
$\cQ_\lambda^d$ and $\cP_\lambda$.  As we will later see, 
these bases have a recursive structure that is naturally related to
the structure of the algorithms 
that work with them.  Here we will show how the bases
can be stored on a quantum computer with a small amount of padding,
and later in this chapter we will show how the subgroup-adapted
bases described here enable efficient implementations of
Clebsch-Gordan and Schur duality transforms.

\subsection{Subgroup Adapted Bases}\label{sec:subgp-adapted}

First we review the basic idea of a subgroup adapted basis.  We
assume that all groups we talk about are finite or compact Lie groups.
Suppose $(\br,V)$ is an irrep of a group $G$ and $H$ is a proper
subgroup of $G$.  We will construct a basis for $V$ via the
representations of $H$.

Begin by restricting the input of $\br$ to $H$ to obtain a
representation of $H$, which we call $(\br|_H,V\da_H)$.   Unlike $V$,
the $H$-representation $V\da_H$ may be reducible.  In fact, if we let
$(\br'_\alpha,V'_\alpha)$ denote the irreps of $H$, then $V\da_H$
will decompose under the action of $H$ as
\be V\da_H \stackrel{H}{\cong} \bigoplus_{\alpha\in\hat{H}}
V'_\alpha \ot \bbC^{n_\alpha}  \ee
 or equivalently, $\br|_H$ decomposes as
\begin{equation}
\br(h) = {\bf r}|_{H}(h) \cong \bigoplus_{\alpha\in\hat{H}}
{\bf r}'_\alpha(h) \ot I_{n_\alpha} \label{eq:restrict}
\end{equation}
where $\hat{H}$ runs over a complete set of inequivalent irreps of
$H$ and $n_\alpha$ is the {\em branching multiplicity} of the irrep
labeled by $\alpha$.  Note that since $\br$ is a unitary
representation, the subspaces corresponding to different irreps of
$H$ are orthogonal.  Thus, the problem of finding an orthonormal
basis for $V$ now reduces to
the problem of (1) finding an orthonormal basis for each irrep of $H$,
$V_\alpha'$ and (2) finding orthonormal bases for the multiplicity spaces
$\bbC^{n_\alpha}$.  The case when all the $n_\alpha$ are either 0 or 1
is known as {\em multiplicity-free branching}.  When this occurs, we
only need to determine which irreps occur in the decomposition of $V$,
and find bases for them.

Now consider a group ${G}$ along with a tower of subgroups
${G}={G}_1\supset {G}_2 \supset \dots
\supset {G}_{k-1} \supset {G}_{k}=\{e\}$ where $\{e\}$
is the trivial subgroup consisting of only the identity element.  For
each $\ccG_i$, denote its irreps by $V_\alpha^i$, for
$\alpha\in\hat{\ccG}_i$.   Any irrep $V_{\alpha_1}^1$ of
$\ccG={G}_1$ decomposes under restriction to $\ccG_2$ into
$\ccG_2$-irreps: say that $V_{\alpha_2}^2$ appears
$n_{\alpha_1,\alpha_2}$ times.  We can then look at these irreps of
${G}_2$, consider their restriction to $\ccG_3$ and decompose
them into different irreps of ${G}_3$. Carrying on in such a
manner down this tower of subgroups will yield a labeling for
subspaces corresponding to each of these restrictions.  Moreover, if
we choose orthonormal bases for the multiplicity spaces, this will
induce an orthonormal basis for $\ccG$. 
This basis is known as a {\em subgroup-adapted basis} and basis
vectors have the form
$\ket{\alpha_2,m_2,\alpha_3,m_3,\ldots,\alpha_k,m_k}$, where
$\ket{m_i}$ is a basis vector for the
($n_{\alpha_{i-1},\alpha_i}$-dimensional) multiplicity space of
$V_{\alpha_i}^i$ in $V_{\alpha_{i-1}}^{i-1}$.

If the branching for each $\ccG_{i+1}\subset \ccG_i$ is
multiplicity-free, then we say that the tower of subgroups is
{\em canonical}.  In this case, the subgroup adapted basis takes the
particularly simple form of $\ket{\alpha_2,\ldots,\alpha_k}$, where
each $\alpha_i\in\hat{\ccG_i}$ and $\alpha_{i+1}$ appears in the
decomposition of $V_{\alpha_i}\da_{\ccG_{i+1}}$.  Often we include
the original irrep label $\alpha=\alpha_1$ as well:
$|\alpha_1,\alpha_2,\dots,\alpha_k\rangle$.  This means that there
exists a basis whose 
vectors are completely determined (up to an arbitrary choice of phase)
by which irreps of $G_1,\ldots,G_k$ they transform according to.
Notice that a basis for
the irrep $V_\alpha$ does not consist of all possible irrep labels
$\alpha_i$, but instead only those which can appear under the
restriction which defines the basis.

The simple recursive structure of subgroup adapted bases makes them
well-suited to performing explicit computations.  Thus, for example,
subgroup adapted bases play a major role in efficient quantum circuits
for the Fourier transform over many nonabelian groups\cite{MRR03}.

\subsection{Explicit orthonormal bases for $\cQ_\lambda^d$ and
$\cP_\lambda$} \label{sec:gz-yy}

In this section we describe canonical towers of subgroups for $\cU_d$
and $\cS_n$, which give rise to subgroup-adapted bases for the irreps
$\cQ_\lambda^d$ and $\cP_\lambda$.  These bases go by many names: for
$\cU_d$ (and other Lie groups) the basis is called the Gel'fand-Zetlin
basis (following \cite{Gelfand:50a}) and we denote it by $Q_\lambda^d$,
while for $\cS_n$ it is called the Young-Yamanouchi basis, or
sometimes Young's orthogonal basis (see
\cite{JK81} for a good review of its properties) and is denoted
$P_\lambda$.  The
constructions and corresponding branching rules are quite simple, but
for proofs we again refer the reader to \cite{GW98}.

{\em The Gel'fand-Zetlin basis for $\cQ_\lambda^d$:}
For $\cU_d$, it turns out that the chain of subgroups $\{1\}=\cU_0\subset
\cU_1 \subset \ldots \subset \cU_{d-1} \subset \cU_d$ is a canonical
tower.  For $c<d$, the subgroup $\cU_c$ is embedded in $\cU_d$ by
$\cU_c := \{U\in\cU_d : U|i\> = |i\> \text{ for } i = c+1,\ldots,
d\}$.  In other words, it corresponds to matrices of the form
\be U\oplus I_{d-c} :=
\l(\begin{array}{c|c} U  & 0\\ \hline
\\ 0 & I_{d-c}\end{array}\r)
,\label{eq:Uc-embed}\ee
 where $U$ is a $c\times c$ unitary matrix.

Since the branching from $\cU_d$ to $\cU_{d-1}$ is multiplicity-free,
we obtain a subgroup-adapted basis $Q_\lambda^d$, which is known as
the Gel'fand-Zetlin (GZ) basis.  Our only free choice in a GZ
basis is the initial choice of basis $\ket{1},\ldots,\ket{d}$ for
$\bbC^d$ which determines the canonical tower of subgroups
$\cU_1\subset \ldots \subset \cU_d$.  Once we have chosen this basis,
specifying $Q_\lambda^d$ reduces to knowing which irreps
$\cQ_\mu^{d-1}$ appear in the decomposition of
$\cQ_\lambda^d\da_{\cU_{d-1}}$.  Recall that the irreps of $\cU_d$ are
labeled by elements of $\cI_{d,n}$ with $n$ arbitrary.  This set can
be denoted by $\bbZ_{++}^d := \cup_n
\cI_{d,n} = \{\lambda\in\bbZ^d :
\lambda_1 \geq \ldots \geq \lambda_d \geq 0\}$.  For
$\mu\in\bbZ_{++}^{d-1}, \lambda\in\bbZ_{++}^d$, we say that $\mu$ {\em
interlaces} $\lambda$ and write $\mu\interlaces\lambda$ whenever $\lambda_1
\geq \mu_1 \geq \lambda_2 \ldots \geq \lambda_{d-1} \geq \mu_{d-1}
\geq \lambda_d$.  In terms of Young diagrams, this means that $\mu$
is a valid partition (i.e. a nonnegative, nonincreasing sequence)
obtained from removing zero or one boxes from each column of
$\lambda$.  For example, if $\lambda=(4,3,1,1)$ (as in
\eq{ferrers-example}), then $\mu\interlaces\lambda$ can be obtained by
removing any subset of the marked boxes below, although if the box
marked $*$ on the second line is removed, then the other marked box on
the line must also be removed.
\be \young(\hfil\hfil\hfil\times,\hfil *\times,\hfil,\times)\ee

Thus a basis vector in $Q_\lambda^d$ corresponds to a sequence of
partitions
$q=(q_d,\ldots, q_1)$ such that $q_d=\lambda$,
$q_1\interlaces q_2 \interlaces \ldots \interlaces q_d$ and
$q_j\in\bbZ^j_{++}$ for $j=1,\ldots,d$.
Again using $\lambda=(4,3,1,1)$ as an example, and choosing $d=5$ (any
$d\geq 4$ is possible), we might have the sequence
\be \begin{array}{cccccccccc}
\Yvcentermath1
\yng(4,3,1,1)& \reverselaces& \Yvcentermath1\yng(3,3,1) &\reverselaces &
\Yvcentermath1\yng(3,3,1)&
\reverselaces& \Yvcentermath1\yng(3,1)& \reverselaces & \Yvcentermath1\yng(2)\\
q_5&&q_4&&q_3&&q_2&&q_1
\end{array} \label{eq:ex-partition-chain}\ee
Observe that it is possible in some steps not to remove any boxes, as
long as $q_j$ has no more than $j$ rows.

In order to work with the Gel'fand-Zetlin basis vectors on a quantum
computer, we will need an efficient way to write them down.
Typically, we think of $d$ as constant and express our resource use in
terms of $n$.  Then an element of $\cI_{d,n}$ can be expressed with
$d\log(n+1)$ bits, since it consists of $d$ integers between $0$ and
$n$.  (This is a crude upper bound on $|\cI_{d,n}|=\binom{n+d-1}{d-1}$,
but for constant $d$ it is good enough for our purposes.)  A
Gel'fand-Zetlin basis vector then requires no more than $d^2\log(n+1)$
bits, since it can be expressed as $d$ partitions of integers no
greater than $n$ into $\leq d$ parts.  (Here we assume that all
partitions have arisen from a decomposition of $(\bbC^d)^{\ot n}$, so
that no Young diagram has more than $n$ boxes.)  Unless otherwise
specified, our algorithms will use this encoding of the GZ basis
vectors.

It is also possible to express GZ basis vectors in a more visually
appealing way by writing numbers in the boxes of a Young diagram.  If
$q_1\interlaces \ldots \interlaces q_d$ is a chain of partitions, then
we write the number $j$ in each box contained in $q_j$ but not
$q_{j-1}$ (with $q_0=(0)$).  For example, the sequence in
\eq{ex-partition-chain} would be denoted
\be \young(1125,233,3,5).\ee
Equivalently, any method of filling a Young diagram with numbers from
$1,\ldots,d$ corresponds to a valid chain of irreps as long as the
numbers are nondecreasing from left to right and are strictly
increasing from top to bottom.  This gives another way of encoding a
GZ basis vector; this time using $n\log d$ bits.  
(In fact, we have an exact formula for $\dim\cQ_\lambda^d$
(\eq{cQ-dim}) and later in this section we will give an algorithm for
efficiently encoding a GZ basis vector in the optimal $\lceil \log
\dim\cQ_\lambda^d\rceil$ qubits.  However, this is not necessary for
most applications.)

{\em Example: irreps of $\cU_2$:}  To ground the above discussion in
an example more familiar to physicists, we show how the GZ basis for
$\cU_2$ irreps corresponds to states of definite angular momentum
along one axis.  An irrep of $\cU_2$ is labeled by two integers
$(\lambda_1,\lambda_2)$ such that $\lambda_1+\lambda_2=n$ and
$\lambda_1 \geq \lambda_2 \geq 0$.  A GZ basis vector for
$\cQ_\lambda^2$ has $\lambda_2+m$ 1's in the first row, followed by
$\lambda_1-(\lambda_2+m)$ 2's in the first row and $\lambda_2$ 2's in
the second row, where $m$ ranges from 0 to $\lambda_1-\lambda_2$.
  This arrangement is necessary to satisfy
the constraint that numbers are strictly increasing from top to bottom
and are nondecreasing 
from left to right.  Since the GZ basis vectors are completely specified
by $m$, we can label the vector $\ket{(\lambda_1,\lambda_2) ;
(\lambda_2 + m)}\in 
Q_\lambda^2$ simply by $\ket{m}$.  For example, $\lambda=(9,4)$ and $m=2$ would
look like
\be \young(111111222,2222).\ee

Now observe that $\dim\cQ_\lambda^2 = \lambda_1 - \lambda_2 +1$, a
fact which is consistent with having angular momentum
$J=(\lambda_1-\lambda_2)/2$.  We claim that $m$ corresponds to the $Z$
component of angular momentum (specifically, the $Z$ component of
angular momentum is $m-J = m - (\lambda_1-\lambda_2)/2$).  To see
this, first note that $\cU_1$ acts 
on a GZ basis vector $\ket{m}$ according to the representation $x \ra
x^{\lambda_2 + m}$, for $x\in\cU_1$; equivalently
$\bq_\lambda^2\l(\l(\begin{smallmatrix}x&0\\0&1\end{smallmatrix}\r)\r)\ket{m}
= x^{\lambda_2+m}\ket{m}$. Since $\bq_\lambda^2(yI_2)\ket{m} =
y^n\ket{m} = y^{\lambda_1+\lambda_2}\ket{m}$, we can find the action
of $e^{i\theta\sigma_z} = \l(\begin{smallmatrix}e^{2i\theta}&0\\0 &
1\end{smallmatrix} \r)
\l(\begin{smallmatrix}e^{-i\theta} & 0 \\ 0 &
e^{-i\theta}\end{smallmatrix}\r)$ on 
$\ket{m}$.  We do this by combining the above arguments to find that
$\bq_\lambda^2(e^{i\theta\sigma_z})\ket{m} =
e^{2i\theta(\lambda_2+m)}e^{-i\theta(\lambda_1+\lambda_2)}\ket{m} =
e^{2i\theta(m-J)}\ket{m}$.  Thus we obtain the desired action of
a $Z$ rotation on a particle with total angular momentum $J$ and
$Z$-component of angular momentum $m$.

{\em Example: The defining irrep of $\cU_d$:}  The simplest
nontrivial irrep of $\cU_d$ is its action on $\bbC^d$.  This
corresponds to the partition $(1)$, so we say that
$(\bq_{(1)}^d,\cQ_{(1)}^d)$ is the {\em defining irrep} of $\cU_d$
with $\cQ_{(1)}^d = \bbC^d$ and $\bq_{(1)}^d(U) = U$.
Let $\ket{1},\ldots,\ket{d}$ be an orthonormal basis for $\bbC^d$
corresponding to the canonical tower of subgroups $\cU_1\subset \cdots
\subset \cU_d$.  It turns out that this is already a GZ basis.  To see
this, note
that $\cQ_{(1)}^d\da_{\cU_{d-1}}\stackrel{\cU_{d-1}}{\cong}
\cQ_{(0)}^{d-1}
\oplus \cQ_{(1)}^{d-1}$.  This is because $\ket{d}$ generates
$\cQ_{(0)}^{d-1}$, a trivial irrep of $\cU_{d-1}$; and
$\ket{1},\ldots,\ket{d-1}$ generate $\cQ_{(1)}^{d-1}$, a defining
irrep of $\cU_{d-1}$.  Another way to say this is that $\ket{j}$ is
acted on according to the trivial irrep of $\cU_1,\ldots,\cU_{j-1}$
and according to the defining irrep of $\cU_j,\ldots,\cU_d$.  Thus
$\ket{j}$ corresponds to the chain of partitions
$\{(0)^{j-1},(1)^{d-j+1}\}$.  We will return to this example several
times in the rest of the chapter.

{\em The Young-Yamanouchi basis for $\cP_\lambda$:}
The situation for $\cS_n$ is quite similar.  Our chain of subgroups is
$\{e\} = \cS_1 \subset \cS_2 \subset \ldots \subset \cS_n$, where for
$m<n$ we define $\cS_m\subset \cS_n$ to be the permutations in $\cS_n$
which leave the last $n-m$ elements fixed.  For example, if $n=3$,
then ${\mathcal S}_3=\{e,(12),(23),(13),(123),(321)\}$, ${\mathcal
S}_2=\{e,(12)\}$, and ${\mathcal S}_1=\{e\}$.  Recall that the
irreps of ${\mathcal S}_n$ can be labeled by $\cI_n=\cI_{n,n}$: the
partitions of $n$ into $\leq n$ parts.

Again, the branching from $\cS_n$ to $\cS_{n-1}$
is multiplicity-free, so to determine an orthonormal basis $P_\lambda$
for the space $\cP_\lambda$ we need only know which irreps occur in the
decomposition of $\cP_\lambda\da_{\cS_{n-1}}$.  It turns out that the
branching rule is given by finding all ways to remove one box from
$\lambda$ while leaving a valid partition.  Denote the set of such
partitions by $\lambda-\Box$.  Formally, $\lambda-\Box := \cI_n \cap
\{\lambda-e_j : j=1,\ldots,n\}$, where we recall that $e_j$ is the
unit vector in 
$\bbZ^n$ with a one in the $j^{\text{th}}$ position and zeroes
elsewhere.  Thus, the general branching rule is
\be \cP_\lambda\da_{\cS_{n-1}} \iso{\cS_{n-1}}
\bigoplus_{\mu\in\lambda-\Box} \cP_\mu.
\label{eq:cP-branching}\ee
For example, if $\lambda=(3,2,1)$, we might have the chain
of partitions:
\be \begin{array}{cccccccccccc}
\Yvcentermath1\yng(3,2,1) & \ra & \Yvcentermath1\yng(2,2,1) & \ra &
\Yvcentermath1\yng(2,2) & \ra & \Yvcentermath1\yng(2,1) & \ra &
\Yvcentermath1\yng(1,1) & \ra & \Yvcentermath1\yng(1)\\
n=6 && n=5 && n=4 && n=3 && n=2 && n=1\end{array}\ee
Again, we can concisely label this chain by writing the number $j$ in
the box that is removed when restricting from $\cS_j$ to $\cS_{j-1}$.
The above example would then be
\be \young(136,24,5).\ee
Note that the valid methods of
filling a Young diagram are slightly different than for the $\cU_d$
case.  Now we use each integer in $1,\ldots,n$ exactly once such that
the numbers are increasing from left to right and from top to bottom.
(The same filling scheme appeared in the description of Young's
natural representation in \sect{weyl-symmetrizer}, but the resulting
basis states are of course quite different.)

This gives rise to a straightforward, but inefficient, method of
writing an element of $P_\lambda$ using $\log n!$ bits.  However, for
applications such as data compression\cite{Hayashi:02b,Hayashi:02c} we
will need an encoding which gives us closer to the optimal $\log
P_\lambda$ bits.  First recall that \eq{cP-dim} gives an exact (and
efficiently computable) expression for
$|P_\lambda|=\dim\cP_\lambda$. Now we would like to efficiently and
reversibly map an element of $P_\lambda$ (thought of as a chain of
partitions $p=(p_n=\lambda,\ldots,p_1=(1))\in P_\lambda$, with $p_j
\in p_{j+1}-\Box$) to an integer in $[|P_\lambda|] :=
\{1,\ldots,|P_\lambda|\}$.  We will construct
this bijection $f_n:P_\lambda \ra [|P_\lambda|]$ by defining an ordering on
$P_\lambda$ and setting $f_n(p):=|\{p'\in P_\lambda : p'\leq p\}|$.
First fix an arbitrary, but easily computable, 
(total) ordering on partitions in $\cI_n$ for each $n$; for example,
lexicographical order.  This
induces an ordering on $P_\lambda$ if we rank a basis vector $p\in
P_\lambda$ first according to $p_{n-1}$, using the order on partitions
we have chosen, then according to $p_{n-2}$ and so on.  We skip $p_n$,
since it is always equal to $\lambda$.  In other words, for $p,p'\in
P_\lambda$, $p>p'$ if $p_{n-1}>p_{n-1}'$ {\em or} $p_{n-1}=p_{n-1}'$
and $p_{n-2}>p_{n-2}'$ {\em or} $p_{n-1}=p_{n-1}'$,
$p_{n-2}=p_{n-2}'$ and $p_{n-3}>p_{n-3}'$, and so on.
Thus $f_n:P_\lambda \ra [|P_\lambda|]$ can be easily verified to be
\be f_n(p) = f_n(p_1,\ldots,p_n) :=
1 + \sum_{k=2}^n\sum_{\substack{\mu\in p_k-\Box\\\mu<p_{k-1}}}
\dim \cP_\mu.
\label{eq:Plambda-relabel}\ee 
Thus $f_n$ is an injective map from $P_\lambda$
to $[|P_\lambda|]$.
Moreover, since there are $O(n^2)$ terms in \eq{Plambda-relabel} and
\eq{cP-dim} gives an efficient way to calculate each
$|P_\lambda|$, this mapping can be performed in time polynomial in $n$.

Of course, the same techniques could be used to efficiently write an
element of $Q_\lambda^d$ in $\lceil \log |Q_\lambda^d| \rceil$ bits,
but unless $d$ is large this usually is not necessary.

\section{Constructing the Schur transform from a series of
Clebsch-Gordan transforms}\label{sec:schur-circuit}

In this section, we will show how the Schur transform on $\nqudits$
can be reduced to a series of CG transforms on $\cU_d$.  The argument
is divided into two parts.  First, we give the theoretical
underpinnings in \sect{cg-Ud} by using Schur duality to relate the $\cU_d$ CG
transform to branching in $\cS_n$.  Then we show how the actual
algorithm works in \sect{schurcg}.

\subsection{Branching rules and Clebsch-Gordan series for $\cU_d$}
\label{sec:cg-Ud}

Recall that CG transform for $\cU_d$ is given by
\be \cQ_\mu^d \ot \cQ_\nu^d \iso{\cU_d}
\bigoplus_{\lambda\in\dom} \cQ_\lambda^d \ot
\bbC^{M_{\mu\nu}^\lambda}.
\label{eq:Ud-CG}\ee
For now, we will work with Littlewood-Richardson coefficients
$M_{\mu\nu}^\lambda$ rather than the more structured space
$\Hom(\cQ_\lambda^d, \cQ_\mu^d \ot\cQ_\nu^d)^{\cU_d}$.
The partitions $\lambda$ appearing on the RHS of \eq{Ud-CG} are
sometimes known as the {\em Clebsch-Gordan series}.  In this section,
we will show (following \cite{GW98}) how the $\cU_d$ Clebsch-Gordan
series is related to the behavior of $\cS_n$ irreps under
restriction.

For integers $k,n$ with $1\leq k\leq n$, embed $\cS_k\times \cS_{n-k}$
as a subgroup of $\cS_n$ in the natural way; as permutations that
leave the sets $\{1,\ldots,k\}$ and $\{k+1,\ldots,n\}$ invariant.  The
irreps of $\cS_k\times \cS_{n-k}$ are $\cP_\mu \hat{\ot} \cP_\nu$,
where $\mu\in\cI_k$ and $\nu\in\cI_{n-k}$.

Under restriction to $\cS_k\times \cS_{n-k}\subset \cS_n$,
the $\cS_n$-irrep $\cP_\lambda$ decomposes as
\be \cP_\lambda \iso{\cS_k\times \cS_{n-k}}
\bigoplus_{\mu\in\cI_k}\bigoplus_{\nu\in\cI_{n-k}}
\cP_\mu \hat{\ot}\cP_\nu \ot \bbC^{N_{\mu\nu}^\lambda},
\label{eq:Sn-restriction}\ee
for some multiplicities $N_{\mu\nu}^\lambda$ (possibly zero).

\begin{claim}
$M_{\mu\nu}^\lambda = N_{\mu\nu}^\lambda$.
\end{claim}
As a corollary, $M_{\mu\nu}^\lambda$ is only nonzero when $|\lambda| =
|\mu| + |\nu|$.
\begin{proof}
Consider the action of $\cS_k\times \cS_{n-k}\times\cU_d$ on
$\nqudits$.  On the one hand, \eq{Sn-restriction} gives
\be \nqudits \iso{\cS_n\times \cU_d}
\bigoplus_{\lambda\in\cI_{d,n}} \cP_\lambda \hat{\ot} \cQ_\lambda^d
\iso{\cS_k\times \cS_{n-k}}
\bigoplus_{\substack{\mu\in\cI_{d,k},\nu\in\cI_{d,n-k}\\
\lambda\in\cI_{d,n}}} \cP_\mu \hat{\ot}\cP_\nu\hat{\ot}\cQ_\lambda^d
\ot \bbC^{N_{\mu\nu}^{\lambda}}.\label{eq:LR-decomp1}\ee
On the other hand, we can apply \eq{Ud-CG} to obtain
\be \nqudits \cong (\bbC^d)^{\ot k} \ot (\bbC^d)^{\ot n-k}
\iso{\cS_k\times \cS_{n-k}}\!\!\!\!
\bigoplus_{\substack{\mu\in\cI_{d,k}\\\nu\in\cI_{d,n-k}}}
(\cP_\mu \ot \cQ_\mu^d)\hat{\ot} (\cP_\nu \ot \cQ_\nu^d)
 \iso{\cU_d}\!\!\!\!\!
\bigoplus_{\substack{\mu\in\cI_{d,k},\nu\in\cI_{d,n-k}\\
\lambda\in\dom}}\!\!\! \cP_\mu \hat{\ot}\cP_\nu \hat{\ot} \cQ_\lambda^d
\ot \bbC^{M_{\mu\nu}^{\lambda}}.\label{eq:LR-decomp2}\ee
Equating \eqs{LR-decomp1}{LR-decomp2} proves the desired equality.  
\end{proof}

This means that the branching rules of $\cS_n$ determine the CG series
for $\cU_d$.\footnote{We can similarly obtain the CG series for $\cS_n$ by
studying the branching from $\cU_{d_1d_2}$ to
$\cU_{d_1}\ot\cU_{d_2}$.  This is a useful tool for studying the
relation between spectra of a bipartite density matrix $\rho^{AB}$ and
of the reduced density matrices $\rho^A$ and
$\rho^B$\cite{CM04,Klyachko04}.} In particular, suppose $k=n-1$.  Then
$\cS_1$ is the trivial group, so restricting to $\cS_{n-1}\times
\cS_{1}$ is equivalent to simply restricting to $\cS_{n-1}$.
According to the branching rule stated in \eq{cP-branching}, this
means that $M_{\lambda,(1)}^{\lambda'}$ is one if $\lambda\in\lambda'
- \Box$ and zero otherwise.  In other words, for the case when one
irrep is the defining irrep, the CG series is
\be \cQ_\lambda^d \ot \cQ_{(1)}^d \cong 
\bigoplus_{\lambda'\in\lambda + \Box} \cQ_{\lambda'}^d.
\label{eq:cg-add-one}\ee
Here $\lambda+\Box$ denotes the set of valid Young diagrams obtained
by {\em adding} one box to $\lambda$.

For example if $\lambda=(3,2,1)$ then
\begin{equation}
\cQ_{(3,2,1)}^3 \otimes \cQ_{(1)}^3 \stackrel{\cU_3}{\cong}
\cQ_{(4,2,1)}^3 \oplus \cQ_{(3,3,1)}^3 \oplus \cQ_{(3,2,2)}^3
\end{equation}
or in Young diagram form
\begin{equation}
\Yvcentermath1 \yng(3,2,1) \otimes \yng(1)
\stackrel{\cU_3}{\cong}
 \yng(4,2,1) \oplus \yng(3,3,1)\oplus \yng(3,2,2)
\end{equation}
Note that if we had $d>3$, then the partition $(3,2,1,1)$ would also
appear.

We now seek to define the CG transform as a quantum circuit.  We
specialize to the case where one of the input irreps is the defining
irrep, but allow the other irrep to be specified by a quantum input.
The resulting CG  transform is defined as:
\be \Ucg = \sum_{\lambda\in\bbZ_{++}^d} \oprod{\lambda}
\ot \Ucg^{\lambda,(1)}.\ee
This takes as input a state of the form $\ket{\lambda}\ket{q}\ket{i}$,
for $\lambda\in\bbZ_{++}^d$, $\ket{q}\in Q_\lambda^d$ and $i\in [d]$.
The output is a superposition over vectors
$\ket{\lambda}\ket{\lambda'}\ket{q'}$, where $\lambda'=\lambda+e_j\in
\bbZ_{++}^d$, $j\in [d]$ and $\ket{q'}\in Q_{\lambda'}^d$.
Equivalently, we could output $\ket{\lambda}\ket{j}\ket{q'}$ or
$\ket{j}\ket{\lambda'}\ket{q'}$, since $(\lambda,\lambda')$,
$(\lambda,j)$ and $(\lambda',j)$ are all trivially related via
reversible classical circuits.

To better understand the input space of $\Ucg$, we introduce the {\em
model representation} $\cQ_*^d :=
\bigoplus_{\lambda\in\bbZ_{++}^d} \cQ_\lambda^d$, with corresponding
matrix $\bq_*^d(U) = \sum_\lambda \oprod{\lambda}\ot
\bq_\lambda^d(U)$.  The model representation (also sometimes called
the {\em Schwinger representation}) is infinite dimensional
and contains each irrep once.\footnote{By contrast, $L^2(\cU_d)$,
which we will not use, contains
$\cQ_\lambda^d$ with multiplicity $\dim\cQ_\lambda^d$.}  Its basis
vectors are of the form $\ket{\lambda,q}$ for $\lambda\in\bbZ_{++}^d$
and $\ket{q}\in Q_\lambda^d$.  Since $\cQ_*^d$ is
infinite-dimensional, we cannot store it on
a quantum computer and in this thesis work only with representations
$\cQ_\lambda^d$ with $|\lambda|\leq n$; nevertheless $\cQ_*^d$ is a
useful abstraction.

Thus $\Ucg$ decomposes $\cQ_*^d \ot \cQ_{(1)}^d$ into irreps.  There
are two important things to notice about this version of the CG
transform.  First is that it operates simultaneously on different
input irreps.  Second is that different input irreps must remain
orthogonal, so in order to to maintain unitarity $\Ucg$ needs to keep
the information of which irrep we started with.  However, since
$\lambda' = \lambda + e_j$, this information requires only storing
some $j\in [d]$.  Thus, $\Ucg$ is a map from $\cQ_*^d \ot \bbC^d$ to
$\cQ_*^d \ot \bbC^d$, where the $\bbC^d$ in the input is the defining
representation and the $\bbC^d$ in the output tracks which irrep we
started with.

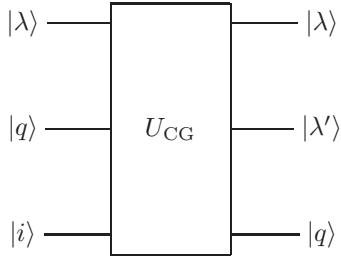
\begin{figure}[ht]
\begin{centering}
\leavevmode\xymatrix{
|\lambda\> & \gnqubit{~~~\Ucg~~~}{dd}\w & |\lambda\>\w \\
|q\> & \gspace{~~~\Ucg~~~}\w & |\lambda'\>\w\\
|i\> & \gspace{~~~\Ucg~~~}\w & |q\>\w\\
}

\caption{Schematic of the Clebsch-Gordan transform.  Equivalently, we
  could replace either the $\lambda$ output or the $\lambda'$ output
  with $j$.}\label{fig:cg}
\end{centering}
\end{figure}

\subsection{Constructing the Schur Transform from Clebsch-Gordan
Transforms}\label{sec:schurcg}

We now describe how to construct the Schur transform out of a
series of Clebsch-Gordan transforms.  Suppose we start with an input
vector $\ket{i_1,\ldots,i_n} \in (\bbC^d)^{\ot n}$, corresponding to
the $\cU_d$-representation $(\cQ_{(1)}^d)^{\ot n}$.  According to
Schur duality (\eq{schur-decomp}), to perform the Schur transform it
suffices to
decompose $(\cQ_{(1)}^d)^{\ot n}$ into $\cU_d$-irreps.  This is
because Schur duality means that the multiplicity space of
$\cQ_\lambda^d$ must be isomorphic to $\cP_\lambda$.  In other words,
if we show that
\be (\cQ_{(1)}^d)^{\ot n} \stackrel{\cU_d}{\cong}
\bigoplus_{\lambda\in\bbZ_{++}^d}
\cQ_\lambda^d \ot \cP_\lambda',
\label{eq:pseudo-schur-decomp}\ee
then we must have $\cP_\lambda' \stackrel{\cS_n}{\cong} \cP_\lambda$
when $\lambda\in\cI_{d,n}$ and $\cP_\lambda' = \{0\}$ otherwise.

\begin{figure}[ht]
\begin{centering}
\includegraphics{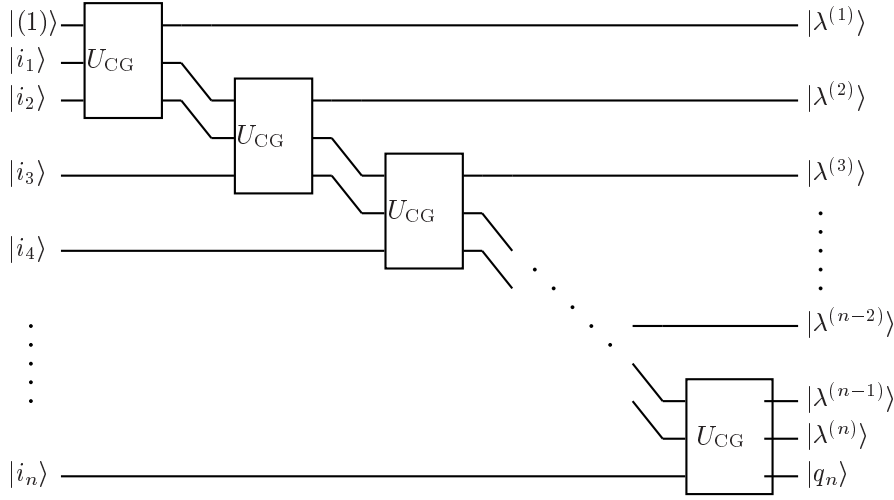}
\caption{Cascading Clebsch-Gordan transforms to produce the Schur transform.
Not shown are any ancilla inputs to the Clebsch-Gordan transforms.  The
structure of inputs and outputs of the Clebsch-Gordan transforms are the same
as in \fig{cg}.} \label{fig:cascade} 
\end{centering}
\end{figure}

To perform the $\cU_d$-irrep decomposition of
\eq{pseudo-schur-decomp}, we simply combine each of
$\ket{i_1},\ldots,\ket{i_n}$
using the CG transform, one at a time.  We start by inputting
$\ket{\lambda^{(1)}}=\ket{(1)}$, $\ket{i_1}$ and $\ket{i_2}$ into $\Ucg$
which outputs $\ket{\lambda^{(1)}}$ and a superposition of different values of
$\ket{\lambda^{(2)}}$ and $\ket{q_2}$.  Here $\lambda^{(2)}$ can be either
$(2,0)$ or $(1,1)$ and $\ket{q_2}\in Q_{\lambda^{(2)}}^d$.  Continuing, we
apply $\Ucg$ to $\ket{\lambda^{(2)}}\ket{q_2}\ket{i_3}$, and output a
superposition of vectors of the form
$\ket{\lambda^{(2)}}\ket{\lambda^{(3)}}\ket{q_3}$, with
$\lambda^{(3)}\in\cI_{d,3}$ and $\ket{q_3}\in Q_{\lambda^{(3)}}^d$. 
Each time we are combining an arbitrary irrep $\lambda^{(k)}$ and
an associated basis vector $\ket{q_k}\in Q_{\lambda^{(k)}}^d$, together
with a vector from the defining irrep $\ket{i_{k+1}}$.  This is repeated
for $k=1,\ldots,n-1$ and the resulting circuit is depicted in
\fig{cascade}.

Finally, we are left with a superposition of states of the form
$\ket{\lambda^{(1)},\ldots,\lambda^{(n)}}\ket{q_n}$, where $\ket{q_n}\in
Q_{\lambda^{(n)}}^d$, $\lambda^{(k)}\in\cI_{d,k}$ and each $\lambda^{(k)}$ is
obtained by adding a single box to $\lambda^{(k-1)}$; i.e. $\lambda^{(k)} =
\lambda^{(k-1)} + e_{j_k}$ for some $j_k\in [d]$.  If we define
$\lambda=\lambda^{(n)}$ and $\ket{q}=\ket{q_n}$, then we have the
decomposition of
\eq{pseudo-schur-decomp} with $\cP_\lambda'$ spanned by the vectors
$\ket{\lambda^{(1)},\ldots, \lambda^{(n-1)}}$ satisfying the
constraints described above.  But this is
precisely the Young-Yamanouchi basis $P_\lambda$ that we have defined
in \sect{gz}!  Since the first $k$ qudits transform under
$\cU_d$ according to $\cQ_{\lambda^{(k)}}^d$, Schur duality implies that
they also transform under $\cS_n$ according to $\cP_{\lambda^{(k)}}$.
Thus we set $\ket{p}=\ket{\lambda^{(1)},\ldots,\lambda^{(n-1)}}$ (optionally
compressing to $\lceil\log|P_\lambda|\rceil$ qubits using the
techniques described in the last section) and obtain the desired
$\ket{\lambda}\ket{q}\ket{p}$.  As a check on this result, note that
each $\lambda^{(k)}$ is invariant under $\bQ(\cU_d)$ since $U^{\ot n}$
acts on the first $k$ qubits simply as $U^{\ot k}$.

If we choose not to perform the $\poly(n)$ steps to optimally compress
$\ket{\lambda^{(1)},\ldots,\lambda^{(n-1)}}$, we could instead have
our circuit output the equivalent $\ket{j_1,\ldots,j_{n-1}}$, which
requires only $n\log d$ qubits and asymptotically no extra running
time.

We can now appreciate the similarity between the $\cU_d$ CG ``add a
box'' prescription and the $\cS_{n-1}\subset \cS_n$ branching rule of
``remove a box.''  Schur duality implies that the representations
$\cQ_{\lambda'}^d$ that are obtained by decomposing $\cQ_\lambda^d \ot
\cQ_{(1)}^d$ are the same as the $\cS_n$-irreps $\cP_{\lambda'}$ that
include $\cP_\lambda$ when restricted to $\cS_{n-1}$.

Define $T_{\text{CG}}(n,d,\eps)$ to be the time complexity (in terms
of number of gates) of performing a single $\cU_d$ CG transform to
accuracy $\eps$ on Young diagrams with $\leq n$ boxes.  Then the total
complexity for the Schur transform is $n\cdot
(T_{\text{CG}}(n,d,\eps/n)+O(1))$, possibly plus a $\poly(n)$ factor
for compressing the $\cP_\lambda$ register to
$\lceil\log\dim\cP_\lambda\rceil$ qubits (as is required for
applications such as data compression and entanglement concentration,
cf.~\sect{schur-qit-apps}).  In the next section we will show that
$T_{\text{CG}}(n,d,\eps)$ is $\poly(\log n,d,\log 1/\eps)$, but first
we give a step-by-step description of the algorithm for the Schur
transform.

\begin{tabbing}
~~~~ \= ~~~ \= ~~~ \= ~~~ \= ~~~\= \kill
\+{\bf Algorithm: Schur transform (plus optional compression)} \\
{\bf Inputs:} (1) Classical registers $d$ and $n$.
 (2) An $n$ qudit quantum register
$\ket{i_1,\ldots,i_n}$.\\ 
{\bf Outputs:} Quantum registers $\ket{\lambda}\ket{q}\ket{p}$, with
$\lambda\in\cI_{d,n}$, $q\in Q_\lambda^d$ and $p\in P_\lambda$.\\
\pushtabs {\bf Runtime:} \=
$n\cdot (T_{\text{CG}}(n,d,\eps/n)+O(1))$ to achieve
accuracy $\eps$.\\
\> (Optionally plus $\poly(n)$ to compress the $\cP_\lambda$ register
to $\lceil \log\dim\cP_\lambda\rceil$ qubits.)\\\poptabs
{\bf Procedure:}\\
{\bf 1.}\> Initialize $\ket{\lambda^{(1)}}:=\ket{(1)}$ and
$\ket{q_1}=\ket{i_1}$.\\
{\bf 2.}\> For $k=1,\ldots,n-1$:\\
{\bf 3.}\>\> Apply $\Ucg$ to
$\ket{\lambda^{(k)}}\ket{q_k}\ket{i_{k+1}}$ to obtain output
$\ket{j_k}\ket{\lambda^{(k+1)}}\ket{q_{k+1}}$, where
$\lambda^{(k+1)}=\lambda^{(k)} + e_{j_k}$.\\
{\bf 4.}\> Output  $\ket{\lambda}:=\ket{\lambda^{(n)}}$,
$\ket{q}:=\ket{q_n}$ and $\ket{p}:=\ket{j_1,\ldots,j_{n-1}}$.\\
{\bf 5.}\> (Optionally use \eq{Plambda-relabel} to reversibly map
$\ket{j_1,\ldots,j_{n-1}}$ to an integer $p\in [\dim\cP_\lambda]$.)
\end{tabbing}

This algorithm will be made efficient in the next
section, where we efficiently construct the CG transform for $\cU_d$,
proving that $T_{\text{CG}}(n,d,\eps) = \poly(\log n, d, \log
1/\eps)$.

\section{Efficient circuits for the Clebsch-Gordan transform}
\label{sec:cg-construct}
We now turn to the actual construction of the circuit for the Clebsch-Gordan
transform described in \sect{cg-Ud}.  To get a feel for the what
will be necessary, we start by giving a circuit for the CG transform
that is efficient when $d$ is constant; i.e. it has complexity
$n^{O(d^2)}$, which is $\poly(n)$ for any constant value of $d$.

First recall that $\dim\cQ_\lambda^d \leq (n+1)^{d^2}$.  Thus, controlled on
$\lambda$, we want to construct a unitary transform on a $D$-dimensional system
for $D=\max_{\lambda\in\cI_{d,n}}\dim\cQ_\lambda^d = \poly(n)$.  There are
classical algorithms\cite{Louck:70a} to compute matrix elements of $\Ucg$ to an
accuracy $\epsilon_1$ in time $\poly(D)\poly\log(1/\epsilon_1)$.  Once we have
calculated all the relevant matrix elements (of which there are only
polynomially many), we can (again in time $\poly(D)\poly\log(1/\epsilon)$)
decompose $\Ucg$ into $D^2\poly\log(D)$ elementary one and two-qubit
operations\cite{Shende:04a,Reck:94a,Barenco:95a,NC00}.  These can in
turn be approximated to accuracy $\epsilon_2$ by 
products of unitary operators from a fixed finite set (such as Clifford
operators and a $\pi/8$ rotation) with a further overhead of
$\poly\log(1/\epsilon_2)$\cite{Dawson05,Kitaev:02a}.  We can either assume
the relevant classical computations (such as decomposing the $D\times D$ matrix
into elementary gates) are performed coherently on a quantum computer, or as
part of a polynomial-time classical Turing machine which outputs the quantum
circuit. In any case, the total complexity is $\poly(n,\log 1/\epsilon)$ if the
desired final accuracy is $\epsilon$ and $d$ is held constant.

The goal of this section is to reduce this running time to $\poly(n,
d, \log(1/\epsilon))$; in fact, we will achieve circuits of size
$\poly(d,\log n, \log (1/\epsilon))$.  To do so, we will reduce the
$\cU_d$ CG transform to two components; first, a $\cU_{d-1}$ CG
transform, and second, a $d\times d$ unitary matrix whose entries can
be computed classically in $\poly(d,\log
n,1/\epsilon)$ steps.  After computing all $d^2$ entries, the second
component can then be implemented with $\poly(d, \log 1/\epsilon)$
gates according to the above arguments.

This reduction from the $\cU_d$ CG transform to the $\cU_{d-1}$ CG
transform is a special case of the Wigner-Eckart Theorem, which
we review in \sect{wigner}.  Then, following
\cite{Biedenharn:68a,Louck:70a}, we
use the Wigner-Eckart Theorem to give an efficient recursive
construction for $\Ucg$ in \sect{cg-recurse}.  Putting everything
together, we obtain a quantum circuit for the Schur transform that is
accurate to within $\epsilon$ and runs in time $n\cdot\poly(\log n, d, \log
1/\epsilon)$, optionally plus an additional $\poly(n)$ time to
compress the $\ket{p}$ register.

\subsection{The Wigner-Eckart Theorem and Clebsch-Gordan transform}
\label{sec:wigner}

In this section, we introduce the concept of an irreducible tensor operator,
which we use to state and prove the Wigner-Eckart Theorem.  Here we will
find that the CG transform is a key part of the Wigner-Eckart Theorem,
while in the next section we will turn this around and use the
Wigner-Eckart Theorem to give a recursive decomposition of the CG
transform.

Suppose $(\br_1,V_1)$ and $(\br_2,V_2)$ are representations of
$\cU_d$.  Recall that $\Hom(V_1,V_2)$ is a representation of $\cU_d$
under the map $T \ra \br_2(U) T \br_1(U)^{-1}$ for
$T\in\Hom(V_1,V_2)$.  If $\bmath{T}=\{T_1,T_2,\ldots\} \subset
\Hom(V_1,V_2)$ is a basis for a $\cU_d$-invariant subspace of
$\Hom(V_1,V_2)$, then we call $\bmath{T}$ a {\em tensor operator}.
Note that a tensor operator $\bT$ is a collection of operators
$\{T_i\}$ indexed by $i$, just as a tensor (or vector) is a collection
of scalars labeled by some index.  For example, the Pauli matrices
$\{\sigma_x,\sigma_y,\sigma_z\}\subset \Hom(\bbC^2,\bbC^2)$ comprise a
tensor operator, since conjugation by $\cU_2$ preserves the subspace
that they span.

Since $\Hom(V_1,V_2)$ is a representation of $\cU_d$, it can be
decomposed into irreps.  If $\bmath{T}$ is a basis for one of these
irreps, then we call it an {\em irreducible tensor operator}.  For
example, the Pauli matrices mentioned above comprise an irreducible
tensor operator, corresponding to the three-dimensional irrep
$\cQ_{(2)}^2$.  Formally, we say that $\bmath{T}^\nu =
\{T^\nu_{q_\nu}\}_{q_\nu \in Q_\nu^d}\subset \Hom(V_1,V_2)$ is an
irreducible tensor operator 
(corresponding to the irrep $\cQ_\nu^d$) if for all $U\in \cU_d$ we
have
\be 
\br_2(U) T^\nu_{q_\nu} \br_1(U)^{-1}
= \sum_{q'_\nu\in Q_\nu^d} \bra{q'_\nu}\bq_\nu^d(U)\ket{q_\nu}
T^\nu_{q'_\nu}. \label{eq:irred-tensor-op}\ee

Now assume that $V_1$ and $V_2$ are irreducible (say $V_1=\cQ_\mu^d$
and $V_2=\cQ_\lambda^d$), since if they are not, we could always
decompose $\Hom(V_1,V_2)$ into a direct sum of homomorphisms from an
irrep in $V_1$ to an irrep in $V_2$.  We can decompose
$\Hom(\cQ_\mu^d, \cQ_\lambda^d)$ into irreps using \eq{hom-decomp} and
the identity $\Hom(A,B) \cong A^* \ot B$ as follows:
\be\begin{split}
\Hom(\cQ_\mu^d, \cQ_\lambda^d) &\stackrel{\cU_d}{\cong}
\bigoplus_{\nu \in \bbZ_{++}^d} \cQ_\nu^d \ot
\Hom(\cQ_\nu^d, \Hom(\cQ_\mu^d, \cQ_\lambda^d))^{\cU_d} \\
&\stackrel{\cU_d}{\cong}
\bigoplus_{\nu \in \bbZ_{++}^d} \cQ_\nu^d \ot
\Hom(\cQ_\nu^d, (\cQ_\mu^d)^* \ot \cQ_\lambda^d)^{\cU_d} \\
&\stackrel{\cU_d}{\cong}
\bigoplus_{\nu \in \bbZ_{++}^d} \cQ_\nu^d \ot
\l((\cQ_\mu^d)^* \ot (\cQ_\nu^d)^* \ot \cQ_\lambda^d\r)^{\cU_d}\\
&\stackrel{\cU_d}{\cong}
\bigoplus_{\nu \in \bbZ_{++}^d} \cQ_\nu^d \ot
\Hom(\cQ_\mu^d \ot \cQ_\nu^d, \cQ_\lambda^d)^{\cU_d}
\end{split}\label{eq:hom-cg-decomp}\ee

Now consider a particular irreducible tensor operator $\bT^\nu
\subset \Hom(\cQ_\mu^d, \cQ_\lambda^d)$ with components
$T^\nu_{q_\nu}$ where $q_\nu$ ranges over $Q_\nu^d$.  We can define a
linear operator $\hat{T}: \cQ_\mu^d \ot \cQ_\nu^d \ra
\cQ_\lambda^d$ by letting
\be \hat{T} \ket{q_\mu}\ket{q_\nu} := T^\nu_{q_\nu}
\ket{q_\mu} \label{eq:Tnu-def}\ee
for all $q_\mu \in Q_\mu^d, q_\nu \in Q_\nu^d$ and extending it to the
rest of $\cQ_\mu^d \ot \cQ_\nu^d$ by linearity.
By construction, $\hat{T} \in \Hom(\cQ_\mu^d \ot \cQ_\nu^d,
\cQ_\lambda^d)$, but we claim that in addition $\hat{T}$ is
invariant under the action of $\cU_d$; i.e. that it lies in
$\Hom(\cQ_\mu^d \ot \cQ_\nu^d, \cQ_\lambda^d)^{\cU_d}$.  To see this,
apply \eqs{irred-tensor-op}{Tnu-def} to show that for any
$U\in\cU_d$, $q_\mu\in Q_\mu^d$ and $q_\nu\in Q_\nu^d$, we have
\be\begin{split}
\bq_\lambda^d(U) \hat{T}
\bigl[\bq_\mu^d(U)^{-1} \ot \bq_\nu^d(U)^{-1}\bigr]
\ket{q_\mu}\ket{q_\nu}
&= \sum_{q_\nu'\in Q_\nu^d}\bra{q'_\nu} \bq_\nu^d(U)^{-1}\ket{q_\nu}
\bq_\lambda^d(U) {T}^\nu_{q'_\nu} \bq_\mu^d(U)^{-1}
\ket{q_\mu} \\
&= \sum_{q_\nu',q_\nu''\in Q_\nu^d}
\bra{q''_\nu} \bq_\nu^d(U)\ket{q'_\nu}
\bra{q'_\nu} \bq_\nu^d(U)^{-1}\ket{q_\nu}
{T}^\nu_{q''_\nu} \ket{q_\mu}
\\ &= T^\nu_{q_\nu} \ket{q_\mu}
= \hat{T} \ket{q_\mu}\ket{q_\nu}.
\end{split}\ee

Now, fix an orthonormal basis for $\Hom(\cQ_\mu^d \ot \cQ_\nu^d,
\cQ_\lambda^d)^{\cU_d}$ and call it $M_{\mu,\nu}^\lambda$.  Then we
can expand $\hat{T}$ in this basis as
\be \hat{T} = \sum_{\alpha\in M_{\mu,\nu}^\lambda}
\hat{T}_\alpha \cdot \alpha,\ee
where the $\hat{T}_\alpha$ are scalars.  Thus
\be \bra{q_\lambda}T^\nu_{q_\nu}\ket{q_\mu}
= \sum_{\alpha\in M_{\mu,\nu}^\lambda}
\hat{T}_{\alpha}
\bra{q_\lambda}\alpha\ket{q_\mu,q_\nu}.
\label{eq:Talpha-def}\ee
This last expression $\bra{q_\lambda}\alpha\ket{q_\mu,q_\nu}$
bears a striking resemblance to the CG transform.  Indeed, note that
the multiplicity space $\Hom(\cQ_\lambda^d, \cQ_\mu^d \ot
\cQ_\nu^d)^{\cU_d}$ from \eq{CG-hom-decomp} is the dual of $\Hom(\cQ_\mu^d
\ot \cQ_\nu^d, \cQ_\lambda^d)^{\cU_d}$ (which contains $\alpha$),
meaning that we can map between the two by taking the transpose.
In fact, taking the conjugate transpose of \eq{Ucg-dag} gives
$\bra{q_\lambda}\alpha = \bra{q_\lambda,\alpha^\dag}\Ucg^{\mu,\nu}$.  Thus
\be \bra{q_\lambda}\alpha\ket{q_\mu,q_\nu}
= \bra{q_\lambda, \alpha^\dag}\Ucg^{\mu,\nu}
\ket{q_\mu,q_\nu}.\ee

The arguments in the last few paragraphs constitute a proof of the
Wigner-Eckart theorem\cite{Messiah62}, which is
stated as follows:\\
\begin{theorem}[Wigner-Eckart]
\sloppypar{For any irreducible tensor operator
 $\bT^\nu = \{T^\nu_{q_\nu}\}_{q_\nu\in Q_\nu^d} \subset
 \Hom(\cQ_\mu^d,\cQ_\lambda^d)$,  there
exist $\hat{T}_\alpha\in\bbC$ for each $\alpha\in
M_{\mu,\nu}^\lambda$ such that for all $\ket{q_\mu}\in \cQ_\mu^d$,
$\ket{q_\nu}\in \cQ_\nu^d$ and $\ket{q_\lambda}\in \cQ_\lambda^d$:}
\be \bra{q_\lambda}T^\nu_{q_\nu} \ket{q_\mu}
= \sum_{\alpha\in M_{\mu,\nu}^\lambda}
\hat{T}_\alpha
\bra{q_\lambda,\alpha^\dag}\Ucg^{\mu,\nu}\ket{q_\mu,q_\nu}.\ee
\end{theorem}

Thus, the action of tensor operators can be related to a component
$\hat{T}_\alpha$ that is invariant under $\cU_d$ and a component
that is equivalent to the CG transform.  We will use this in the next
section to derive an efficient quantum circuit for the CG transform.

\subsection{A recursive construction of the Clebsch-Gordan transform}
\label{sec:cg-recurse}
In this section we show how the $\cU_d$ CG transform (which here we call
$\Ucg^{[d]}$) can be efficiently reduced to the $\cU_{d-1}$ CG
transform (which we call $\Ucg^{[d-1]}$).  Our strategy, following
\cite{Biedenharn:68a}, will be to express $\Ucg^{[d]}$ in terms of
$\cU_{d-1}$ tensor operators and then use the Wigner-Eckart Theorem to
express it in terms of $\Ucg^{[d-1]}$.  After we have explained this as a
relation among operators, we describe a quantum circuit for
$\Ucg^{[d]}$ that uses $\Ucg^{[d-1]}$ as a subroutine.

First, we express $\Ucg^{[d]}$ as a $\cU_d$ tensor operator.  For
$\mu\in\bbZ_{++}^d$, $\ket{q}\in Q_\mu^d$ and $i\in [d]$, we can expand
$\Ucg^{[d]} \ket{\mu}\ket{q}\ket{i}$ as
\be \Ucg^{[d]} \ket{\mu}\ket{q}\ket{i}
= \ket{\mu}
\sum_{\substack{j\in[d]\st\\ \mu + e_j\in\bbZ_{++}^d}}
\sum_{q'\in Q_{\mu+e_j}^d}
C^{\mu,j}_{q,i,q'}\ket{\mu + e_j}\ket{q'}.\ee
for some coefficients $C^{\mu,j}_{q,i,q'}\in \bbC$.
Now define operators $T^{\mu,j}_i: \cQ_\mu^d \ra \cQ_{\mu+e_j}^d$ by
\be T^{\mu,j}_i = \sum_{q\in Q_\mu^d}\;
\sum_{q'\in Q_{\mu+e_j}^d}
C^{\mu,j}_{q,i,q'} \ket{q'}\bra{q},\ee
so that $\Ucg^{[d]}$ decomposes as
\be \Ucg^{[d]} \ket{\mu}\ket{q}\ket{i}
= \ket{\mu}
\sum_{\substack{j\in[d]\st\\ \mu + e_j\in\bbZ_{++}^d}}
\ket{\mu + e_j}T^{\mu,j}_i \ket{q}.
\label{eq:CG-to-tensor}\ee
Thus $\Ucg^{[d]}$ can be understood in terms of the maps
$T_i^{\mu,j}$, which are irreducible tensor operators in
$\Hom(\cQ_\mu^d, \cQ_{\mu+e_j}^d)$  corresponding to the irrep
$\cQ_{(1)}^d$.  (This is unlike the notation of the last section in
which the superscript denoted the irrep corresponding to the tensor
operator.)

The plan for the rest of the section is to decompose the $T_i^{\mu,j}$
operators under the action of $\cU_{d-1}$, so that we can apply the
Wigner-Eckart theorem.  This involves decomposing three different
$\cU_d$ irreps into $\cU_{d-1}$ irreps: the input space $\cQ_\mu^d$,
the output space $\cQ_{\mu+e_j}^d$ and the space $\cQ_{(1)}^d$
corresponding to the subscript $i$.  Once we have done so,
the Wigner-Eckart Theorem gives an expression for $T_i^{\mu,j}$ (and
hence for $\Ucg^{[d]}$) in terms of $\Ucg^{[d-1]}$ and a small number
of coefficients, known as {\em reduced Wigner coefficients}.  These
coefficients can be readily calculated, and in the next section we
cite a formula from \cite{Biedenharn:68a} for doing so.

First, we examine the decomposition of $\cQ_{(1)}^d$, the $\cU_d$-irrep
according to which the $T_i^{\mu,j}$ transform.  Recall that
$\cQ_{(1)}^d \stackrel{\cU_{d-1}}{\cong} 
\cQ_{(0)}^{d-1} \oplus \cQ_{(1)}^{d-1}$.  In terms of the tensor
operator we have defined, this means that $T^{\mu,j}_d$ is an
irreducible $\cU_{d-1}$ tensor operator corresponding to the trivial
irrep $\cQ_{(0)}^{d-1}$ and $\{T^{\mu,j}_1, \ldots, T^{\mu,j}_{d-1}\}$
comprise an irreducible $\cU_{d-1}$ tensor operator corresponding to
the defining irrep $\cQ_{(1)}^{d-1}$.

Next, we would like to decompose $\Hom(\cQ_\mu^d, \cQ_{\mu+e_j}^d)$
into maps between irreps of $\cU_{d-1}$.  This is 
 slightly
more complicated, but can be derived from the $\cU_{d-1}\subset
\cU_d$ branching rule introduced in \sect{gz-yy}.  Recall that
$\cQ_\mu^d \stackrel{\cU_{d-1}}{\cong} \bigoplus_{\mu'\interlaces \mu}
\cQ_{\mu'}^{d-1}$, and similarly $\cQ_{\mu+e_j}^d
\stackrel{\cU_{d-1}}{\cong} \bigoplus_{\mu''\interlaces \mu+e_j}
\cQ_{\mu''}^{d-1}$.  This is the moment that we anticipated in
\sect{gz-yy} when we chose our set of basis vectors
$Q_\mu^d$ to respect these decompositions.   As a result,  a vector
$\ket{q}\in Q_\mu^d$ can be expanded as
$q = (q_{d-1},q_{d-2},\ldots,q_1) = (\mu',q_{(d-2)})$ with
$q_{d-1}=\mu'\in\bbZ_{++}^{d-1}$, $\mu'\interlaces \mu$ and $\ket{q_{(d-2)}}
= \ket{q_{d-2},\ldots,q_1}\in Q_{\mu'}^{d-1}$.  In other words, we
will separate vectors in $Q_\mu^d$ into a $\cU_{d-1}$ irrep label
$\mu'\in\bbZ_{++}^{d-1}$ and a basis vector from $\cQ_{\mu'}^{d-1}$.

This describes how to decompose the spaces $\cQ_\mu^d$ and
$\cQ_{\mu+e_j}^d$.  To extend this to decomposition of
$\Hom(\cQ_\mu^d, \cQ_{\mu+e_j}^d)$, we use the 
canonical isomorphism $\Hom(\bigoplus_x A_x, \bigoplus_y B_y) \cong
\bigoplus_{x,y} \Hom(A_x,B_y)$, which holds for any sets of vector
spaces $\{A_x\}$ and $\{B_y\}$.  Thus
\begin{subequations}\label{eq:wigner-op-decomp}
\be \Hom(\cQ_\mu^d, \cQ_{\mu+e_j}^d)  \stackrel{\cU_{d-1}}{\cong}
\bigoplus_{\mu'\interlaces\mu} \;\; \bigoplus_{\mu''\interlaces\mu+e_j}
\Hom(\cQ_{\mu'}^{d-1}, \cQ_{\mu''}^{d-1}).
\label{eq:wigner-op-decomp1}\ee
Sometimes we will find it convenient to denote the
$\cQ_{\mu'}^{d-1}$ subspace of $\cQ_\mu^d$ by
$\cQ_{\mu'}^{d-1} \subset \cQ_\mu^d$, so that \eq{wigner-op-decomp1}
becomes
\be \Hom(\cQ_\mu^d, \cQ_{\mu+e_j}^d)  \stackrel{\cU_{d-1}}{\cong}
\bigoplus_{\mu'\interlaces\mu}\;\; \bigoplus_{\mu''\interlaces\mu+e_j}
\Hom(\cQ_{\mu'}^{d-1}\subset\cQ_\mu^d,
\cQ_{\mu''}^{d-1}\subset\cQ_{\mu+e_j}^d).
\label{eq:wigner-op-decomp2}\ee
\end{subequations}

According to \eq{wigner-op-decomp} (either version), we can decompose
$T^{\mu,j}_i$ as 
\be T^{\mu,j}_i =
\sum_{\mu'\interlaces \mu}\;\;\sum_{\mu''\interlaces \mu+e_j}
\ket{\mu''}\bra{\mu'} \ot T^{\mu,j,\mu',\mu''}_i.
\label{eq:tensor-decomp}\ee
Here $T^{\mu,j,\mu',\mu''}_i \in
\Hom(\cQ_{\mu'}^{d-1}\subset\cQ_\mu^d,
\cQ_{\mu''}^{d-1}\subset\cQ_{\mu+e_j}^d)$ and we have implicitly
decomposed $\ket{q}\in Q_\mu^d$ into
$\ket{\mu'}\ket{q_{(d-2)}}$.

The next step is to decompose the representions in
\eq{wigner-op-decomp} into irreducible 
components.  In fact, we are not
interested in the entire space $\Hom(\cQ_{\mu'}^{d-1},
\cQ_{\mu''}^{d-1})$, but only the part that is equivalent to
$\cQ_{(1)}^{d-1}$ or $\cQ_{(0)}^{d-1}$, depending on whether
$i\in[d-1]$ or $i=d$ (since $T^{\mu,j,\mu',\mu''}_i$ transforms
according to $\cQ_{(1)}^{d-1}$ if $i\in\{1,\ldots,d-1\}$ and
according to $\cQ_{(0)}^{d-1}$ if $i=d$).  This knowledge of how
$T^{\mu,j,\mu',\mu''}_i$ transforms under $\cU_{d-1}$ will give us two
crucial simplifications: first, we can greatly reduce the range of
$\mu''$ for which $T^{\mu,j,\mu',\mu''}_i$ is nonzero, and second, we
can apply the Wigner-Eckart theorem to describe
$T^{\mu,j,\mu',\mu''}_i$ in terms of $\Ucg^{[d-1]}$.

The simplest case is $\cQ_{(0)}^{d-1}$, when $i=d$: according to
Schur's Lemma 
the invariant component of $\Hom(\cQ_{\mu'}^{d-1},
\cQ_{\mu''}^{d-1})$ is zero if $\mu'\neq\mu''$ and consists of the
matrices proportional to $I_{\cQ_{\mu'}^{d-1}}$ if $\mu'=\mu''$.  In
other words $T_d^{\mu,j,\mu',\mu''}=0$ unless $\mu'=\mu''$, in which
case $T_d^{\mu,j,\mu',\mu'} := \hat{T}^{\mu,j,\mu',0}
I_{\cQ_{\mu'}^{d-1}}$ for some scalar $\hat{T}^{\mu,j,\mu',0}$.  (The
final superscript 0 will later be convenient when we want a single
notation to encompass both the $i=d$ and the $i\in\{1,\ldots,d-1\}$
cases.)

The $\cQ_{(1)}^{d-1}$ case, which occurs when $i\in\{1,\ldots,d-1\}$, is more
interesting.  We will simplify the $T^{\mu,j,\mu',\mu''}_i$ operators
(for $i=1,\ldots,d-1$) in two stages: first using the branching rules
from \sect{gz-yy} to reduce the number of nonzero terms and then by
applying the Wigner-Eckart theorem to find an exact expression for
them. Begin by recalling from \eq{hom-cg-decomp} that the multiplicity of
$\cQ_{(1)}^{d-1}$ in the isotypic decomposition of
$\Hom(\cQ_{\mu'}^{d-1}, \cQ_{\mu''}^{d-1})$ is 
given by $\dim\Hom(\cQ_{\mu'}^{d-1} \ot \cQ_{(1)}^{d-1},
\cQ_{\mu''}^{d-1})^{\cU_{d-1}}$.  According to the $\cU_{d-1}$ CG
``add a box''  prescription (\eq{cg-add-one}), this is one if
 $\mu'\in\mu''-\Box$ 
 and zero otherwise.  Thus if $i\in [d-1]$, then
$T^{\mu,j,\mu',\mu''}_i$ is zero unless $\mu'' = \mu' + e_{j'}$
for some $j'\in [d-1]$.  Since we need not consider all possible
$\mu''$, we can define $T^{\mu,j,\mu',j'}_i := T^{\mu,j,\mu',\mu' +
e_{j'}}_i$.
This notation can be readily extended to cover the case when
$i=d$; define $e_0=0$, so that the only nonzero operators for
$i=d$ are of the form $T^{\mu,j,\mu',0}_d := T^{\mu,j,\mu',\mu'}_d =
\hat{T}^{\mu,j,\mu',0} I_{\cQ_{\mu'}^{d-1}}$.  Thus, we can 
replace \eq{tensor-decomp} with
\be T^{\mu,j}_i =
\sum_{\mu'\interlaces \mu}\;\;\sum_{j'=0}^{d-1}
\ket{\mu'+e_{j'}}\bra{\mu'} \ot T^{\mu,j,\mu',\mu'+e_{j'}}_i.
\label{eq:tensor-decomp2}\ee

Now we show how to apply the Wigner-Eckart theorem to the $i\in[d-1]$
case.  The operators $T^{\mu,j,\mu',j'}_i$
map $\cQ_{\mu'}^{d-1}$ to $\cQ_{\mu'+e_{j'}}^{d-1}$ and comprise an
irreducible $\cU_{d-1}$ tensor operator corresponding to the irrep
$\cQ_{(1)}^{d-1}$. This means
we can apply the Wigner-Eckart Theorem and since the multiplicity of
$\cQ_{\mu'+e_{j'}}^{d-1}$ in $\cQ_{\mu'}^{d-1}\ot \cQ_{(1)}^{d-1}$ is
one, the sum over the multiplicity label $\alpha$ has only a single
term.  The theorem 
implies the existence of a set of scalars
$\hat{T}^{\mu,j,\mu',j'}$ such that for any $\ket{q}\in
Q_{\mu'}^{d-1}$ and $\ket{q'} \in Q_{\mu'+e_{j'}}^{d-1}$,
\be \bra{q'} T_i^{\mu,j,\mu',j'} \ket{q} =
\hat{T}^{\mu,j,\mu',j'}
\bra{\mu',\mu'+e_{j'},q'} \Ucg^{[d-1]} \ket{\mu',q,i}.
\label{eq:wigner-tensor-reduction}\ee
Sometimes the matrix elements of $\Ucg$ or
$T^{\mu,j,\mu',j'}_i$ are called {\em Wigner coefficients} and
the $\hat{T}^{\mu,j,\mu',j'}$ are known as {\em reduced Wigner
coefficients}.  

Let us now try to interpret these equations operationally.  
\eq{CG-to-tensor} reduces the $\cU_d$ CG transform to a $\cU_d$ tensor
operator, \eq{tensor-decomp2} decomposes this tensor operator into
$d^2$ different $\cU_{d-1}$ tensor operators (weighted by the
$\hat{T}^{\mu,j,\mu',j'}$ coefficients) and
\eq{wigner-tensor-reduction} turns this into a $\cU_{d-1}$ CG
transform followed by a $d\times d$ unitary matrix.  The coefficients
for this matrix are the $\hat{T}^{\mu,j,\mu',j'}$, which we will see
in the next section can be efficiently computed by conditioning on
$\mu$ and $\mu'$.

\begin{figure}[ht]
\begin{center}\leavevmode
\xymatrix@R=4pt@C=4pt{
{|\mu\>} & ~~~ & \gnqubit{~\hat{T}^{[d]}~}{dd}\ar@{-}[ll] &
*{~~~} & |\mu\>\ar@{-}[ll] \\
{|\mu'\>} & \nw & \gspace{~\hat{T}^{[d]}~}\w &\nw& |\mu+e_j\>\w\\
{|\mu'+e_{j'}\>} & \nw & \gspace{~\hat{T}^{[d]}~}\w
&\nw& |\mu'+e_{j'}\>\w
}
\xymatrix@R=4pt@C=4pt{
& |\mu\> & \nw & \nw & \b\w\ar@{-}[d] & \nw & \b\w\ar@{-}[d] & |\mu\>\w \\
*{~\Huge\mbox{$\cong$}~} & |\mu'\> & *+={\oplus}\w & *+{|e_{j'}\>}\w & 
\op{\hat{T}^{[d]}_{\mu,\mu'}}\w & *+{|e_j\>}\w & *+={\oplus}\w
& |\mu+e_j\>\w\\
& |\mu'+e_{j'}\> & \b\w\ar@{-}[u] & \nw & \b\w\ar@{-}[u] & \nw & \nw &
|\mu'+e_{j'}\>\w
}
\caption{The reduced Wigner transform $\hat{T}^{[d]}$ can be
expressed as a $d\times d$ rotation whose coefficients are controlled
by $\mu$ and $\mu'+e_{j'}$.}
\label{fig:reduced-wigner}
\end{center}
\end{figure}
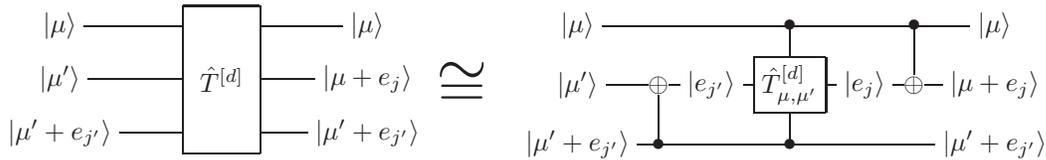

Now we spell this recursion out in more detail.  Suppose we wish to
apply $\Ucg^{[d]}$ to
$\ket{\mu}\ket{q}\ket{i}=\ket{\mu}\ket{\mu'}|q_{(d-2)}\>\ket{i}$,
for some $i\in\{1,\ldots,d-1\}$.  Then 
\eq{wigner-tensor-reduction} indicates that we should first apply
$\Ucg^{[d-1]}$ to $\ket{\mu'}|q_{(d-2)}\>\ket{i}$ to obtain output
that is a superposition over states
$\ket{\mu'+e_{j'}}\ket{j'}|q'_{(d-2)}\>$ for $j'\in\{1,\ldots,d-1\}$ and
$|q'_{(d-2)}\>\in Q_{\mu'+e_{j'}}^{d-1}$.  Then, controlled by
$\mu$ and $\mu'$, we want to map the $(d-1)$-dimensional $\ket{j'}$
register into the $d$-dimensional $\ket{j}$ register, which will then
tell us the output irrep $\cQ_{\mu+e_j}^d$.
According to \eq{wigner-tensor-reduction}, the coefficients of this
$d\times(d-1)$ matrix are given by the reduced Wigner coefficients
$\hat{T}^{\mu,j,\mu',j'}$,
 so we will denote the overall matrix
$\hat{T}^{[d]}_{\mu,\mu'} := \sum_{j,j'}
\hat{T}^{\mu,j,\mu'+e_{j'},j'} \ket{j}\!\bra{j'}$.\footnote{The reason
 why $\mu'+e_{j'}$ appears in the superscript rather than $\mu'$ is
 that after applying $\hat{T}^{[d]}_{\mu,\mu'}$ we want to keep a
 record of $\mu'+e_{j'}$ rather than of $\mu'$.  This is further
 illustrated in \fig{reduced-wigner}.}
  The resulting circuit is depicted in \fig{CG-decomp}: a $\cU_{d-1}$
CG transform is followed by the $\hat{T}^{[d]}$ operator, which is
defined to be
\be\hat{T}^{[d]} = \sum_{\mu'\interlaces\mu}\sum_{j,j'}
\hat{T}^{\mu,j,\mu',j'} \oprod{\mu} \ot \ket{\mu+e_j}\bra{\mu'}
\ot \oprod{\mu' + e_{j'}}.
\label{eq:red-wig-def}\ee
Then \fig{reduced-wigner} shows how $\hat{T}^{[d]}$ can be expressed
as a $d\times (d-1)$ matrix $\hat{T}^{[d]}_{\mu,\mu'}$ that is
controlled by $\mu$ and $\mu'$.  In fact, once we consider the $i=d$
case in the next paragraph, we will find that
$\hat{T}^{[d]}_{\mu,\mu'}$ is actually a $d\times d$ unitary matrix.
In the next section, we will then show how the individual reduced
Wigner coefficients $\hat{T}^{\mu,j,\mu',j'}$ can be efficiently
computed, so that ultimately $\hat{T}^{[d]}_{\mu,\mu'}$ can be
implemented in time $\poly(d,\log 1/\eps)$.

Now we turn to the case of $i=d$.   The circuit is much simpler, but
we also need to explain how it 
works in coherent superposition with the $i\in[d-1]$ case.  Since
$i=d$ corresponds to the trivial representation of $\cU_{d-1}$, the
$\Ucg^{[d-1]}$ operation is not performed.  Instead, $\ket{\mu'}$ and
$|q_{(d-2)}\>$ are left untouched and the $\ket{i}=\ket{d}$
register is relabeled as a $\ket{j'}=\ket{0}$ register.  We can
combine this relabeling operation with $\Ucg^{[d-1]}$ in the
$i\in[d-1]$ case by defining
\be \tilde{U}_{\text{CG}}^{[d-1]} :=
\l(\ket{0}\bra{d} \ot \sum_{\mu'\in\bbZ_{++}^{d-1}} \oprod{\mu'}\r)
\ot I_{\cQ_{\mu'}^{d-1}} + \Ucg^{[d-1]}.
\label{eq:tildeUcg-def}\ee
This ends up mapping $i\in\{1,\ldots,d\}$ to $j'\in\{0,\ldots,d-1\}$
while mapping $\cQ_{\mu'}^{d-1}$ to $\cQ_{\mu'+e_{j'}}^{d-1}$.
Now we can interpret the sum on $j'$ in the above definitions of
$\hat{T}^{[d]}$ and $\hat{T}^{[d]}_{\mu,\mu'}$ as ranging over
$\{0,\ldots,d-1\}$, so that $\hat{T}^{[d]}_{\mu,\mu'}$ is a $d\times
d$ unitary matrix.  We thus obtain the circuit in \fig{CG-decomp} with
the implementation of $\hat{T}^{[d]}$ depicted in \fig{reduced-wigner}.

\begin{figure}[ht]
\begin{centering}
\setlength{\unitlength}{3947sp}%
\begin{picture}(5250,1599)(376,-1198)
\thinlines
{\put(1201,-1186){\framebox(750,1125){}}}
{\put(1201,-211){\line(-1, 0){450}}}
{\put(1201,-586){\line(-1, 0){450}}}
{\put(1201,-1036){\line(-1, 0){450}}}
{\put(2401,-211){\line(-1, 0){450}}}
{\put(2176,-586){\line(-1, 0){225}}}
{\put(4501,-1036){\line(-1, 0){2550}}}
{\put(3301,-586){\line(-1, 0){300}}}
{\put(3301,-211){\line(-1, 0){450}}}
{\put(3301,-736){\framebox(750,1125){}}}
{\put(4501,-211){\line(-1, 0){450}}}
{\put(4501,239){\line(-1, 0){450}}}
{\put(4501,-586){\line(-1, 0){450}}}
{\put(3301,239){\line(-1, 0){2550}}}
\put(1426,-736){\makebox(0,0)[lb]{$\tilde{U}_{\text{CG}}^{[d-1]}$}}
\put(3486,-286){\makebox(0,0)[lb]{$\hat{T}^{[d]}$}}
\put(376,-286){\makebox(0,0)[lb]{$|\mu'\>$}}
\put(4651,-661){\makebox(0,0)[lb]{$|u'+e_{j'}\>$}}
\put(4651,-1111){\makebox(0,0)[lb]{$|q'_{(d-2)}\>$}}
\put(4651,164){\makebox(0,0)[lb]{$|\mu\>$}}
\put(451,-1111){\makebox(0,0)[lb]{$|i\>$}}
\put(376,-661){\makebox(0,0)[lb]{$|q\>$}}
\put(2326,-661){\makebox(0,0)[lb]{$|\mu'+e_{j'}\>$}}
\put(2476,-286){\makebox(0,0)[lb]{$|\mu'\>$}}
\put(4651,-286){\makebox(0,0)[lb]{$|\mu+e_j\>$}}
\put(5300,-1020){\makebox(0,0)[lb]{$$\Bigg\}$$}}
\put(5500,-850){\makebox(0,0)[lb]{$|q\>$}}
\put(376,164){\makebox(0,0)[lb]{$|\mu\>$}}
\end{picture}
\caption{The $\cU_d$ CG transform, $\Ucg^{[d]}$, is decomposed into a
$\cU_{d-1}$ CG transform $\tilde{U}_{\text{CG}}^{[d-1]}$ (see 
\eq{tildeUcg-def}) and a reduced Wigner operator
$\hat{T}^{[d]}$.  In \fig{reduced-wigner} we show how to reduce
the reduced Wigner operator to a $d\times d$ matrix
conditioned on $\mu$ and $\mu'+e_{j'}$.
}
\label{fig:CG-decomp}
\end{centering}
\end{figure}

We have now reduced the problem of performing the CG transform
$\Ucg^{[d]}$ to the problem of computing reduced Wigner coefficients
$\hat{T}^{\mu,j,\mu',j'}$.

\subsection{Efficient Circuit for the Reduced Wigner Operator}

The method of Biedenharn and Louck\cite{Biedenharn:68a} allows us to
compute reduced Wigner coefficients for the cases we are interested
in.  This will allow us to construct an efficient circuit to implement the
controlled-$\hat{T}$ operator to accuracy $\epsilon$ using an overhead which
scales like $\poly(\log n, d, \log(\epsilon^{-1}))$.

To compute $\hat{T}^{\mu,j,\mu',j'}$, we first introduce the vectors
$\tilde{\mu} := \mu + \sum_{j=1}^d (d-j)e_j$ and
$\tilde{\mu}' := \mu' + \sum_{j=1}^{d-1} (d-1-j)e_j$.  
Also
define $S(j-j')$ to be 1 if $j\geq j'$ and $-1$ if $j<j'$.  Then
according to Eq.~(38) in
Ref~\cite{Biedenharn:68a},
\begin{equation}
\hat{T}^{\mu,j,\mu',j'} = 
\l\{\begin{array}{ll}
S(j-j')
\l[\frac{\prod_{s\in [d-1]\backslash j} 
(\tilde{\mu}_j - \tilde{\mu}'_s)
\prod_{t\in [d]\backslash j'} 
(\tilde{\mu}'_{j'} - \tilde{\mu}_t + 1)
}{
\prod_{s\in [d]\backslash j}
(\tilde{\mu}'_j - \tilde{\mu}'_s)
\prod_{t\in [d-1]\backslash j'} 
(\tilde{\mu}'_{j'} - \tilde{\mu}'_t + 1)
}\r]^{\frac{1}{2}}
& \mbox{ if $j'\in\{1,\ldots,d-1\}$.}\\
S(j-d)
\l[\frac{\prod_{s\in [d-1]\backslash j} 
(\tilde{\mu}_j - \tilde{\mu}'_s)
}{
\prod_{s\in [d]\backslash j}
(\tilde{\mu}'_j - \tilde{\mu}'_s)
}\r]^{\frac{1}{2}}
& \mbox{ if $j'=0$.}
\end{array}\r. 
\end{equation}
The elements of the partitions here are of size $O(n)$, so
the total computation necessary is $\poly(d, \log n)$.  Now how do we
implement the $\hat{T}^{[d]}$ transform given this
expression?

As in the introduction to this section, note that any unitary gate of
dimension $d$ can be 
implemented using a number of two qubit gates polynomial in
$d$\cite{Reck:94a,Barenco:95a,NC00}.  The method of this
construction is to take a unitary gate of dimension $d$ with {\em
known} matrix elements and then convert this into a series of unitary
gates which act non-trivially only on two states.  These two state
gates can then be constructed using the methods described in
\cite{Barenco:95a}.  In order to modify this for our work, we
calculate, to the specified accuracy $\epsilon$, the elements of
the $\hat{T}^{[d]}$ operator, conditional on the $\mu$ and
$\mu'+e_{j'}$ inputs, perform the decomposition into two qubit gates
as described in \cite{Reck:94a,Barenco:95a} {\em online}, and then,
conditional on this calculation perform the appropriate controlled
two-qubit gates onto the space where $\hat{T}^{[d]}$ will act.
Finally this classical computation must be undone to reset any garbage
bits created during the classical computation.   To produce
an accuracy $\epsilon$ we need a classical computation of size ${\rm
poly}(\log(1/\epsilon))$ since we can perform the appropriate
controlled rotations with bitwise accuracy.

Putting everything together as depicted in figures \ref{fig:CG-decomp}
and \ref{fig:reduced-wigner} gives a $\poly(d,\log n,\log 1/\epsilon)$
algorithm to reduce $\Ucg^{[d]}$ to $\Ucg^{[d-1]}$.  Naturally this
can be applied $d$ times to yield a $\poly(d,\log n,\log 1/\epsilon)$
algorithm for $\Ucg^{[d]}$.  (We can end the recursion either at
$d=2$, using the construction in \cite{BCH04}, or at $d=1$, where
the CG transform simply consists of the map $\mu\ra \mu+1$ for
$\mu\in\bbZ$, or even at $d=0$, where the CG transform is completely
trivial.)  We summarize the CG algorithm as follows.

\begin{tabbing}
~~~~ \= ~~~~ \= ~~~ \= ~~~ \= ~~~\= \kill
\+{\bf Algorithm: Clebsch-Gordan transform} \\
\parbox[t]{5.6in}{{\bf Inputs:} (1) Classical registers $d$ and
$n$. (2) Quantum 
registers $\ket{\lambda}$ (in any superposition over different
$\lambda\in\cI_{d,n}$), $\ket{q}\in \cQ_\lambda^d$ (expressed as a
superposition of GZ basis elements) and $\ket{i}\in\bbC^d$.}\\
\parbox[t]{5.6in}{{\bf Outputs:} (1) Quantum registers $\ket{\lambda}$
(equal to the 
input), $\ket{j}\in\bbC^d$ (satisfying $\lambda+e_j\in\cI_{d,n+1}$)
and $\ket{q'}\in\cQ_{\lambda+e_j}^d$.}\\
{\bf Runtime:} $d^3\poly(\log n,\log 1/\eps)$ to achieve accuracy
$\eps$.\\
{\bf Procedure:}\\
{\bf 1.} \> If $d=1$\\
{\bf 2.} \> Then output $\ket{j}:=\ket{i}=\ket{1}$ and
$\ket{q'}:=\ket{q}=\ket{1}$ (i.e. do nothing).\\
{\bf 3.} \> Else\\
{\bf 4.} \> \> Unpack $\ket{q}$ into
$\ket{\mu'}\ket{q_{(d-2)}}$, 
such that $\mu'\in\cI_{d,m}$, $m\leq n$, $\mu'\interlaces\mu$ and
$\ket{q_{(d-2)}}\in\cQ_{\mu'}^{d-1}$.\\ 
{\bf 5.} \> \> If $i<d$\\
{\bf 6.} \> \> \parbox[t]{5in}{Then perform the CG transform with inputs
$(d-1,m,\ket{\mu'},\ket{q_{(d-2)}},\ket{i})$ and outputs
$(\ket{\mu'},\ket{j'},\ket{q'_{(d-2)}})$.}\\
{\bf 7.} \> \> Else (if $i=d$)\\
{\bf 8.} \> \> \> \parbox[t]{5.5in}{Replace $\ket{i}=\ket{d}$ with
$\ket{j'}:=\ket{0}$ and set $\ket{q'_{(d-2)}}:=\ket{q'_{(d-2)}}$.}\\
{\bf 9.} \> \> End. (Now $i\in\{1,\ldots,d\}$ has been replaced by
$j\in\{0,\ldots, d-1\}$.)\\
{\bf 10.} \> \> Map $\ket{\mu'}\ket{j'}$ to
$\ket{\mu'+e_{j'}}\ket{j'}$.\\
{\bf 11.} \> \> \parbox{5.5in}{Conditioned on $\mu$ and $\mu'+e_j'$,
calculate the gate sequence necessary to implement $\hat{T}^{[d]}$,
which inputs $\ket{j'}$ and outputs $\ket{j}$.}\\
{\bf 12.} \> \> Execute this gate sequence, implementing
$\hat{T}^{[d]}$.\\
{\bf 13.} \> \> Undo the computation from {\bf 11}.\\
{\bf 14.} \> \> Combine $\ket{\mu'+e_{j'}}$ and $\ket{q'_{(d-2)}}$ to
form $\ket{q'}$.\\
{\bf 15.} \> End.
\end{tabbing}

Finally, in \sect{schur-circuit} we described
how $n$ CG transforms can be used to perform the Schur transform, so
that $\Usch$ can be implemented in time $n\cdot \poly(d,\log n,\log
1/\epsilon)$, optionally plus an additional $\poly(n)$ time to
compress the $\ket{p}$ register.

\chapter{Relations between the Schur transform and the $\cS_n$ QFT}
\label{chap:sch-Sn}

This final chapter is devoted to algorithmic connections
between the Schur transform and the quantum Fourier transform on
$\cS_n$.  In \sect{GPE} we describe {\em generalized phase estimation},
which is a reduction from measuring in the Schur basis (a weaker
problem than the full Schur transform) to the $\cS_n$ QFT.  Then in
\sect{QFT-from-Sch} we show a reduction in the other direction, from
the $\cS_n$ QFT to the Schur transform.  The goal of these reductions
is not so much to perform new tasks efficiently, since efficient
implementations of the QFT already exist, but to help clarify the
position of the Schur transform {\em vis-a-vis} known algorithms.

\section{Generalized phase estimation}\label{sec:GPE}
The last chapter developed the Schur transform based on the
$\cU_d$ CG transform.  Can we instead build the Schur transform out of
operations on $\cS_n$?  This section explores that possibility.  We
will see that using the $\cS_n$ QFT allows us to efficiently measure a
state in the Schur basis, a slightly weaker task than performing the
full Schur transform.  Our algorithm for this measurement generalizes
the quantum circuits used to estimate the phase of a black-box unitary
transform\cite{Shor94,Kitaev:02a} (see also \cite{Klappenecker03}) to
a nonabelian setting; hence we 
call it generalized phase estimation (GPE).  As we will see, our
techniques actually extend to measuring the irrep labels in reducible
representations of any group for which we can efficiently perform
group operations and a quantum Fourier transform.

The main idea behind GPE is presented in \sect{GPE-lambda}, where we
show how it can be used to measure $\ket{\lambda}$ (and optionally
$\ket{p}$ as well) in the Schur basis.
Here the techniques are completely general and we
show how similar results hold for any group.  We specialize to 
Schur basis measurements in \sect{GPE-Ud}, where we show that GPE can
be extended to also measure the $\cQ_\lambda^d$ register, thereby making a
complete Schur basis measurement possible based only on the $\cS_n$
QFT.  We conclude in \sect{GPE-CG} with an alternate interpretation
of GPE, which shows its close connection with the $\cS_n$ CG
transform. 

\subsection{Using GPE to measure $\lambda$ and $\cP_\lambda$}
\label{sec:GPE-lambda}
Let $\ccG$ be an arbitrary finite group over which there exists an
efficient circuit for the quantum Fourier transform\cite{MRR03},
$\Uqft$.  Fix a set of inequivalent irreps $\hat{\ccG}$, where
$\mu\in\hat{\ccG}$ corresponds to the irrep
$(\br_\mu,V_\mu)$.
$\Uqft$ then maps the group algebra $\bbC[\ccG]$ to
$\bigoplus_{\mu\in\hat{\ccG}} V_\mu \ot V_\mu^*$, and is
explicitly given by \eq{uqft-explicit}.

Now suppose $(\rho,V)$ is a representation of $\ccG$ for which we can
efficiently perform the controlled-$\rho$ operation, $C_\rho =
\sum_{g\in\ccG} \oprod{g} \ot \rho(g)$.  To specialize to the Schur
transform we will choose $V=\nqudits$ and $\rho=\bP$, but everything in
the section can be understood in terms of arbitrary $\ccG$ and $(\rho,V)$.
Let the multiplicity of the
irrep $V_\nu$ in $V$ be given by $m_\nu$, so that $V$
decomposes as
\be V \iso{\ccG} \bigoplus_{\nu\in\hat{\ccG}}
\bbC^{m_\nu} \ot V_\nu.
\label{eq:GPE-V-decomp}\ee
This induces a basis for $V$, analogous to the Schur basis, given by
$\ket{\nu, \alpha, k}_{\rm V}$, where
$\nu\in\hat{\ccG}$, $\alpha\in[m_\nu]$ and $k\in[d_\nu]$, where $d_\nu
:= \dim V_\nu$. 
For any $\lambda\in\hat{\ccG}$, define the
projector onto the $V_\lambda$-isotypic subspace in terms of this basis as
\be \Pi_\lambda = \oprod{\lambda}
\ot I_{m_\lambda} \ot I_{d_\lambda}.\ee
Note that this becomes \eq{Pi-lambda-def} for the special case of
$(\rho,V)=(\bP,\nqudits)$.

\begin{figure}[ht]
\begin{center}\leavevmode
\xymatrix@R=4pt@C=4pt{
    |\mu\> & *-{~}\w & \nw &
    \gnqubit{\Uqft^\dag}{dd}\w\a{gCA} & {} &{} & {} & {} & {} &
\gnqubit{\Uqft}{dd}\a{gIA} & \nw & \nt & \nt\a{gLA} & \n & \nt \\  
{\ket{i}} & \n   &\n   & \gspace{\Uqft^\dag}\w & *-{~}\w &
    *-{~}\w & \b\w\a{gFB} & *-{~}\w & *-{~}\w & \gspace{\Uqft}\w & \nw
    &\nw &\nw\a{gLC} &\nw &\nt
\\  \ket{j} & \n   &\n   & \gspace{\Uqft^\dag}\w & & &
    & & &\gspace{\Uqft} & \nw & \nw &\nw &\nw &\nt\a{gNC}
\\  {} & {}
\\  \ket{\psi} & \n   &\n   &\n   &\n   &\n   & \op{\rho(g)}\w\a{gFE}
& \nw &\nw &\nw &\nw &\nw &\nw &\nw & *-{~~~~} \a{gNE}
\ar@{-}"gFE";"gFB"
\save "gLC"!R;"gLA"+<0ex,+2ex>."gNE"!D(1.1)**\frm{\}} \restore
\save "gNC"!R*-{~~~~~~~~~~~~~~|\phi_{\rm out}\>} \restore
}
\caption{Quantum circuit $\hat{C}_\rho$ used in generalized phase
estimation.}
\label{fig:GPE-basic}
\end{center}
\end{figure}
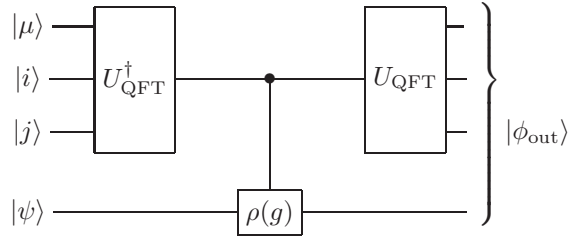

The problem is that, as with the Schur basis, there is no immediately
obvious way to measure or otherwise access the register labeling the
irreps.  We are given no information about the isomorphism in
\eq{GPE-V-decomp} or about how to implement it.
However, by using the Fourier transform along with the
controlled-$\rho$ operator, it is possible to efficiently perform the
projective measurement $\{\Pi_\lambda\}_{\lambda\in\hat{\ccG}}$.  To do
so, we define the operator 
\be \hat{C}_\rho = (\Uqft \ot I_V) C_\rho (\Uqft^\dag \ot I_V)
\label{eq:hatC-def}\ee
acting on $\bigoplus_\mu V_\mu \ot V_\mu^* \ot V$.  This
is represented in \fig{GPE-basic}.

The procedure for performing the projective measurement
$\{\Pi_\lambda\}_{\lambda\in\hat{\ccG}}$ is as follows:
\begin{tabbing}
~~~~~ \= ~~~ \= ~~~ \= ~~~ \= ~~~\= \kill
{\bf Algorithm: Generalized Phase Estimation}\+ \\
{\bf Inputs:} A state $\ket{\psi}\in V$.\\
\pushtabs{\bf Outputs:} ~\= (1) Classical variable $\lambda$ with
probability $p_\lambda:=\bra{\psi}\Pi_\lambda\ket{\psi}$\\
\> (2) The state $\Pi_\lambda\ket{\psi}/\sqrt{p_\lambda}$.
\\\poptabs
{\bf Runtime:}
\parbox[t]{5.4in}{$2T_{\text{QFT}}+T_{C_\rho}$ where $T_{\text{QFT}}$
(resp. $T_{C_\rho}$) is the running time for the QFT on $G$
(resp. the controlled-$\rho$ operation).}\\
{\bf Procedure:}\\
{\bf 1.}\>
\parbox[t]{5.6in}{Create registers $\ket{\mu}\ket{i}\ket{j}$ (see 
\fig{GPE-basic}) with $\mu$
corresponding to the trivial representation $V_0$ and
$\ket{i}=\ket{j}=\ket{1}\in V_0$.}\\
{\bf 2.}\> Apply $\hat{C}_\rho$.  This involves three steps.\\
\>{\bf a)}\> Apply $\Uqft^\dag$ to $\ket{\mu}\ket{i}\ket{j}$,
obtaining the uniform superposition $|G|^{-1/2}\sum_{g\in
G}\ket{g}$.\\
\>{\bf b)}\> Perform $C_\rho = \sum_g \oprod{g} \ot \rho(g)$.\\
\>{\bf c)}\> Apply $\Uqft$ to the first register.\\
\>\parbox[t]{5.6in}{The output $\ket{\psi_{\text{out}}}$ is a 
superposition of $\ket{\lambda}\ket{i'}\ket{j'}\ket{v}$ with
$\lambda\in\hat{\ccG}$, $\ket{i'}\in V_\lambda$, $\ket{j'}\in
V_\lambda^*$ and $\ket{v}\in V$.}\\
{\bf 3.}\> Measure $\lambda$.\\
{\bf 4.}\> \parbox[t]{5.6in}{Optionally perform $\hat{C}_\rho^\dag$.
This is only necessary if we need the residual state
$\Pi_\lambda\ket{\psi}$.}
\end{tabbing}

To analyze this circuit, expand $\ket{\psi}$ in the
$\ket{\mu}\ket{\alpha}\ket{k}_{\rm V}$ basis as
\be \ket{\psi} = \sum_{\mu\in\hat{\ccG}}
\sum_{\alpha=1}^{m_\mu}\sum_{k=1}^{d_\mu}
c_{\mu,\alpha,k}\ket{\mu,\alpha,k}_{\rm V}.\ee
\eq{GPE-V-decomp} means that $\rho(g)$ acts on $\ket{\psi}$ according 
to
\be \rho(g)\ket{\psi}= \sum_{\mu\in\hat{\ccG}}
\sum_{\alpha=1}^{m_\mu}\sum_{k=1}^{d_\mu}
c_{\mu,\alpha,k}\ket{\mu,\alpha}\br_\mu(g)\ket{k}_{\rm V}.
\label{eq:GPE-R-action}\ee

Now examine the $\bbC[\ccG]$ register.  The
initial $\Uqft^\dag$ in $\hat{C}_\rho$ maps the trivial irrep to the
uniform superposition of group elements
$\smfrac{1}{|\ccG|}\sum_{g\in\ccG} \ket{g}$.  This is analogous to the
initialization step of phase estimation on abelian
groups\cite{Shor94,Kitaev:02a}.
Thus the output of the circuit in \fig{GPE-basic} is
\be \ket{\phi_{\text{out}}}  =
\sum_{g\in\ccG}\sum_{\nu\in\hat{\ccG}}
\sum_{\lambda\in\hat{\ccG}}\sum_{i,j=1}^{d_\lambda}
\frac{\sqrt{d_\lambda}}{|\ccG|}
\lbm\br_\lambda(g)\rbm_{i,j}
\ket{\lambda,i,j}\ot\rho(g)\ket{\psi}.
\label{eq:GPE-output1}\ee
We can simplify this using \eq{GPE-R-action} and the orthogonality
relations for irrep matrix elements\cite{GW98} to reexpress
\eq{GPE-output1} as
\be
    |\phi_{\rm out}\> =
    \sum_{\lambda \in \hat{\mathcal G}} \sum_{\alpha=1}^{m_\lambda}
    \sum_{i,j=1}^{d_\alpha} \frac{c_{\lambda,\alpha,i}}
    {\sqrt{d_\lambda}} |\lambda,i,j \> \otimes
    |\lambda,\alpha,j\>_{\rm V}
\,.\label{eq:GPE-output2}\ee
The output $|\phi_{\rm out}\>$ has several interesting properties
which we can now exploit.  Measuring the first register (the irrep
label index) produces outcome $\lambda$ with the correct probability
$\sum_{j=1}^{m_\lambda} \sum_{k=1}^{d_\lambda} |c_{\lambda,j,k}|^2$.
Remarkably, this is achieved independent of the basis in which ${\bf
C}_\rho$ is implemented.  As mentioned above, this reduces to
measuring the irrep label $\lambda$ in the Schur basis when
$\ccG=\cS_n$ and $(\rho,V)=(\bP,\nqudits)$.  In this case, the circuit
requires running time $\poly(n)$ for the $\cS_n$ QFT\cite{Beals97} and
$O(n\log d)$ time for the controlled permutation $C_\bP$, comparable
to the efficiency of the Schur transform given in the last chapter.

This circuit also allows us to perform arbitrary instruments on the
irrep spaces $V_\lambda$; for example, we could perform a complete
measurement, or could perform a unitary rotation conditioned on
$\lambda$.  This is because \eq{GPE-output2} has extracted the irrep
basis vector from $\ket{\psi}$ into the $\ket{i}$ register.  We can
perform an arbitrary instrument on this $V_\lambda$ register, and then
return the information to the $V$ register by performing
$\hat{C}_\rho^\dag$.

To put this more formally, suppose we want to perform an instrument
with operation elements 
\be \sum_{\lambda\in\hat{\ccG}} \oprod{\lambda}
\ot I_{m_\lambda} \ot A_\lambda^{(x)} 
\label{eq:GPE-instrument}\ee
on $V$, where $x$ labels the outcomes of the instrument and the
normalization condition is that $\sum_x (A_\lambda^{(x)})^\dag
A_\lambda^{(x)} = I_{d_\lambda}$ for each $\lambda$.
Then this can be effected by performing the instrument
\be \hat{C}_\rho^\dag
\l(\sum_{\lambda\in\hat{\ccG}} \oprod{\lambda}
\ot A_\lambda^{(x)} \ot I_{d_\lambda} \ot I_V \r)
\hat{C}_\rho.
\label{eq:GPE-instr-construct}\ee
This claim can be verified by explicit calculation and use of the
orthogonality relations, but we will give a simpler proof in
\sect{GPE-CG}.

To recap, so far we have shown how GPE can be used to efficiently:
\bi \item
measure the $\ket{\lambda}$ and $\ket{p}$ registers, or perform
general instruments of the form of \eq{GPE-instrument}, in the Schur
basis of $\nqudits$ using $\poly(n)+O(n\log d)$ gates; and
\item
perform instruments of the form of \eq{GPE-instrument} for any
group $\ccG$ and representation $(\rho,V)$ such that the QFT on $\ccG$
and the controlled-$\rho$ operation can be implemented efficiently.
\ei

\subsection{Using GPE to measure $\cQ_\lambda^d$}\label{sec:GPE-Ud}
In this section we specialize to the case of the Schur basis and show
how GPE can be adapted to measure the $\ket{q}$ register. This allows
us to perform a complete measurement in the
$\ket{\lambda}\ket{p}\ket{q}_{\rm Sch}$ basis, or more generally, to
perform instruments with operation elements
\be \sum_{\lambda\in\cI_{d,n}}\sum_{q\in Q_\lambda^d}
 \Usch\l(\oprod{\lambda} \ot \oprod{q} \ot A_{\lambda,q}^{(x)} 
\r)\Usch^\dag.
\label{eq:GPE-instrument2}\ee
Here $Q_\lambda^d$ is the GZ basis defined in \sect{gz-yy}.  We will
find that the running time is $d\poly(n,\log d,\log 1/\epsilon)$, which is
comparable to the running time of the circuits in
\sect{schur-circuit}, but has slightly less dependence on $d$ and
slightly more dependence on $n$.\footnote{If $d$ is much larger than
$n$, then it is always possible (even with the CG-based Schur
transform) to reduce the time for a Schur basis measurement to
$\poly(n,\log 1/\eps) + O(n\log d)$.  This is because, given a string
$\ket{i_1,\ldots,i_n}\in\nqudits$, we can first measure the type in
time $O(n\log d)$ and then unitarily map $\ket{i_i,\ldots,i_n}$ to
$\ket{i_i',\ldots,i_n'}$, where $i_j'\in [n]$ and $i_j'=i_k'$ iff
$i_j=i_k$.  Measuring $\ket{i_i',\ldots,i_n'}$ in the Schur basis then
requires $\poly(n,\log 1/\epsilon)$ time, and the measured value of
$\ket{q}$ can be translated to the proper value by replacing each
instance $i_j'$ in the Young tableau with $i_j$.  Moreover, the final
answer can be used to uncompute the type, so this modification also
works when implementing $\Usch$ rather than simply a Schur basis
measurement.}  More importantly, it gives a conceptually independent
method for a Schur basis measurement.

The main idea is that we can measure $\ket{q}\in Q_\lambda^d$ by
measuring the irrep label $q_c$ for each subgroup $\cU_c\subset
\cU_d$, $c=1,\ldots,d-1$.  We can measure $q_c$ by performing GPE in
a way that only looks at registers in states
$\ket{1},\ldots,\ket{c}$.  As these measurements
commute\cite{Biedenharn:63a}---in fact, 
they are simultaneously diagonalized by the GZ
basis\cite{Gelfand:50a}---we can perform them sequentially without
worrying about the disturbance that they cause.
After performing this modified GPE $d-1$ times, we can extract
the register $\ket{q}$ in addition to the $\ket{\lambda}\ket{p}$ that
we get from the first application of GPE.  

We now describe this modification of GPE in more detail.  To do so, we
will need to consider performing GPE on a variable number of qubits.
Define $\Ugpe^{(d,n)}$ by
\be \Ugpe^{(d,n)} = \sum_{\lambda\in\cI_{d,n}}
\ket{\lambda}\otimes
(\Usch^{(d,n)})^\dag \l(\ket{\lambda}\bra{\lambda} \otimes I_{\cQ_\lambda^d}
\otimes I_{\cP_\lambda} \r) \Usch^{(d,n)}
= \sum_{\lambda\in\cI_{d,n}} \ket{\lambda}\ot \Pi_\lambda^{(d,n)}
\ee
This coherently extracts the $\ket{\lambda}$ register from $\nqudits$.
Here we have also explicitly written out the dependence of
$\Pi_\lambda^{(d,n)}$ and $\Usch^{(d,n)}$ on $d$ and $n$.  Also, we
have expressed $\Ugpe^{(d,n)}$ as an isometry to avoid writing out the
ancilla qubits initialized to zero, but it is of course a reversible
unitary transform.  For example, we use GPE to construct
$\Pi_\lambda^{(d,n)}$ by performing $\Ugpe^{(d,n)}$, measuring
$\lambda$ and then undoing $\Ugpe^{(d,n)}$:
\be \Pi_\lambda^{(d,n)} = (\Ugpe^{(d,n)})^\dag
\l(\oprod{\lambda} \ot I_d^{\ot n}\r) \Ugpe^{(d,n)}.
\label{eq:proj-from-GPE}\ee

The observables we want to measure correspond to determining the irrep
label of $\cU_c$.  For any $\cU_c$-representation $(\bq,\cQ)$, we
define the $\cQ_\mu^c$-isotypic subspace of $\cQ$ to be the direct sum
of the irreps in the decomposition of $\cQ$ that are isomorphic to
$\cQ_\mu^c$ (cf.~\eq{rep-space-decomp}).  The projector onto this
subspace can be given explicitly (though we will not need the exact
formula) in terms of $\bq$ as follows:
\be \pi_\mu^c(\bq)= \dim\cQ^c_\mu \int_{U\in\cU_c} dU
\l(\tr \bq_\mu^c(U)\r)^* \bq(U),\ee
where $dU$ is a Haar measure on $\cU_c$.  Define the
$\cU_c$-representation $(\bQ_d^n,\nqudits)$ by $\bQ_d^n(U)=(U\oplus
I_{d-c})^{\ot n}$, where $(U\oplus I_{d-c})$ is the embedding of
$\cU_c$ in $\cU_d$ given by \eq{Uc-embed}.  Our goal is to perform the
projective measurements
$\{\pi_\mu^c(\bQ_d^n)\}_{\mu\in\bbZ^c_{++},|\mu|\leq n}$ for
$c=1,\ldots,d$.  Since for each $c$, $\pi_\mu^c(\bQ_d^n)$ is diagonal
in the GZ basis $Q_\lambda^d$, the projectors commute and can be
measured simultaneously.

For the special case of $|\lambda|=m=n$ and $c=d$, $\bQ_d^n$
is the same as the $\bQ$ defined in \eq{bQ-def} and we have
$\Pi_\lambda^{(d,n)} = \pi_\lambda^d(\bQ_d^n)$.   In this case,
\eq{proj-from-GPE} tells us how to perform the projective measurement
$\{\pi_\lambda^d(\bQ_d^n)\}_{\lambda\in\cI_{d,n}}$.  We now need to
extend this to measure
$\{\pi_\mu^c(\bQ_d^n)\}_{\mu\in\cI_{c,m},m\leq n}$ for any $c\in[d]$.

The first step in doing so is to measure the number of positions in
$\ket{i_1,\ldots,i_n}$ where $i_j\in \{1,\ldots, c\}$.  Call this
number $m$.  Though we will measure $m$, we will not identify which
$j$ have $i_j\in[c]$.  Instead we will coherently separate them by
performing the unitary operation
$\Usel^{(c)}$ which implements the isomorphism:
\be
(\bbC^d)^{\otimes n}
\cong \bigoplus_{m=0}^n (\bbC^c)^{\otimes m} \otimes 
(\bbC^{d-c})^{\otimes n-m} \otimes \bbC^{\binom{n}{m}}
\ee
It is straightforward to implement $\Usel^{(c)}$ in time linear in the
size of the input (i.e. $O(n\log d)$), though we will have to pad
quantum registers as in the original Schur transform.

We can use $\Usel^{(c)}$ to construct $\pi_\mu^c(\bQ_d^n)$ in terms of
$\pi_\mu^c(\bQ_c^m)$ as follows:
\be \pi_\mu^c(\bQ_d^n) = (\Usel^{(c)})^\dag \l( \oprod{m} \otimes
\pi_\mu^c(\bQ_c^m) \otimes I_{d-c}^{\otimes n-m} \otimes I_{\binom{n}{m}}
\r)\Usel^{(c)}.
\label{eq:Q_n-sel-Q_m}\ee
If $m=|\mu|$, then we can use \eq{proj-from-GPE} to construct the
projection on the RHS of \eq{Q_n-sel-Q_m}, obtaining
\be \pi_\mu^c(\bQ_d^n) = (\Usel^{(c)})^\dag \l(\oprod{m} \ot
\l[\l(\Ugpe^{(c,m)}\r)^\dag\l(\oprod{\mu}\ot I_c^{\ot m}\r)
\Ugpe^{(c,m)}\r] 
\otimes I_{d-c}^{\otimes n-m} \otimes I_{\binom{n}{m}}\r)\Usel^{(c)}.
\label{eq:one-GZ-invariant}\ee

This gives a prescription for measuring $\pi_\mu^c(\bQ_d^n)$, whose
output $\mu$ corresponds to the component $q_c$ of the GZ basis
element.  First measure $m=|\mu|$ by counting the number of qudits
that have values in $\{1,\ldots,c\}$.  Then select only those $m$
qudits using $\Usel^{(c)}$ and perform GPE on them to find $\mu$.  

Finally, measuring the commuting observables
$\{\pi_\mu^c(\bQ_d^n)\}_{\mu\in\bbZ_{++}^c}$ for $c=1,\ldots,d$ yields
a complete von Neumann measurement of the $\cQ_\lambda^d$
register with operation elements as follows:
\be (\Usch^{(d,n)})^\dag
\l(\oprod{\lambda}\ot\oprod{q}\ot I_{\cP_\lambda}\r)
\Usch^{(d,n)} = \prod_{c=1}^d \pi_{q_c}^c(\bQ_d^n),\ee
with $q$ ranging over $Q_\lambda^d$.

Combined with the results of the last section, we now have an
algorithm for performing a complete measurement in the Schur basis, or
more generally an arbitrary instrument with operation elements
\be (\Usch^{(d,n)})^\dag
\l(\sum_{\lambda\in\cI_{d,n}}
\oprod{\lambda}\ot\oprod{q}\ot A_{\lambda,q}^{(x)}\r)
\Usch^{(d,n)},
\label{eq:general-GPE-instrument}\ee
where the $A_{\lambda,q}^{(x)}$ are arbitrary operators on
$\cP_\lambda$ and $x$ labels the measurement outcomes.  In particular,
if we let $x$ range over triples $(\lambda,q,p)$ with $\lambda\in\cI_{d,n}$, $q\in
Q_\lambda^d$ and $p\in P_\lambda$; and set
$A_{\lambda',q'}^{(\lambda,q,p)}=\delta_{\lambda,\lambda'}
\delta_{q,q'}\oprod{p}$, then 
\eq{general-GPE-instrument} corresponds to a complete von Neumann
measurement in the Schur basis.  The general algorithm is as follows:

\begin{tabbing}
~~~~ \= ~~~~ \= ~~~ \= ~~~ \= ~~~\= \kill
{\bf Algorithm: Complete Schur basis measurement using GPE}\+ \\
{\bf Input:} A state $\ket{\psi}\in \nqudits$.\\
{\bf Output:} \parbox[t]{5.4in}{
$\sum_{\lambda\in\cI_{d,n}}\sum_{q\in Q_\lambda^d}\sum_x
\ket{\lambda}\ket{q}\ket{x}
 (\Usch^{(d,n)})^\dag
\l(\sum_{\lambda\in\cI_{d,n}}
\oprod{\lambda}\ot\oprod{q}\ot A_{\lambda,q}^{(x)}\r)
\Usch^{(d,n)}\ket{\psi}$\\
corresponding to the coherent output of the instrument in
\eq{general-GPE-instrument}}\\
{\bf Runtime:} $d \cdot O(T_{\text{QFT}}(\cS_n) + n\log d)$\\
{\bf Procedure:}\\
{\bf 1.}\> For $c=1,\ldots,d-1$:\\
{\bf 2.}\>\> Apply $\Usel^{(c)}$ to $\ket{\psi}$, outputting
superpositions of $\ket{m}\ket{\alpha_c^m}\ket{\beta_{d-c}^{n-m}}
\ket{\gamma}_{\binom{n}{m}}$.\\
{\bf 3.}\>\> Perform $\sum_{m=0}^n \oprod{m}\ot \Ugpe^{(c,m)}$ on
$\ket{m}\ket{\alpha_c^m}$ to output $\ket{m}\sum_{\mu_c\in\cI_{c,m}}
\ket{\mu_c}\pi_{\mu_c}^c(\bQ_d^n)\ket{\alpha_c^m}$.\\
{\bf 4.}\>\> Apply $(\Usel^{(c)})^\dag$.\\
{\bf 5.}\> Set $\ket{q} := \ket{\mu_1}\ldots\ket{\mu_{d-1}}$.\\
(Steps 6--10 are based on \eq{GPE-instr-construct}.)\\
{\bf 6.}\> Add registers $\ket{\mu}\!=\!\ket{(n)}$ and
$\ket{i}\!=\!\ket{j}\!=\!\l|\young(12)\cdots\young(n)\r\rangle$
corresponding to the trivial irrep of $\cS_n$.\\
{\bf 7.}\> Perform $\hat{C}_{\bf P}$ ($:=(\Uqft \ot I)C_{\bf
P}(\Uqft^\dag \ot I)$) on $\ket{\mu}\ket{i}\ket{j}\ket{\psi}$ to
output
$\ket{\lambda}\ket{i'}\ket{j'}\ket{\psi'}$.\\
{\bf 8.}\> Perform the instrument
$\l\{\sum_{\lambda\in\cI_{d,n}}\sum_{q\in Q_\lambda^d}
\oprod{q} \ot \oprod{\lambda} \ot A_{\lambda,q}^{(x)} 
\ot I_{\cP_\lambda} \ot I_d^{\ot n}\r\}_x$.\\
{\bf 9.} \> Apply $\hat{C}_{\bf P}^\dag$ to output
$\ket{\mu}\ket{i}\ket{j}\ket{\psi''}$.\\
{\bf 10.} \> The registers $\ket{\mu}\ket{i}\ket{j}$ are always in the
state $\ket{(n)}\l|\young(12)\cdots\young(n)\r\rangle^{\ot 2}$
 and can be discarded.\\
{\bf 11.} \> Reverse steps 1--5.\\
\end{tabbing}


{\em Generalizations to other groups:}  Crucial to this procedure is
not only that $\cU_d$ and $\cS_n$ form a dual reductive pair
(cf.~\sect{dual-reductive}), but that both groups have GZ bases, and their
canonical towers of subgroups also form dual reductive pairs.  These
conditions certainly exist for other groups (e.g. $\cU_{d_1}\times\cU_{d_2}$
acting on polynomials of $\bbC^{d_1+d_2}$), but it is an open problem
to find useful applications of the resulting algorithms.

\subsection{Connection to the Clebsch-Gordan transform}
\label{sec:GPE-CG}
In this section, we explain how GPE can be thought of in terms of the
CG transform on $\ccG$, or on $\cS_n$ when we specialize to the case of
the Schur basis.  The goal is to give a simple
representation-theoretic interpretation of the measurements described
in \sect{GPE-lambda} as well as pointing out relations between the QFT
and the CG transform.

We begin with a quick review of GPE.  We have two registers,
$\bbC[\ccG]$ and $V$, where $(\rho,V)$ is a representation of $\ccG$.
Assume for simplicity that the isomorphism in \eq{GPE-V-decomp} is an
equality:
\be V = \bigoplus_{\nu\in\hat{\ccG}} V_\nu \ot \bbC^{m_\nu}.\ee
This means that the controlled-$\rho$ operation $C_\rho$ is given by
\be C_\rho=\sum_{g\in\ccG} \ket{g}\bra{g}\otimes \rho(g)
= \sum_{g\in\ccG} \ket{g}\bra{g}\otimes \sum_{\nu\in\hat{\ccG}}
\oprod{\nu} \otimes \br_\nu(g) \otimes I_{m_\nu}.\ee
The GPE prescription given in \sect{GPE-lambda} began with
initializing $\bbC[\ccG]$ with the trivial irrep (or equivalently a
uniform superposition over group elements).  We will relax this
condition and analyze the effects of $\hat{C}_\rho$ (the
Fourier-transformed version of $C_\rho$, cf.~\eq{hatC-def}) on
arbitrary initial states in $\bbC[\ccG]$ in order to see how it acts
like the CG transform.

There are two equivalent ways of understanding how $\hat{C}_\rho$ acts
like the CG transform; in terms of representation spaces or in terms
of representation matrices.  We first present the explanation based on
representation matrices.  Recall from \sect{QFT} that $\bbC[\ccG]$ can
be acted on by either the left or the right representation 
($\bL(h)\ket{g}=\ket{hg}$ and $\bR(h)\ket{g}=\ket{gh^{-1}}$).  The
controlled-$\rho$ operation acts on these matrices as follows
\begin{align}
C_\rho \l(\bL(g) \ot I_V\r) C_\rho^\dag &= \bL(g) \ot \rho(g)
\label{eq:Crho-left}\\
C_\rho^\dag \l(\bR(g) \ot I_V\r) C_\rho &= \bR(g) \ot \rho(g)
\label{eq:Crho-right}\end{align}
The proofs of these claims are straightforward\footnote{
Here we prove \eqs{Crho-left}{Crho-right}:
\begin{align}
C_\rho \l(\bL(h) \ot I_V\r) C_\rho^\dag &=
\l(\sum_{g_1\in\ccG}\oprod{hg_1}\ot\rho(hg_1)\r)
\l(\sum_{g_2\in\ccG}\ket{hg_2}\bra{g_2}\ot I_V\r)
\l(\sum_{g_3\in\ccG}\oprod{g_3}\ot\rho(g_3^{-1})\r)
\non\\&= \sum_{g\in\ccG}\ket{hg}\bra{g} \ot \rho(h) = 
\bL(h) \ot \rho(h) \\
C_\rho^\dag \l(\bR(h) \ot I_V\r) C_\rho &=
\!\l(\sum_{g_1\in\ccG}\oprod{g_1h^{-1}}\ot\rho(hg_1^{-1})\r)
\!\!\!\l(\sum_{g_2\in\ccG}\ket{g_2h^{-1}}\bra{g_2}\ot I_V\r)
\!\!\!\l(\sum_{g_3\in\ccG}\oprod{g_3}\ot\rho(g_3)\r)
\non\\&=
 \sum_{g\in\ccG}\ket{gh^{-1}}\bra{g}\! \ot\! \rho(h) 
=\bR(h) \ot \rho(h)
\end{align}
}.  If we combine them, we find that
\be C_\rho(\bL(g_1)\bR(g_2) \ot \rho(g_2))C_\rho^\dag
 = \bL(g_1)\bR(g_2) \ot \rho(g_1).
\label{eq:Crho-both}\ee

Let us examine how $C_\rho$ transforms the left and right
representations, any observable on $\bbC[\ccG]$ can be constructed out
of them.  Focus for now on the action of $C_\rho$ on the left
representation.  \eq{Crho-left} says that conjugation by $C_\rho$ maps
the left action on $\bbC[\ccG]$ to the tensor product action on
$\bbC[\ccG] \ot V$.  To see how this acts on the representation
spaces, we conjugate each operator by $\Uqft$, replacing $C_\rho$ with
$\hat{C}_\rho$, $\bL$ with $\hat{\bL}$ and $\bR$ with $\hat{\bR}$.
Then $\hat{C}_\rho$ couples an irrep $V_\mu$ from
$\bbC[\ccG]\cong\bigoplus_\mu V_\mu\ot V_\mu^*$ with an irrep $V_\nu$
from $V$ and turns this into a sum of irreps $V_\lambda$.  This
explains how GPE can decompose $V$ into irreps: if initialize $\mu$ to
be the trivial irrep, then only $\lambda=\nu$ appears in the output
and measuring $\lambda$ has the effect of measuring the irrep label of
$V$.  The right representation is acted on by $\hat{C}_\rho$ in the
opposite manner; we will see that conjugating by $\hat{C}_\rho$
corresponds to the inverse CG transform, mapping $V_\mu^*$ in the
input to $V_\lambda^* \ot V_\nu$ in the output.

To make this concrete, Fourier transform each term in \eq{Crho-both}
to obtain
\begin{flalign} 
\hat{C}_\rho^\dag&
\l(\sum_{\mu\in\hat{\ccG}} \oprod{\mu} \ot \br_\mu(g_1)
\ot \br_\mu(g_2^{-1}) \ot
\sum_{\nu\in\hat{\ccG}} \oprod{\nu} \ot \br_\nu(g_1)
\ot I_{m_\nu}\r)\hat{C}_\rho 
\label{eq:GPE-rep-matrices1}\\\qquad=&
\l(\Uqft\ot I_V\r) C_\rho^\dag
\l(\bL(g_1)\bR(g_2) \ot \rho(g_1)\r) C_\rho
\l(\Uqft^\dag\ot I_V\r)
\\=& \l(\Uqft\ot I_V\r)
\l(\bL(g_1)\bR(g_2) \ot \rho(g_2)\r)
\l(\Uqft^\dag\ot I_V\r)
\\=&
\sum_{\lambda\in\hat{\ccG}}
\oprod{\lambda} \ot \br_\lambda(g_1) \ot \br_\lambda(g_2^{-1})
\ot \sum_{\nu\in\hat{\ccG}} \oprod{\nu} \ot \br_\nu(g_2)\ot I_{m_\nu}
\end{flalign}
To understand this we need to work backwards.  Measuring an observable
on the $V_\lambda$ register of the final state corresponds to measuring
that observable on the $V_\lambda$-isotypic subspace of the original
$V_\mu\ot V_\nu$ inputs.  On the other hand, the initial $V_\mu^*$
register splits into $V_\lambda^*$ and $V_\nu$ registers.  We can see
an example of this in \eq{GPE-output2}, where the $V_\mu$ register
($\ket{i})$ has been transferred to $V_\lambda$, while $V_\lambda^*$
and $V_\nu$ are in the maximally entangled state
$\ket{\Phi_\lambda} = d_\lambda^{-1/2}\sum_{j=1}^{d_\lambda} \ket{jj}$
corresponding to the trivial irrep that $V_\mu^*$ was initialized to.

\begin{figure}[ht]
\begin{center}
\leavevmode
\xymatrix@R=4pt@C=4pt{
&& V_\nu\a{nu-right}\\
V_\mu^*\a{mu-star}\ar[rru] \ar[rrd] & *{~~~~~~} & \\
V_\mu\a{mu} \ar[rrd] & & V_\lambda^*\\
& & V_\lambda \\
V_\nu\a{nu-left}\ar[rru]
}
\caption{Performing $\hat{C}_\rho$ combines $V_\mu$ and $V_\nu$ to
form $V_\lambda$ and splits $V_\mu^*$ into $V_\lambda^*$ and $V_\nu$.
Here $V_\mu\ot V_\mu^*$ and $V_\lambda\ot V_\lambda^*$ come from the
decomposition of $\bbC[\ccG]$ and $V_\nu$ comes from the decomposition
of $V$.}
\label{fig:GPE-rep-maps}
\end{center}
\end{figure}
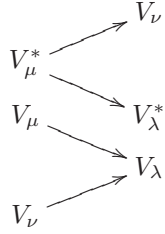

Thus $\hat{C}_\rho$ corresponds to a CG transform from $V_\mu\ot
V_\nu$ to $V_\lambda$ and an inverse CG transform from $V_\mu^*$ to
$V_\lambda^* \ot V_\nu$.  These maps are sketched in
\fig{GPE-rep-maps}.

We can verify the two maps separately by replacing
$\hat{C}_\rho$ in \eq{GPE-rep-matrices1} with $\Ucg$ or $\Ucg^\dag$
acting on the appropriate registers and checking that the
representation matrices transform appropriately.
There are two details here which still need
to be explained.  First, our description of the CG transform has not
accounted for the multiplicity spaces that are generated.  Second, we
have not explained how the inverse CG transform always creates the
correct irreps $V_\lambda^* \ot V_\nu$ when $\lambda$ is a label
output by the first CG transform.  To explain both of these, we track
the representation spaces through a series of transformations
equivalent to $C_\rho$.  As with $C_\rho$, we will begin and end with
$\bbC[\ccG] \ot V$, but we will show how the component irreps transform
along the way.  Here, $\iso{U}$ is used to mean that the unitary
operation $U$ implements the isomorphism; all of the isomorphisms
respect the action of the group $\ccG$.

\begin{eqnarray}
\bbC[G]\otimes V
& \stackrel{\Uqft}{\cong} &
\l(\bigoplus_{\mu\in\hat{G}} \ccG_\mu \otimes \ccG_\mu^*\r) \otimes 
\l(\bigoplus_{\nu\in\hat{G}} \ccG_\nu \otimes \bbC^{m_\nu}\r)\\
& \stackrel{\Ucg}{\cong} &
\bigoplus_{\mu,\nu,\lambda\in\hat{G}} \ccG_\lambda \otimes 
\Hom(\ccG_\lambda, \ccG_\mu\otimes \ccG_\nu)^\ccG
 \otimes \ccG_\mu^* \otimes \bbC^{m_\nu} \\
& \cong & \label{eq:GPE-hom-reverse}
\bigoplus_{\mu,\nu,\lambda\in\hat{G}} \ccG_\lambda \otimes 
\Hom(\ccG_\mu^*, \ccG_\lambda^* \otimes \ccG_\nu)^\ccG
 \otimes \ccG_\mu^* \otimes \bbC^{m_\nu} \\
& \stackrel{\Ucg^\dag}{\cong} &
\bigoplus_{\nu,\lambda\in\hat{G}} \ccG_\lambda \otimes \ccG_\lambda^*
 \otimes \ccG_\nu \otimes \bbC^{m_\nu} \\
& \stackrel{\Uqft^\dag}{\cong} &
\bbC[G] \otimes 
\bigoplus_{\nu\in\hat{G}} \ccG_\nu \otimes \bbC^{m_\nu}
= \bbC[G] \otimes V
\end{eqnarray}
The isomorphism in \eq{GPE-hom-reverse} is based on repeated
application of the identity $\Hom(A,B)\cong A^* \ot B$.  This
equivalence between 
$\Hom(\ccG_\lambda, \ccG_\mu\otimes \ccG_\nu)^\ccG$ and
$\Hom(\ccG_\mu^*, \ccG_\lambda^* \otimes \ccG_\nu)^\ccG$
is the reason that a CG transform followed by an inverse CG transform
on different registers can yield the correct representations in the
output.

\subsubsection{Application: using $\Uqft$ to construct $\Ucg$ }
So far the discussion in this section has been rather abstract: we
have shown that $\hat{C}_\rho$ acts in a way analogous to $\Ucg$, but
have not given any precise statement of a connection.  To give the
ideas in this section operational meaning, we now show how $\Uqft$ can
be used to perform $\Ucg$ on an arbitrary group.  This idea is
probably widely known, and has been used for the dihedral group in
\cite{Kup03}, but a presentation of this form has not appeared before
in the literature.

\begin{figure}[ht]
\begin{center}
\leavevmode
\xymatrix@R=4pt@C=4pt{
|\mu\> & \nw & \gnqubit{\Uqft^\dag}{dd}\w &&&& \gnqubit{\Uqft}{dd} &
\nw &  |\lambda\>\w \\
V_\mu & \nw & \gspace{\Uqft^\dag}\w & \nw & \b\w\a{control}& \nw &
\gspace{\Uqft}\w & \nw & V_\lambda\w\\ 
V_\mu^* & *{~} & \gspace{\Uqft^\dag}\w &&&& \gspace{\Uqft} & \nw &
V_\lambda^*\w\\
{\ket{\Phi_\mu}} \ar@{-}[ur]\ar@{-}[dr]\\
V_\mu & *{~} &\nw &\nw &\nw &\nw &\nw&\nw & V_\mu \w 
\\
|\nu\> & \nw & \gnqubit{\Uqft^\dag}{dd}\w &&&& \gnqubit{\Uqft}{dd} &
\nw &  |\nu\>\w \\
V_\nu & \nw & \gspace{\Uqft^\dag}\w & \nw & \op{\bL(g)}\w\a{target}&
\nw & \gspace{\Uqft}\w & \nw & V_\nu\w\\ 
V_\nu^* & \nw & \gspace{\Uqft^\dag}\w &&&& \gspace{\Uqft} & \nw &
V_\nu^*\w
\ar@{-}"control";"target"
}
\caption{Using $\Uqft$ to construct $\Ucg$ for an arbitrary group.
The inputs to the CG transform are put in the $\ket{\mu}$, $V_\mu$,
$\ket{\nu}$ and $V_\nu$ registers.  The $V_\nu^*$ register is not
affected by the circuit, but is included so that the QFT will have
valid inputs.  The output of the CG transform is in the
$\ket{\lambda}$ and $V_\lambda$ registers.  $\ket{\nu}$ saves the
irrep label of one of the inputs and $(V_\mu \ot V_\nu \ot
V_\lambda^*)^{\ccG}$ is the multiplicity space.  }
\label{fig:GPE-to-CG}
\end{center}
\end{figure}
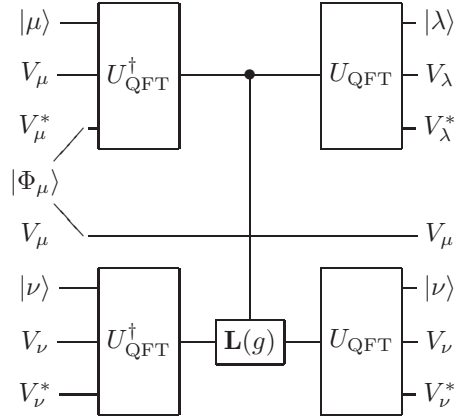

The algorithm for $\Ucg$ is  depicted in \fig{GPE-to-CG}
and is described as follows:
\begin{tabbing}
~~~~ \= ~~~~ \= ~~~ \= ~~~ \= ~~~\= \kill
{\bf Algorithm: Clebsch-Gordan transform using GPE}\+ \\
{\bf Input:}
$\ket{\mu}^{A_1}\ket{v_\mu}^{A_2}\ket{\nu}^{B_1}\ket{v_\nu}^{B_2}$, 
where $\mu,\nu\in\hat{G}$, $\ket{v_\mu}\in V_\mu$ and
$\ket{v_\nu}\in V_\nu$.\\
{\bf Output:}
\parbox[t]{5.5in}{$\ket{\lambda}^{A_1}\ket{v_\lambda}^{A_2}\ket{\nu}^{B_1} 
\ket{\alpha}^C$ with $\lambda\in\hat{G}$, $\ket{v_\lambda}\in
V_\lambda$ the irrep of the combined space and
$\ket{\alpha}\in\l(V_\mu \ot V_\nu \ot V_\lambda^*\r)^G$ the
multiplicity label.}\\
{\bf Runtime:} $4 T_{\text{QFT}} + T_{C_{\bL}}$ where $T_{C_{\bf
L}}$ is the time of the controlled-$\bL$ operation.\\
{\bf Procedure:}\\
{\bf 1.}\>\parbox[t]{5.5in}{Add states $\ket{\Phi_\mu}^{A_3A_4}$ 
and $\ket{v_\nu^*}^{B_3}$, where $\ket{\Phi_\mu}$ is the unique state
(up to phase) in the one-dimensional space $(V_\mu^* \ot V_\mu)^\ccG$
(cf.~\eq{unique-invariant-bipartite}) and
$\ket{v_\nu^*}\in V_\nu^*$ is arbitrary.}\\
{\bf 2.}\>\parbox[t]{5.5in}{Perform the inverse QFT on $A_1A_2A_3$
(yielding output $A$) and 
on $B_1B_2B_3$ (yielding output $B$); i.e.
$$ \Uqft^{A_1A_2A_3 \ra A} \ot \Uqft^{B_1B_2B_3 \ra B}.$$
Registers $A$ and $B$ now contain states in $\bbC[\ccG]$.}\\
{\bf 3.}\> Apply $C_\bL^{AB}$, mapping $\ket{g_1}^A\ket{g_2}^B$ to
$\ket{g_1}^A\ket{g_1g_2}^B$.\\
{\bf 4.}\> Perform the QFT on $A$ and $B$, yielding output $A_1A_2A_3$
and $B_1B_2B_3$.\\
{\bf 5.}\> Discard the register $B_3$, which still contains the state
$\ket{v_\nu^*}$.\\
{\bf 6.}\> \parbox[t]{5.5in}{$A_1$ now contains the combined irrep
label, which we call $\lambda$.  The irrep space $V_\lambda$ is in
$A_2$, while the multiplicity space $(V_\mu \ot V_\nu \ot
V_\lambda^*)^{\ccG}$ is in $A_4B_2A_3$, which we relabel as $C$.}
\end{tabbing}

The representations are transformed in the same way as in
\fig{GPE-rep-maps} with the addition of $V_\nu^*$ which is left
unchanged.  This is because we only act on the second register using
left multiplication, which acts as the identity on $V_\nu^*$.

Thus, we can perform the CG transform efficiently whenever we can
efficiently perform the QFT, with the small caveat that efficiently
manipulating the multiplicity space may take additional effort.  One
application of this reduction is to choose $\ccG=\cU_d$ and thereby
replace the $\cU_d$ CG construction of \sect{cg-construct} with a CG
transform based on the $\cU_d$ QFT.  Unfortunately, no fast quantum
algorithms are known for the $\cU_d$ and is is not immediately clear
how quantizing $\bbC[\cU_d]$ should correspond to cutting off the
irreps that appear in the decomposition $\bigoplus_\lambda
\cQ_\lambda^d \ot (\cQ_\lambda^d)^*$.  However, QFTs are known
for discrete matrix groups such as $GL_n(\mathbb{F}_q)$ (running in
time $q^{O(n)}$)\cite{MRR03} and  classical fast Fourier transforms are
known for $\cU_d$ and other compact groups\cite{MR97}.

A problem related to the $\cU_d$ QFT was also addressed in
\cite{Zalka04}, which sketched an algorithm for implementing
$\bq_\lambda^2(U)$ in time polylogarithmic in $|\lambda|$, though some
crucial details about efficiently integrating Legendre functions
remain to be established.  In contrast, using the 
Schur transform to construct $\bq_\lambda^2(U)$ would require
$\poly(|\lambda|)$ gates.  The idea behind \cite{Zalka04} is to embed
$\cQ_\lambda^2$ in $\cQ_*^d := \bigoplus_\mu \cQ_\mu^2$, where $\mu\in
\bbZ_{++}^2$ and $|\mu|$ can be exponentially large.  Then $\cQ_*^d$
corresponds to functions on the 2-sphere which can be discretized and
efficiently rotated, though some creative techniques are necessary to
perform this unitarily.  It is possible that this approach could
ultimately yield efficient implementations of $\Ucg$ and $\Uqft$ on
$\cU_d$.

\section{Deriving the $\cS_n$ QFT from the Schur
transform} 
\label{sec:QFT-from-Sch}

We conclude the chapter by showing how $\Usch$ can be used to
construct $\Uqft$.  Of course, an efficient algorithm for $\Uqft$
already exists\cite{Beals97}, but the circuit we present here
appears to be quite different.  Some of the mathematical principles
behind this connection are in Thm.~9.2.8 of \cite{GW98} and I am
grateful to Nolan Wallach for a very helpful conversaton on this
subject.

The algorithm is based on the embedding of $\cS_n$ in $[n]^n$ given by
$s\ra (s(1),\ldots,s(n))$.  This induces a map from $\bbC[\cS_n]\ra
(\bbC^n)^{\otimes n}$.  More precisely, if $1^n$ denotes the weight
$(1,\ldots,1)$ with $n$ ones, then we have a unitary map between
$\bbC[\cS_n]$ and $(\bbC^n)^{\otimes n}(1^n)$.  This is the natural
way we would represent a permutation on a computer (quantum or
classical): as a string of $n$ distinct numbers from $\{1,\ldots,
n\}$.  Similarly, we can embed $\cS_n$ in $\cU_n$ by letting a
permutation $s$ denote the unitary matrix $\sum_{i=1}^n
\ket{s(i)}\bra{i}$. 

Using this embedding, the algorithm for $\Uqft$ is as follows:
\begin{tabbing}
~~~~ \= ~~~~ \= ~~~ \= ~~~ \= ~~~\= \kill
{\bf Algorithm: $\cS_n$ QFT using the Schur transform}\+ \\
{\bf Input:} $\bbC[\cS_n]$\\
{\bf Output:} $\bigoplus_{\lambda\in\cI_n} \cP_\lambda^* \ot
\cP_\lambda$.\\
{\bf Runtime:} $\poly(n,\log 1/\eps)$.\\
{\bf Procedure:}\\
{\bf 1.}\> Embed $\bbC[\cS_n]$ in $(\bbC^n)^{\ot n}(1^n)$.\\
{\bf 2.}\> Perform $\Usch^{(n,n)}$ on $(\bbC^n)^{\ot n}(1^n)$ to
output $\ket{\lambda}\ket{q}\ket{p}$.\\
{\bf 3.}\> Output $\ket{\lambda}$ as the irrep label, $\ket{q}$ as
the state of $\cP_\lambda^*$ and $\ket{p}$ for $\cP_\lambda$.
\end{tabbing}

First we need to argue that setting $\ket{q}$ to be the
$\cP_\lambda^*$ output is well-defined.  Note that
$\ket{q}\in\cQ_\lambda^n(1^n)$, so if $\ket{q}$ is a GZ basis vector,
then its branching pattern $(q_1,\ldots,q_n)$ satisfies $q_i \in
q_{i+1}-\Box$, and thus $\ket{q_1,\ldots,q_n}\in P_\lambda$.

Now to prove that this algorithm indeed performs a Fourier transform
on $\cS_n$, we examine a series of isomorphisms.  The Fourier
transform relates $\bbC[\cS_n]$ to $\bigoplus_\lambda \cP_\lambda
\otimes \cP_\lambda$. Since weights are determined by the action of
the unitary group, restricting \eq{schur-decomp} on both sides to the
$1^n$ weight space gives the relation
\be (\bbC^n)^{\otimes n}(1^n)  \iso{\cU_d\times\cS_n}
\bigoplus_{\lambda\in\cI_n} \cQ_\lambda^n(1^n) \hat{\ot}\cP_\lambda
\ee
Thus we have the isomorphisms:
$$\begin{CD}
\bbC[\cS_n]  @>{\text{embed}}>{(1)}>  (\bbC^n)^{\otimes n}(1^n) \\
 @V{(2)}V{\Uqft}V @V{(3)}V{\Usch}V\\
\bigoplus_{\lambda\in\cI_n} \cP_\lambda \otimes \cP_\lambda
@>{(4)}>>
\bigoplus_{\lambda\in\cI_n} \cQ_\lambda^n(1^n) \otimes \cP_\lambda
\end{CD}$$

Our goal is to understand the isomorphism (4) by examining how the
other isomorphisms act on representation matrices.  First we look at
how (1) relates $\bP,\bQ$ with $\bL, \bR$.  Note that
$\bQ(\cS_n)$ and $\bP(\cS_n)$ act on $(\bbC^n)^{\otimes
n}(1^n)$ according to
\be
\bQ(\pi)\bigotimes_{j=1}^n \ket{s(j)}= 
\bigotimes_{j=1}^n \ket{\pi(s(j))} 
\qquad\text{and}\qquad
\bP(\pi)\bigotimes_{j=1}^n \ket{s(j)}= 
\bigotimes_{j=1}^n \ket{s(\pi^{-1}(j))} 
\ee
And from the definition of multiplying permutations, 
$\bL$ and $\bR$ act on $\bbC[\cS_n]$ according to
\be \bL(\pi)\bigotimes_{j=1}^n \ket{s(j)}= 
\bigotimes_{j=1}^n \ket{\pi(s(j))} 
\qquad\text{and}\qquad
\bR(\pi)\bigotimes_{j=1}^n \ket{s(j)}= 
\bigotimes_{j=1}^n \ket{s(\pi^{-1}(j))}, 
\ee
if we write permutations as elements of $[n]^n$.
Thus the embedding map relates $\bL$ and $\bR$ to $\bQ$ and
$\bP$ respectively.

This means that for any $\pi_1,\pi_2\in\cS_n$, the isomorphism (4)
maps $\sum_\lambda \ket{\lambda}\bra{\lambda} \otimes
\bp_\lambda(\pi_1) \otimes \bp_\lambda(\pi_2)$ to $\sum_\lambda
\ket{\lambda}\bra{\lambda} \otimes
\left.\bq_\lambda^n(\pi_1)\right|_{\cQ_\lambda^n(1^n)} \otimes
\bp_\lambda(\pi_2)$.  This proves that $\cQ_\lambda^n(1^n) \iso{\cS_n}
\cP_\lambda$ (cf.~Thm~9.2.8 of \cite{GW98}).

Moreover, it is straightforward to verify that the GZ basis of
$\cQ_\lambda^n(1^n)$ corresponds to the same chain of partitions that
labels the GZ basis of $\cP_\lambda$; one need only look at which
weights appear in the restriction to $\cS_{n-1}\subset\cU_{n-1}$.
This establishes that the representation matrices are the same, up to
an arbitrary phase difference for each basis vector.  The existence of
this phase means that we have constructed a slightly different Fourier
transform than \cite{Beals97}, and it is an interesting open question
to calculate this phase difference and determine its significance.

\bibliography{ref}
\bibliographystyle{thesis}


\end{document}